%% file: report.tex
\documentclass[a4paper,12pt,twoside,openright]{report}

\input{./Style/Packages.tex}
\input{./Style/Colors.tex}


\author{Larissa Bruna Streher}
\date{\Large{May, 2017}}
\title{Large-eddy simulations of the flow around a NACA0012 airfoil at different angles of attack}

\input{corporate/stroemungsmechanik}

\def\drafttype{false}   

\def\FigPath{./Images}

\begin{document}

\renewcommand{\arraystretch}{1.2}

\ifpdf
   \DeclareGraphicsExtensions{.pdf,.png,.jpg,}
\fi

\DeclareGraphicsRule{.eps.gz}{eps}{.pdf}{`gunzip -c #1}

\pagestyle{empty}

\includepdf[pages=-]{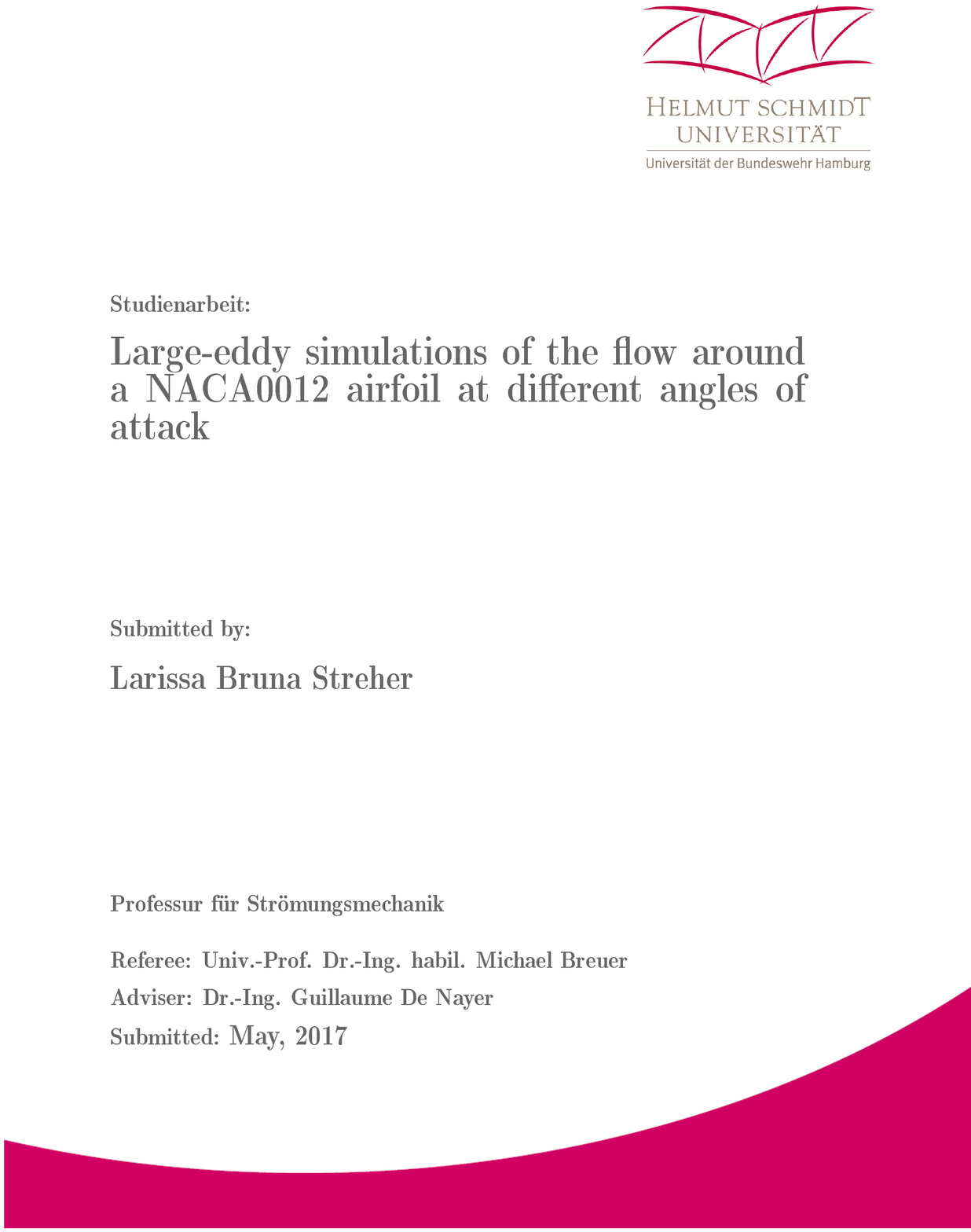}

\cleardoublepage
\include{abstract}

\cleardoublepage
\pagestyle{empty}
\include{thanks}

\cleardoublepage
\pagestyle{empty}

\pagestyle{fancy}
\fancyhf{}
\fancyhead{}
\fancyfoot{}
\fancyhead[RO]{\rightmark}
\fancyhead[LO]{\leftmark}
\fancyfoot[C]{\thepage}


\pagenumbering{roman}

\setcounter{page}{1}

\setcounter{tocdepth}{3}      
\setcounter{secnumdepth}{3}   

\tableofcontents
\listoffigures
\listoftables
\include{nomenclature}

\pagestyle{fancy}


\include{introduction}
\include{chapter1}

\include{chapter2}

\include{chapter3}
\include{conclusion}

\pagestyle{fancy}

\bibliographystyle{acm}

\bibliography{report}




\label{end} \end{document}

%% file: Style/Packages.tex
\usepackage[utf8x]{inputenc}
\usepackage[T1]{fontenc}

\usepackage{ae,aecompl} 

\usepackage{amsmath} 
\usepackage{amsfonts}
\usepackage{amssymb} 

\usepackage{array}
\usepackage{multirow} 

\usepackage[a4paper]{geometry} 
\usepackage{xspace} 
\usepackage{path} 
\usepackage{xcolor} 
\usepackage{colortbl} 
\usepackage{enumitem} 
\usepackage{fancyhdr} 
\usepackage{anysize} 
\usepackage{indentfirst}
\usepackage[small,hang]{caption} 
\usepackage{graphicx} 
\usepackage{float} 
\usepackage{pifont} 
\usepackage{textcomp} 
\usepackage{tocbibind} 
\usepackage{pdfpages} 
\usepackage{diagbox} 
\usepackage{ifthen} 
\usepackage{ifpdf} 
\usepackage{rotating}

\usepackage{eurosym} 

\ifpdf               
\usepackage{corporate/design_pdf}
\else
\usepackage{corporate/design_ps}
\fi

\ifpdf
\else
\fi
\usepackage{subfigure}

\ifpdf
	\usepackage[
		    pagebackref, 
                    ]{hyperref}
        \hypersetup{pdftitle={Large-eddy simulations around a NACA0012 airfoil at different angles of attack},
		    pdfsubject={LES NACA0012},
		    pdfauthor={Larissa Bruna Streher},
                    linktocpage=true, 
		    bookmarks=true,
		    bookmarksopen=true,
		    bookmarksnumbered=true, 
		    pdfpagelabels=true,
		    pdfpagemode=None, 
		    plainpages=false, 
	            colorlinks=true, 
                    citecolor=blue,
		    urlcolor=Magenta,
		    filecolor=green,
		    linkcolor=red
		}
                    
\else
        \usepackage[bookmarks]{hyperref}
         \hypersetup{pdftitle={Large-eddy simulations around a NACA0012 airfoil at different angles of attack},
 		    pdfsubject={LES NACA0012},
 		    pdfauthor={Larissa Bruna Streher},
                    linktocpage=true, 
                    breaklinks=true, 
 		    bookmarks=true,
 		    bookmarksopen=true,
 		    bookmarksnumbered=true, 
 		    pdfpagelabels=true,
 		    pdfpagemode=None, 
 	            colorlinks=true, 
                    citecolor=blue,
 		    urlcolor=Magenta,
 		    filecolor=green,
 		    linkcolor=red
                    }
\fi             

%% file: Style/Colors.tex
\definecolor{GreenYellow}   {cmyk}{0.15,0,0.69,0}
\definecolor{Yellow}        {cmyk}{0,0,1,0}
\definecolor{Goldenrod}     {cmyk}{0,0.10,0.84,0}
\definecolor{Dandelion}     {cmyk}{0,0.29,0.84,0}
\definecolor{Apricot}       {cmyk}{0,0.32,0.52,0}
\definecolor{Peach}         {cmyk}{0,0.50,0.70,0}
\definecolor{Melon}         {cmyk}{0,0.46,0.50,0}
\definecolor{YellowOrange}  {cmyk}{0,0.42,1,0}
\definecolor{Orange}        {cmyk}{0,0.61,0.87,0}
\definecolor{BurntOrange}   {cmyk}{0,0.51,1,0}
\definecolor{Bittersweet}   {cmyk}{0,0.75,1,0.24}
\definecolor{RedOrange}     {cmyk}{0,0.77,0.87,0}
\definecolor{Mahogany}      {cmyk}{0,0.85,0.87,0.35}
\definecolor{Maroon}        {cmyk}{0,0.87,0.68,0.32}
\definecolor{BrickRed}      {cmyk}{0,0.89,0.94,0.28}
\definecolor{Red}           {cmyk}{0,1,1,0}
\definecolor{OrangeRed}     {cmyk}{0,1,0.50,0}
\definecolor{RubineRed}     {cmyk}{0,1,0.13,0}
\definecolor{WildStrawberry}{cmyk}{0,0.96,0.39,0}
\definecolor{Salmon}        {cmyk}{0,0.53,0.38,0}
\definecolor{CarnationPink} {cmyk}{0,0.63,0,0}
\definecolor{Magenta}       {cmyk}{0,1,0,0}
\definecolor{VioletRed}     {cmyk}{0,0.81,0,0}
\definecolor{Rhodamine}     {cmyk}{0,0.82,0,0}
\definecolor{Mulberry}      {cmyk}{0.34,0.90,0,0.02}
\definecolor{RedViolet}     {cmyk}{0.07,0.90,0,0.34}
\definecolor{Fuchsia}       {cmyk}{0.47,0.91,0,0.08}
\definecolor{Lavender}      {cmyk}{0,0.48,0,0}
\definecolor{Thistle}       {cmyk}{0.12,0.59,0,0}
\definecolor{Orchid}        {cmyk}{0.32,0.64,0,0}
\definecolor{DarkOrchid}    {cmyk}{0.40,0.80,0.20,0}
\definecolor{Purple}        {cmyk}{0.45,0.86,0,0}
\definecolor{Plum}          {cmyk}{0.50,1,0,0}
\definecolor{Violet}        {cmyk}{0.79,0.88,0,0}
\definecolor{RoyalPurple}   {cmyk}{0.75,0.90,0,0}
\definecolor{BlueViolet}    {cmyk}{0.86,0.91,0,0.04}
\definecolor{Periwinkle}    {cmyk}{0.57,0.55,0,0}
\definecolor{CadetBlue}     {cmyk}{0.62,0.57,0.23,0}
\definecolor{CornflowerBlue}{cmyk}{0.65,0.13,0,0}
\definecolor{MidnightBlue}  {cmyk}{0.98,0.13,0,0.43}
\definecolor{NavyBlue}      {cmyk}{0.94,0.54,0,0}
\definecolor{RoyalBlue}     {cmyk}{1,0.50,0,0}
\definecolor{Blue}          {cmyk}{1,1,0,0}
\definecolor{Cerulean}      {cmyk}{0.94,0.11,0,0}
\definecolor{Cyan}          {cmyk}{1,0,0,0}
\definecolor{ProcessBlue}   {cmyk}{0.96,0,0,0}
\definecolor{SkyBlue}       {cmyk}{0.62,0,0.12,0}
\definecolor{Turquoise}     {cmyk}{0.85,0,0.20,0}
\definecolor{TealBlue}      {cmyk}{0.86,0,0.34,0.02}
\definecolor{Aquamarine}    {cmyk}{0.82,0,0.30,0}
\definecolor{BlueGreen}     {cmyk}{0.85,0,0.33,0}
\definecolor{Emerald}       {cmyk}{1,0,0.50,0}
\definecolor{JungleGreen}   {cmyk}{0.99,0,0.52,0}
\definecolor{SeaGreen}      {cmyk}{0.69,0,0.50,0}
\definecolor{Green}         {cmyk}{1,0,1,0}
\definecolor{ForestGreen}   {cmyk}{0.91,0,0.88,0.12}
\definecolor{PineGreen}     {cmyk}{0.92,0,0.59,0.25}
\definecolor{LimeGreen}     {cmyk}{0.50,0,1,0}
\definecolor{YellowGreen}   {cmyk}{0.44,0,0.74,0}
\definecolor{SpringGreen}   {cmyk}{0.26,0,0.76,0}
\definecolor{OliveGreen}    {cmyk}{0.64,0,0.95,0.40}
\definecolor{RawSienna}     {cmyk}{0,0.72,1,0.45}
\definecolor{Sepia}         {cmyk}{0,0.83,1,0.70}
\definecolor{Brown}         {cmyk}{0,0.81,1,0.60}
\definecolor{Tan}           {cmyk}{0.14,0.42,0.56,0}
\definecolor{Gray}          {cmyk}{0,0,0,0.50}
\definecolor{Black}         {cmyk}{0,0,0,1}
\definecolor{White}         {cmyk}{0,0,0,0}

%% file: corporate/stroemungsmechanik.tex
\addresshead{Helmut-Schmidt-Universität}
\address{Universität der Bundeswehr Hamburg\\Postfach 70 08 22\\22008 Hamburg}

\faculty{Maschinenbau}
\department{Professur für Strömungsmechanik}
\professor{Univ.-Prof. Dr.-Ing. habil. Michael Breuer}

%% file: abstract.tex
\addcontentsline{toc}{chapter}{Abstract}
\chapter*{Abstract\markboth{ABSTRACT}{}}
\subsection*{Large-eddy simulations of the flow around a NACA0012 airfoil at different angles of attack}
\par The purpose of this work is to investigate the flow around a fixed NACA0012 airfoil profile at different angles of attack using wall-resolved LES. The profile has a chord length of $c=0.1\,m$ and is exposed to a flow at a Reynolds number of $Re=100{,}000$. The background is that in the next step the coupled problem should be considered, dealing with the flutter problem.
\par The software ANSYS ICEM CFD is utilized in order to generate the  meshes, which are then applied for the simulation with the in-house software FASTEST-3D. This is based on a finite-volume approach and a three sub-steps predictor-corrector scheme as spatial and temporal discretizations, respectively. The turbulence is modeled by the large-eddy simulation technique and the subgrid-scales are modeled according to Smagorinsky.
\par An analysis of the meshes concerning the dimensionless wall distances is performed based on the spatial and time-averaged results of the velocities, the spatial and time-averaged streamlines, the instantaneous velocities, the spatial and time-averaged Reynolds stresses and the aerodynamic coefficients. A selection of the less computational time demanding grids that can correctly approximate the solutions of the test case is made. In a near future these will be applied to investigate the fluid-structure interaction and particularly the flutter phenomenon of the NACA0012 profile within a turbulent flow.

%% file: thanks.tex
\addcontentsline{toc}{chapter}{Acknowledgements}
\chapter*{Acknowledgements}

\par Foremost, I would like to thank Univ.\--Prof.\ Dr.\--Ing.\ habil.\ Michael Breuer for the introduction in the computational fluid dynamics area and the support through the development of this project.
\par I also would like to express my sincere gratitude to my advisor Dr.\--Ing.\ \mbox{Guillaume} De Nayer for the continuous support of my research project, for his patience, motivation, enthusiasm and immense knowledge. His guidance helped me in all the time of research and writing of this work.
\par Finally, I must express my very profound gratitude to my parents and to my boyfriend for providing me with unfailing support and continuous encouragement throughout my years of study and through the process of researching and writing this work. This accomplishment would not have been possible without them.

%% file: nomenclature.tex
\chapter*{Nomenclature\markboth{NOMENCLATURE}{}}
\label{nomenclature}
\addcontentsline{toc}{chapter}{Nomenclature}

\noindent The abbreviations, notations and variables used in this report are explained below with the corresponding units. Notations and variables not included in this section are explained in the context of the report.

\section*{Abbreviations}

\begin{tabular}{ll}
CFD	& Computational Fluid Dynamics \\
CPU & Central Processing Unit \\
DNS	& Direct Numerical Simulation \\
LES	& Large-Eddy Simulation \\
FLOPs & Floating Point Operations per Second \\
FSI	& Fluid-Structure Interaction \\
FVM & Finite-Volume Method \\
IDW & Inverse Distance Weighting \\
NACA & National Advisory Committee for Aeronautics \\
RANS	& Reynolds-Averaged Navier Stokes \\
e.g.    & exempli gratia \\
i.e.	& id est \\
Eq. & Equation\\
Fig.	& Figure \\
\end{tabular}

\section*{Notations}

\begin{tabular}{ll}
\(A_i\) & Index notation of vector A\\
\(B_{ij}\) & Second-order tensor, index notation of B \\ 
\(\overline{C}\) & Large turbulence scale of C\\
\(D^{t_n}\) & D in the time step $t_n$  \\
\(E^*\) & Dimensionless value of E  \\
\({<}{F}{>}\) & Spatial and time-averaged F \\

\end{tabular}

\section*{Variables}

\begin{tabular}{p{1.9cm}p{10.7cm}p{2cm}}
\(c\) & Airfoil chord length & $(m)$ \\
\(C_D\) & Drag coefficient &  \\
\(C_f\) & Friction coefficient &  \\
\(C_L\) & Lift coefficient &  \\
\(C_p\) & Pressure coefficient &  \\
\(C_s\) & Smagorinsky constant &  \\
\(l_k\) & Kolmogorov length  & $(m)$ \\
\(L_z\) & Span-wise length & $(m)$ \\
\(Ma\) & Mach number & \\
\(N\) & Number of nodes & \\
\(N_{L_z}\) & Number of nodes in the span-wise direction & \\
\(N_R\) & Number of nodes in the domain radius & \\
\(N_{SS}\) & Number of nodes in the suction side & \\
\(N_W\) & Number of nodes in the wake & \\
\(n_{avg}\) & Time step to start the averaging process &  \\
\(n_i\) & Normal vector &  \\
\(p\) & Pressure & $(Pa)$ \\
\(p^{\,corr}\) & Pressure correction & $(Pa)$ \\
\(p^{\,pred}\) & Predicted pressure & $(Pa)$ \\
\(p'\) & Pressure fluctuation & $(Pa)$ \\
\(p_{t_0}\) & Initial pressure & $(Pa)$ \\
\(p_{\infty}\) & Free stream pressure & $(Pa)$ \\
\(q\) & Stretching factor & \\
\(q_R\) & Stretching factor of the domain radius  & \\
\(q_W\) & Stretching factor of the wake & \\
\(q_\Phi\) & Source term of the transport variable $\Phi$ & \\
\(R\) & Domain radius & $(m)$ \\
\(R_{u'_iu'_j}\) & Two-point correlation of the velocity fluctuation & \\
\(Re\) & Reynolds number &  \\
\(S_{ij}\) & Strain rate tensor & $(s^{-1})$ \\	
\(t\) & Time & $(s)$ \\	
\(t^*\) & Dimensionless time &  \\
\(t_n\) & Time step number  &  \\
\(U_{conv}\) & Mean convection velocity & $(m{\cdot}s^{-1})$ \\
\(u_i\) & Velocity vector in Cartesian coordinates & $(m{\cdot}s^{-1})$ \\
\(u_i^{\,pred,\,j}\) & Predicted velocity vector in Cartesian coordinates & $(m{\cdot}s^{-1})$ \\
\(u^*_i\) & Dimensionless velocity vector in Cartesian coordinates & \\
\end{tabular}

\section*{Variables}

\begin{tabular}{p{1.9cm}p{10.7cm}p{2cm}}
\(u_{i,\,avg}\) & Time-averaged velocity vector in Cartesian coordinates & $(m{\cdot}s^{-1})$ \\
\(u_{i,\,in}\) & Inlet velocity vector in Cartesian coordinates & $(m{\cdot}s^{-1})$ \\
\(u_{i,\,sym}\) & Velocity vector in Cartesian coordinates at the symmetry boundary & $(m{\cdot}s^{-1})$ \\
\(u_{i,\,t_0}\) & Initial velocity vector in Cartesian coordinates & $(m{\cdot}s^{-1})$ \\
\(u_{n,\,w}\) & Velocity normal to the wall & $(m{\cdot}s^{-1})$ \\	
\(u_{t,\,w}\) & Velocity tangential to the wall & $(m{\cdot}s^{-1})$ \\
\(u_{\tau}\) & Shear stress velocity & $(m{\cdot}s^{-1})$ \\
\(V_{cell}\) & Cell volume & $(m^3)$ \\
\(W\) & Wake length & $(m)$ \\
\(x_i\) & Cartesian coordinates &$(m)$\\
\(x_i^*\) & Dimensionless Cartesian coordinates &\\
\(y^+\) & Dimensionless wall distance & \\
\(y_{first\,cell}\) & First cell wall distance & $(m)$ \\
\end{tabular}

\section*{Greek variables}

\begin{tabular}{p{1.9cm}p{10.7cm}p{2cm}}
\(\alpha\) & Angle of attack & $(^\circ)$ \\
\(\Gamma_{\Phi}\) & Diffusive flux of the transport variable $\Phi$  & \\
\(\Delta\) & Filter cutoff length & $(m)$ \\
\(\Delta t\) & Time step	&$(s)$\\
\(\Delta y\) & Distance of the midpoint of the first cell & $(m)$ \\
\(\theta\) & Due to fluid-structure interaction induced rotation angle  &  $(^\circ)$\\
\(\mu\) & Dynamic viscosity & $(Pa{\cdot}s)$ \\
\(\mu_T\) & Eddy viscosity & $(Pa{\cdot}s)$ \\
\(\nu\)& Kinematic viscosity &$(m^{2}{\cdot}s^{-1})$\\
\(\rho\) & Density	&$(kg{\cdot}m^{-3})$\\
\(\sigma\) & Standard deviation &  \\
\(\tau_{ij}^{mol}\) & Molecular-dependent momentum transport  & $(Pa)$ \\
\(\tau_{ij}^{SGS}\) & Subgrid-scale stress tensor & $(Pa)$ \\
\(\tau_{ij}^{turb^*}\) & Dimensionless Reynolds stress tensor & \\
\(\tau_w\) & Wall shear stress & $(Pa)$\\
\(\Phi\) &  Transport variable&  \\
\(\omega_z\) & Vorticity in the span-wise direction  & $(s^{-1})$ \\
\end{tabular}

%% file: introduction.tex
\chapter*{Introduction\markboth{INTRODUCTION}{}}
\label{introduction}
\setcounter{page}{1}
\pagenumbering{arabic}
\addcontentsline{toc}{chapter}{Introduction}
\par Flows play an important role in nature and technical applications. Most of the current available technologies were only possible due to preliminary observations of phenomena in nature. These were, then, studied aiming the acquirement of knowledge and its consequent application in technical areas, for instance, in the development of airplanes. Firstly machines imitating the flight of birds were invented, then balloons filled with gases lighter than air, until the current state-of-the-art airplanes were achieved. 
\par The knowledge of fluid dynamics is based on the equations proposed by Navier and Stokes in the 19$^{th}$ century. These describe fluid motions by five coupled non-linear partial differential equations: conservations of mass, momentum and energy. Due to the complexity of these laws only very simple problems subjected to simplifications are able to be analytically solved. Therefore, the more complicated flows have to be numerically approximated using discretization and computers, according to a discipline called Computational Fluid Dynamics (CFD).
\par CFD is based on the discretization of the governing equations and the fluid domain aiming at the approximation of the flow through the use of initial and boundary conditions. The most utilized spatial discretization technique is the finite-volume method (FVM), while the temporal discretization approach is chosen according to each problem and can be either explicit or implicit. 
\par As per Moore's law, the computer performance, evaluated according to the number of floating point operations per second, i.e$.$, FLOPs, increases in an exponential pace. This growth enables the calculation of increasingly complicated fluid problems utilizing also more accurate and therefore more computational time allowing turbulence approaches, such as Direct Numerical Simulation and Large-Eddy Simulation. Nowadays, for instance, even simulations of the flow around an aircraft including the calculation of the forces generated by the fluid flow up to coupled fluid-structure interactions (FSI) were enabled as a result of the computer performance growth and the introduction of supercomputers.
\par The generation of instantaneous forces that may induce the airfoils to vibrate and possibly fail is a phenomena called flutter, which is a fluid-structure interaction observed by airplanes. This effect will be investigated in the near future by LES of a symmetrical NACA0012 airfoil profile with a chord length of $c=0.1\,m$ exposed to a Reynolds number of $Re=100{,}000$. 
\par In order to achieve accurate results regarding the vibration of the airfoil, firstly wall-resolved LES on a fixed NACA0012 profile at various angles of attack are performed. For this purpose several high quality meshes are generated and evaluated, as described in the following work. Firstly, in Chapter \ref{chap:numerical_methodology}, the applied numerical methodology with a brief discussion of the Navier-Stokes equations, the utilized discretization methods and turbulence approach is mentioned. Secondly, in Chapter \ref{chap:test_case}, the test case is presented regarding the geometry, the computational setup and the utilized initial and boundary conditions. Then, in Chapter \ref{chap:results_and_discussion}, the results obtained by LES are discussed and the grids are evaluated and compared. Finally, the achieved conclusions are summarized and a selection of the meshes that will be used to analyze the flutter effect is performed.

%% file: chapter1.tex
\chapter{Numerical methodology}
\label{chap:numerical_methodology}

\section{Navier-Stokes Equations}
\label{sec:navier_stokes}
In the $19^{th}$ century the Navier-Stokes equations were independently proposed by the French engineer Claude Louis  M. H. Navier and the Irish mathematician George Gabriel Stokes to describe the behavior of viscous fluids based on its velocity $u_i$ and its thermodynamic properties: Temperature $T$, pressure $p$ and density $\rho$. 
\par They are based on three conservation laws: The conservation of mass, momentum and energy, which are used in conjunction with the thermal equation of state. In the case of the flow around a NACA0012 profile only the conservation of mass and of momentum are required, since the flow is incompressible, submitted to a constant temperature and thus constant fluid properties.  
     
\subsection{Conservation of mass}
\label{subsec:conservation_mass}
The conservation of mass can be derived either via Lagrange or via Euler, as summarized by Breuer \cite{Breuer_2013}. While the first one is based on the principle that in a closed system the total mass of a time-dependent infinitesimal fluid volume must stay constant, the second one states that the difference between the inflowing and outflowing mass in a rigid control volume must be equal to the change of mass per time.
\par Both of these approaches result, for an incompressible fluid, in Eq.\ (\ref{eq:conservation_mass}). 
\begin{equation}
\label{eq:conservation_mass}
\frac{\partial u_i}{\partial x_i}=0
\end{equation}

\subsection{Conservation of momentum}
\label{subsec:conservation_momentum}
The conservation of momentum is based on Newton's second law, which states that the forces acting on a fluid control volume must be in equilibrium, i.e$.$, the time rate of change of momentum in j-direction is proportional to both the external forces $($volume and surface forces$)$ working in j-direction and the molecular-dependent momentum transport $\tau_{ij}^{mol}$. 
\par The dimensionless equation of momentum for an incompressible fluid submitted to a constant temperature is described by Eq.\ (\ref{eq:conservation_momentum}).
\begin{equation}
\label{eq:conservation_momentum}
\frac{\partial u_j}{\partial t}+\frac{\partial \left(u_i u_j\right)}{\partial x_i}=-\frac{\partial p}{\partial x_j} -\frac{1}{Re} \frac{\partial \tau_{ij}^{mol}}{\partial x_i}
\end{equation} 
\par $Re$ and $\tau_{ij}^{mol}$ are the Reynolds number and the molecular-dependent momentum transport tensor, respectively. The last one is proportional to the dynamic viscosity $\mu$ and the strain rate tensor $S_{ij}$, as per Eqs.\ (\ref{eq:molecule_transport}) and (\ref{eq:deformation_tensor}).
\begin{equation} \label{eq:molecule_transport}
\tau_{ij}^{mol}=-\mu \left(\frac{\partial u_i}{\partial x_j}+\frac{\partial u_j}{\partial x_i}\right)=-2 \mu S_{ij}
\end{equation}  
\begin{equation} \label{eq:deformation_tensor}
S_{ij}=\frac{1}{2}\left(\frac{\partial u_i}{\partial x_j}+\frac{\partial u_j}{\partial x_i}\right)
\end{equation}
\par The dimensionless variables are acquired by dividing the non-dimensionless ones by a reference value. In this study these variables are not identified by a special marker in order to simplify the display of the equations. 

\section{Discretization}
\label{sec:discretization}
Since the conservation equations are non-linear and coupled, the solutions for complex problems are numerically approximated using spatial and temporal discretization methods.

\subsection{Spatial discretization}
\label{subsec:spatial_discretization}
Equations (\ref{eq:conservation_mass}) and (\ref{eq:conservation_momentum}) are discretized according to the finite-volume method (FVM), which have the advantage of automatically fulfilling the conservation principle. The equations are integrated over the volume and the convective and diffusive terms are then converted to surface integrals by Gauss' divergence theorem. Equation (\ref{eq:spatial_discretization}) shows this process for the general form of the incompressible conservation laws, where $\Phi=1$ and $\Phi=u_i$ stands for the conservation of mass and momentum, respectively.
\begin{equation}
\label{eq:spatial_discretization}
\int_V\frac{\partial \left( \Phi \right)}{\partial t}\mathrm{d}V+\int_S\left( u_j \Phi - \Gamma_{\Phi} \frac{\partial \Phi}{\partial x_j}\right)\cdot n_j \mathrm{d}S=\int_V q_{\Phi}\mathrm{d}V
\end{equation}
\par In order to numerically approximate the volume and surface integrals, the domain is divided into several small control volumes. In the case of the software FASTEST-3D, these must have hexahedral forms. A second-order accurate midpoint rule is then applied to all control volumes in order to approximate the volume and surface integrals. While the former does not require interpolation methods, since all flow values are stored at the cell center and therefore already know, the latter requires an interpolation process in order to calculate the values on the cell faces. This process occurs according to a linear interpolation from the neighboring cells and is also second-order accurate.

\subsection{Temporal discretization}
\label{subsec:temporal_discretization}
\par Since large-eddy simulation is used to describe the turbulent flow field, it is necessary to work with very small time steps. Therefore, an explicit temporal discretization is preferred.  

\par The explicit method demands a small time step $\Delta t$ in order to be stable. However, it is easier to implement on high-performance computers (vectorization and parallelization) and requires a smaller numerical effort per time step. It uses only known variable values from old time steps, e.g$.$, $t_n$, $t_{n-1}$, $t_{n-2}$, in order to calculate the current ones, i.e$.$, $t_{n+1}$.

\par An explicit low-storage three sub-steps Runge-Kutta scheme, described by Breuer et al$.$ \cite{Breuer_2012}, is used by FASTEST-3D to predict the velocities $u_i^{\,pred}$ for $t_{n+1}$, according to \mbox{Eq\ (\ref{eq:predictor_step})}. $u_i^{t_n}$ stands for the velocity at the last time step and $f\left(t_n,u_i^{t_n},p^{\,pred}\right)$ represents a function of the time step, pressure and velocity, as well as of the geometry of the control volume. 
\begin{eqnarray}
\label{eq:predictor_step}
u_i^{\,pred,\,1}&=&u_i^{t_n}+\,\frac{\Delta t}{3}\;f\left(t_n,u_i^{t_n},p^{\,pred}\right)\nonumber\\
u_i^{\,pred,\,2}&=&u_i^{t_n}+\,\frac{\Delta t}{2}\;f\left(t_n,u_i^{\,pred,\,1},p^{\,pred}\right)\nonumber\\
u_i^{\,pred}&=&u_i^{t_n}+\;\Delta t\; f\left(t_n,u_i^{\,pred,\,2},p^{\,pred}\right)
\end{eqnarray}
\par These velocities are calculated applying the values of the pressure $p^{\,pred}$ (either a presumed value or an approximation based on the value at the last time step $p^{t_n}$) in the conservation of momentum. However, they may not fulfill the continuity equation. Hence, a corrector step must be executed in order to guarantee its fulfillment. 

\par The velocity field only satisfies both conservation laws when a correct pressure field is available. Thus, a Poisson equation (see Eq.\ (\ref{eq:Poisson})), deduced from the divergence of the momentum conservation, is used to calculate a pressure correction $p^{\,{corr}}$.
\begin{equation}
\label{eq:Poisson}
\frac{\partial}{\partial x_i} \left[\frac{\quad\partial p^{\,{corr}}}{\partial x_i}\right]=\frac{\rho}{\Delta t} \frac{\quad\partial u_i^{\,pred}}{\partial x_i}
\end{equation}
\par The pressure correction $p^{\,{corr}}$ is then used to estimate a new pressure $p^{t_{n+1}}$ and  velocity $u_i^{t_{n+1}}$, according to Eqs.\ (\ref{eq:neue_Druck}) and (\ref{eq:neue_Geschwindigkeit}), respectively. The latter, however, does not necessarily precisely fulfill the continuity equation for $t_{n+1}$. Therefore, the correction step must be repeated until a convergence criterion is achieved.
\begin{equation}
\label{eq:neue_Druck}
p^{t_{n+1}}=p^{\,pred}+p^{\,{corr}}
\end{equation}
\begin{equation}
\label{eq:neue_Geschwindigkeit}
u_i^{t_{n+1}}=u_i^{\,pred}-\frac{\Delta t}{\rho}\frac{\quad\partial p^{\,{corr}}}{\partial x_i}
\end{equation}

\section{Turbulence modeling} 
\label{sec:turbulence_modelling}
The Reynolds number is a relation between the inertial and viscous forces acting on a fluid, according to Eq.\ (\ref{eq:Reynolds_number}). $c$ and $u_{in}$ represent the chord length and the inflow velocity respectively, i$.$e$.$, the free-stream velocity.
\begin{equation}
Re=\frac{u_{in}\;\rho\;c}{\mu}
\label{eq:Reynolds_number}
\end{equation}
\par At larger Reynolds numbers, the inertial forces overcome the viscous ones, resulting in a turbulent flow. This is characterized by its vorticity, three-dimensionality, as well as its chaotic and transient behavior.
\par The turbulence can be described as an energy cascade, in which the largest eddies (scales) decay into smaller ones, transferring their turbulent kinetic energy. When the eddies achieve the Kolmogorov length $l_k$ (smallest scale), the turbulent kinetic energy dissipates into thermal energy.
\par In order to simulate turbulent flows, the energy cascade can be either directly solved by the Direct Numerical Simulation, partially modeled as done in Large-Eddy Simulation or fully modeled according to Reynolds-Averaged Navier-Stokes Equations. 
\par DNS directly applies the Navier-Stokes equations for the whole turbulence spectrum, requiring small enough control volumes in order to resolve even the Kolmogorov length scale $l_k$. Despite of being the most accurate approach, it requires a small time step in order to be stable, a high computational effort and is restricted to Reynolds numbers in the order of magnitude of $Re\leq \mathcal{O}\left(10^4\right)$. Therefore, its mainly usage is as a comparison base for other turbulence approaches in a process called a-priori test. This aims at the validation of LES or RANS simulations mainly when experimental data are not available.
\par The RANS approach tries to model the entire turbulence spectrum utilizing a time-averaging process, in which all quantities are divided into a statistically stationary value plus a fluctuation. This process, however, results in six new unknown terms for fluids submitted to a constant temperature. Various methods are available to model these terms and are divided into first-moment-closure eddy viscosity models and second-moment-closure Reynolds stress model, as described by Breuer \cite{Breuer_2002}. This turbulence approach has the advantages of being computationally less demanding, although it is less accurate than DNS.
\par The LES approach is the technique that stays between both previously described approaches and has been increasingly applied due to the higher available computer performance. Because of lower required computational time when compared to DNS and higher accuracy when compared to RANS, this method is utilized in the present thesis to perform the simulation of the NACA0012 airfoil profile.
\par Large-eddy simulation is based on the division of the turbulence spectrum into small and large scales. The former constitutes the low-energy scale, is short-lived, dissipative, universal (independent from geometry and boundary conditions) and nearly homogeneous and isotropic. Therefore this scale, as well as its influence on the large scales, is modeled. The latter constitutes the high-energy scale, which is strongly problem-dependent. Thus, it is predicted directly by the spatial filtered Navier Stokes equations. 
\par Due to the direct usage of the conservation laws for the large eddies, the simulation time step and the grid resolution must be sufficiently small/fine in order to resolve the smallest eddies of the large scale. Even though this increases the accuracy, it also raises the numerical effort compared to RANS.
\par The governing equations for LES are acquired by the spatial filtering of the conservation laws. In the case of FASTEST-3D, as described by Durst and Schäfer \cite{Durst_1996}, this is an implicit filter, which has the advantage of coupling filtering and numerical method. However, it is susceptible to aliasing error, which arises from the non-linearity of the equations.
\par The filtering process plus some mathematical procedures result in the LES equations, which are subject to the commutation error caused by the approximation of the filtered partial derivatives.
 \begin{eqnarray}\label{eq:LES}
\frac{\partial {\overline{u}}_i}{\partial x_i}&=&0 \\
\label{eq:LES2}
\frac{\partial {\overline{u}}_j}{\partial t}+\frac{\partial \left( {\overline{u}}_i{\overline{u}}_j\right)}{\partial x_i}&=&-\frac{\partial \overline{p}}{\partial x_j}-\frac{1}{Re} \frac{\partial {\overline {\tau}}_{ij}^{mol}}{\partial x_i}-\frac{\partial \tau_{ij}^{SGS}}{\partial x_i}
\end{eqnarray}
\par Equations (\ref{eq:LES}) and (\ref{eq:LES2}) are respectively the dimensionless conservation of mass and momentum for the large turbulent scales (characterized by an overline) of an incompressible fluid subjected to constant temperature. ${\overline{u}_i}$, $\overline{p}$ and ${\overline{\tau}}_{ij}^{mol}$ represent, correspondingly, the large scale velocity, pressure and molecular-dependent transport. The latter is a function of the large scale strain rate tensor ${\overline {S}}_{ij}$, according to Eqs.\ (\ref{eq:molecule_dependent_transport_LES}) and (\ref{eq:strain_rate_tensor_LES}). 
\begin{eqnarray}
\label{eq:molecule_dependent_transport_LES}
{\overline{\tau}}_{ij}^{mol}&=&-2\mu {{\overline {S}}_{ij}} \\
\label{eq:strain_rate_tensor_LES}
{\overline{S}}_{ij}&=&\frac{1}{2}{\left({\frac{\partial {\overline {u}}_i}{\partial x_j}+\frac{\partial {\overline {u}}_j}{\partial x_i}}\right)}
\end{eqnarray}
\par The conservation of momentum (\ref{eq:LES2}) comprises a new term, the subgrid-scale stress tensor $\tau_{ij}^{SGS}$. It is produced by the filtering process of the non-linear convective term, i.e$.$, ${\overline {u_i u_j}}$, and symbolizes the following effects: The interaction of the large eddies that results in the production of small scales, the interaction between large and small scales and finally the interaction between small scales that culminates in the formation of large eddies in a process called backscatter. This term is represented by Eq.\ (\ref{eq:sub_grid_scale_stress}) and cannot be directly calculated. Therefore, it constitutes the closure problem of the turbulence.
\begin{equation}
\label{eq:sub_grid_scale_stress}
\tau_{ij}^{SGS}=\overline{u_i u_j}-{\overline {u}}_i {\overline {u}}_j={\overline{{\overline u}_i{\overline u}_j}}-{\overline u}_i {\overline u}_j+{\overline{{\overline u}_i{u'}_j}}+{\overline{{\overline u}_j{u'}_i}}+{\overline {u'_i u'_j}}
\end{equation}
\par In order to solve the governing equations, i.e$.$, to model the subgrid-scale stress tensor, various models have been proposed, as summarized by Breuer \cite{Breuer_2002}.
\par Concerning the subgrid-scale models, FASTEST-3D uses the Smagorinsky model \cite{Smagorinsky_1963} in order to calculate the subgrid-scale stress of the flow. This model is based on the algebraic Boussinesq approximation (see Eq.\ (\ref{eq:Boussinesq_approximation})) and states an analogy between the molecular-dependent transport for laminar flows and the subgrid-scale tensor, as per \mbox{Eq.\ (\ref{eq:Smagorinsky_equation})}.
\begin{eqnarray}
\label{eq:Boussinesq_approximation}
\tau _{ij}^{mol}&=&-2 \mu {S}_{ij} \\
\label{eq:Smagorinsky_equation}
\tau _{ij}^{SGS}&=&-2 \mu_T {\overline S}_{ij}
\end{eqnarray} 
\par The proportionality factor is the eddy viscosity $\mu_T$. It depends on the turbulence structure and may vary in space and time. Therefore, is not a fluid property. $\mu_T$ can be estimated by the Smagorinsky model, which is based on the assumption that the modeled scales are isotropic, i.e$.$, the turbulence is in local equilibrium.
\par Due to a dimension analysis, Smagorinsky \cite{Smagorinsky_1963} stated that the eddy viscosity is proportional to a characteristic length $l_c$, velocity $v_c$ and density $\rho_c$ (this is assumed constant and equal to one, i.e$.$, $\rho_c =1$), according to Eq.\ (\ref{eq:eddy_viscosity_1}).
\begin{equation}
\label{eq:eddy_viscosity_1}
\mu_T\sim\rho_c\,l_c\,v_c\sim l_c\,v_c
\end{equation}
\par The characteristic velocity $v_c$ is approximated as a function of the characteristic length $l_c$ and the strain rate tensor of the large eddies ${\overline S}_{ij}$, while the characteristic length $l_c$ is estimated with help of the Van-Driest damping function, which, in turn, is needed to evaluate the subgrid-scale stress tensor near fixed walls. $l_c$ is then proportional to the Smagorinsky constant $C_s$, the filter cutoff length $\Delta$ and the dimensionless wall distance $y^+$, as demonstrated in Eq$.$ (\ref{eq:eddy_viscosity_2}).
\begin{equation}
\label{eq:eddy_viscosity_2}
\mu_T=C_s^2\Delta^2\sqrt{2{\overline S}_{ij}{\overline S}_{ij}}{\left[1-\exp{\left(\frac{-y^+}{25}\right)}^3\right]}
\end{equation}
\par The Smagorinsky constant $C_s$ is empirically determined and a standard value of \mbox{$C_s=0.1$} is used for the NACA0012 airfoil simulations.
\par Since the filtering process and numerical method are coupled, the filter cutoff length $\Delta$ is defined as a function of each cell volume $V_{cell}$, according to Eq.\ (\ref{eq:cutoff_length}).
\begin{equation}
\label{eq:cutoff_length}
\Delta=\left(V_{cell}\right)^{\frac{1}{3}}
\end{equation}
\par The dimensionless wall distance is defined as a function of the shear velocity $u_\tau$, the distance of the first cell middle point to the airfoil geometry $\Delta y$ and the kinematic viscosity $\nu$, as stated by Eq.\ (\ref{eq:dimensionless_wall_distance}).
\begin{equation}
y^+=\frac{u_\tau\,\Delta y}{\nu} \label{eq:dimensionless_wall_distance} \\
\end{equation}
\par The shear velocity $u_\tau$ is a function of the fluid density $\rho$ and the wall shear stress $\tau_w$, according to Eq.\ (\ref{eq:shear_velocity}). The latter, i.e$.$, $\tau_w$, is proportional to the dynamic viscosity $\mu$ and the velocity gradient $\partial u/\partial y$ near the wall $w$, as demonstrated in Eq.\ (\ref{eq:wall_shear_stress}).
\begin{eqnarray}
u_\tau&=&\sqrt{\frac{\tau_w}{\rho}} \label{eq:shear_velocity} \\
\tau_w&=&\mu \left.\frac{\partial u}{\partial y}\right |_w \approx \frac{\Delta u_i}{\Delta y}=\frac{u_{first\,cell,\,avg,i}-0}{\Delta y} \label{eq:wall_shear_stress}
\end{eqnarray}
\par In the present work, the dimensionless wall distance is approximated as a function of the time-averaged velocities in the first cell close to the wall $u_{first\,cell,\,avg,\,i}$, the distance of the first cell middle point to the airfoil geometry $\Delta y$ and the kinematic viscosity $\nu$, as stated by Eq.\ (\ref{eq:dimensionless_wall_distance_used}).
\begin{equation}
\label{eq:dimensionless_wall_distance_used}
y^+=\sqrt{{\sqrt{\sum_{i=1}^{3}u_{first\,cell,\,avg,i}^2}}\cdot\frac{\Delta y}{\nu}}
\end{equation}

\section{Initial and boundary conditions}\markboth{CHAPTER 1.$\quad$NUMERICAL METHODOLOGY}{1.4.$\quad$IN. AND BOUND. CONDs.}
\label{sec:initial_and_boundary_conditions}
Since the discretized Navier-Stokes equations constitute an initial boundary value problem, these conditions have to be specified in order to solve CFD problems. Initial conditions for the whole domain and boundary conditions for all boundaries are required, since the conservation laws are parabolic in time and elliptic in space. 
\par The utilized initial and boundary conditions are thoroughly described in Section \ref{sec:simulation_parameters}.

%% file: chapter2.tex
\chapter{Flow around a NACA0012 airfoil}
\label{chap:test_case}
\par The flow around a NACA0012 airfoil profile at a Reynolds number of $Re=100{,}000$ is investigated using wall-resolved LES and the Smagorinsky model. The in-house software FASTEST-3D is used to perform the simulations on seven different grids using FVM for the spatial discretization and a predictor-corrector procedure based on a three sub-steps Runge-Kutta scheme in combination with a pressure Poisson equation for the temporal discretization. 

\section{Description of the geometrical model and the test section}
\label{sec:geometrical_model_test_section}
The NACA0012 profile stands for the National Advisory Committee for Aeronautics four digit series airfoil profiles. The first two digits, i.e$.$, $00$, represent a symmetrical airfoil and the last two stand for a profile with a maximal thickness of twelve percent of the chord length $c$, located at thirty percent of the chord \cite{Jacobs_1935}. Because of its symmetrical profile it does not produce lift for an angle of attack of $\alpha=0^\circ$ and therefore is commonly used for stability purposes in airplanes, such as in horizontal and vertical stabilizers.
\par The simulated airfoil profile has a chord length of $c=0.1\,m$ with a sharp trailing edge geometry. Its profile is described by NASA \cite{NASA_2016} according to Eq$.$ (\ref{eq:NACA0012}) with a global origin of the coordinate system located at the leading edge. $y_N$ and $x_N$ represent the $y$ and $x$ coordinates of the NACA0012 airfoil, respectively. 

\begin{eqnarray}
\nonumber
\label{eq:NACA0012}
y_{N}&=&\pm 0.594689181c\left[0.298222773\sqrt{\frac{x_{N}}{c}}-0.127125232\frac{x_{N}}{c}-0.357907906\left(\frac{x_{N}}{c}\right)^2\right.\\
&& \left. +0.291984971\left(\frac{x_{N}}{c}\right)^3-0.105174606\left(\frac{x_{N}}{c}\right)^4\right]
\end{eqnarray}
  
\par The computational domain is established with help of artificial boundaries, which are carefully chosen according to the work of Almutari \cite{Almutari_2010}. This thesis presents the results of various large-eddy simulations for a NACA0012 profile for diverse configurations of the Reynolds number and angle of attack, validated by a-posteriori studies against the results of DNS simulations achieved by Jones et al. \cite{Jones_2008}, as well as by the experimental results of Rinoiei and Takemura \cite{Takemura_2004}.

\par
\begin{figure}[H]
	\centering
	\centering
	\includegraphics[scale=0.85,draft=\drafttype]{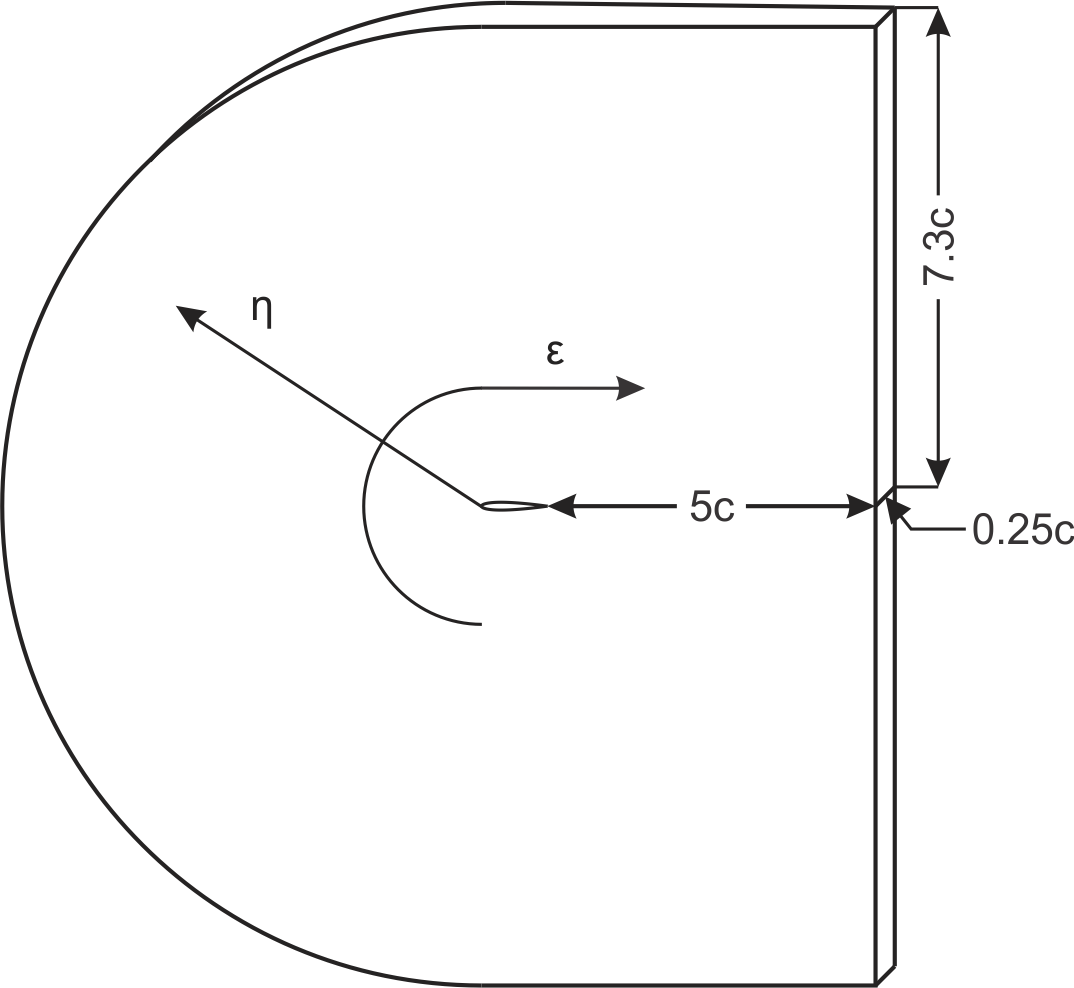}
	\caption{\label{fig:domain_3d}Fluid domain of the NACA0012 airfoil profile exposed to a Reynolds number of $Re=100{,}000$.}
\end{figure}

 Figure \ref{fig:domain_3d} illustrates the simulated fluid domain. $\eta$ and $\varepsilon$ are the curvilinear coordinates utilized in order to simplify the description of the cartesian meshes (see \mbox{Section \ref{sec:computational_mesh}}). A wake length of $W=5c$, a domain radius of $R=7.3c$ and span-wise length of $L_z=0.25c$ characterize the computational domain. This is narrow due to the usage of periodic boundary conditions in the homogeneous span-wise direction, which reduces the required numerical effort. The span-wise length $L_z$ is established based on the work of Almutari \cite{Almutari_2010}, which presents the evaluation of the the two-point velocity correlation within this direction.

\section{Computational meshes} \markboth{CHAPTER 2.$\quad$FLOW AROUND A NACA0012 AIRFOIL}{2.2$\quad$COMPT. MESHES}
\label{sec:computational_mesh}
As described in Subsection \ref{subsec:spatial_discretization}, the computational domain must be spatially discretized in order to numerically solve the governing Eqs.\ (\ref{eq:LES}) and (\ref{eq:LES2}) for turbulent flows. This procedure is done with help of the software ANSYS ICEM CFD and consists of the following steps:

\begin{itemize}
\item Generation of the two-dimensional geometry (NACA0012 profile and boundaries);
\item Generation of a two-dimensional C-grid composed of twelve geometrical blocks;
\item Extrusion of the two-dimensional geometry and blocks;
\item Definition of the edge parameters, regarding the dimensionless wall distance $y^+$ and the stretching factor $q$;
\item Analysis of the mesh quality regarding the internal angles of each cell and its volume;
\item Exportation of the high-quality mesh for the software FASTEST-3D with subsequent generation of a $.grd$ and a $.tbc$ file. The former contains information of the grid itself while the latter holds data regarding the boundary conditions.
\end{itemize}  

\par The grids generated by ANSYS ICEM CFD are composed of only one cell in the span-wise direction. A FORTRAN code is, then, used in order to generate a grid with 64 cells in the span-wise direction. Thereafter, a process called $mapping$ is applied in the three-dimensional grid in order to re-check if all cells have a positive volume and to divide the previously generated geometrical blocks into smaller, as cubic as possible, parallel blocks. Each one is calculated by a different processor with information exchange occurring only at its faces, so that the computation of all blocks can be performed at the same time in a process called $parallel\,computing$.

\par Regarding the generation of grids for different angles of attack, i.e$.$, $\alpha=5^\circ$ and $\alpha=11^\circ$, an inverse distance weighting IDW interpolation method, which is described in the work of Sen et al. \cite{Sen_2017},  is used to adapt the previous generated mesh with $\alpha =0^\circ$.

\par Seven different C-block structured hexahedral grids with varying numbers of nodes, dimensionless wall distances $y^+$ and angles of attack $\alpha$ are generated. A summary of the main parameters for each grid is presented in Tables \ref{table:mesh_nodes} and \ref{table:grids}.

\begin{table}[!htbp]
		\centering
	\begin{tabular}{p{2.5cm} p{2cm} p{3cm} p{0.8cm} p{0.8cm} p{0.8cm} p{0.8cm}}
		\hline
		\centering{\bf{Mesh resolution}} & \centering{\bf{Control volumes}} & \centering{\bf{Number of parallel blocks}} & \centering{\bf{$N_W$}} & \centering{\bf{$N_R$}} & \centering{\bf{$N_{L_z}$}} & \centering{\bf{$N_{SS}$}} \tabularnewline 
		\hline
		\centering{{Fine}} & \centering{28{,}646{,}400} & \centering{376} & \centering{501} & \centering{301} & \centering{65} & \centering{274} \tabularnewline
		\centering{Medium} & \centering{7{,}545{,}600} & \centering{98} & \centering{251} & \centering{151} & \centering{65} & \centering{148} \tabularnewline
		\hline
	\end{tabular}
	\caption{\label{table:mesh_nodes}Number of control volumes, parallel blocks and nodes of the generated meshes for a NACA0012 profile at a Reynolds number of $Re=100{,}000$.}
\end{table}

\par Table \ref{table:mesh_nodes} illustrates the grids according to the number of control volumes, parallel blocks, i.e$.$, number of processors required to perform the simulations, and nodes in the direction of the wake $N_W$, in the direction of the radius $N_R$, in the span-wise direction $N_{L_z}$ and on the suction side of the airfoil $N_{SS}$. Since the NACA0012 profile is symmetric, the number of nodes on the pressure side of the airfoil is the same as on the suction side.

\par The $fine\,\,mesh$ is generated according to the work of Almutari \cite{Almutari_2010} and is composed of more than 28 million cells. The mapping process produced 376 parallel blocks with 92.23\% load-balancing efficiency and an average number of 76,187 cells per parallel block. The wall distance of the first cell, i.e$.$, $y_{first\,cell}=4.4\cdot 10^{-6}\,m$, is based on the work of Kasibhotla \cite{Kasibhotla_2014}, which presents the results of NACA0012 simulations for Reynolds numbers of $Re=10^5$ and $Re=10^6$ using wall-resolved LES. Due to the fine resolution of the mesh and the small wall distance of the first cell, the required time step to simulate the flow on this mesh is in the order of $\Delta t=\mathcal{O}(10^{-8})\,s$, resulting in large computational times. 

\par Aiming at a reduction in CPU-time a $medium\,\,mesh$ was generated. This is composed of nearly a quarter of the control volumes of the fine mesh, i.e$.$, approximately 7 million cells. The mapping process generated 98 parallel blocks with 89.12\% load-balancing efficiency. Therefore, the quality of the blocking and the parallelization is lower than for the fine mesh. This lower load-balancing efficiency indicates that the parallel blocks do not possess the same size, i.e$.$, are not really cubic, which affects the calculation time. The average number of control volumes per processor is 76,995 and the simulation is stable also for larger time steps when compared to the fine resolution mesh. 

\par Table \ref{table:grids} illustrates the differences between the generated grids, i.e$.$, the resolution, the angle of attack $\alpha$, which is stipulated according to Almutari \cite{Almutari_2010}, and the first cell wall distance $y_{first\,cell}$. The latter, i.e$.$,  $y_{first\,cell}$, is varied in order to get stable simulations for time steps in the order of $\Delta t=\mathcal{O}(10^{-6})\,s$ and is further evaluated regarding the resolution of the viscous sublayer (see Section \ref{sec:analysis_dimensionless_wall_distance}). 

\begin{table}[!htbp]
	\centering
	\begin{tabular}{c c c l}
		\hline
		\centering{\bf{Mesh resolution}} & \centering{\bf{Angle of attack}} & \centering{\bf{First cell wall distance}} & \centering{\bf{Mesh name}} \tabularnewline
		\hline
		\centering{Fine} & \centering{$0^\circ$} & \centering{$4.4\cdot 10^{-6}\,m$} & \centering{$\;$$f-0-y^+_{min}$} \tabularnewline \hline
		\multicolumn{1}{c}{\multirow{6}{*}{Medium}} & \multirow{2}{*}{$0^\circ$} & \centering{$9.0\cdot 10^{-6}\,m$} & \centering{$\;$$m-0-y^+_{med}$} \tabularnewline
		& & \centering{$1.8\cdot 10^{-5}\,m$} & \centering{$\;$$m-0-y^+_{max}$} \tabularnewline \cline{2-4}
		& \multirow{2}{*}{$5^\circ$} & \centering{$9.0\cdot 10^{-6}\,m$} & \centering{$\;$$m-5-y^+_{med}$} \tabularnewline
		& & \centering{$1.8\cdot 10^{-5}\,m$} & \centering{$\;$$m-5-y^+_{max}$} \tabularnewline \cline{2-4}
		& \multirow{2}{*}{$11^\circ$} & \centering{$9.0\cdot 10^{-6}\,m$} & \centering{$\;$$m-11-y^+_{med}$} \tabularnewline
		& & \centering{$1.8\cdot 10^{-5}\,m$} & \centering{$\;$$m-11-y^+_{max}$} \tabularnewline
		\hline	
	\end{tabular}
	\caption{\label{table:grids} Generated meshes according to the angle of attack and the first cell wall distance for a NACA0012 profile at a Reynolds number of $Re=100{,}000$.}
\end{table}

\subsection{Fine mesh}
\par A fine resolution mesh is generated only for an angle of attack of $\alpha=0^\circ$ regarding its very small time step, i.e., $\Delta t=6\cdot10^{-8}\,s$, and therefore enormous amount of required CPU-time. Its quality is evaluated in the software ANSYS ICEM CFD based on the angles and determinants of the Jacobi-Matrix, which are used to evaluate the shape and volume of each cell. Target values of respectively $90^\circ$ and $1$ are aimed in order to have only orthogonal cells with a perfect  shape. The achieved values are $\alpha \geq 81^\circ$ and $\det\geq 0.97$, which indicates a high-quality mesh.

\par The computational domain is firstly divided into twelve geometrical blocks. The edge parameters are cautiously calculated in order to get a mesh with smooth transitions between the blocks. The stretching factor $q$ used for the wake length and domain radius are $q_W=1.06$ and $q_R=1.02$, respectively. The values of the first cell wall distance and the stretching factor $q_R$ are carefully established in order to guarantee the resolution of the viscous sublayer. 

\par
\begin{figure}[H]
	\subfigure[Parallel block distribution.]{\includegraphics[width=0.5\textwidth]{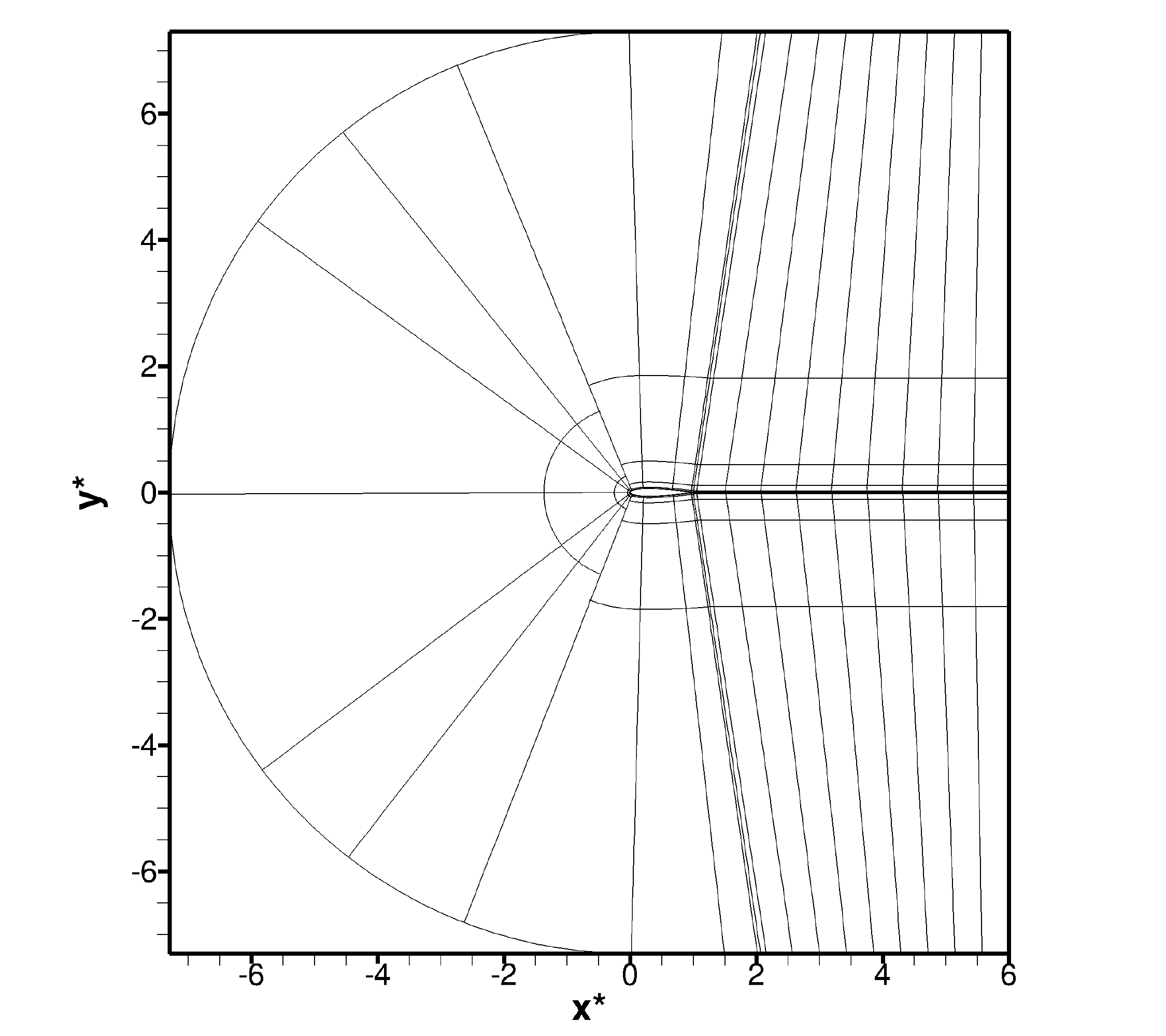}\label{fig:fine_mesh_blocking_a}}\hfill
	\subfigure[Parallel block distribution near the airfoil.]{\includegraphics[width=0.5\textwidth]{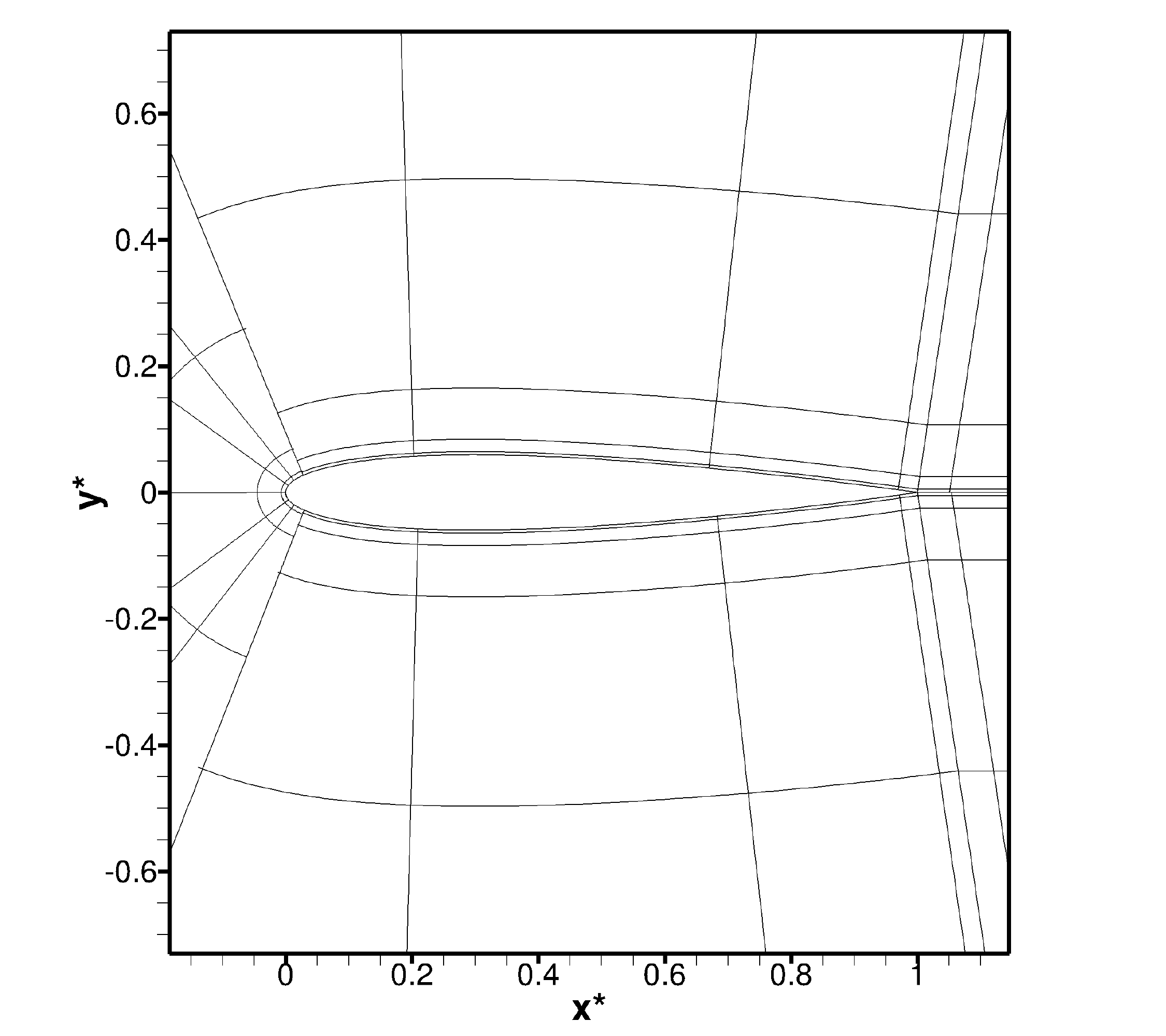}\label{fig:fine_mesh_blocking_b}}
	\caption{Blocking of the $f-0-y^+_{min}$ grid for the NACA0012 profile at a Reynolds number of $Re=100{,}000$. }
	\label{fig:fine_mesh_blocking}
\end{figure}

 The generated mesh contains only one cell in the z-direction, so that a FORTRAN code is used in order to obtain a mesh with 64 cells in the span-wise direction. The mapping process is then executed in order to achieve a high-performance computation through the division of the geometrical blocks into 376 smaller parallel ones. Since each block is computed by a single processor, a number of 376 processors is required to start a simulation. The blocking configuration is illustrated by Fig.\ \ref{fig:fine_mesh_blocking}. Figure \ref{fig:fine_mesh_blocking_a} shows an overall view of the blocking, while Fig.\ \ref{fig:fine_mesh_blocking_b} emphasizes the block concentration near the airfoil profile.
\par A higher control volumes density is present near the airfoil, in the wake and orthogonal to the trailing edge. The first two ones are necessary in order to resolve the boundary layer and the wake, respectively. The latter, however, is not desired but is the result of the meshing process while establishing the edge parameters of the wake length. Since the wake, the top and the bottom edges are parallel, they need to have the same edge parameter specification in order to assure the generation of a mesh with control volume interior angles as close as possible to $90^\circ$ . 

\par
\begin{figure}[H]
	\subfigure[Overall view of the mesh.]{\includegraphics[width=0.5\textwidth]{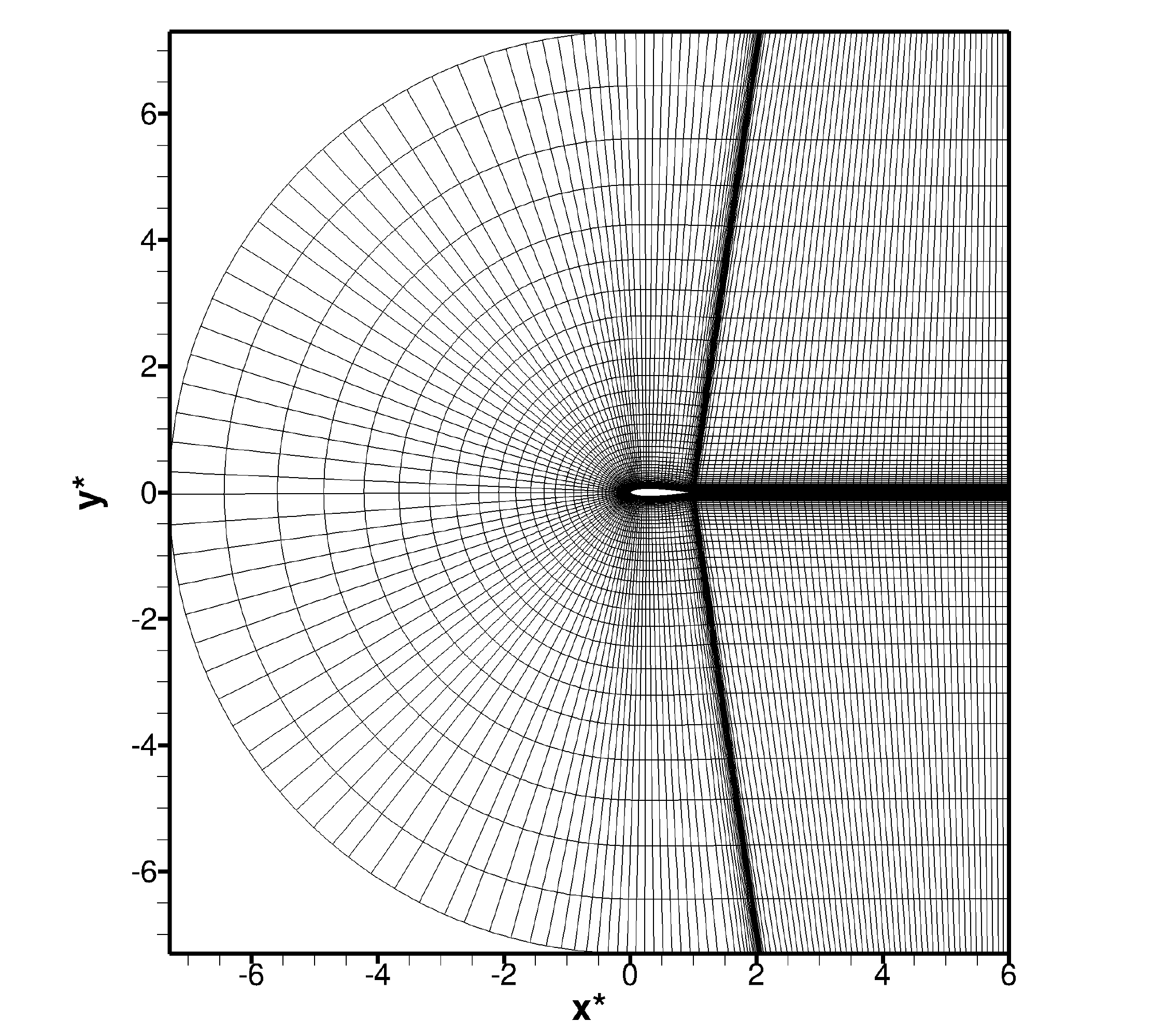}\label{fig:fine_mesh_mesh_a}}\hfill
	\subfigure[Mesh near the airfoil.]{\includegraphics[width=0.5\textwidth]{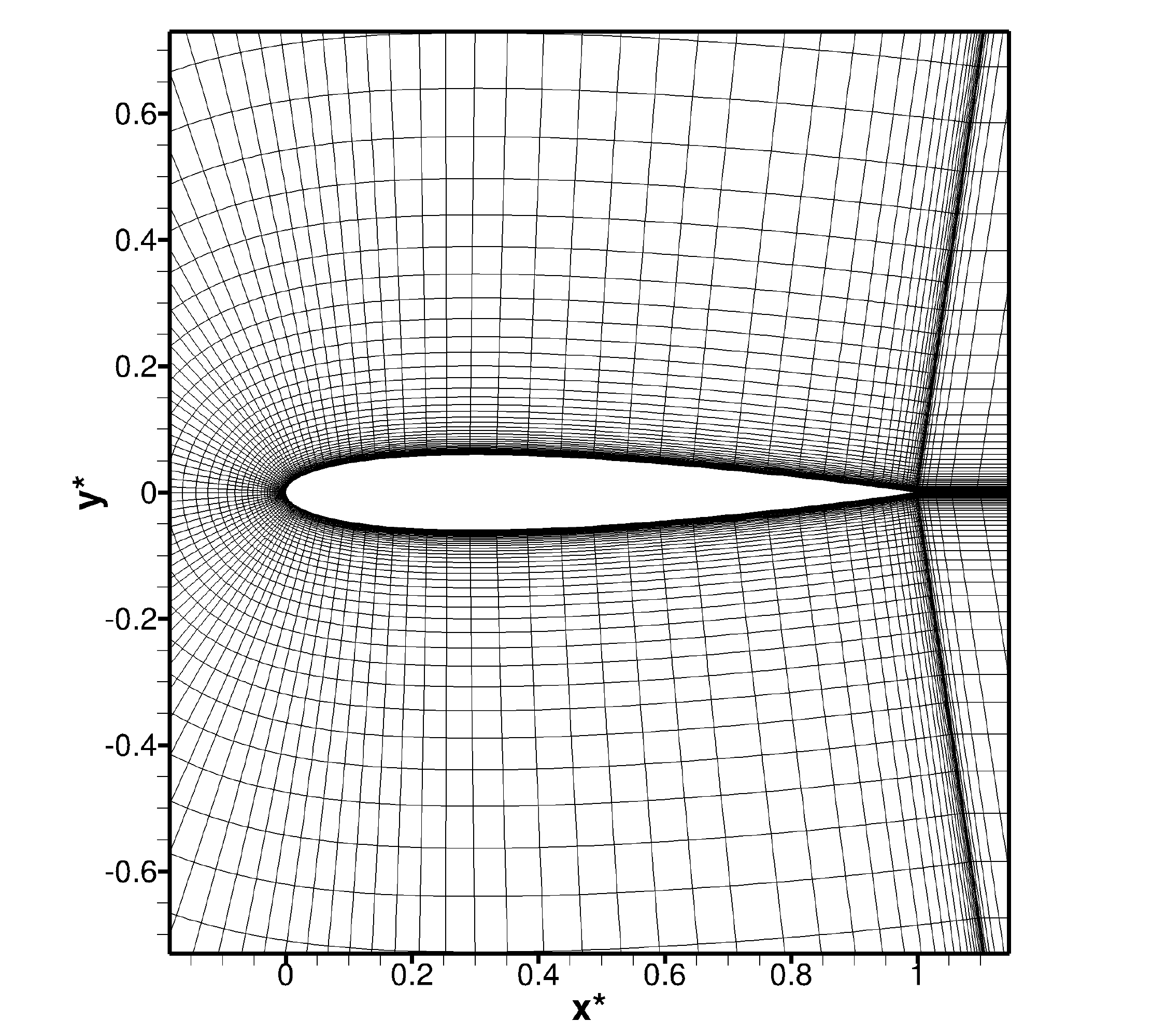} \label{fig:fine_mesh_mesh_b}}
	\caption{Fine resolution mesh $f-0-y^+_{min}$ for the NACA0012 profile at a Reynolds number of $Re=100{,}000$. }
	\label{fig:fine_mesh_mesh}
\end{figure}

 Figures \ref{fig:fine_mesh_mesh_a} and \ref{fig:fine_mesh_mesh_b} illustrate the fine grid. In order to achieve a good quality image, only one out of five mesh lines is delineated in the $\varepsilon$-axis and one out of four lines or one out of ten lines in the $\eta$-axis for the front and the wake domain, respectively.
\par The simulation of this mesh requires a high computational effort that is the result of the fine resolution, the small first cell wall distance and the number of processors needed to perform a parallel computation. Moreover, the utilized supercomputer, SuperMUC, needs to have 376 available processors to run this simulation, increasing the time spent in the queue.  Hence, the medium resolution meshes are generated in order to simulate the flow around this profile for angles of attack of $\alpha=0^\circ$, $\alpha=5^\circ$ and $\alpha=11^\circ$. 

\subsection{Medium mesh}
\label{subsec:medium_mesh} 

\par A total of six medium resolution meshes are generated for diverse combinations of the angle of attack and the first cell wall distance, as illustrated in Table \ref{table:grids}. Despite the same number of nodes per edge, the distribution of them varies, that is, the stretching factors $q_R$ and $q_W$ are adjusted for the first cell wall distances $y^+_{med}$ and $y^+_{max}$.
\par Firstly, the number of nodes per edge from the fine resolution mesh with only one cell in the span-wise direction is modified in order to get a grid with approximately half of the nodes, as per the last four columns of the Table \ref{table:mesh_nodes}. Secondly, the first cell wall distance and the stretching factor are adjusted, resulting in two different meshes for $\alpha=0^\circ$: One with $y_{first\,cell}=9\cdot 10^{-6}\,m$, $q_R=1.06$ and $q_W=1.06$ and the other with $y_{first\,cell}=1.8\cdot 10^{-5}\,m$, $q_R=1.06$ and $q_W=1.06$. Both of them have angles between the grid lines greater than 81.5$^\circ$ and determinants of the Jacobi-Matrix greater than 0.945.
\par These meshes are then adapted in order to achieve angles of attack of $\alpha=5^\circ$ and $\alpha=11^\circ$. The six resulting grids passed through a FORTRAN code in order to obtain 64 cells in the span-wise direction and through the mapping process, which divided the 12 previous geometrical blocks into 98 parallel ones.
\par
\begin{figure}[H]
	\subfigure[Parallel blocks distribution: $m-0-y^+_{max}$.]{\includegraphics[width=0.46\textwidth]{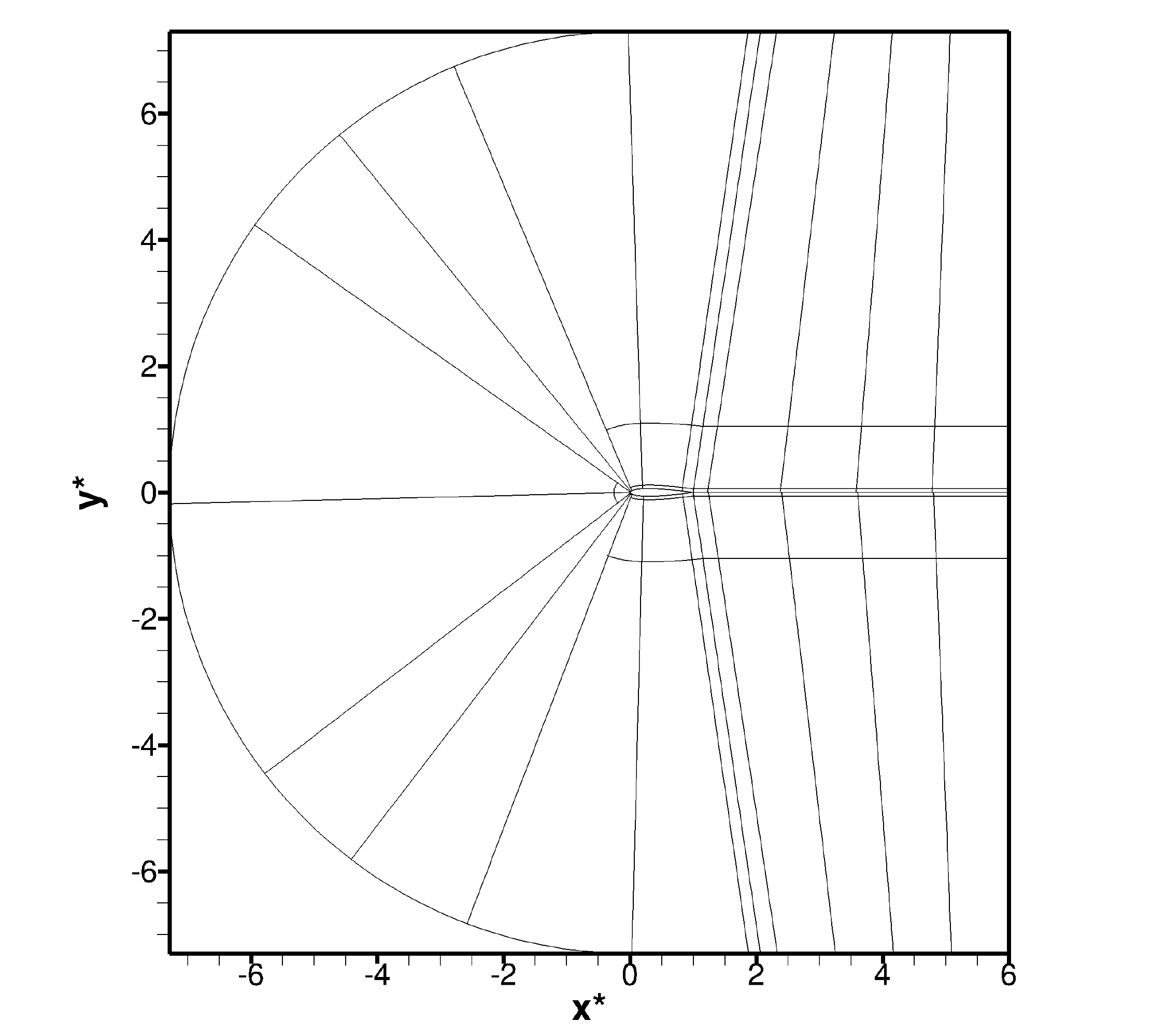}\label{fig:medium_mesh_angle_0_blocking_a}}\hfill
	\subfigure[Parallel blocks near the airfoil: $m-0-y^+_{max}$.]{\includegraphics[width=0.46\textwidth]{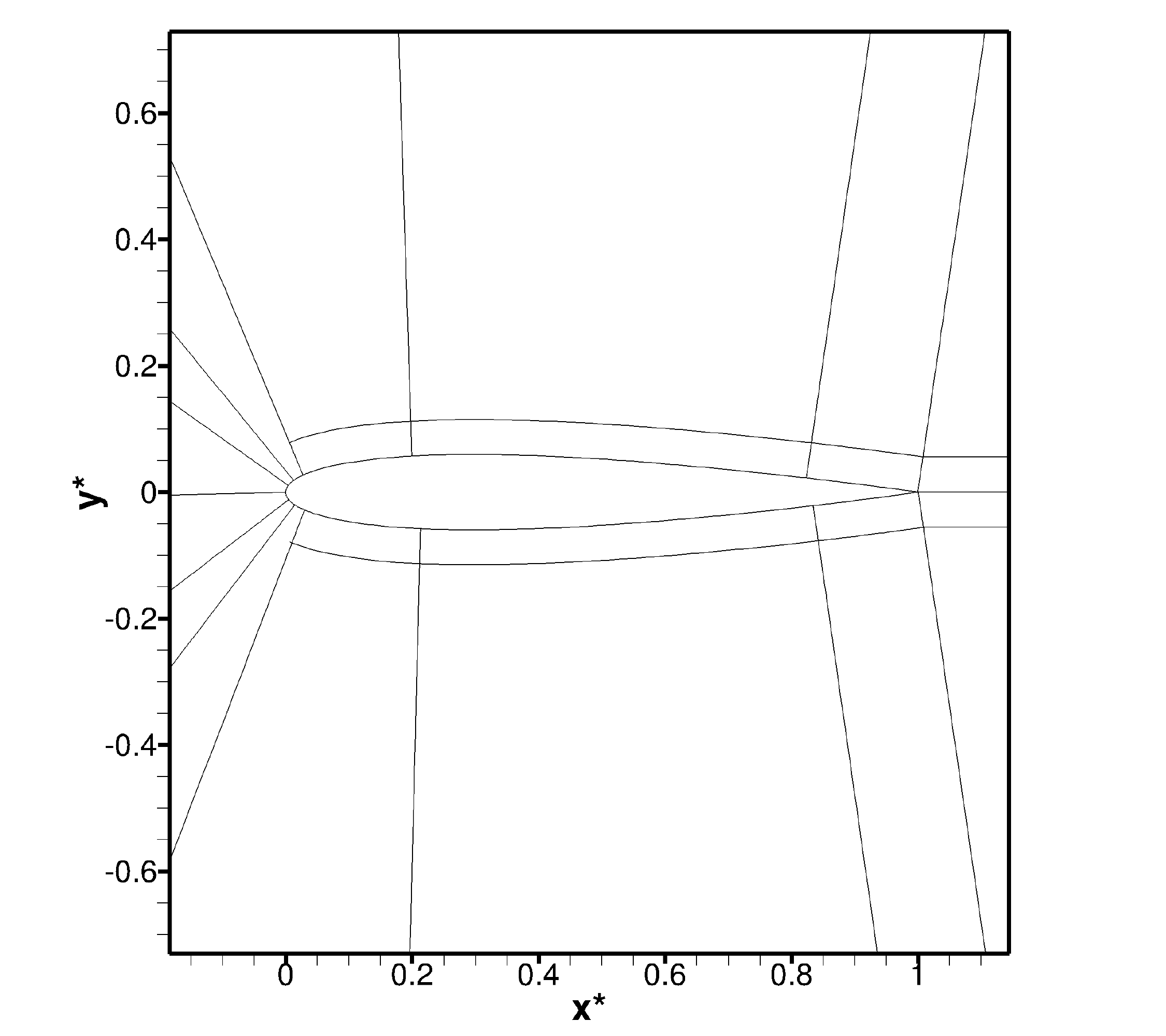}\label{fig:medium_mesh_angle_0_blocking_b}}
	\subfigure[Parallel blocks distribution: $m-5-y^+_{max}$.]{\includegraphics[width=0.46\textwidth]{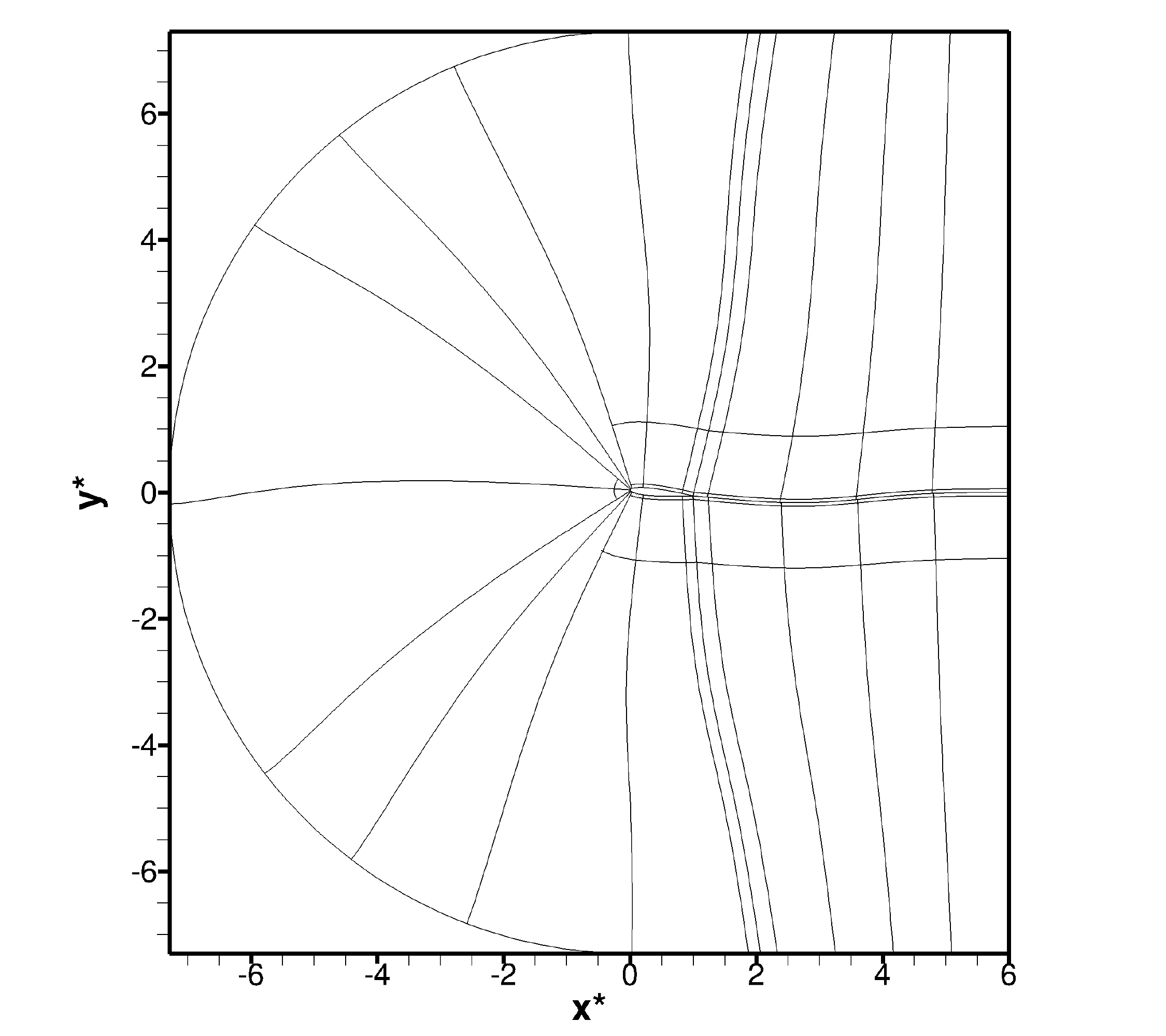}\label{fig:medium_mesh_angle_5_blocking_a}}\hfill
	\subfigure[Parallel blocks near the airfoil: $m-5-y^+_{max}$.]{\includegraphics[width=0.46\textwidth]{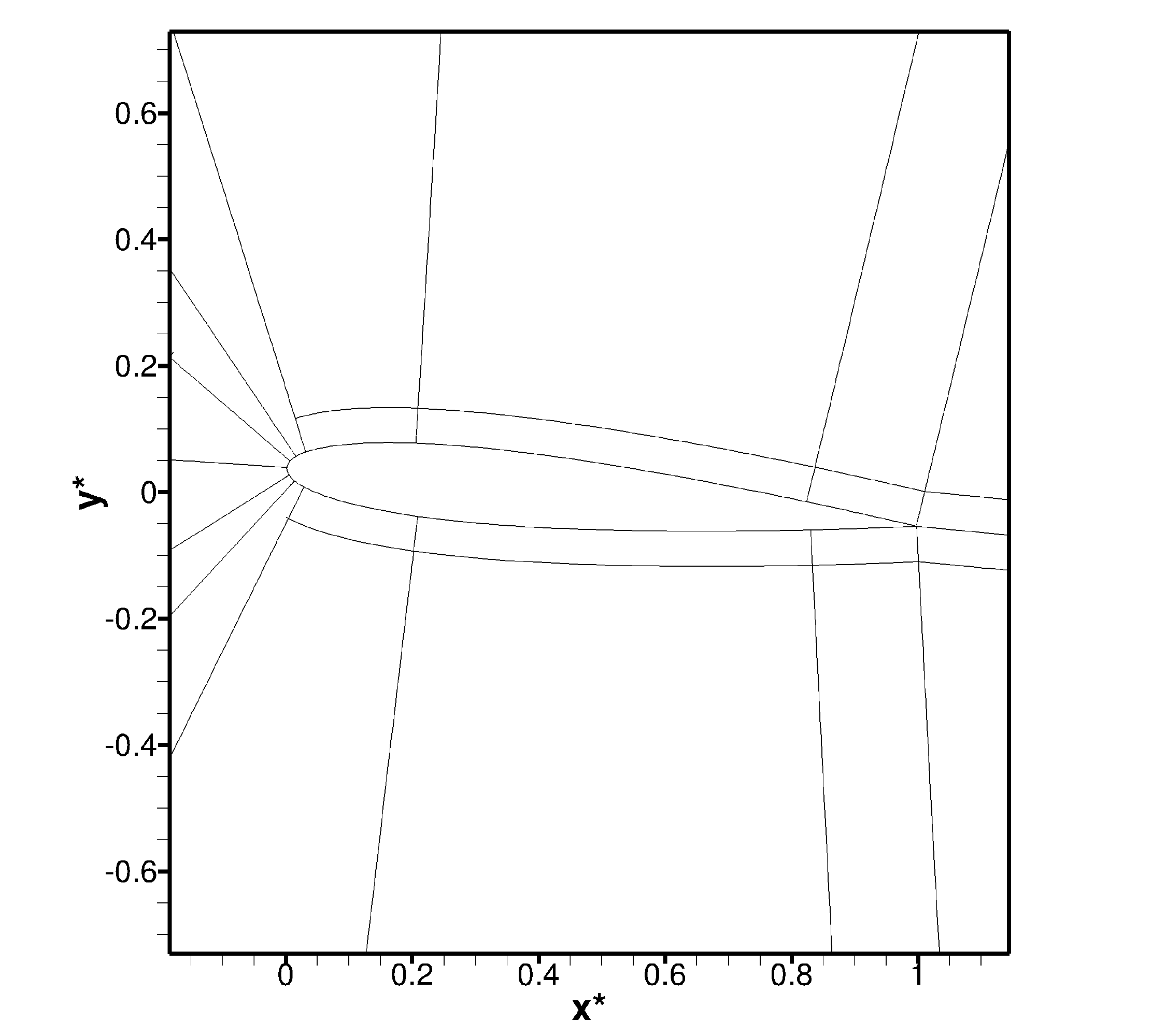}\label{fig:medium_mesh_angle_5_blocking_b}}
	\subfigure[Parallel blocks distribution: $m-11-y^+_{max}$.]{\includegraphics[width=0.46\textwidth]{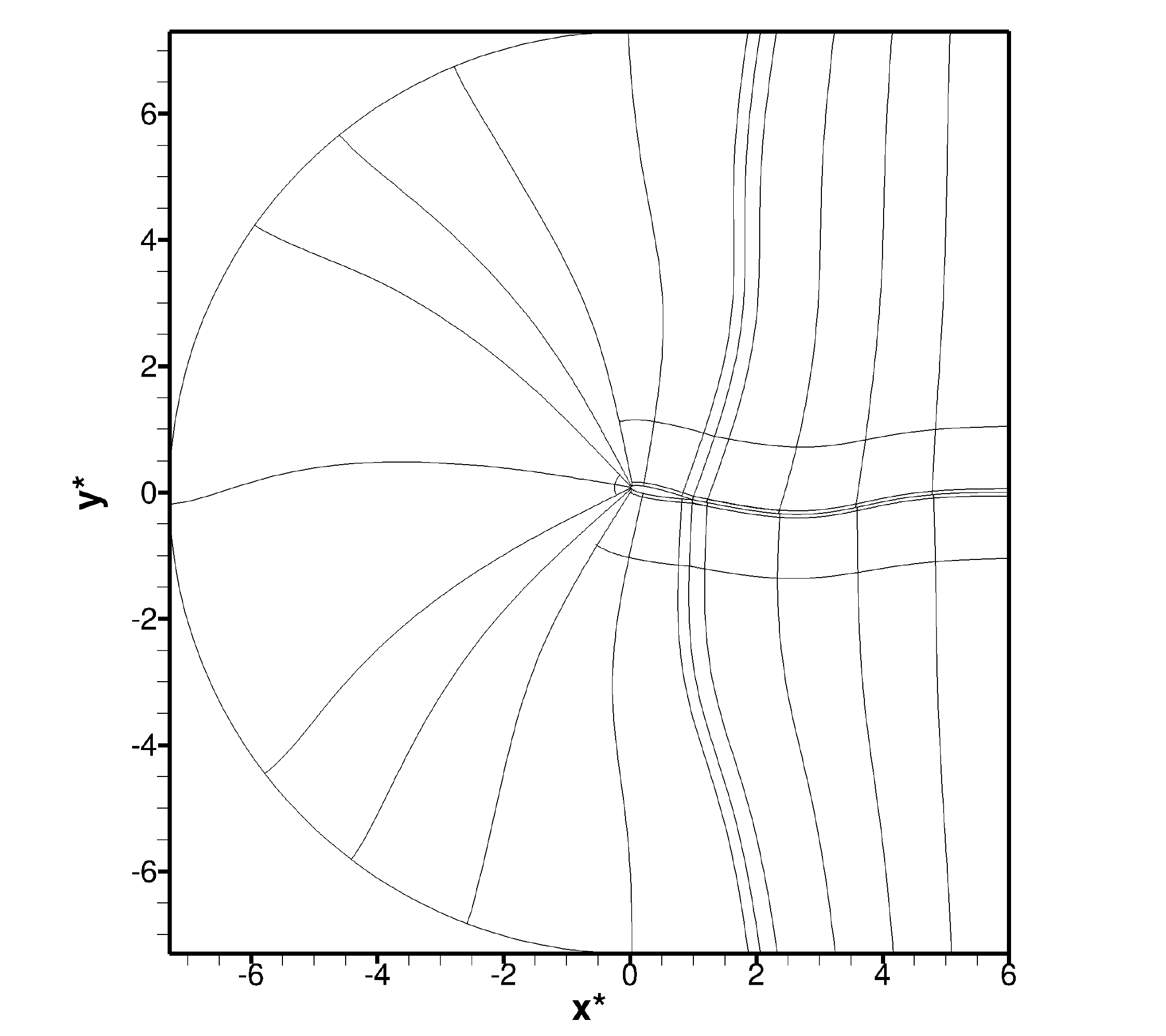}\label{fig:medium_mesh_angle_11_blocking_a}}\hfill
	\subfigure[Parallel blocks near the airfoil: $m-11-y^+_{max}$.]{\includegraphics[width=0.46\textwidth]{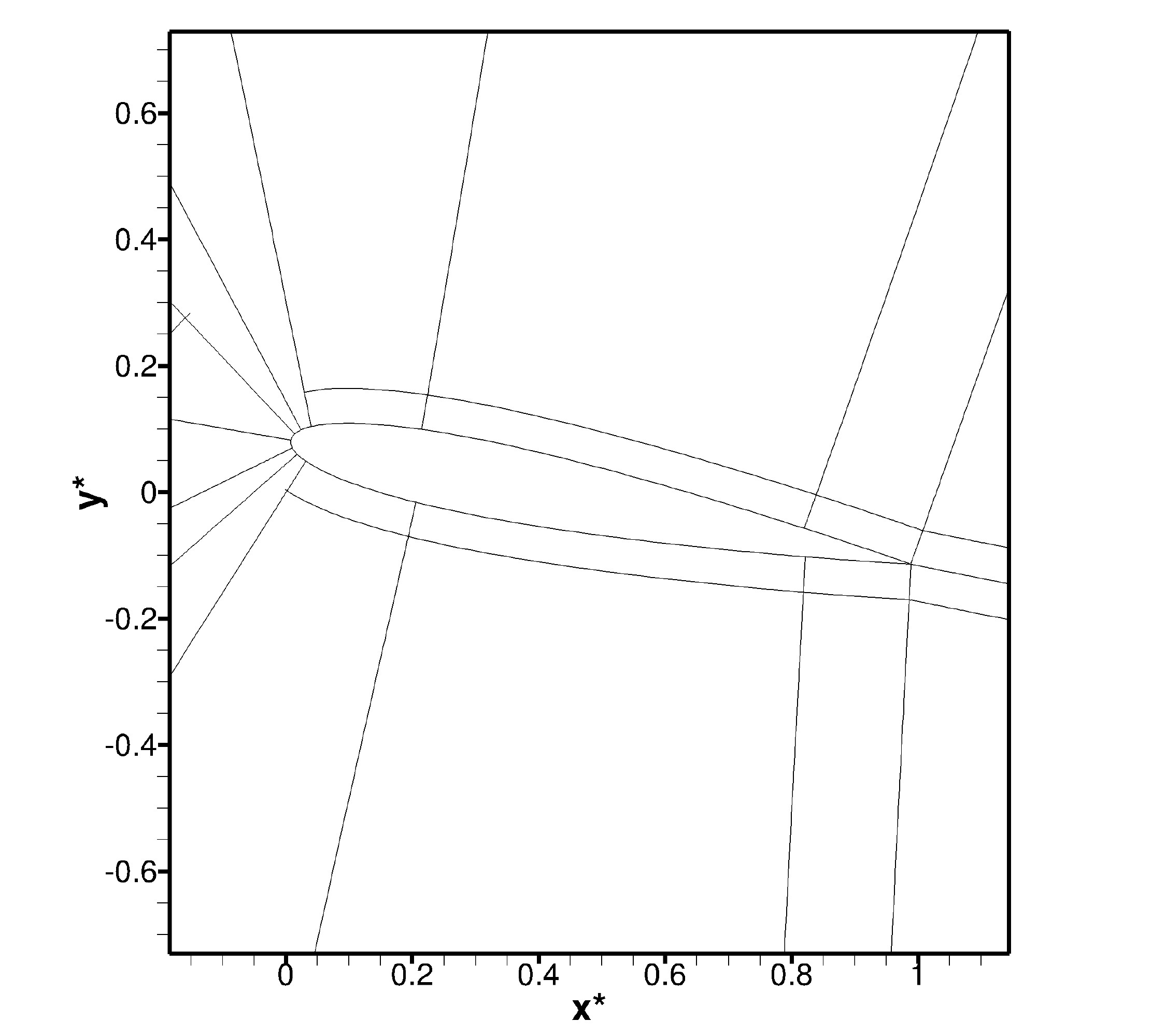}\label{fig:medium_mesh_angle_11_blocking_b}}
	\caption{Blocking of the $m-0-y^+_{max}$, $m-5-y^+_{max}$ and $m-11-y^+_{max}$ meshes.}
	\label{fig:medium_mesh_blocking}
\end{figure}

 Figures \ref{fig:medium_mesh_blocking} illustrates the blocking of the $m-0-y^+_{max}$, $m-5-y^+_{max}$ and $m-11-y^+_{max}$ meshes, i$.$e$.$, the medium meshes with a first cell wall distance of $y_{first\,cell}=1.8\cdot 10^{-5}\,m$ and angles of attack of $\alpha=0^\circ$, $\alpha=5^\circ$ and $\alpha=11^\circ$, respectively. Since the number of cells of the medium grid is smaller than the one of the fine resolution mesh, the required number of blocks is also smaller, regarding the same target of 70,000 control volumes per block. This implicates in a lower block density near the airfoil profile, in comparison with the fine resolution grid. The adapted grids acquire also curved form edges, which helps maintaining the cell angles as close as possible to $90^\circ$, as well as the orthogonality to the airfoil profile.

\par As in the fine resolution grid, the control volumes density is also higher near the airfoil, in the wake and orthogonal to the trailing edge. For the case of the adapted meshes, i.e$.$, $\alpha=5^\circ$ and $\alpha=11^\circ$, the wake edges are also curved, which improves the resolution of this area, since the von K\'{a}rm\'{a}n vortex street follows the same pattern of the curved wake edges. 

\par
\begin{figure}[H]
	\subfigure[Overall view of the $m-0-y^+_{max}$ mesh.]{\includegraphics[width=0.46\textwidth]{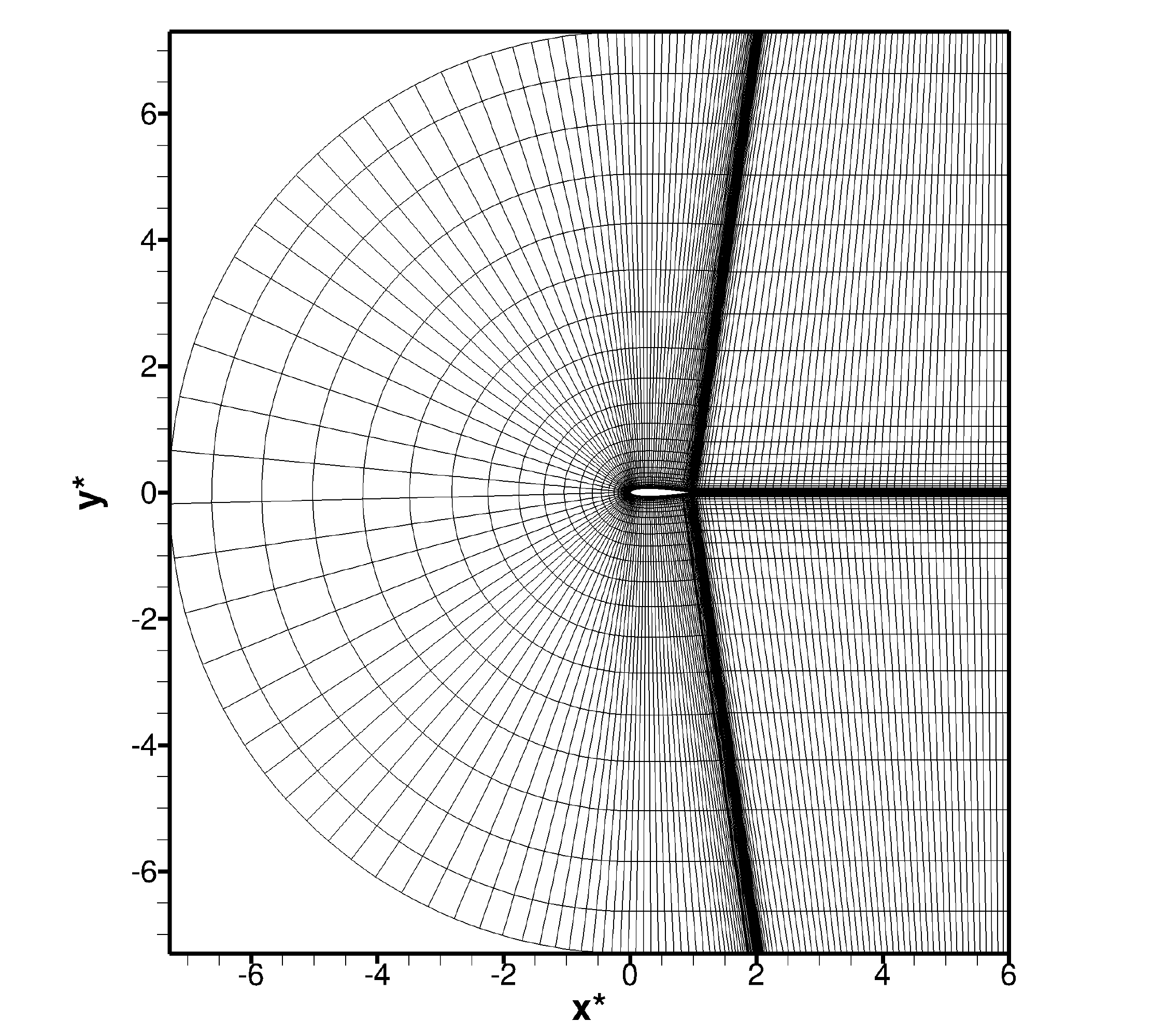}\label{fig:medium_mesh_angle_0_mesh_a}}\hfill
	\subfigure[Mesh $m-0-y^+_{max}$ near the airfoil.]{\includegraphics[width=0.46\textwidth]{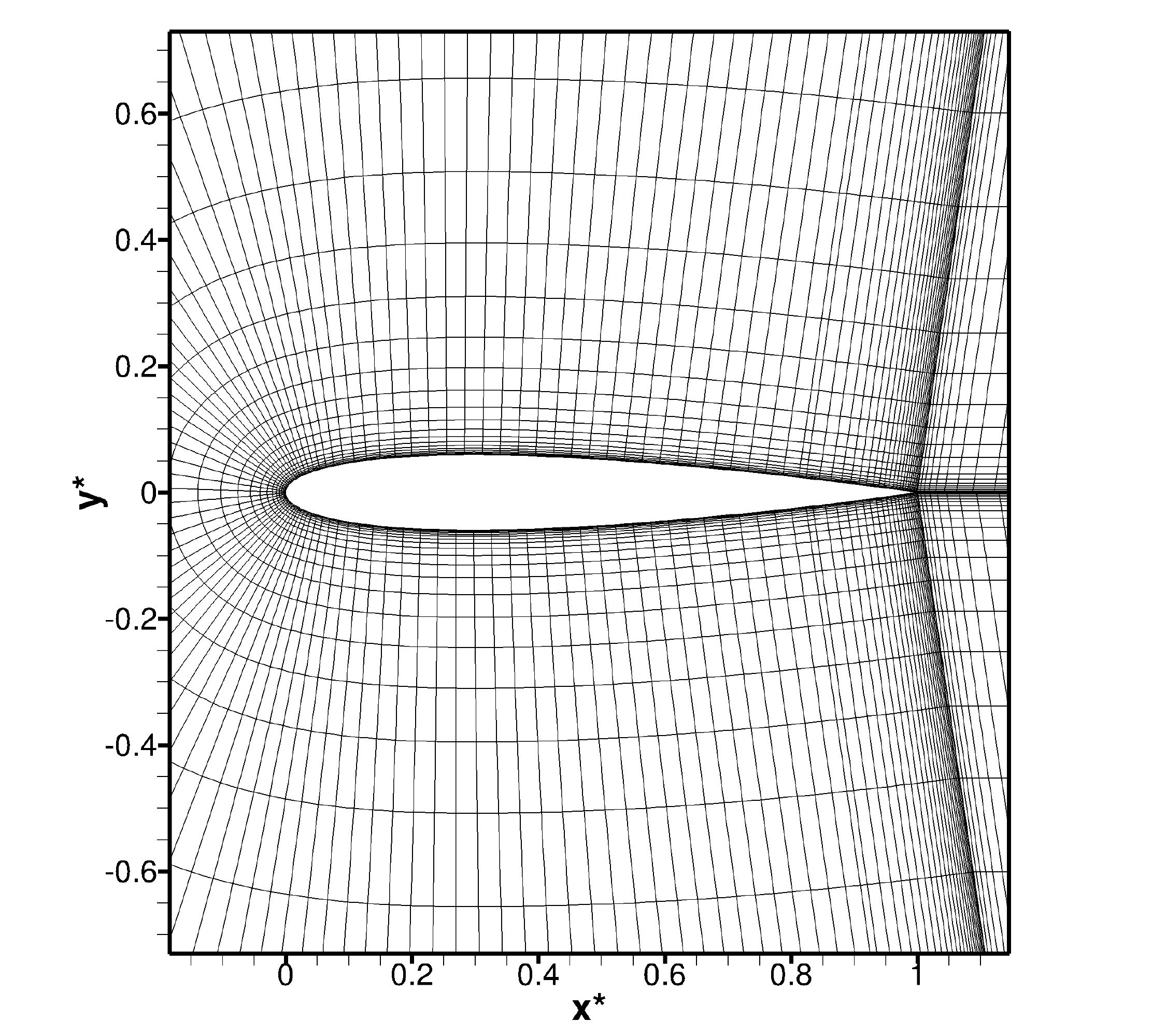} \label{fig:medium_mesh_angle_0_mesh_b}}
	\subfigure[Overall view of the $m-5-y^+_{max}$ mesh .]{\includegraphics[width=0.46\textwidth]{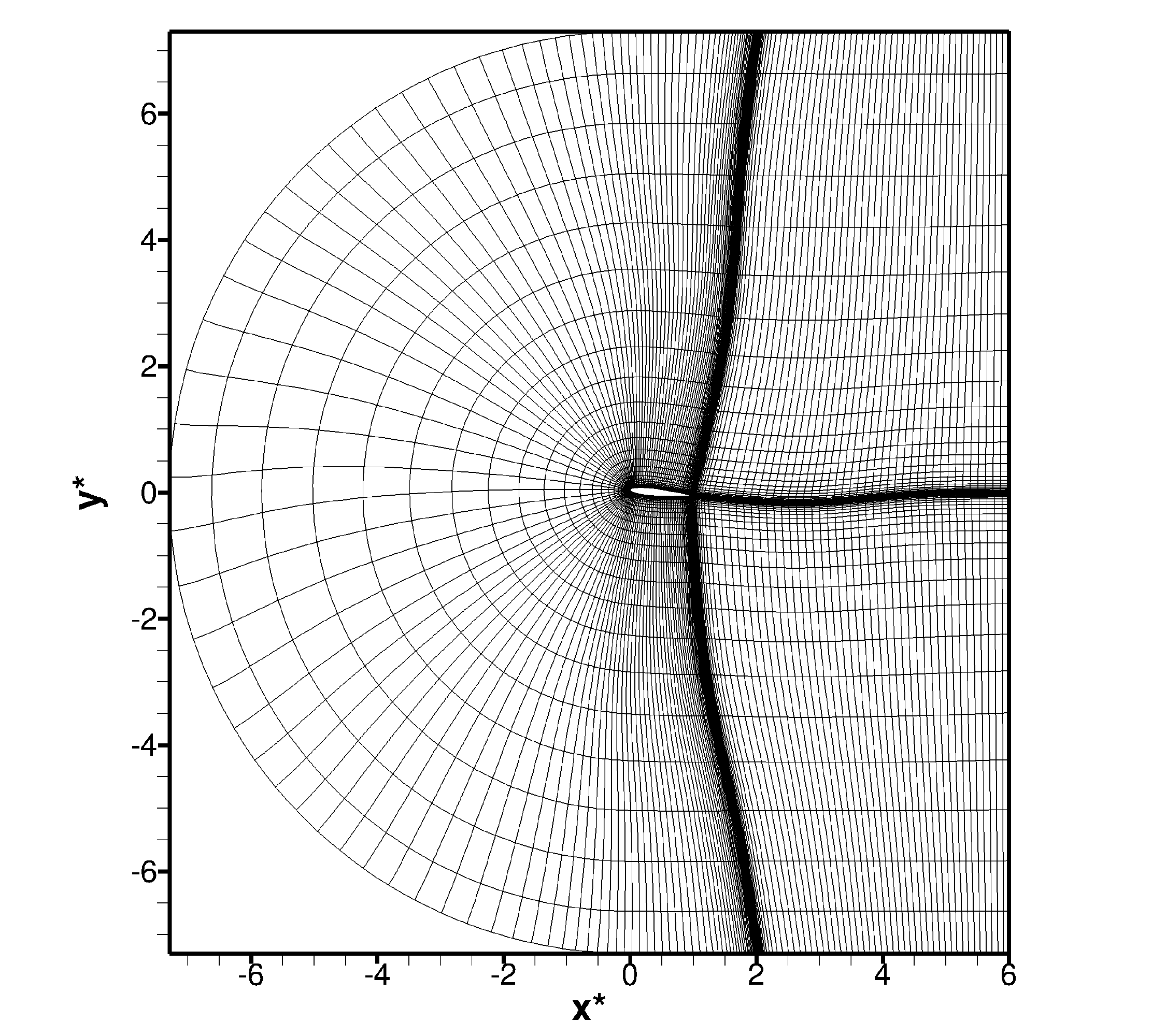}\label{fig:medium_mesh_angle_5_mesh_a}}\hfill
	\subfigure[Mesh $m-5-y^+_{max}$ near the airfoil.]{\includegraphics[width=0.46\textwidth]{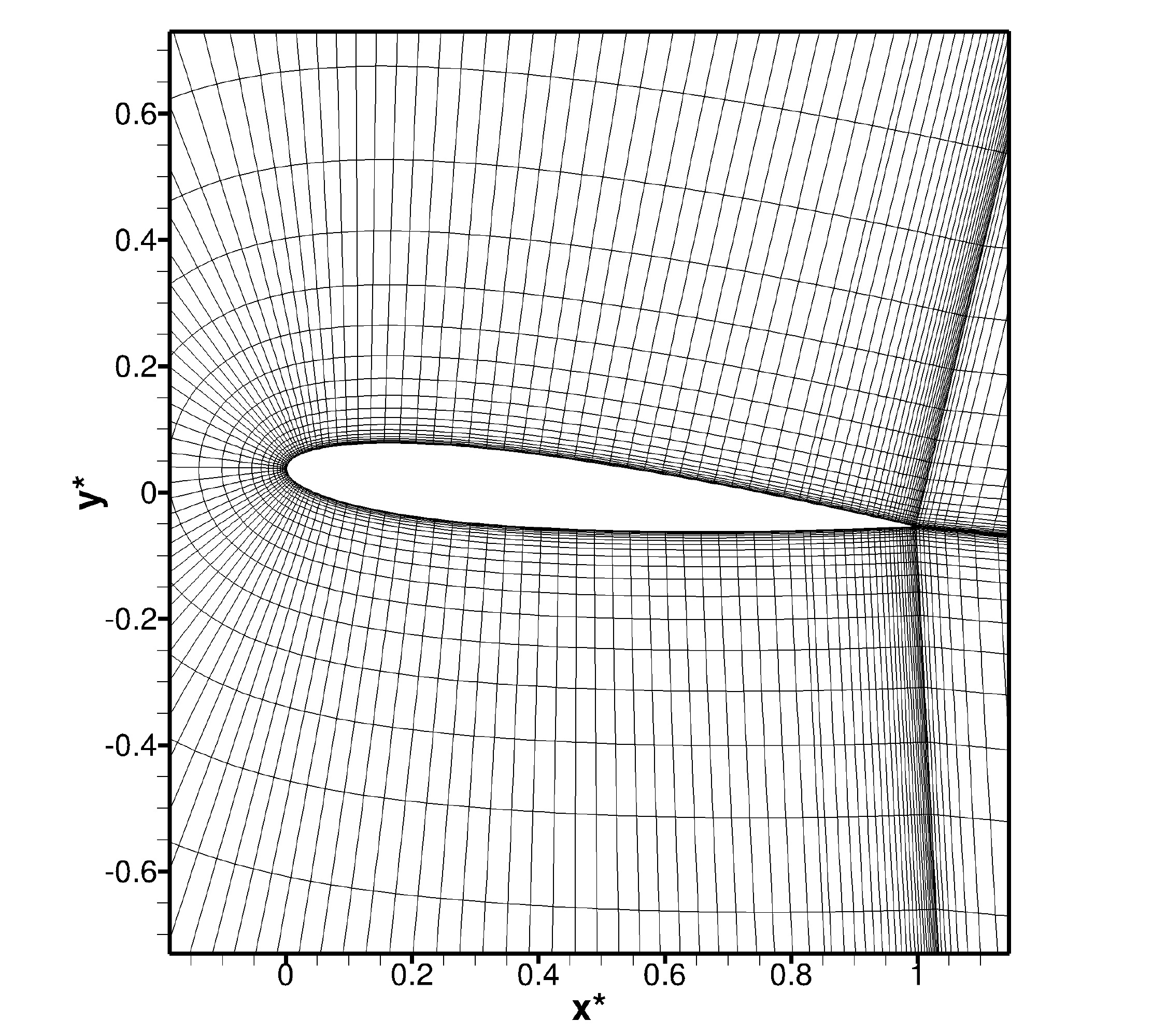} \label{fig:medium_mesh_angle_5_mesh_b}}
	\subfigure[Overall view of the  $m-11-y^+_{max}$ mesh.]{\includegraphics[width=0.46\textwidth]{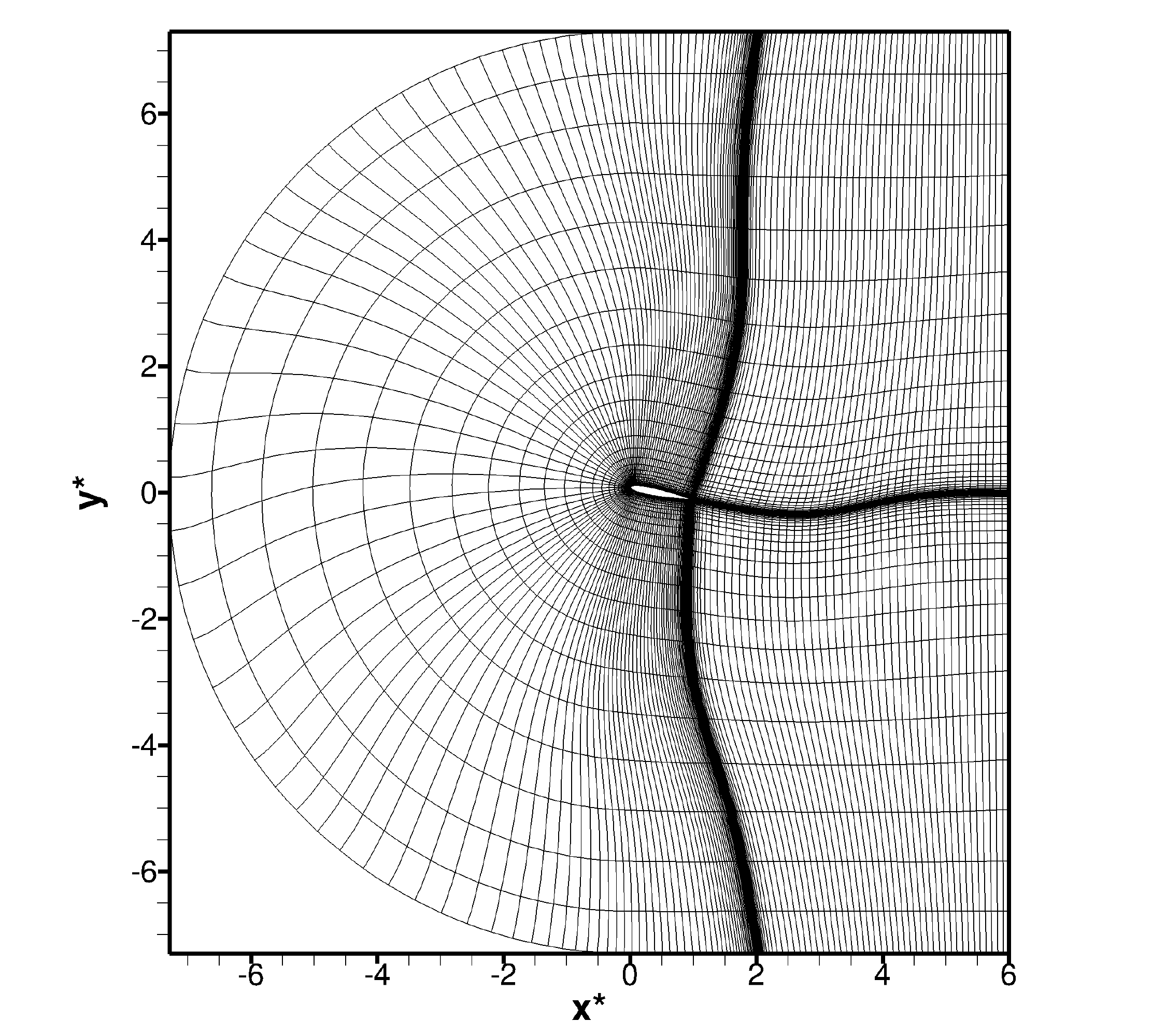}\label{fig:medium_mesh_angle_11_mesh_a}}\hfill
	\subfigure[Mesh $m-11-y^+_{max}$ near the airfoil.]{\includegraphics[width=0.46\textwidth]{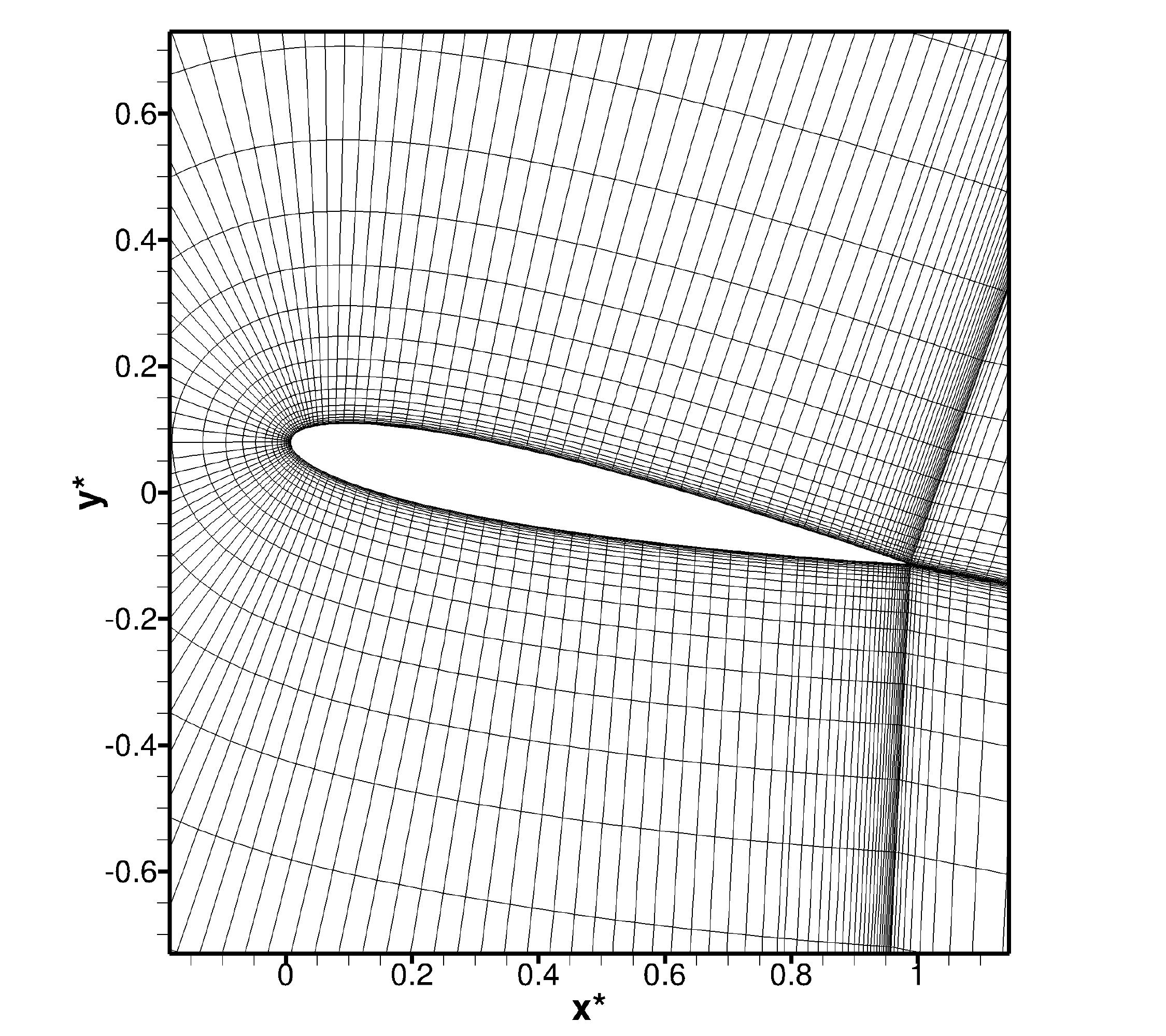} \label{fig:medium_mesh_angle_11_mesh_b}}
	\caption{Meshes of the $m-0-y^+_{max}$, $m-5-y^+_{max}$ and $m-11-y^+_{max}$ grids.}
	\label{fig:medium_mesh_mesh}
\end{figure}

 Figure \ref{fig:medium_mesh_mesh} illustrates the grids of the $m-0-y^+_{max}$, $m-5-y^+_{max}$ and $m-11-y^+_{max}$ meshes, i$.$e$.$, the medium resolution meshes with a first cell wall distance of $y_{first\,cell}=1.8\cdot 10^{-5}\,m$ and angles of attack of $\alpha=0^\circ$, $\alpha=5^\circ$ and $\alpha=11^\circ$, respectively. In order to achieve a good quality image only one out of five mesh lines is delineated in the $\varepsilon$-axis and only one out of two lines or one out of five lines in the $\eta$-axis for the front and the wake domain, respectively.

\par The meshes and blocking of the medium resolution mesh with a first cell wall distance of $y_{first\,cell}=9\cdot 10^{-6}\,m$ are very similar to the ones with a first cell wall distance of $y_{first\,cell}=1.8\cdot 10^{-5}\,m$, therefore they are not displayed. The main reason to vary this parameter is to increase the time step size in order to reduce the computational effort (see Section \ref{sec:simulation_parameters}).

\markboth{CHAPTER 2.$\quad$FLOW AROUND A NACA0012 AIRFOIL}{2.3$\quad$INIT. COND.}
\section{Initial conditions}\markboth{CHAPTER 2.$\quad$FLOW AROUND A NACA0012 AIRFOIL}{2.3$\quad$INIT. COND.}
\label{subsec:initial_conditions}
\par The conservation laws are parabolic in time. Therefore, initial conditions for the whole fluid domain are required. In the current simulation initial conditions for the velocity vector and for the pressure are utilized, according to Eqs.\ (\ref{eq:initial_velocity_i}) through (\ref{eq:initial_pressure}). The initial velocity tensor is characterized only by a velocity in the main flow direction, i.e$.$, the chord-wise direction, and is equal to the inlet velocity tensor (see \mbox{Section \ref{sec:boundary_conditions}}).
\begin{eqnarray}
u_{1,\,t_0}&=&14.7\,m{\cdot}s^{-1} \label{eq:initial_velocity_i}\\
u_{2,\,t_0}&=&0\,m{\cdot}s^{-1} \label{eq:initial_velocity_j}\\
u_{3,\,t_0}&=&0\,m{\cdot}s^{-1} \label{eq:initial_velocity_k}\\
p_{t_0}&=&0\,Pa \label{eq:initial_pressure}
\end{eqnarray}

\section{Boundary conditions}\markboth{CHAPTER 2.$\quad$FLOW AROUND A NACA0012 AIRFOIL}{2.4$\quad$BOUND. COND.}
\label{sec:boundary_conditions}
\par Due to the elliptic character of the incompressible governing equations, boundary conditions for all boundaries are mandatory.
\par In the case of simulations of the flow around the NACA0012 profile artificial boundary conditions, which must not influence the domain itself, are established at some boundaries in order to define the fluid domain. 
\subsection{Inlet}
\label{subsubsec:inflow}
\par The inlet velocity tensor is given in the Dirichlet form, as stated by Eqs.\ (\ref{eq:inlet_velocity_1}) through (\ref{eq:inlet_velocity_3}). This variable corresponds to the velocity tensor of the cells located in the inlet boundary and is stored in the control volumes midpoints. It defines the test case, i.e$.$, is correlated to the simulated Reynolds number of $Re=100{,}000$.
\begin{eqnarray}
u_{1,\,in}&=&14.7\,m{\cdot}s^{-1} \label{eq:inlet_velocity_1} \\
u_{2,\,in}&=&0\,m{\cdot}s^{-1} \label{eq:inlet_velocity_2} \\
u_{3,\,in}&=&0\,m{\cdot}s^{-1} \label{eq:inlet_velocity_3} \\
\end{eqnarray}
\par A pressure boundary condition (Eq. (\ref{eq:inlet_pressure})) in the Neumann form is utilized for the pressure correction $p^{\,corr}$, according to the work of Münsch \cite{Muensch_2015}.
\begin{equation}
\left. \frac{\quad \partial p^{\,corr}}{\partial x_1}\right |_{in}=0
\label{eq:inlet_pressure}
\end{equation}

\subsection{Symmetry boundary}
\label{subsubsec:symmetry_boundary_condition}
\par A symmetry boundary can be used if both geometry and flow behavior have mirror symmetry, which is never the case of turbulent flows. Since the edges located at the top and the bottom of the fluid domain are situated sufficiently far from the airfoil wake, these can be approximated as symmetric. These edges are characterized by zero velocities normal to the boundary, according to Eqs.\ (\ref{eq:symmetry_1}) through (\ref{eq:symmetry_3}). 
\begin{eqnarray}
u_{1,\,sym}&\neq&0\,m{\cdot}s^{-1} \label{eq:symmetry_1} \\ 
u_{2,\,sym}&=&0\,m{\cdot}s^{-1} \label{eq:symmetry_2} \\ 
u_{3,\,sym}&\neq&0\,m{\cdot}s^{-1} \label{eq:symmetry_3}
\end{eqnarray}
\par Furthermore, only a flux of the normal to the boundary velocity $u_{2}$ is available. Therefore, only a normal stress, i.e$.$, $\tau_{22}$, caused by this flux is present in the symmetry edges, conform Eqs.\ (\ref{eq:symmetry_flux_1}) through (\ref{eq:symmetry_flux_3}).
\begin{eqnarray}
\tau_{12,\,sym}&=&-\;\;\mu\left.\frac{\partial u_1}{\partial x_2}\right|_{sym}=0 \label{eq:symmetry_flux_1} \\ 
\tau_{22,\,sym}&=&-2\mu\left.\frac{\partial u_2}{\partial x_2}\right|_{sym}\neq0 \label{eq:symmetry_flux_2} \\ 
\tau_{32,\,sym}&=&-\;\;\mu\left.\frac{\partial u_3}{\partial x_2}\right|_{sym}=0 \label{eq:symmetry_flux_3}
\end{eqnarray}

\subsubsection{No-slip wall}
\label{subsubsec:wall}
\par The no-slip wall condition states that a fluid adheres to the wall and consequently moves with its velocity. For a fixed wall, that results in a zero velocity for the tangential to the wall components of the velocity, according to Eqs.\ (\ref{eq:wall_velocity_1}) and (\ref{eq:wall_velocity_2}). If the wall is also impermeable, which is the case of the NACA0012 airfoil, the velocity normal do the wall is also zero (see Eq.\ (\ref{eq:wall_velocity_3})).
\begin{eqnarray}
u_{t_1,\,w}&=&0\,m{\cdot}s^{-1} \label{eq:wall_velocity_1} \\
u_{t_2,\,w}&=&0\,m{\cdot}s^{-1} \label{eq:wall_velocity_2} \\
u_{n,\,w}&=&0\,m{\cdot}s^{-1} \label{eq:wall_velocity_3}
\end{eqnarray}   
\par Furthermore, the normal stress $\tau_{nn}$ caused by the gradient of the normal velocity $u_n$ is zero, according to Eq.\ (\ref{eq:wall_flux_2}). The shear stresses $\tau_{t_1n}$ and $\tau_{t_2n}$, caused due to the gradients of the tangential velocities $u_{t_1}$ and $u_{t_2}$, are available, conform to Eqs.\ (\ref{eq:wall_flux_1}) and (\ref{eq:wall_flux_3}).
\begin{eqnarray}
\tau_{t_1n,\,w}&=&-\;\mu\left.\frac{\partial u_{t_1}}{\partial n}\right|_{w}\neq0 \label{eq:wall_flux_1} \\ 
\tau_{nn,\,w}&=&-2\mu\left.\frac{\partial u_n}{\partial n}\right|_{w}=0 \label{eq:wall_flux_2} \\ 
\tau_{t_2n,\,w}&=&-\;\mu\left.\frac{\partial u_{t_2}}{\partial n}\right|_{w}\neq0 \label{eq:wall_flux_3}
\end{eqnarray}
\par The no-slip wall also states a boundary condition in the form of Neumann for the pressure, according to Eq.\ (\ref{eq:wall_pressure}).
\begin{equation}
\left.\frac{\partial p}{\partial n}\right |_w=0 \label{eq:wall_pressure}
\end{equation}

\subsection{Outlet}
\label{subsubsec:outlet}
\par The outlet boundary must be located where the eddies can pass through at most undisturbed and without reflection, so that this edge has no or only a minor influence on the domain, as described by Breuer \cite{Breuer_2013}. 
\par The large-eddy simulation of the NACA0012 profile requires the usage of a transport equation, i.e$.$, a convective equation, for the outlet boundary (see Eq.\ (\ref{eq:outlet_boundary})). This considers the effects of the present vortex system through the usage of a convective \mbox{velocity $U_{conv}$}.
\begin{equation}
\label{eq:outlet_boundary}
\frac{\partial u_i}{\partial t}+U_{conv}\left.\frac{\partial u_i}{\partial x_1}\right|_{out}=0
\end{equation}
\par The convective velocity $U_{conv}$ is parallel to the stream-wise direction and therefore normal to the outlet boundary. In the present case $U_{conv}$ is set to the free stream velocity, i.e$.$, $U_{conv}=u_{1,\,in}=14.7\,m{\cdot}s^{-1}$.
\par Due to the extrapolation of the velocities on the outlet boundary, a zero gradient of the pressure correction is available (see Eq.\ (\ref{eq:outlet_pressure})), as stated by Münsch \cite{Muensch_2015}. 
\begin{equation}
\left.\frac{\quad \partial p^{\,corr}}{\partial x_1}\right |_{out}=0 \label{eq:outlet_pressure}
\end{equation}

\subsection{Periodic boundary}
\label{subsubsec:periodic_boundary_condition}
\par The periodic boundary condition allows the reduction of the fluid domain and consequently the decrease of the computational effort. It is based on the simultaneous usage of the boundary values on two correspondent domain boundaries, as stated by Breuer \cite{Breuer_2002}.
\par The usage of this type of boundary is, however, only possible in homogeneous directions, which are characterized by the non-variation of the statistical flow. In the case of the NACA0012 simulations, the span-wise direction is homogeneous and characterized by $u_i(x_1,x_2,x_3,t)=u_i(x_1,x_2,x_3+\Delta x_3,t)$ and $p(x_1,x_2,x_3,t)=p(x_1,x_2,x_3+\Delta x_3,t)$. 
\par The period length must be carefully chosen in order to guarantee the capture of the largest eddies. This can be assured by the usage of the turbulent fluctuation two-point correlation, which is calculated along the homogeneous direction for the half of the period length and must tend to zero, conform the work of Breuer \cite{Breuer_2013}. A zero two-point correlation indicates that the searched variables are not correlated for the utilized period length and therefore periodic boundaries can be applied.
\par The two-point correlation of the velocity fluctuation for the span-wise direction is calculated according to Eq.\ (\ref{eq:two_point_correlation}). The investigated period length is $L_z$ (see Eq.\ (\ref{eq:period_length})). The velocities $<u_i'>$ represent the time-averaged values of the turbulent fluctuations ($<\,>$ stands for the time-averaged values and $'$ stands for the fluctuations). 
\begin{equation}
R_{u_i'u_j'}=\frac{<u_i'(x_1,x_2,x_3,t)\;u_j'(x_1,x_2,x_3+\Delta x_3,t)>}{\sqrt{<(u_i'(x_1,x_2,x_3,t))^2>}\;\sqrt{<(u_j'(x_1,x_2,x_3+\Delta x_3,t))^2>}} \label{eq:two_point_correlation}
\end{equation}
\begin{equation}
\Delta x_3=\frac{L_z}{2} \label{eq:period_length}
\end{equation}
\par For the current simulation, the span-wise direction is homogeneous and a period length of $L_z=0.25c$ is applied. This is based on the work of Almutari \cite{Almutari_2010} and in the work of Schmidt \cite{Schmidt_2016}. The former presents the calculation of the two-point correlation for the span-wise direction of a NACA0012 profile. The latter presents the utilized period  length, i.e$.$, $L_z=0.25c$, based on simulation results available in the work of Visbal et al. \cite{Visbal_2009}.

\section{Computational setup}\markboth{CHAPTER 2.$\quad$FLOW AROUND A NACA0012 AIRFOIL}{2.5$\quad$COMPT. SETUP}
\label{sec:simulation_parameters}
\par The flow parameters of the NACA0012 test case are described in Table \ref{table:flow_parameters}. The temperature is presumed to be constant and the fluid incompressible. The inflow velocities in the y-direction and z-direction are assumed to be $u_{2,\,in}=0\,m{\cdot}s^{-1}$ and $u_{3,\,in}=0\,m{\cdot}s^{-1}$, respectively. The Reynolds number is calculated using the free-stream conditions and the chord length, according to Eq.\ (\ref{eq:Reynolds_number}). No turbulence generator is applied at the inlet.
\begin{table}[!htbp]
	\centering
	\begin{tabular}{l l}
		\hline
		Temperature & $T=300\,K$ \tabularnewline 
		Inflow velocity & $u_{1,\,in}=14.7\,m{\cdot}s^{-1}$ \tabularnewline
		Fluid density & $\rho_f=1.225\,kg{\cdot}m^{-3}$ \tabularnewline
		Dynamic fluid viscosity & $\displaystyle{\mu_f=18.27\cdot10^{-6}\,Pa{\cdot}s}$ \tabularnewline
		Reynolds number & $\displaystyle{Re=\frac{u_{1,\,in}\;\rho_f\;c}{\mu_f}=100{,}000}$ \tabularnewline
		\hline	
	\end{tabular}
	\caption{\label{table:flow_parameters}Fluid parameters.}
\end{table}
\par The simulation parameters are described in Table \ref{table:simulation_parameters}. The temporal discretization adopts a three sub-steps low-storage Runge-Kutta scheme as predictor and a pressure Poisson equation as corrector. The Smagorinsky model is applied with a constant \mbox{$C_s=0.1$}. 
\begin{table}[!htbp]
	\centering
	\begin{tabular}{l l}
		\hline
		Spatial discretization & FVM \tabularnewline
		Temporal discretization & Predictor-corrector scheme \tabularnewline
		Turbulence approach & LES \tabularnewline
		Sub-grid scale model & Smagorinsky \tabularnewline
		Damping function & Van-Driest \tabularnewline
		Smagorinsky constant & $C_s=0.1$ \tabularnewline 
		Wall function & None \tabularnewline
		\hline	
	\end{tabular}
	\caption{\label{table:simulation_parameters}Simulation parameters.}
\end{table}
\par No wall function is utilized in order to fully resolve the viscous sublayer, so that the required dimensionless wall distance must be contained within $0 \leq y^+_{visc} \leq 5$. Therefore, first cell wall distances of $y_{first\,cell}=4.4\cdot 10^{-6}\,m$, $y_{first\,cell}=9.0\cdot 10^{-6}\,m$ and $y_{first\,cell}\nobreak=\nobreak1.8\nobreak\cdot\nobreak 10^{-6}\,m$ are utilized. The required time step varies due to this distance and the mesh resolution, so that the generated grids demand different values in order to be stable. In general, the larger the first cell wall distance, the larger the time step. 
\par The presentation of the predicted results for the flow around the NACA0012 airfoil presented in Sections \ref{sec:simulation_analysis} and \ref{sec:reynolds_stresses} are based on time-averaged and in the span-wise direction spatial-averaged velocities and Reynolds stresses. In order to achieve these quantities a spatial-averaging process in the span-wise direction and a time-averaging process, which starts after an initialization phase of the flow ($t^*> t^*_{init}\geq 50)$, are required.
\par The spatial-averaging in the span-wise direction is used due to the geometry of the NACA0012 profile and the fluid domain. Indeed, the computational domain is obtained with the extrusion in the span-wise direction of a two-dimensional geometry. It is based on the arithmetical average of the cell velocities in the span-wise direction.
\par Table \ref{table:time_step} illustrates the seven generated grids with the respective time step sizes $\Delta t$, the dimensionless duration of the initialization phase $t^*_{init}$ and the in Chapter \ref{chap:results_and_discussion} analyzed dimensionless duration of the averaging process $t^*_{avg}$. 

\begin{table}[H]
	\centering
	\begin{tabular}{c l r r}
		\hline
		\multicolumn{1}{c}{\bf{Grid}} & \multicolumn{1}{c}{\bf{Time step}} & \multicolumn{1}{c}{\bf{$t^*_{init}$}} & \multicolumn{1}{c}{\bf{$t^*_{avg}$}} \tabularnewline
		\hline
		$\,f-\,0\,-y^+_{min}$ & $\Delta t=8\cdot10^{-8}\,s$ & $53$ & $40$ \tabularnewline
		$m-\,0\,-y^+_{med}$ & $\Delta t=2\cdot10^{-7}\,s$ & $59$ & $322$ \tabularnewline
		$m-\,0\,-y^+_{max}$ & $\Delta t=1\cdot10^{-6}\,s$  & $73$ & $1009$ \tabularnewline
		$m-\,5\,-y^+_{med}$ & $\Delta t=1\cdot10^{-7}\,s$ & $50$ & $152$ \tabularnewline
		$m-\,5\,-y^+_{max}$ & $\Delta t=1\cdot10^{-6}\,s$ & $51$ & $1039$ \tabularnewline
		$m-11-y^+_{med}$ & $\Delta t=9\cdot10^{-8}\,s$ & $50$ & $76$ \tabularnewline
		$m-11-y^+_{max}$ & $\Delta t=7\cdot10^{-7}\,s$ & $51$ & $454$ \tabularnewline
		\hline
	\end{tabular}
	\caption{\label{table:time_step}Time step, dimensionless duration of the initialization phase $t^*_{init}$ and dimensionless duration of the averaging process $t^*_{avg}$ for the NACA0012 test case.}
\end{table}

\par The time step number to start the time-averaging process $n_{avg}$ is calculated according to Eqs.\ (\ref{eq:iteration_averaging}) and (\ref{eq:dimensionless_time}), with the dimensionless time to initialize this process varying within \mbox{$50\leq t^*_{init} \leq 73$}. 
\begin{eqnarray}
\label{eq:iteration_averaging}
n_{avg}&=&\frac{c\;\;t^*}{u_{1,\,in}\;\;\Delta t}\,, \\
\label{eq:dimensionless_time}
t^*&=&\frac{t\;\,u_{1,\,in}}{c}\,.
\end{eqnarray}

%% file: chapter3.tex
\chapter{Results and discussion}
\label{chap:results_and_discussion}

\par The dimensionless wall distance $y^+$ is firstly calculated in order to guarantee that this is small enough to resolve the viscous sublayer. Secondly, the velocity and Reynolds stresses of the time and spatial-averaged results are analyzed. In case these are not available, a preliminary analysis with the instantaneous values is executed. Finally, the simulation results of the various grids are compared in order to establish the independence of the meshes.
\par The results are presented in a dimensionless form, which is calculated according to \mbox{Eqs.\ (\ref{eq:dimensionless_coordinates}), (\ref{eq:dimensionless_velocities}) and (\ref{eq:dimensionless_reynolds_stresses})}. The former stands for the dimensionless coordinates $x_i^*$, the second for the dimensionless velocities $u_i^*$ and the latter for the dimensionless Reynolds stresses $\tau_{ij}^{turb^*}$. $<{u_i'u_j'}>$ represents the mean value (time and spatial averaged) of the velocity fluctuations $u_i'u_j'$. $<u_i>$ and $<\tau_{ij}^{turb}>$ represent the spatial and time-averaged velocities and Reynolds stresses, respectively. 
\begin{eqnarray}
x_i^*&=&\frac{x_i}{c} \label{eq:dimensionless_coordinates}\\ 
u_i^*&=&\frac{<u_i>}{u_{1,\,in}} \label{eq:dimensionless_velocities}\\
\tau_{ij}^{turb^*}&=&\frac{<\tau_{ij}^{turb}>}{{u_{1,\,in}^2}}=\frac{<{u_i' u_j'}>}{{u_{1,\,in}^2}} \label{eq:dimensionless_reynolds_stresses}
\end{eqnarray}

\section{Dimensionless wall distance analysis}\markboth{CHAPTER 3.$\quad$RESULTS AND DISCUSSION}{3.1$\quad$DIM. WALL DIST. ANALYSIS}
\label{sec:analysis_dimensionless_wall_distance}
The dimensionless wall distance $y^+$ is computed according to Eq.\ (\ref{eq:dimensionless_wall_distance}) with a kinematic viscosity of $\nu_f=1.49\cdot10^{-5}\,m^2{\cdot}s^{-1}$ by the software TECPLOT 360. If the time-averaged velocities $u_{i,\,avg}$ are not available, the instantaneous velocities are used instead. 

\subsection{Fine mesh}
\label{subsec:dimensionless_wall_distance_fine_mesh}
\par Since the time step to simulate the fine resolution grid is in the order of $\Delta t=\mathcal{O}(10^{-8})\,s$ and therefore it demands a lot of computational time to reach a fully developed state, a preliminary analysis of the dimensionless wall distance is performed when the flow was not yet fully developed and consequently not yet time-averaged.
\par
\begin{figure}[H]
	\centering
	\subfigure[Unsteady dimensionless wall distance.]{\includegraphics[width=0.495\textwidth]{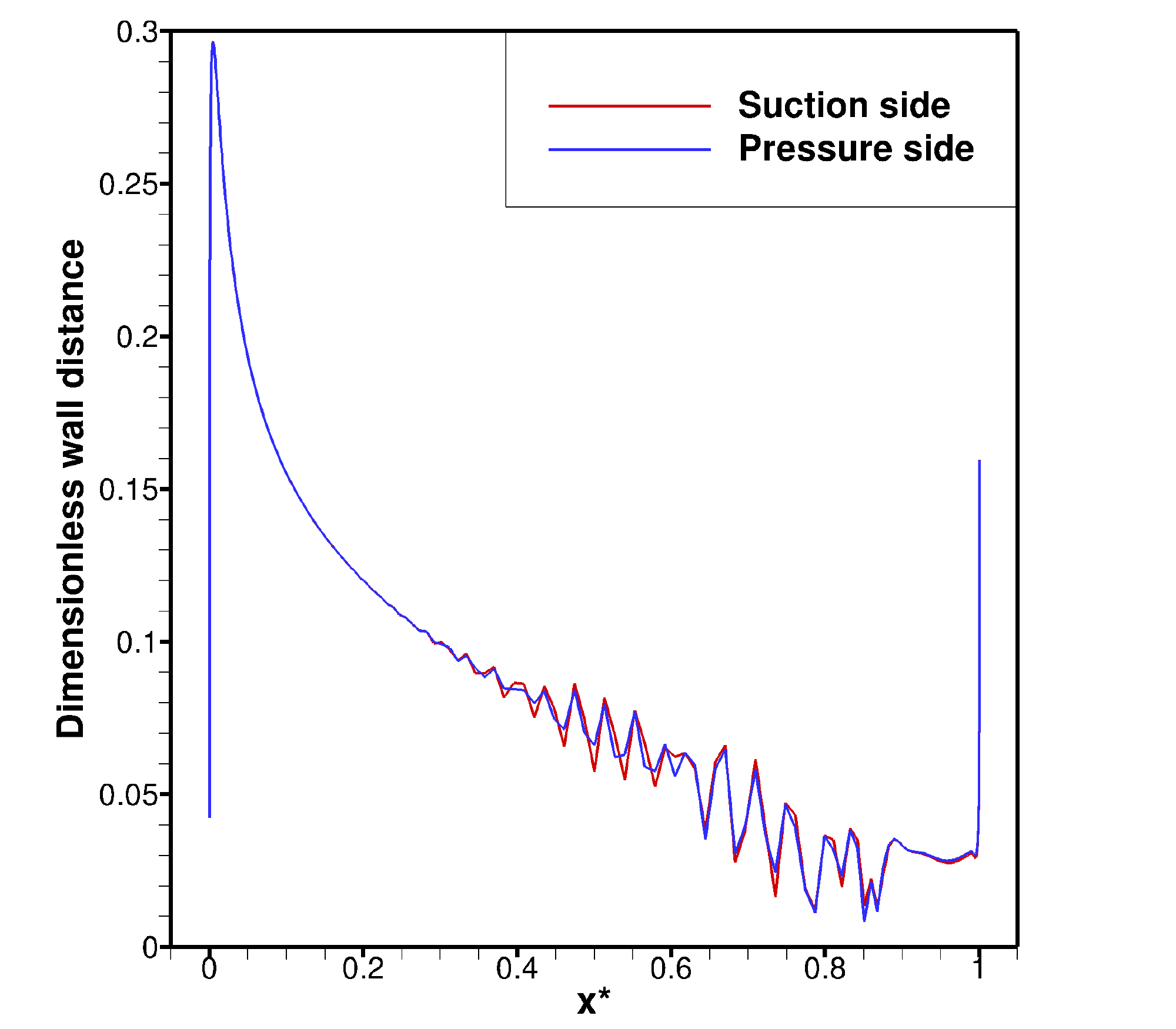}\label{fig:y_plus_unsteady_fine_grid}}\hfill
	\subfigure[First cell instantaneous velocities.]{\includegraphics[width=0.495\textwidth]{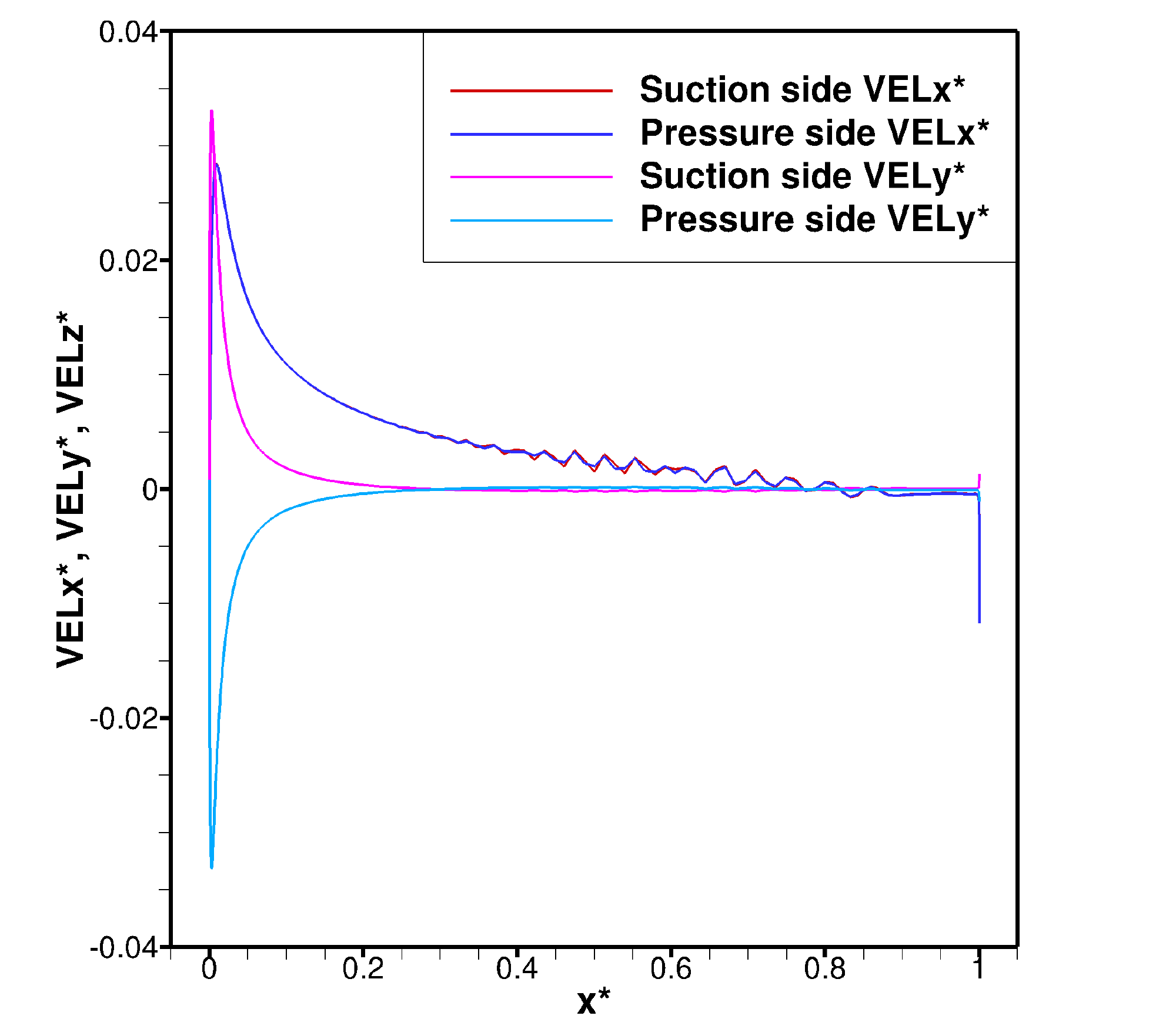} \label{fig:u_unsteady_fine_grid}}\hfill
	\subfigure[Instantaneous velocity in x-direction.]{\includegraphics[width=0.495\textwidth]{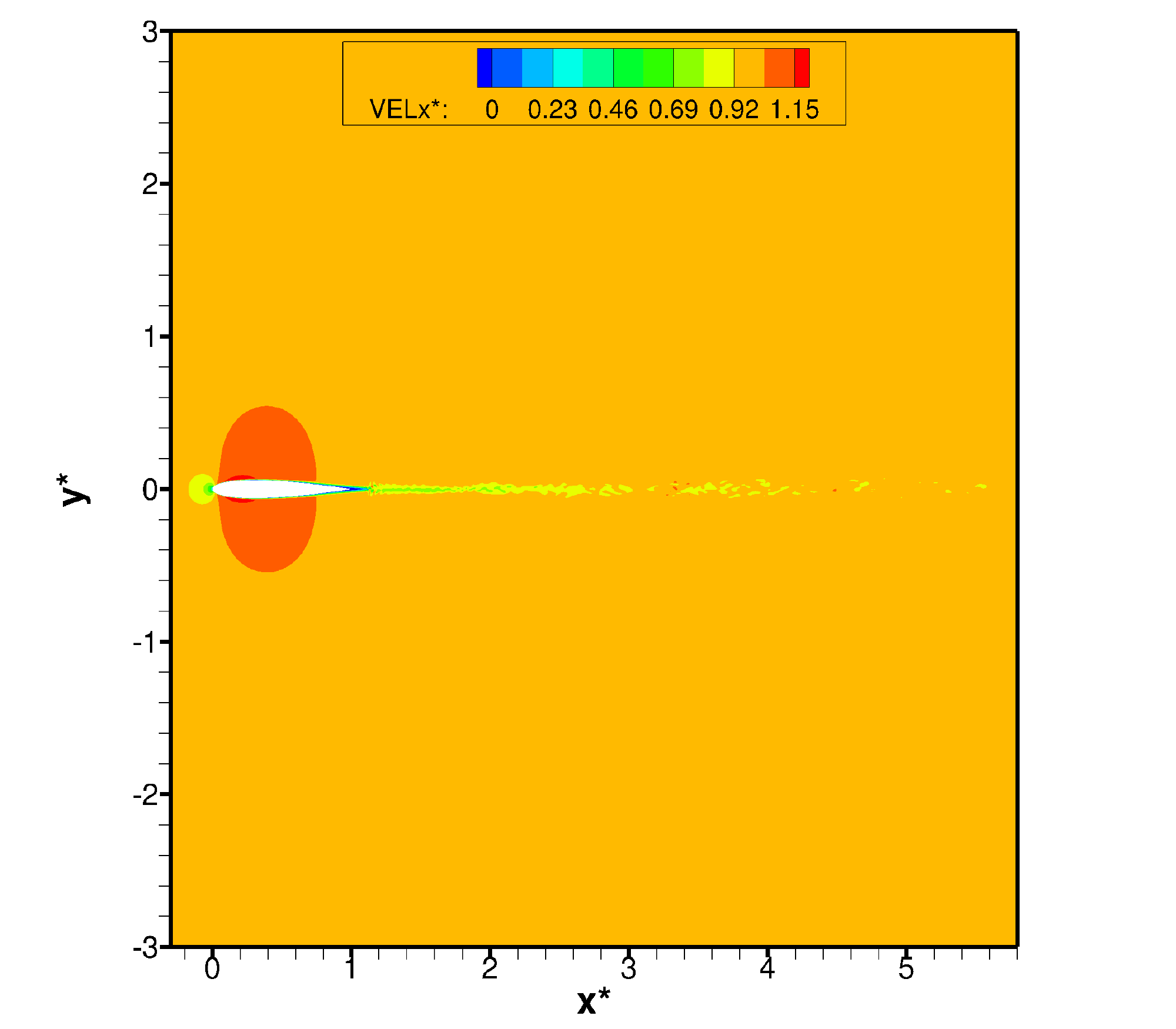} \label{fig:flow_y_plus_unsteady_fine_grid}}
	\caption{Unsteady dimensionless wall distances and velocities of the  $f-0-y^+_{min}$ mesh for $t^*=40$ (results are spatial-averaged in the span-wise direction).}
	\label{fig:fine_mesh_y+_u}
\end{figure}

 Figures \ref{fig:y_plus_unsteady_fine_grid} and \ref{fig:u_unsteady_fine_grid} represent respectively the unsteady dimensionless wall distances and the instantaneous dimensionless first cell velocities of the fine resolution grid as functions of the dimensionless coordinate $x^*$. These values are spatial-averaged but not time-averaged, since the flow has not yet achieved a developed state. This condition can be assured through Fig.\ \ref{fig:flow_y_plus_unsteady_fine_grid}, since the von K\'arm\'an vortex street is not yet totally formed.
\par The instantaneous dimensionless wall distance of the fine resolution mesh is similar for the pressure and the suction side, which is expected due to the symmetrical profile at an angle of attack of $\alpha=0^\circ$. 
\par As a result of the adopted wall distance and stretching factor, the viscous sublayer ($0\leq y^+_{visc} \leq 5$) is fully resolved. This is assured by the fact, that the dimensionless wall distances vary within $0.009 \leq y^+ \leq 0.3$. The first peak of this value, as shown in Fig.\ \ref{fig:y_plus_unsteady_fine_grid}, is due to the flow acceleration caused by the curvature of the profile: Since the velocity and the dimensionless wall distance are proportional, as the former increases, the latter also rises. Then, the dimensionless wall distance suffers a reduction also caused due to the airfoil curvature. Both the dimensionless wall distances and the first cell velocities show strong fluctuations due to the absence of a time-averaging process.

\subsection{Medium mesh}
\label{subsec:dimensionless_wall_distance_medium_mesh}

\par The computed dimensionless wall distances of the six medium resolution meshes are analyzed aiming at the study of the viscous sublayer resolution.

\subsubsection{Angle of attack $\alpha=0^\circ$}
\label{subsubsec:dimensionless_wall_distance_angle_0}
\par
\begin{figure}[H]
	\centering
	\subfigure[Time-averaged dimensionless wall distance.]{\includegraphics[width=0.495\textwidth]{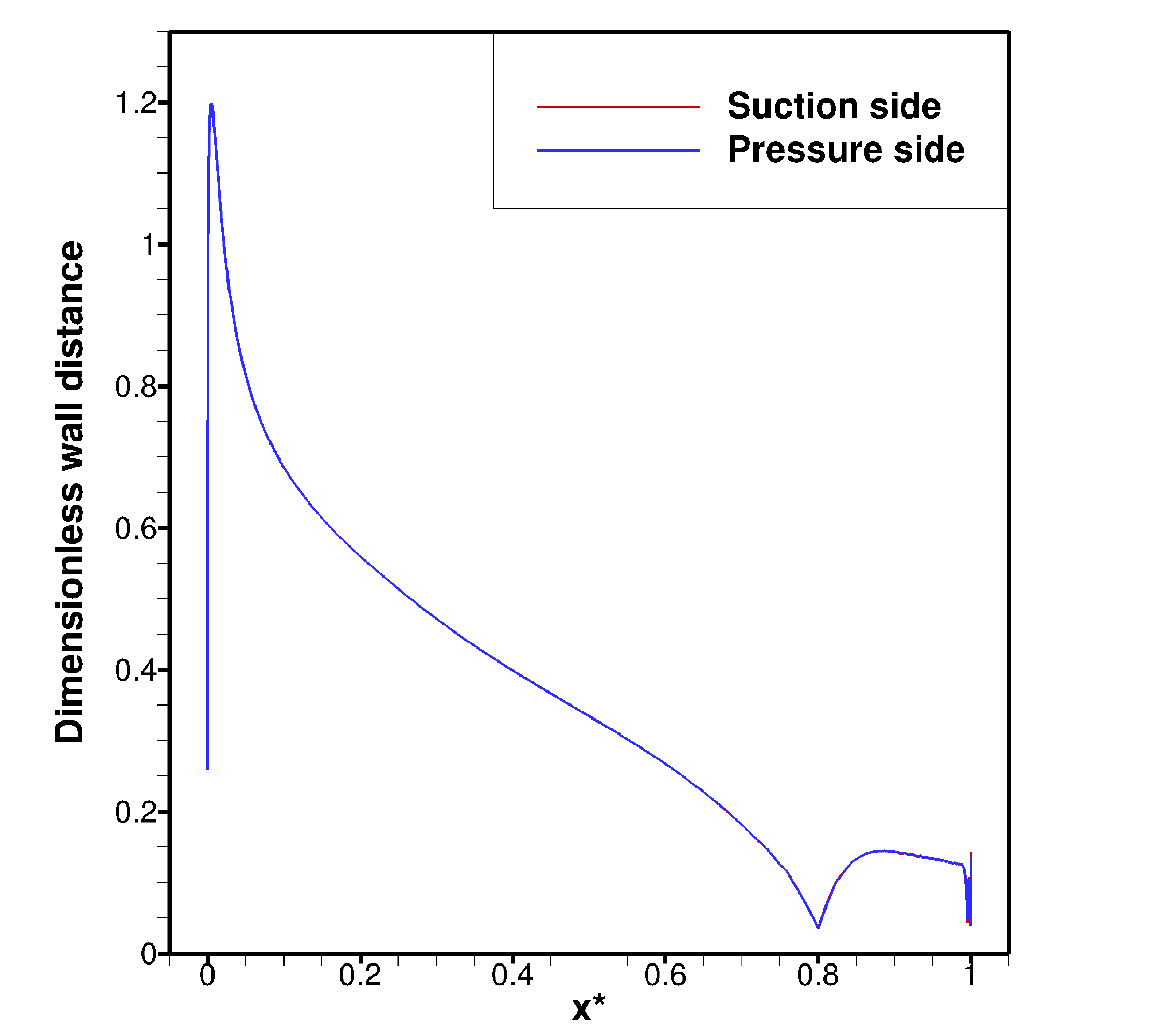}\label{fig:y_plus_avg_medium_grid_e-5}}\hfill
	\subfigure[First cell time-averaged velocities.]{\includegraphics[width=0.495\textwidth]{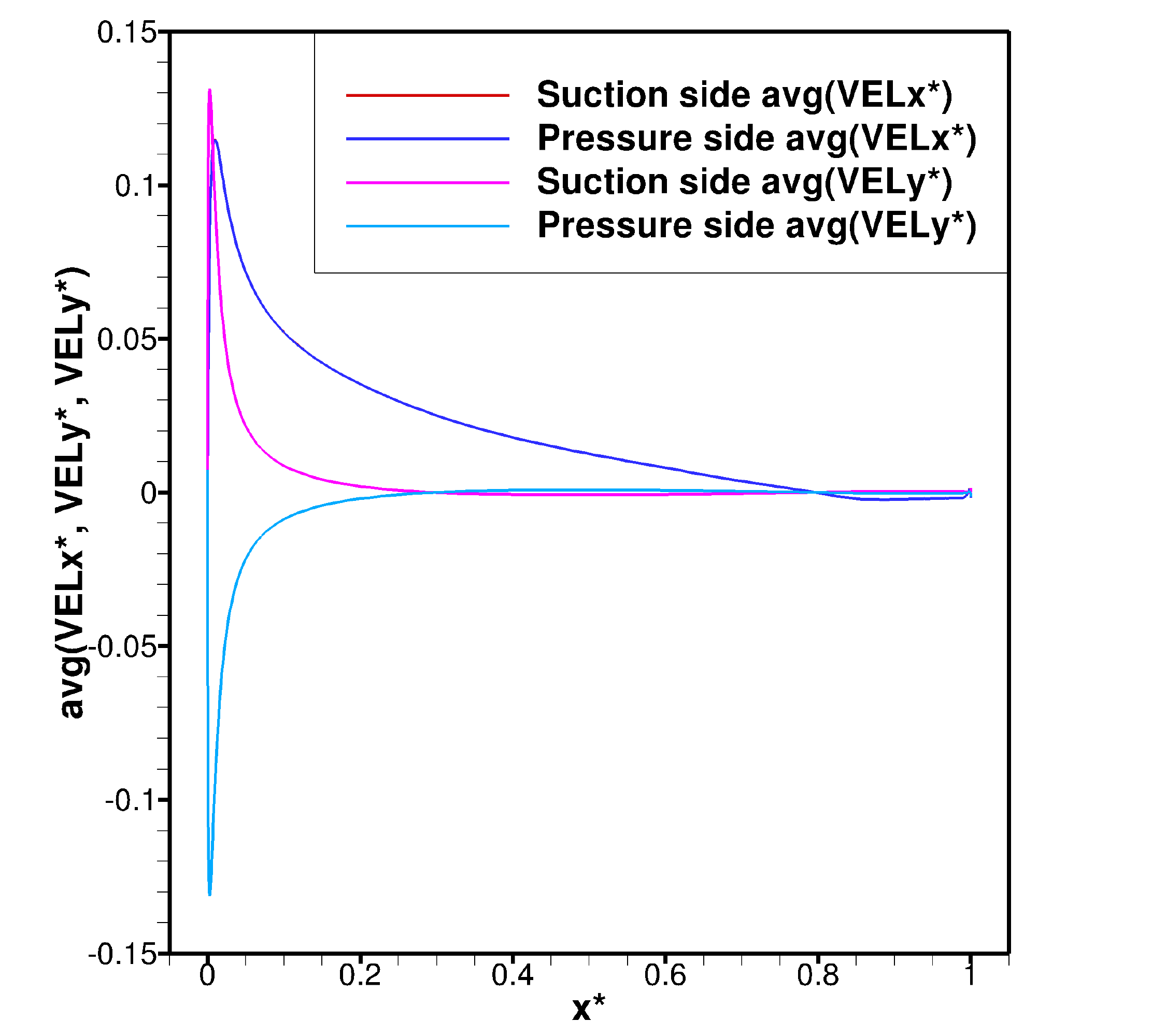} \label{fig:u_avg_medium_grid_e-5}}\hfill
	\caption{Time-averaged dimensionless wall distances and first cell velocities of the  $m-0-y^+_{max}$ mesh for $t^*_{avg}=1009$ (results are spatial-averaged in the span-wise direction).}
	\label{fig:medium_mesh_e-5_angle_0_y+_u_avg}
\end{figure}

\begin{figure}[H]
	\centering
	\subfigure[Time-averaged dimensionless wall distance.]{\includegraphics[width=0.495\textwidth]{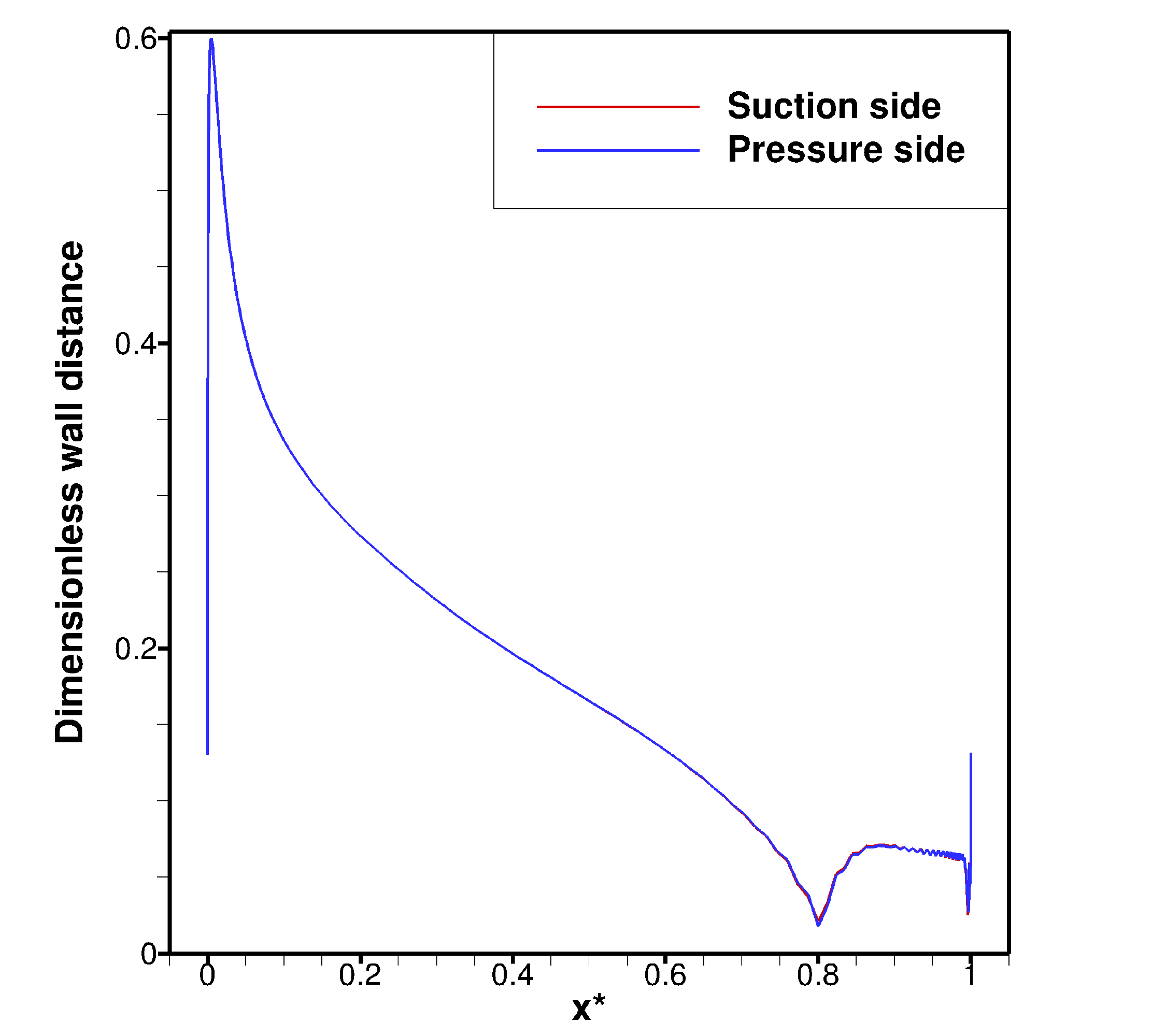}\label{fig:y_plus_avg_medium_grid_e-6}}\hfill
	\subfigure[First cell time-averaged velocities.]{\includegraphics[width=0.495\textwidth]{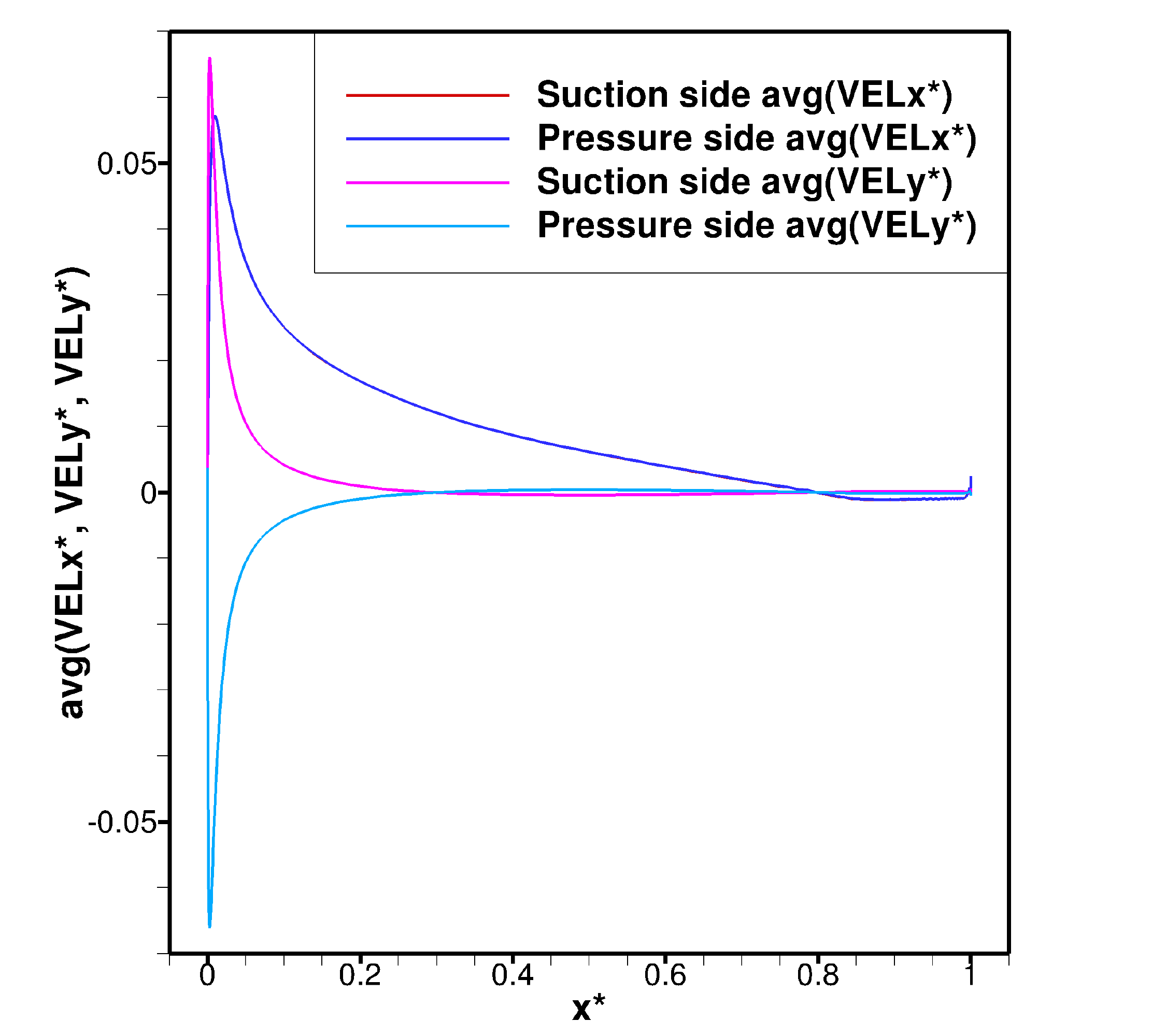} \label{fig:u_avg_medium_grid_e-6}}\hfill
	\caption{Time-averaged dimensionless wall distances and first cell velocities of the  $m-0-y^+_{med}$ mesh for $t^*_{avg}=322$ (results are spatial-averaged in the span-wise direction).}
	\label{fig:medium_mesh_e-6_angle_0_y+_u_avg}
\end{figure}

 The time-averaged values of the dimensionless wall distances for the $m-0-y^+_{max}$ and $m-0-y^+_{med}$ meshes are illustrated in Figs.\ \ref{fig:y_plus_avg_medium_grid_e-5} and \ref{fig:y_plus_avg_medium_grid_e-6}, respectively. Because of the curvature of the profile, the distributions for the suction and pressure side are similar. As illustrated in Figs.\ \ref{fig:u_avg_medium_grid_e-5} and \ref{fig:u_avg_medium_grid_e-6} for respectively the $m-0-y^+_{max}$ and $m-0-y^+_{med}$ meshes, the main flow velocity primarily increases in both suction and pressure sides. Then, it decreases and at about $0.8c$ a recirculation region is generated, increasing the main flow velocity in the negative chord-wise direction and consequently increasing the dimensionless wall distance.
\par The dimensionless wall distances vary within approximately $0.05\leq y^+ \leq 1.2$ for the $m-0-y^+_{max}$ mesh. Therefore, the viscous sublayer can be resolved. Despite of the presence of strong fluctuations on results of the fine resolution grid, the velocities, as well as the dimensionless wall distances have also the same pattern for both $f-0-y^+_{min}$ and $m-0-y^+_{max}$ meshes, i.e$.$, the flow is similar for both pressure and suction sides and is firstly accelerated (near the leading edge) and then decelerated.
\par When compared to the data of the $m-0-y^+_{med}$ (illustrated in Figs.\ \ref{fig:y_plus_avg_medium_grid_e-6} and \ref{fig:u_avg_medium_grid_e-6}) the same pattern is observed. Since the dimensionless averaging time is smaller for this grid ($t^*_{avg}=322$), small fluctuations are still present between $0.8c \leq x \leq 1c$. Due to the fact that the dimensionless wall distance is directly proportional to the wall distance of the first cell and that this influences the velocities within these control volumes, expected discrepancies in the dimensionless wall distance also exists. These are summarized in Table \ref{table:dimensionless_wall_distance_angle_0}, for both the fine and medium resolution meshes with an angle of attack of $\alpha=0^\circ$.
\begin{table}[H]
	\centering
	\begin{tabular}{l c c}
		\hline
		\multicolumn{1}{c}{\bf{Mesh}} & \bf{First cell wall distance} & \bf{Dimensionless wall distance} \tabularnewline \hline
		$\;f-0-y^+_{min}$ & $y_{first\,cell}=4.4\cdot 10^{-6}\,m$ & $0.009\leq y^+ \leq 0.296$ \tabularnewline
		$m-0-y^+_{med}$ & $y_{first\,cell}=9.0\cdot 10^{-6}\,m$ & $0.014\leq y^+ \leq 0.599$ \tabularnewline
		$m-0-y^+_{max}$ & $y_{first\,cell}=1.8\cdot 10^{-5}\,m$ & $0.043\leq y^+ \leq 1.196$ \tabularnewline
		\hline	
	\end{tabular}
	\caption{\label{table:dimensionless_wall_distance_angle_0}Dimensionless wall distance variation for the grids of a NACA0012 profile with an angle of attack of $\alpha=0^\circ$.}
\end{table}

\par The highest values of the dimensionless wall distances are, as expected, for the medium resolution grid with $y_{first\,cell}=1.8\cdot 10^{-5}\,m$, i$.$e$.$, $m-0-y^+_{max}$. Since this is still located in the viscous sublayer ($0\leq y^+_{visc} \leq 5$) and the first cell velocity, as well as the dimensionless wall distance follow the same pattern as the other two meshes, the first cell wall distance of $y_{first\,cell}=1.8\cdot 10^{-5}\,m$ is sufficient to accurately simulate the NACA0012 test case for an angle of attack of $\alpha=0^\circ$. Further investigations of the velocity field and the Reynolds stresses (see Sections \ref{sec:simulation_analysis} \ref{sec:reynolds_stresses}) are, however, performed in order to guarantee the independence of this mesh. 

\subsubsection{Angle of attack $\alpha=5^\circ$}
\label{subsubsec:dimensionless_wall_distance_angle_5}

\par
\begin{figure}[H]
	\centering
	\subfigure[Time-averaged dimensionless wall distance.]{\includegraphics[width=0.495\textwidth]{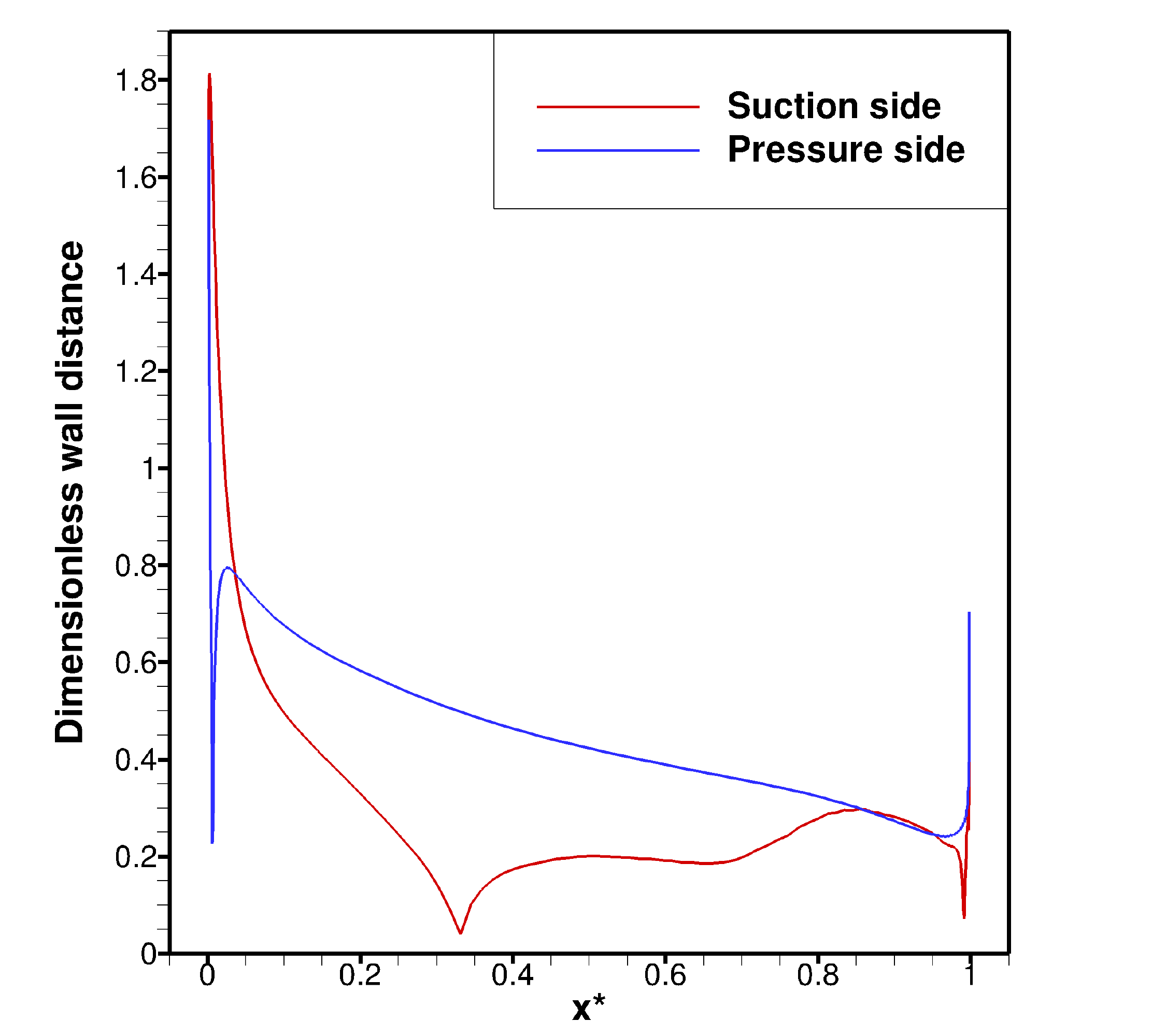}\label{fig:y_plus_avg_medium_grid_e-5_angle_5}}\hfill
	\subfigure[First cell time-averaged velocities.]{\includegraphics[width=0.495\textwidth]{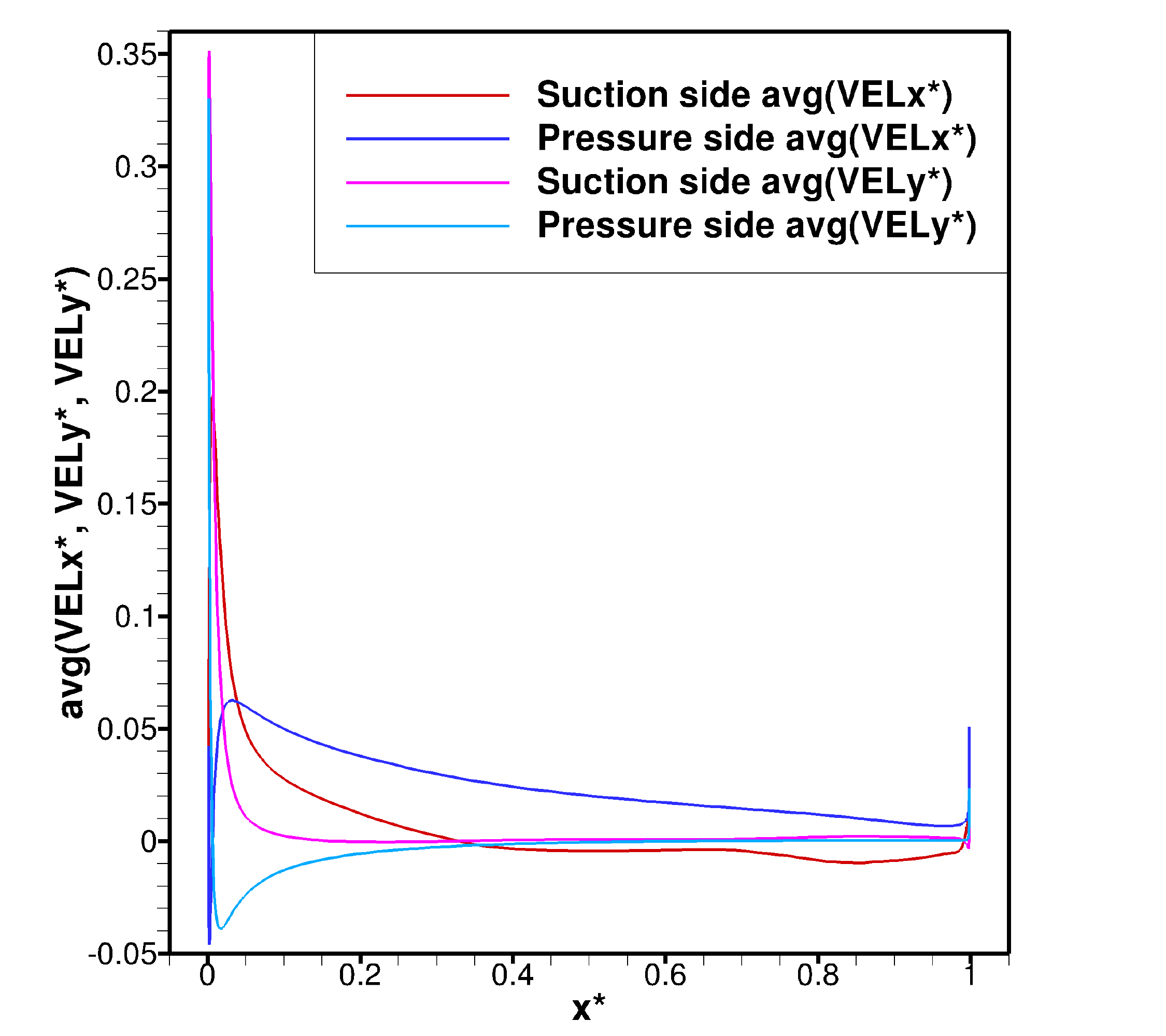} \label{fig:u_avg_medium_grid_e-5_angle_5}}\hfill
	\caption{Time-averaged dimensionless wall distances and first cell velocities of the  $m-5-y^+_{max}$ mesh for $t^*_{avg}=1039$ (results are spatial-averaged in the span-wise direction).}
	\label{fig:medium_mesh_e-5_angle_5_y+_u_avg}
\end{figure}

\begin{figure}[H]
	\centering
	\subfigure[Time-averaged dimensionless wall distance.]{\includegraphics[width=0.495\textwidth]{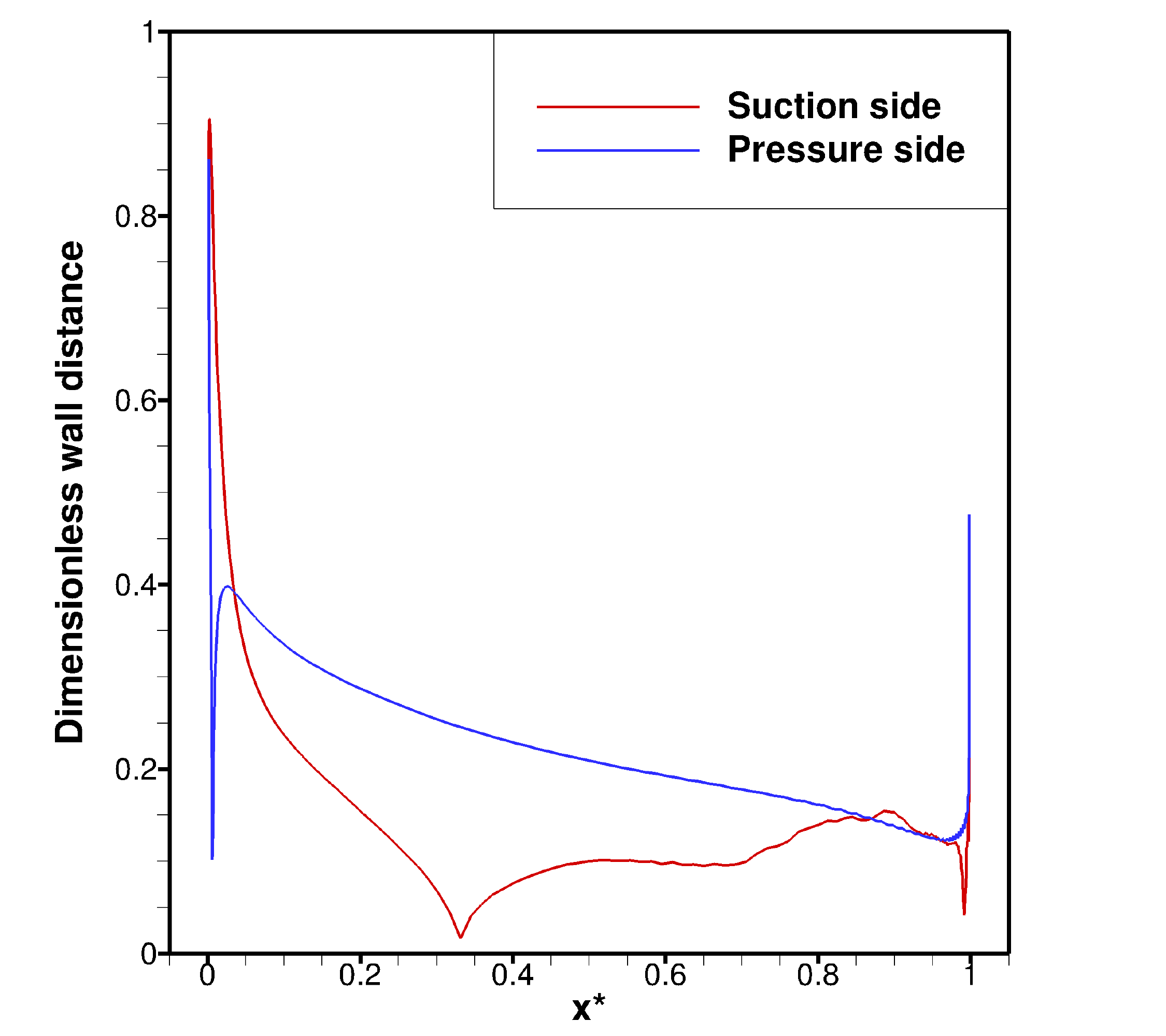}\label{fig:y_plus_avg_medium_grid_e-6_angle_5}}\hfill
	\subfigure[First cell time-averaged velocities.]{\includegraphics[width=0.495\textwidth]{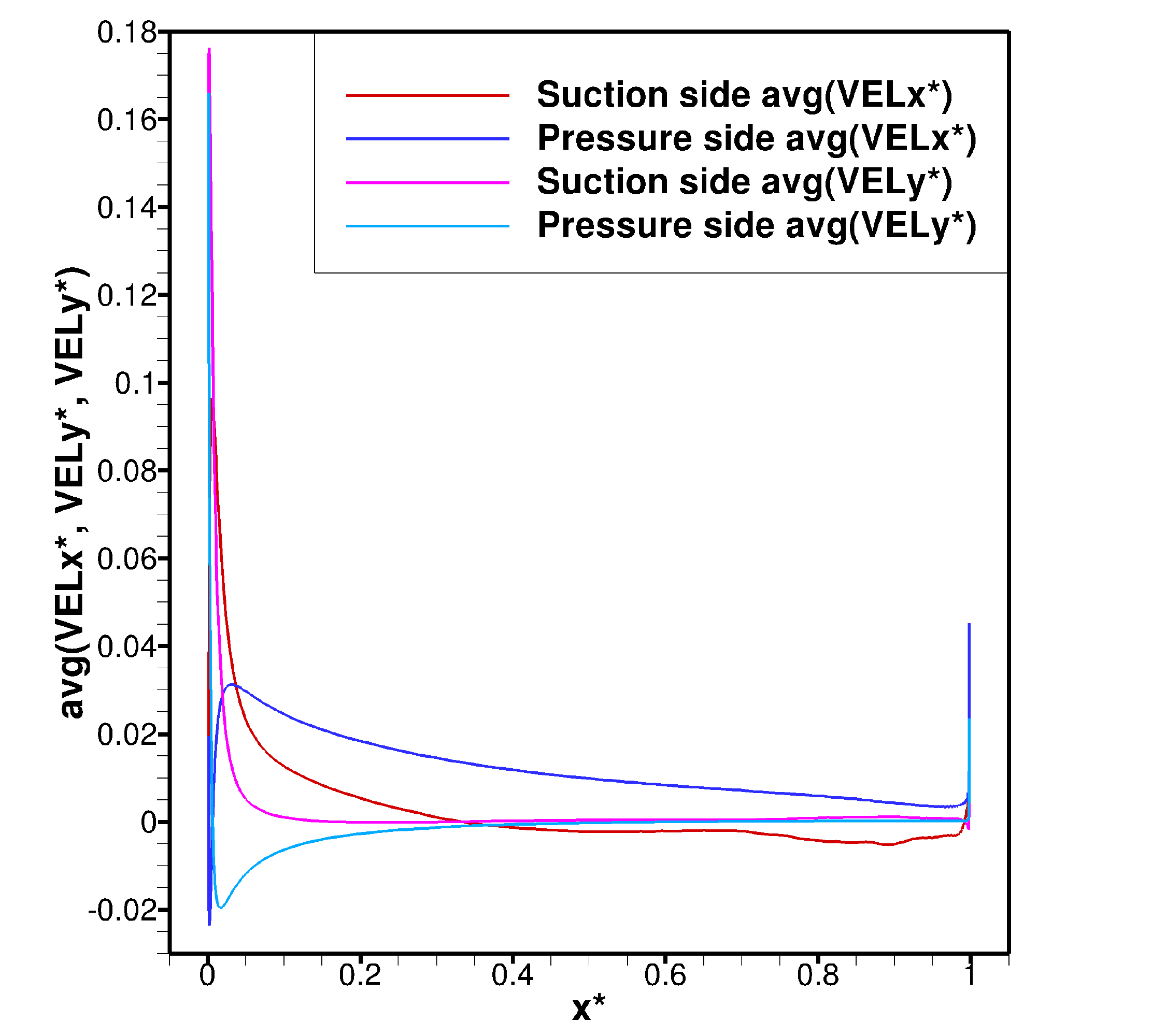} \label{fig:u_avg_medium_grid_e-6_angle_5}}\hfill
	\caption{Time-averaged dimensionless wall distances and first cell velocities of the  $m-5-y^+_{med}$ mesh for $t^*_{avg}=152$ (results are spatial-averaged in the span-wise direction).}
	\label{fig:medium_mesh_e-6_angle_5_y+_u_avg}
\end{figure}

 The time-averaged dimensionless wall distances for the grids with an angle of attack of $\alpha=5^\circ$ are illustrated in Figs$.$ \ref{fig:y_plus_avg_medium_grid_e-5_angle_5} and \ref{fig:y_plus_avg_medium_grid_e-6_angle_5}. The former stands for the $m-5-y^+_{max}$ mesh and is calculated with the time-averaged simulation data in a dimensionless time of $t^*_{avg}=1039$. The latter illustrates the $m-5-y^+_{med}$ grid  and is computed for a dimensionless time of $t^*_{avg}=152$. The discrepancy between both of the dimensionless times is caused by the difference of the utilized time steps (see Eq.\ (\ref{eq:dimensionless_time})). The smaller the time step $\Delta t$, the larger the computational time to achieve the same dimensionless averaging time $t^*_{avg}$.

\par As the symmetrical profile is subjected to an angle of attack of $\alpha=5^\circ$ the flow in the suction and pressure side do not follow the same pattern, as demonstrated in Figs$.$ \ref{fig:u_avg_medium_grid_e-5_angle_5} and \ref{fig:u_avg_medium_grid_e-6_angle_5}. 
On the suction side the flow is accelerated near the leading edge and followed by a deceleration until approximately $0.33c$, where a separation occurs and a recirculation region starts. On the pressure side the flow is firstly decelerated due to the angle of the profile, secondly suffers an acceleration as the flow starts following the airfoil and finally is decelerated due to the profile curvature, without suffering a boundary layer separation. This difference in velocity leads to a difference in the static pressure, i.e$.$, higher on the pressure side and lower on the suction side, resulting in a lift force. 
\par As the time-averaged dimensionless wall distance depends on the values of the time-averaged velocities, the former follows the same pattern as the latter, resulting in two different curves for the dimensionless wall distance on the suction and on the pressure side. Due to the smaller available dimensionless time for the $m-5-y^+_{med}$ mesh, i$.$e$.$ $t^*_{avg}=152$, some small fluctuations are still present in the dimensionless wall distances and first cell velocities of this mesh, according to Figs$.$ \ref{fig:y_plus_avg_medium_grid_e-6_angle_5} and \ref{fig:u_avg_medium_grid_e-6_angle_5}.
  
\par The range of the dimensionless wall distances is compared in Table \ref{table:dimensionless_wall_distance_angle_5}. As expected, the higher values of the dimensionless wall distances occurs for the $m-5-y^+_{max}$ grid. Although it is approximately twice the value of the $m-5-y^+_{med}$ grid, it is still located in the viscous sublayer ($0\leq y^+_{visc}\leq 5$). Therefore, the resolution of this grid should be sufficient to accurately resolve this sublayer. Further analysis of the velocity field and the Reynolds stresses are also performed in order to guarantee the independence of this mesh (see Sections \ref{sec:simulation_analysis} and \ref{sec:reynolds_stresses}).
\begin{table}[H]
 	\centering
 	\begin{tabular}{l c c}
 		\hline
 		\multicolumn{1}{c}{\bf{Mesh}} & \bf{First cell wall distance} & \bf{Dimensionless wall distance} \tabularnewline \hline
 		$\; \; m-5-y^+_{med}$ & $y_{first\,cell}=9.0\cdot 10^{-6}\,m$ & $0.028\leq y^+ \leq 0.90$ \tabularnewline
 		$\; \; m-5-y^+_{max}$ & $y_{first\,cell}=1.8\cdot 10^{-5}\,m$ & $0.032\leq y^+ \leq 1.81$ \tabularnewline
 		\hline	
 	\end{tabular}
 	\caption{\label{table:dimensionless_wall_distance_angle_5}Dimensionless wall distance variation for the grids of a NACA0012 profile with an angle of attack of $\alpha=5^\circ$.}
\end{table}
 
\subsubsection{Angle of attack $\alpha=11^\circ$}
\label{subsubsec:dimensionless_wall_distance_angle_11}
 
\par
\begin{figure}[H]
 	\centering
 	\subfigure[Time-averaged dimensionless wall distance.]{\includegraphics[width=0.495\textwidth]{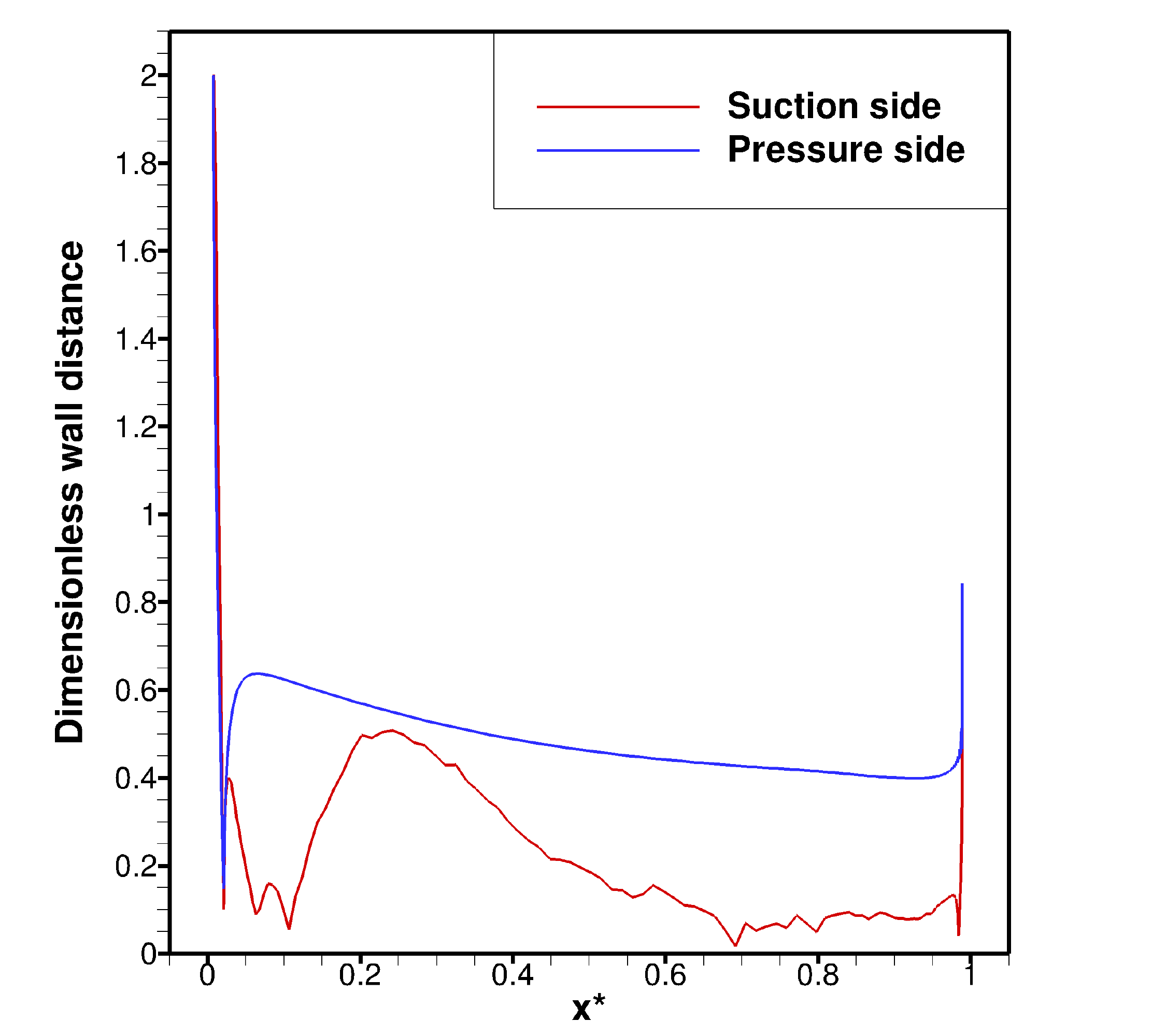}\label{fig:y_plus_avg_medium_grid_e-5_angle_11}}\hfill
 	\subfigure[First cell time-averaged velocities.]{\includegraphics[width=0.495\textwidth]{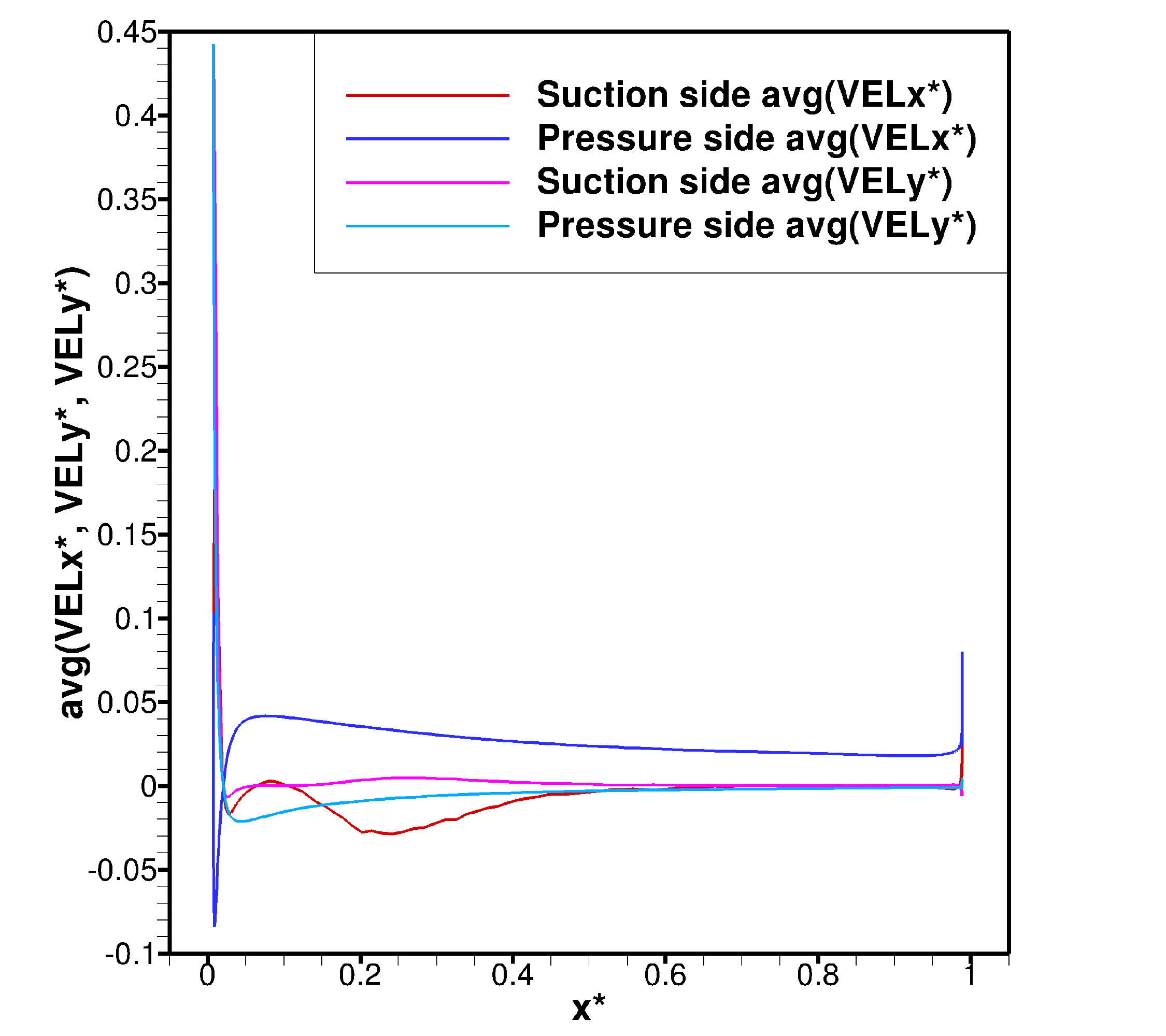} \label{fig:u_avg_medium_grid_e-5_angle_11}}\hfill
 	\caption{Time-averaged dimensionless wall distances and first cell velocities of the $m-\nobreak 11-\nobreak y^+_{max}$ mesh for $t^*_{avg}=454$ (results are spatial-averaged in the span-wise direction).}
 	\label{fig:medium_mesh_e-5_angle_11_y+_u_avg}
 \end{figure}

\begin{figure}[H]
 	\centering
 	\subfigure[Time-averaged dimensionless wall distance.]{\includegraphics[width=0.495\textwidth]{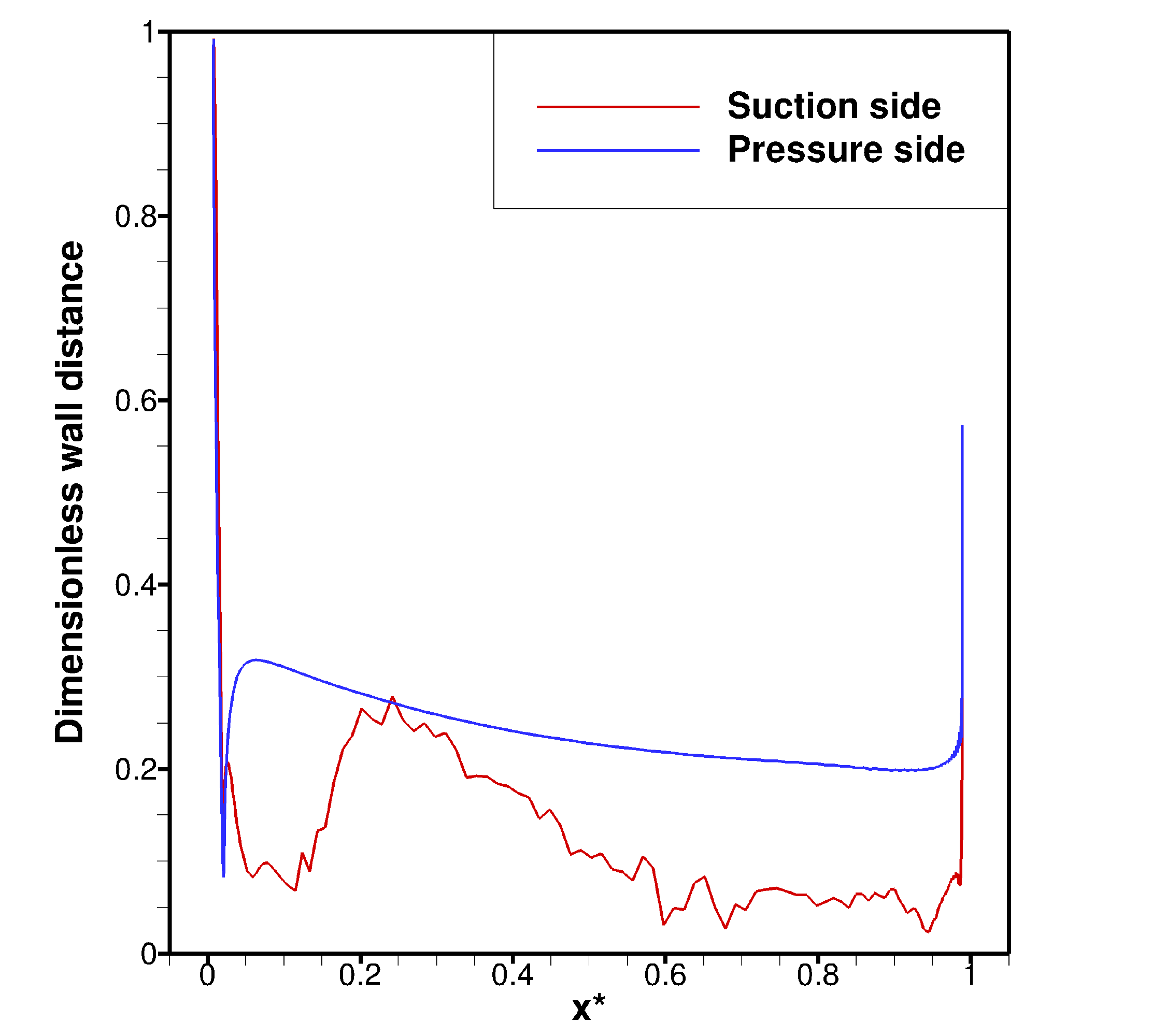}\label{fig:y_plus_avg_medium_grid_e-6_angle_11}}\hfill
 	\subfigure[First cell time-averaged velocities.]{\includegraphics[width=0.495\textwidth]{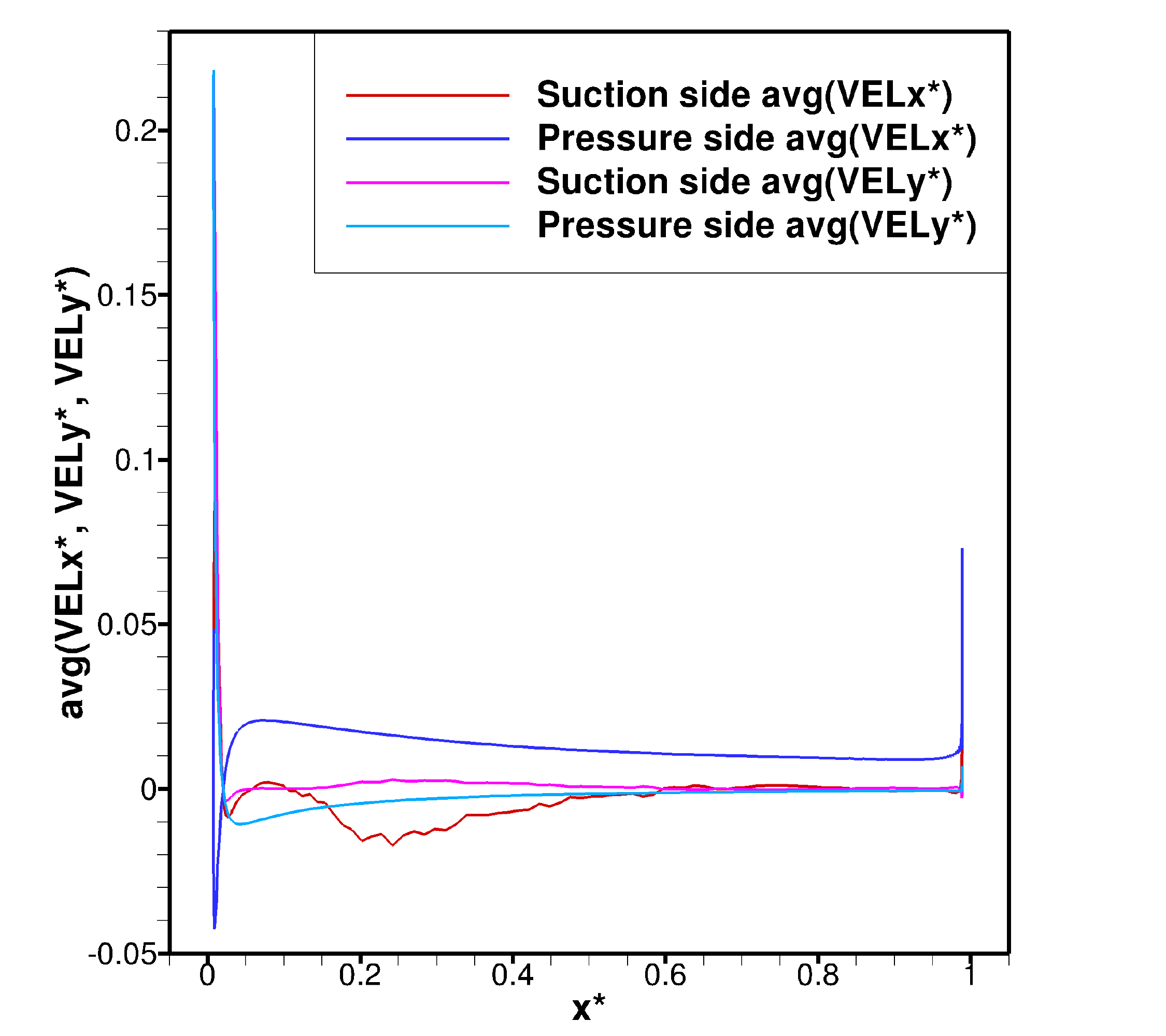} \label{fig:u_avg_medium_grid_e-6_angle_11}}\hfill
 	\caption{Time-averaged dimensionless wall distances and first cell velocities of the  $m-\nobreak11-\nobreak y^+_{med}$ mesh for $t^*_{avg}=76$ (results are spatial-averaged in the span-wise direction).}
 	\label{fig:medium_mesh_e-6_angle_11_y+_u_avg}
 \end{figure}

 The time-averaged dimensionless wall distances for the $m-11-y^+_{max}$ and $m-11-y^+_{med}$ are illustrated in Figs$.$ \ref{fig:y_plus_avg_medium_grid_e-5_angle_11} and \ref{fig:y_plus_avg_medium_grid_e-6_angle_11}, respectively. The former is calculated with the data at a dimensionless time of $t^*_{avg}=454$, while the latter is computed at $t^*_{avg}=76$. This difference in the averaging dimensionless time is caused by the dissimilar required time steps $\Delta t$ for both grids. Therefore, the $m-11-y^+_{med}$ mesh requires a larger computational time to achieve the same dimensionless averaging time $t^*_{avg}$ than the $m-11-y^+_{max}$ grid.
\par Fluctuations are present in the time-averaged dimensionless wall distances, as well as in the first cell velocities, according to Figs.\ \ref{fig:medium_mesh_e-5_angle_11_y+_u_avg} and \ref{fig:medium_mesh_e-6_angle_11_y+_u_avg} for the $m-11-y^+_{max}$ and $m-11-y^+_{med}$ meshes, respectively. This indicates that although the time and spatial-averaging processes are already executed, the achieved averaging dimensionless time is not yet enough to damp all fluctuations. Nevertheless, the studies of the dimensionless wall distance and the first cell velocities are performed.
\par Since the symmetrical profile is subjected to an angle of attack of $\alpha=11^\circ$ the flow does not follow the same pattern on the pressure and suction sides. On the former the flow is firstly decelerated due to the angle of the profile, then accelerated as it starts following the profile and finally suffers a slight deceleration due to the profile curvature. On the latter the flow is accelerated at the leading edge and followed by a deceleration, until a flow separation occurs, leading to the formation of a laminar separation bubble between approximately $0.02c < x < 0.77c$ for the $m-11-y^+_{max}$ mesh and between approximately $0.02c < x < 0.66c$ for the $m-11-y^+_{med}$ grid (see Sections \ref{sec:simulation_analysis} and \ref{sec:flow_summary}). This phenomenon, described in the work of Almutari \cite{Almutari_2010}, occurs mainly for Reynolds numbers between $10^4 < Re < 10^6$, and is characterized by a detachment of the boundary layer followed by a reattachment. 
\par As the dimensionless wall distance is directly proportional to the velocities, the same pattern occurs: While the pressure side does not contain fluctuations, the suction side shows strong fluctuations, which may interfere the precision of the analysis. Nevertheless, a preliminary study is executed as shown in Table \ref{table:dimensionless_wall_distance_angle_11}.
\begin{table}[H]
	\centering
	\begin{tabular}{l c c}
		\hline
		\multicolumn{1}{c}{\bf{Mesh}} & \bf{First cell wall distance} & \bf{Dimensionless wall distance} \tabularnewline \hline
		$\;\;m-11-y^+_{med}$ & $y_{first\,cell}=9.0\cdot 10^{-6}\,m$ & $0.023\leq y^+ \leq 0.974$ \tabularnewline
		$\;\;m-11-y^+_{max}$ & $y_{first\,cell}=1.8\cdot 10^{-5}\,m$ & $0.018\leq y^+ \leq 1.997$ \tabularnewline
		\hline	
	\end{tabular}
	\caption{\label{table:dimensionless_wall_distance_angle_11}Dimensionless wall distance variation for the grids of a NACA0012 profile with an angle of attack of $\alpha=11^\circ$.}
\end{table}
\par The dimensionless wall distances for the $m-11-y^+_{max}$ mesh varies within approximately $0.018 \leq y^+ \leq 1.997$, with a highest value of more than twice the uppermost value of the $m-11-y^+_{med}$ grid. Although this value is within the viscous sublayer, it is already high, since only two cells are used to resolve this portion. Nevertheless, this high dimensionless wall distance is only reached on a peak near the leading edge, otherwise these distances are within the range of approximately $0.018<y^+<0.845$. Consequently, further analysis of the velocity field and the Reynolds stresses are required in order to assure the grid-independence of the meshes for an angle of attack of $\alpha=11^\circ$.  
 
\section{Analysis of the time-averaged velocity field}\markboth{CHAPTER 3.$\quad$RES. AND DIS.}{3.2$\quad$ANALYSIS OF THE TIME-AVG. VEL. FIELD}
\label{sec:simulation_analysis}

\par The fluid velocity within the computational domain is analyzed for all seven meshes. The results are compared for the grids with the same angle of attack in order to acquire information about the independence of the mesh. 
\par Firstly, the whole fluid domain is analyzed in spite of studying the flow behavior on the suction and pressure side, as well as in the wake. Secondly, an analysis focused on the flow close to the profile is executed in order to assure that the viscous sublayer is resolved. Finally, the separations are located, which lead to either the detachment of the boundary layer or the formation of separation bubbles. 
\par The development of laminar separation bubbles is a common phenomenon in flows around airfoils for Reynolds numbers of $Re=100{,}000$ \cite{Almutari_2010}. This is characterized by a detachment followed by a reattachment of the boundary layer and may modify the effective shape of the airfoil. Therefore, it may negatively influence the aerodynamic performance of the profile.

\subsection{Angle of attack $\alpha=0^\circ$}
\label{subsec:velocity_analysis_angle_0}
\par The dimensionless velocity distribution at different dimensionless averaging times for the meshes $f-0-y^+_{min}$, $m-0-y^+_{med}$ and $m-0-y^+_{max}$ is analyzed. The study of the former grid is carried out based on instantaneous values, since no time-averaged velocity is available due to its high required computational time.
\par
\begin{figure}[H]
	\centering
	\subfigure[Unsteady velocity of the $f-0-y^+_{min}$ mesh for $t^*=40$.]{\includegraphics[width=0.49\textwidth]{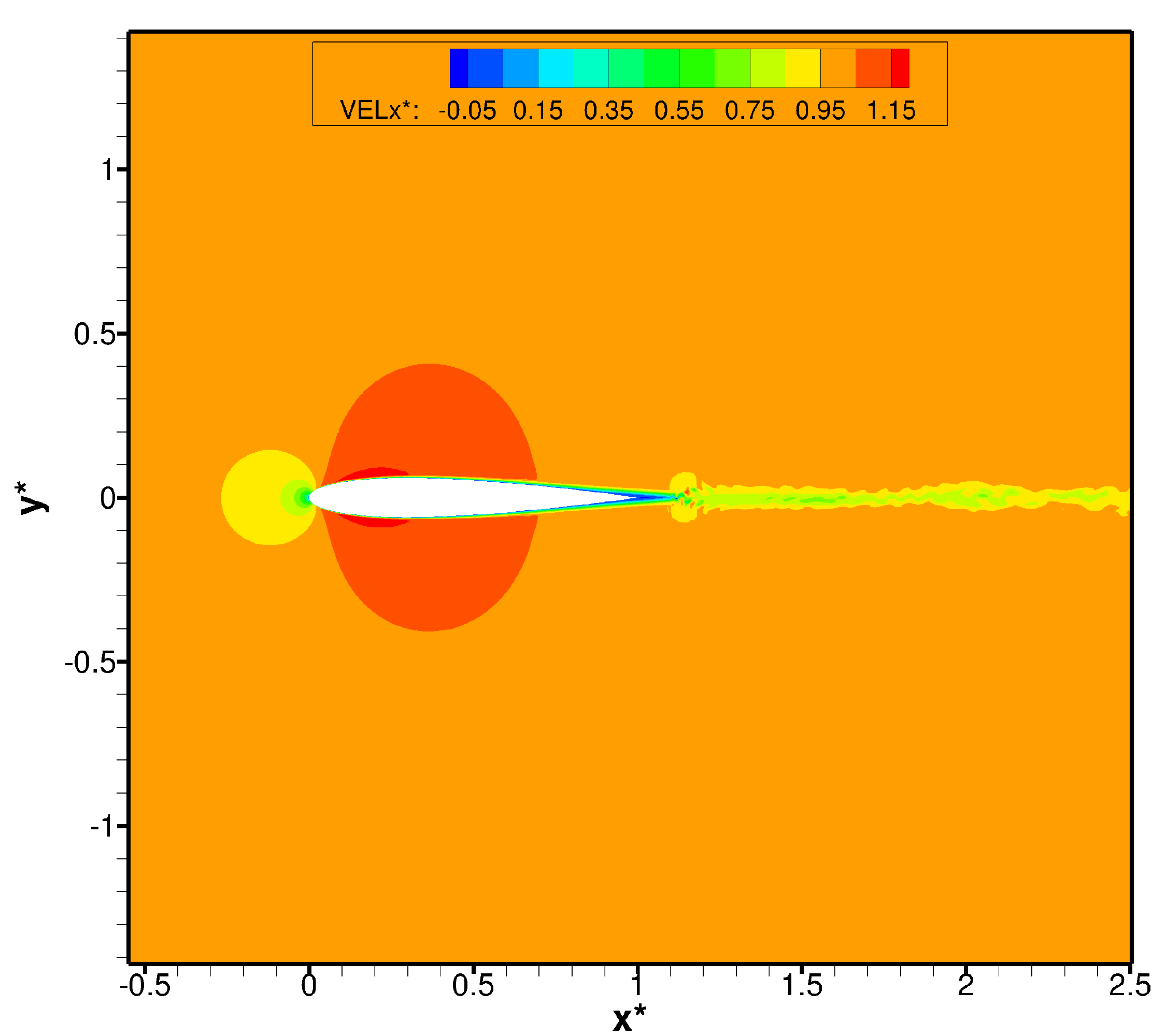}\label{fig:flow_domain_fine_grid}}\hfill
	\subfigure[Time-averaged velocity of the $m-0-y^+_{med}$ mesh for $t^*_{avg}=322$.]{\includegraphics[width=0.49\textwidth]{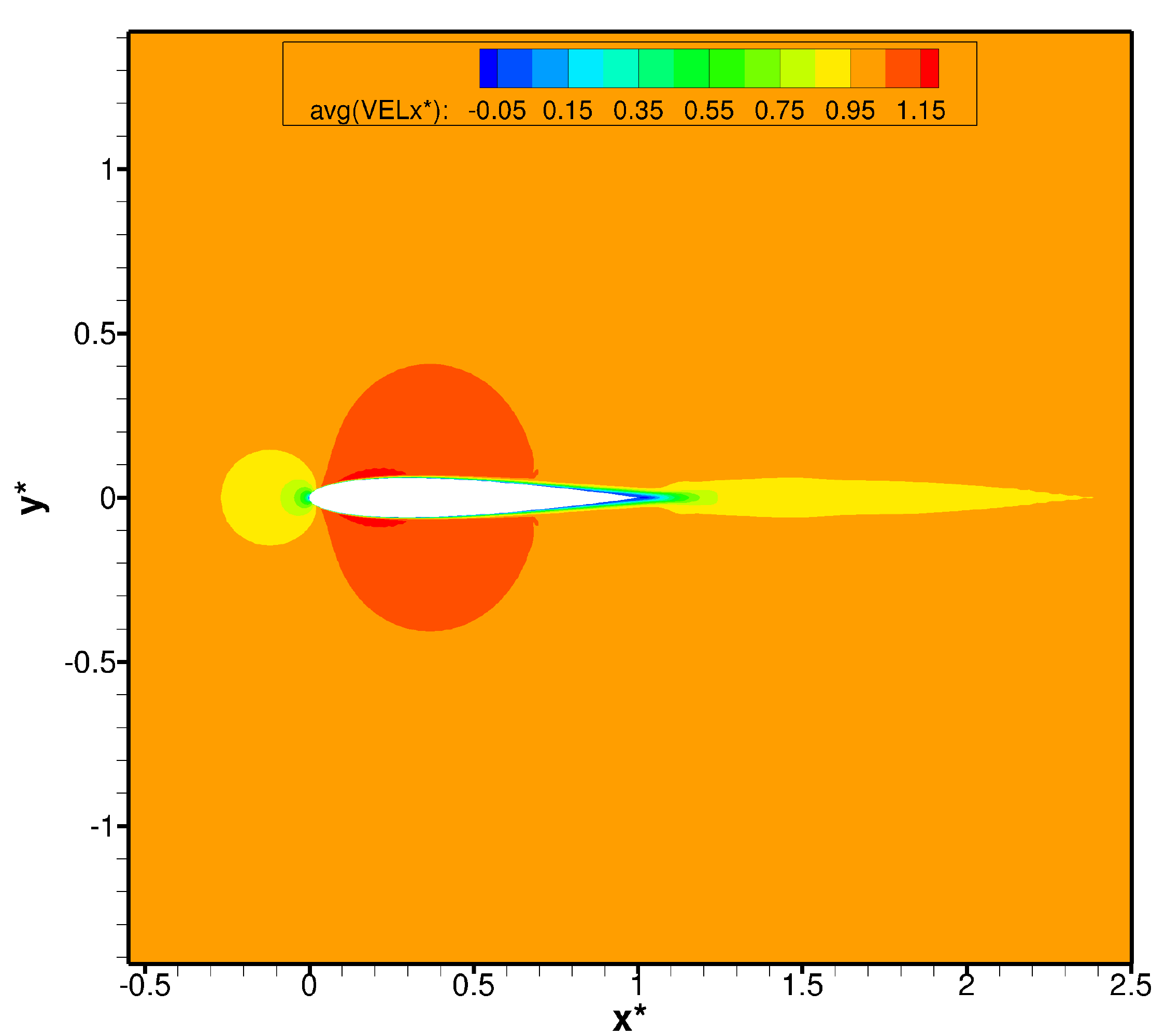} \label{fig:flow_domain_medium_grid_e-6_angle_0}}\hfill
\end{figure}

\begin{figure}[H]
	\centering
	\subfigure[Time-averaged velocity of the $m-0-y^+_{max}$ mesh for $t^*_{avg}=1009$.]{\includegraphics[width=0.49\textwidth]{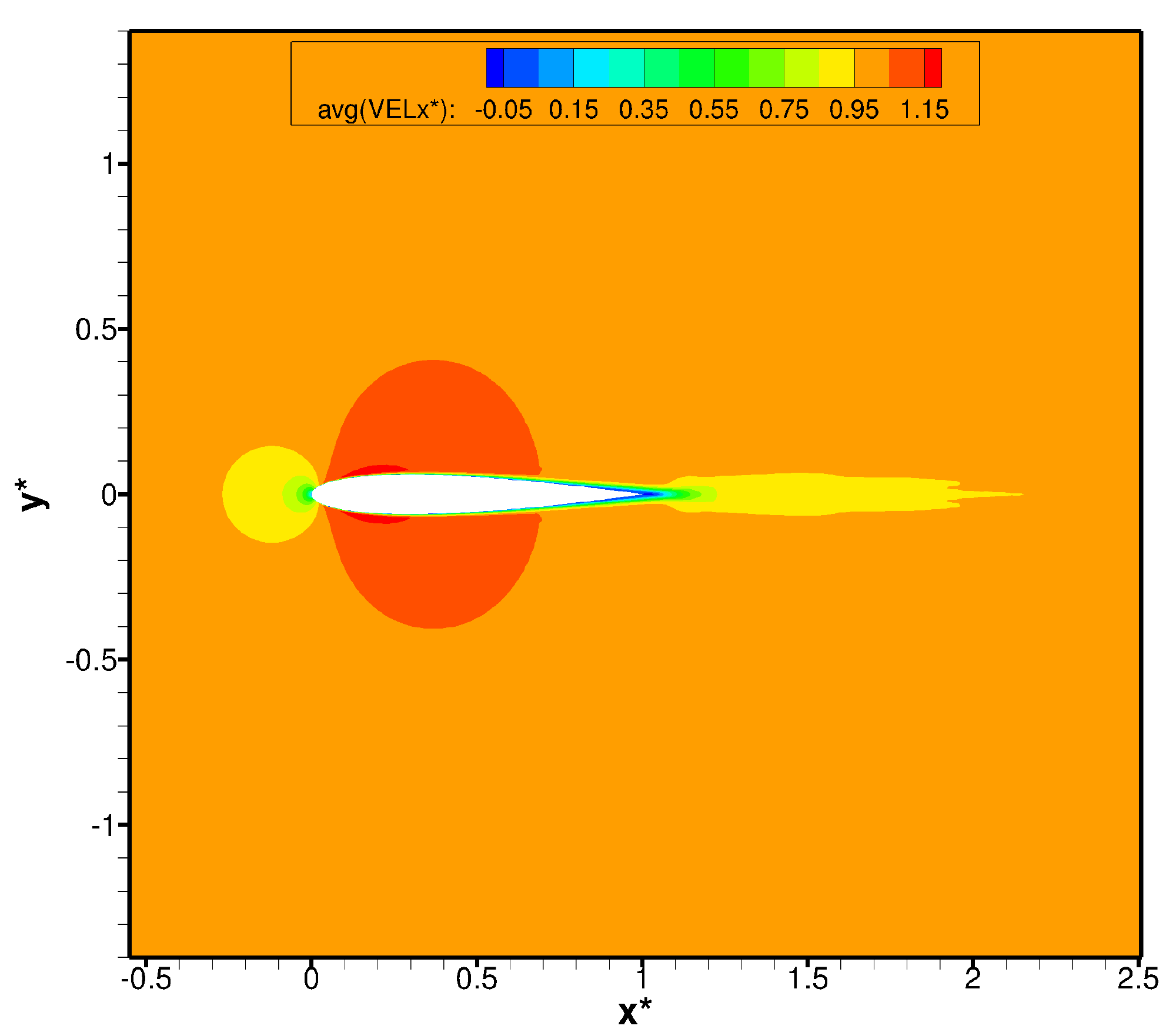} \label{fig:flow_domain_medium_grid_e-5_angle_0}}
	\caption{Velocity distribution in the main flow direction for the meshes with $\alpha=0^\circ$ (results are spatial-averaged in the span-wise direction).}
	\label{fig:velocities_angle_o}
\end{figure}

 Figures \ref{fig:flow_domain_fine_grid}, \ref{fig:flow_domain_medium_grid_e-6_angle_0} and \ref{fig:flow_domain_medium_grid_e-5_angle_0} illustrate respectively the instantaneous velocity distribution in the main flow direction of the $f-0-y^+_{min}$, $m-0-y^+_{med}$ and $m-0-y^+_{max}$ meshes. Due to the symmetrical profile with an angle of attack of $\alpha=0^\circ$, the flow on the suction and pressure side follow the same pattern, which is caused by the airfoil curvature. Firstly, the flow is accelerated until a dimensionless velocity of $u_1^*\approx1.17$ is achieved. Then, it is subsequently decelerated, which leads to a back-flow and consequent boundary layer detachment on both suction and pressure sides (see Section \ref{sec:flow_summary}). This, in turn, generates a repeating pattern of swirling vortex in the wake, which is called von K\'arm\'an vortex street.
\par
\begin{figure}[H]
	\centering
	\subfigure[Unsteady velocity profiles of the $f-0-y^+_{min}$ mesh for $t^*=40$.]{\includegraphics[width=0.49\textwidth]{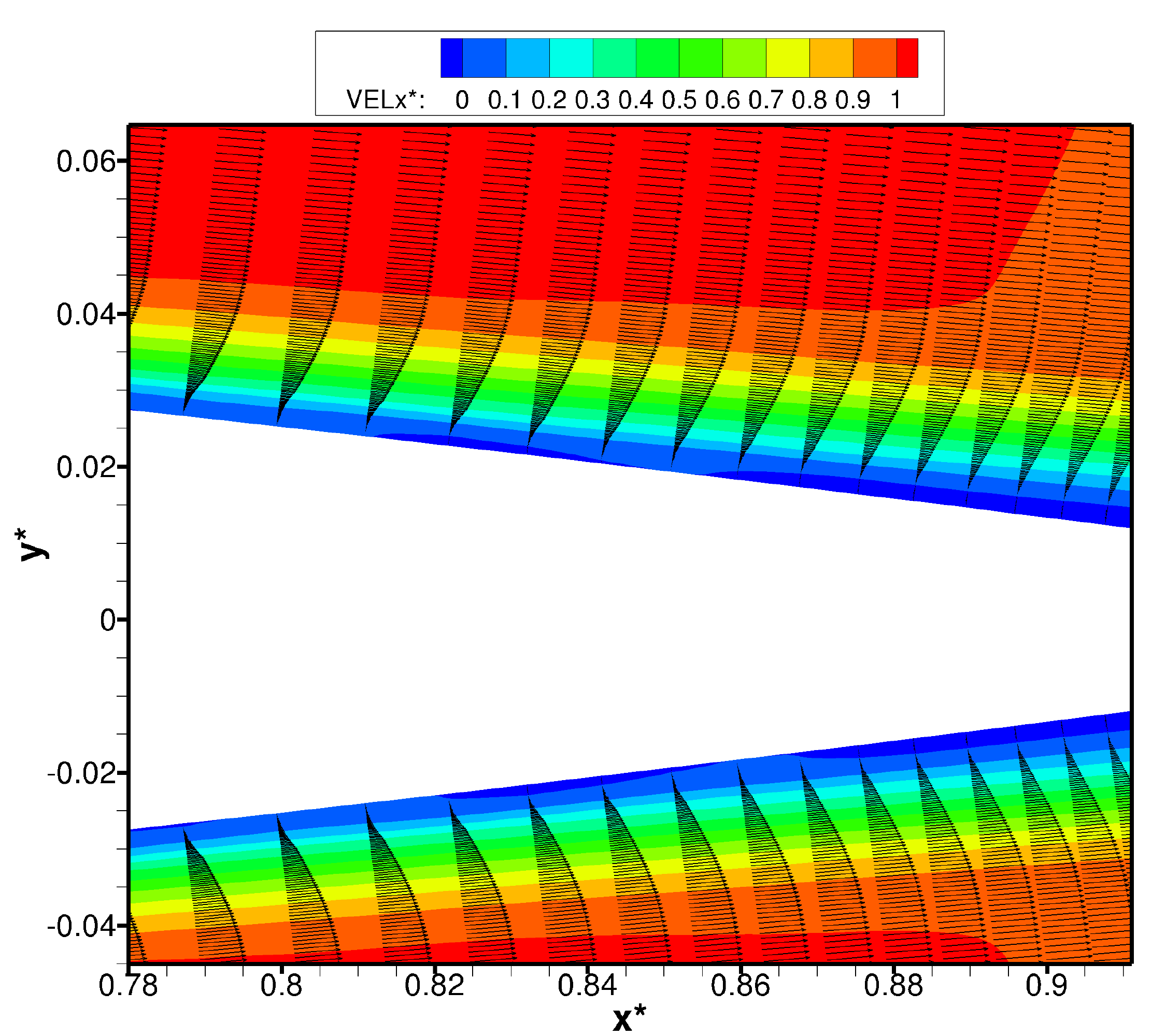}\label{fig:flow_near_profiel_fine_grid_e-6_angle_0}}\hfill
	\subfigure[Time-averaged velocity profiles of the $m-\nobreak 0-\nobreak y^+_{med}$ mesh for $t^*_{avg}=322$.]{\includegraphics[width=0.49\textwidth]{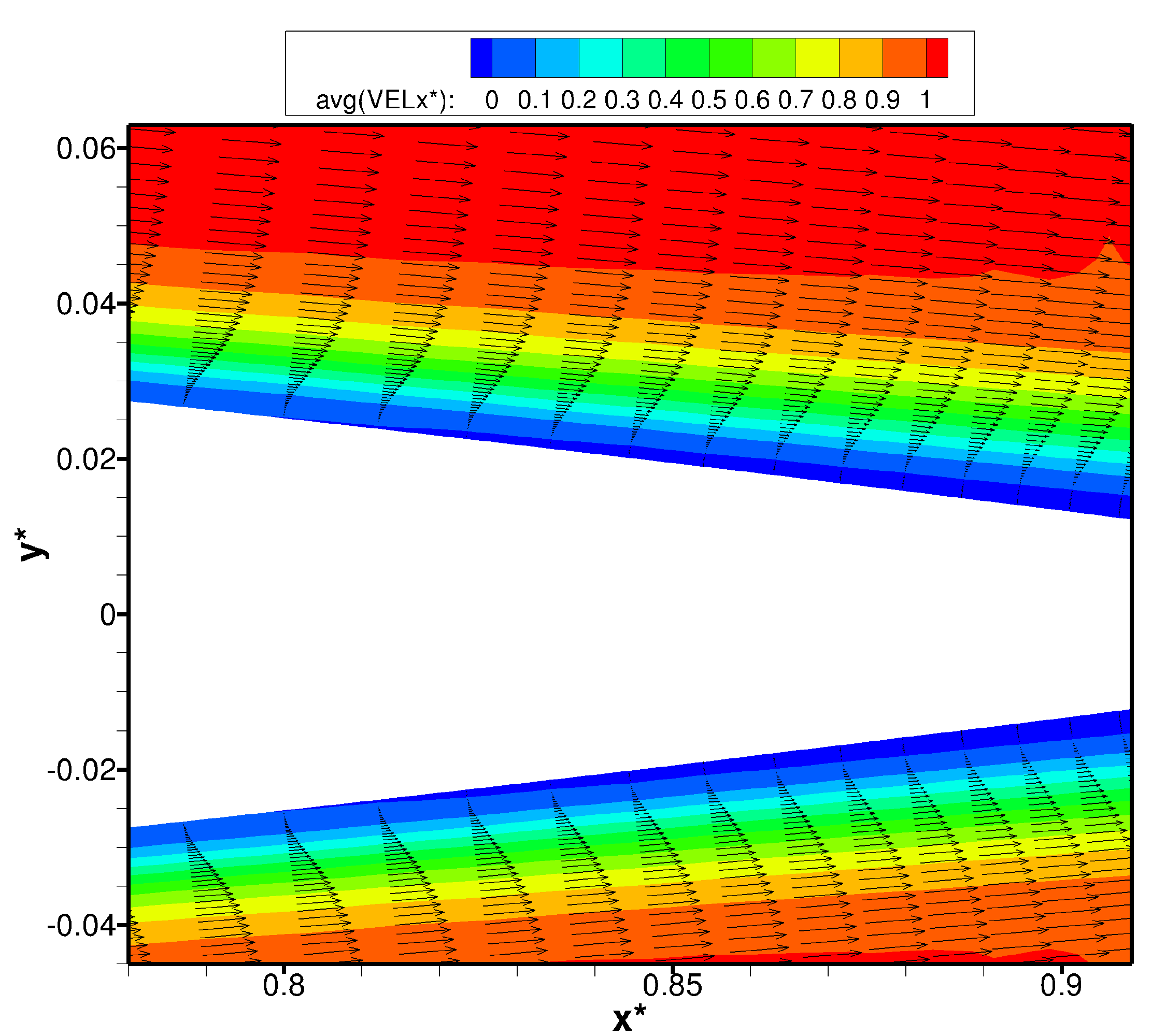} \label{fig:flow_near_profiel_medium_grid_e-6_angle_0}}\hfill
\end{figure}

\begin{figure}[H]
	\centering
	\subfigure[Time-averaged velocity profiles of the $m-\nobreak 0-\nobreak y^+_{max}$ mesh for $t^*_{avg}=1009$.]{\includegraphics[width=0.49\textwidth]{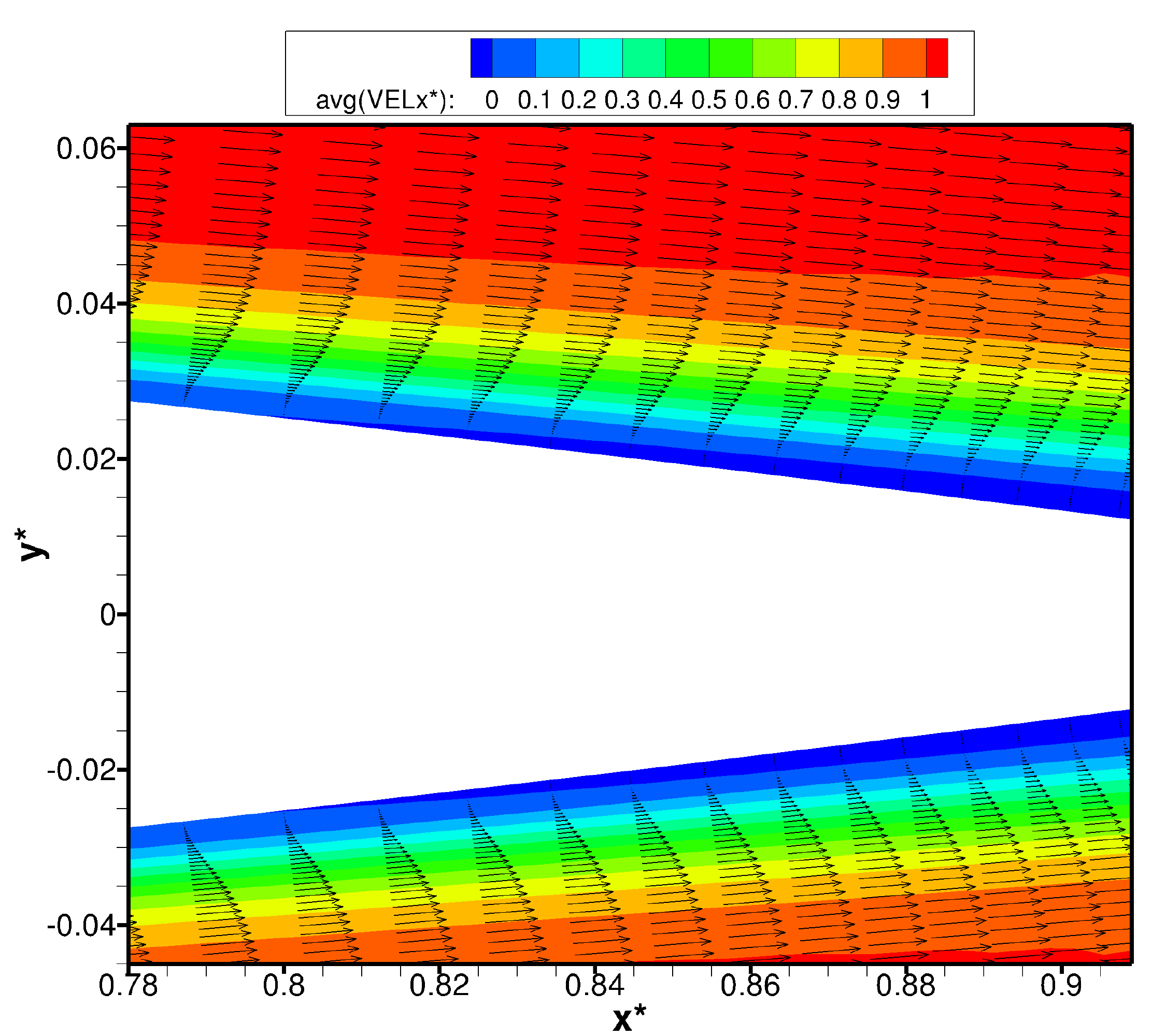} \label{fig:flow_near_profile_medium_grid_e-5_angle_0}}
	\caption{Velocity profiles in the main flow direction for the meshes with $\alpha=0^\circ$ (results are spatial-averaged in the span-wise direction).}
	\label{fig:velocity_profiles_angle_o}
\end{figure}

 In order to provide a more thorough analysis of the medium resolution meshes, the resolution of the boundary layer is analyzed where the flow is separated ($x=0.8\,c$), as shown in Figs$.$ \ref{fig:flow_near_profiel_fine_grid_e-6_angle_0}, \ref{fig:flow_near_profiel_medium_grid_e-6_angle_0}  \ref{fig:flow_near_profile_medium_grid_e-5_angle_0} for the $f-0-y^+_{min}$, $m-0-y^+_{med}$ and $m-0-y^+_{max}$ meshes, respectively.
\par A detachment of the boundary layer that leads to a back-flow and consequent formation of the von K\'arm\'an vortex street starts at approximately $x=0.8c$ for both pressure and suction sides of the $m-0-y^+_{max}$ and $m-0-y^+_{med}$ meshes. However, for the $f-0-y^+_{min}$ grid, this detachment occurs at about $x=0.854c$ for the suction side and at about $x=0.863c$ for the pressure side. Moreover, another two recirculation zones are also present on the unsteady results of the $f-0-y^+_{min}$ mesh.
\par The differences of the velocities for the fine and the medium resolution meshes are caused by the fact, that at the analyzed dimensionless time the flow of the $f-0-y^+_{min}$ grid is not yet fully developed and consequently not yet time-averaged. Therefore, even the results on the medium mesh with the greater first cell wall distance, i.e$.$, \mbox{$m-0-y^+_{max}$}, are assumed to be grid-independent. This result emphasizes the previous study of the dimensionless wall distance (see Section \ref{sec:analysis_dimensionless_wall_distance}), since both of them indicate a grid-independence for all three meshes. Nevertheless, studies of the Reynolds stresses (see Section \ref{sec:reynolds_stresses}) and the flow separation (see Section \ref{sec:flow_summary}) are also executed. 
 
\subsection{Angle of attack $\alpha=5^\circ$}
\label{subsec:velocity_analysis_angle_5}
\par The dimensionless velocity field for the $m-5-y^+_{med}$ and $m-5-y^+_{max}$ meshes is investigated for dimensionless averaging times of $t^*_{avg}=152$ and $t^*_{avg}=1039$, respectively. The divergence in the analyzed dimensionless averaging times is caused due to the larger required computational time to simulate the $m-5-y^+_{med}$ grid. 
\par Although the NACA0012 profile is symmetrical, the angle of attack of $\alpha=5^\circ$ causes an unsymmetrical flow. On the suction side the flow is accelerated, while on the pressure side it suffers a deceleration. This velocity difference causes a gradient of the pressure between both airfoil sides and therefore, a lift force is generated (see Section \ref{sec:drag_lift_coefficients}). 
\par
\begin{figure}[H]
	\centering
	\subfigure[Time-averaged velocity of the $m-5-y^+_{med}$ mesh for $t^*_{avg}=152$.]{\includegraphics[width=0.49\textwidth]{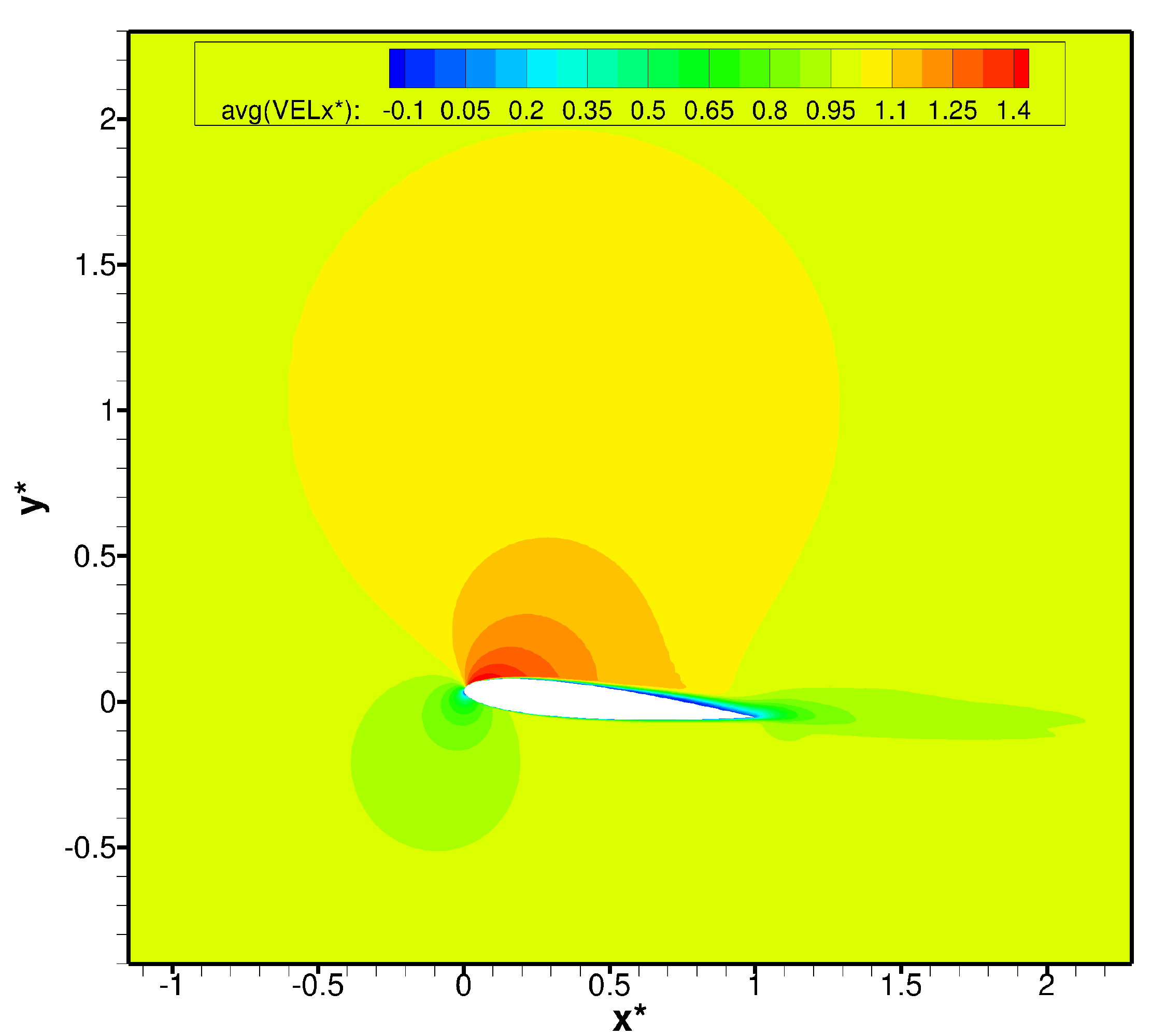}\label{fig:flow_domain_medium_grid_e-6_angle_5}}\hfill
	\subfigure[Time-averaged velocity of the $m-5-y^+_{max}$ mesh for $t^*_{avg}=1039$.]{\includegraphics[width=0.49\textwidth]{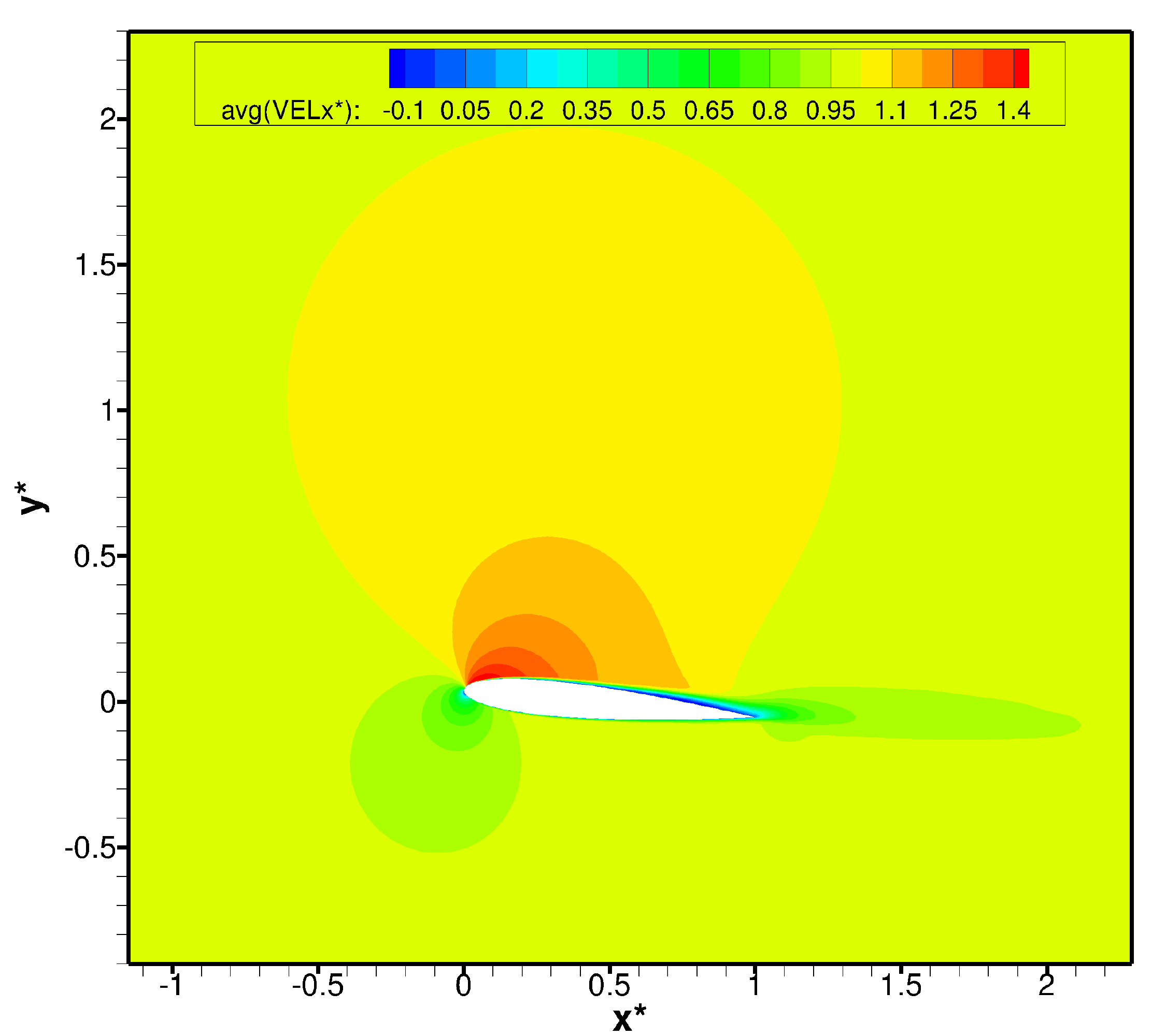} \label{fig:flow_domain_medium_grid_e-5_angle_5}}\hfill
	\caption{Time-averaged velocity distribution in the main flow direction for the meshes with an angle of attack of $\alpha=5^\circ$ (results are spatial-averaged in the span-wise direction).}
	\label{fig:velocities_angle_5}
\end{figure}

 Figures \ref{fig:flow_domain_medium_grid_e-6_angle_5} and \ref{fig:flow_domain_medium_grid_e-5_angle_5} illustrate the spatial and time-averaged dimensionless velocity fields in the main flow direction of the $m-5-y^+_{med}$ and $m-5-y^+_{max}$ meshes, respectively. A detachment of the boundary layer and a small divergence on the velocity fields in the wake are recognizable. The latter is, however, caused by the difference of the investigated dimensionless averaging times and therefore is not influenced by the grid.
\par
\begin{figure}[H]
	\centering
	\subfigure[Time-averaged velocity profiles of the $m-\nobreak 5-\nobreak y^+_{med}$ mesh for $t^*_{avg}=152$.]{\includegraphics[width=0.49\textwidth]{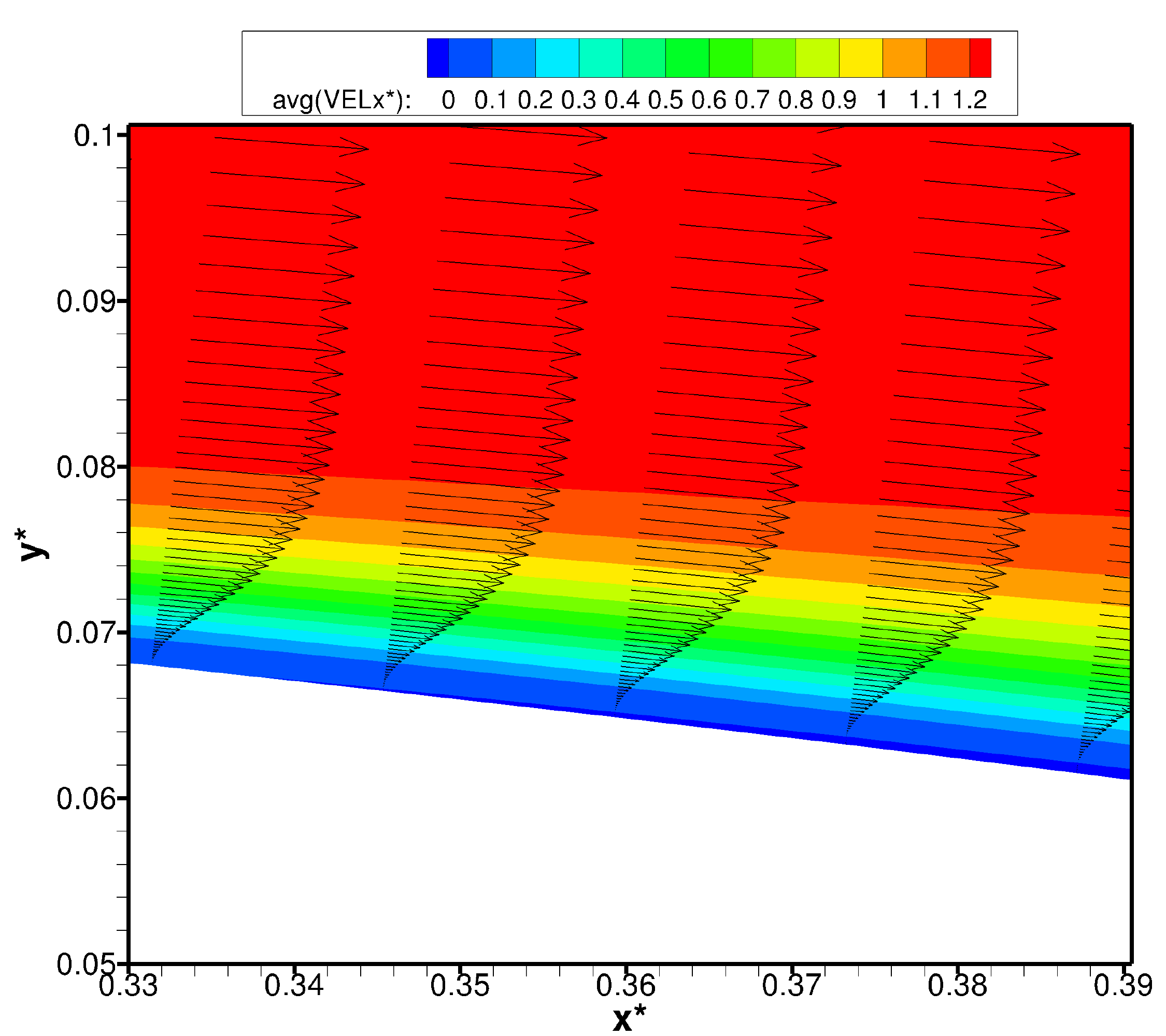}\label{fig:flow_near_profile_medium_grid_e-6_angle_5}}\hfill
	\subfigure[Time-averaged velocity profiles of the $m-\nobreak 5-\nobreak y^+_{max}$ mesh for $t^*_{avg}=1039$.]{\includegraphics[width=0.49\textwidth]{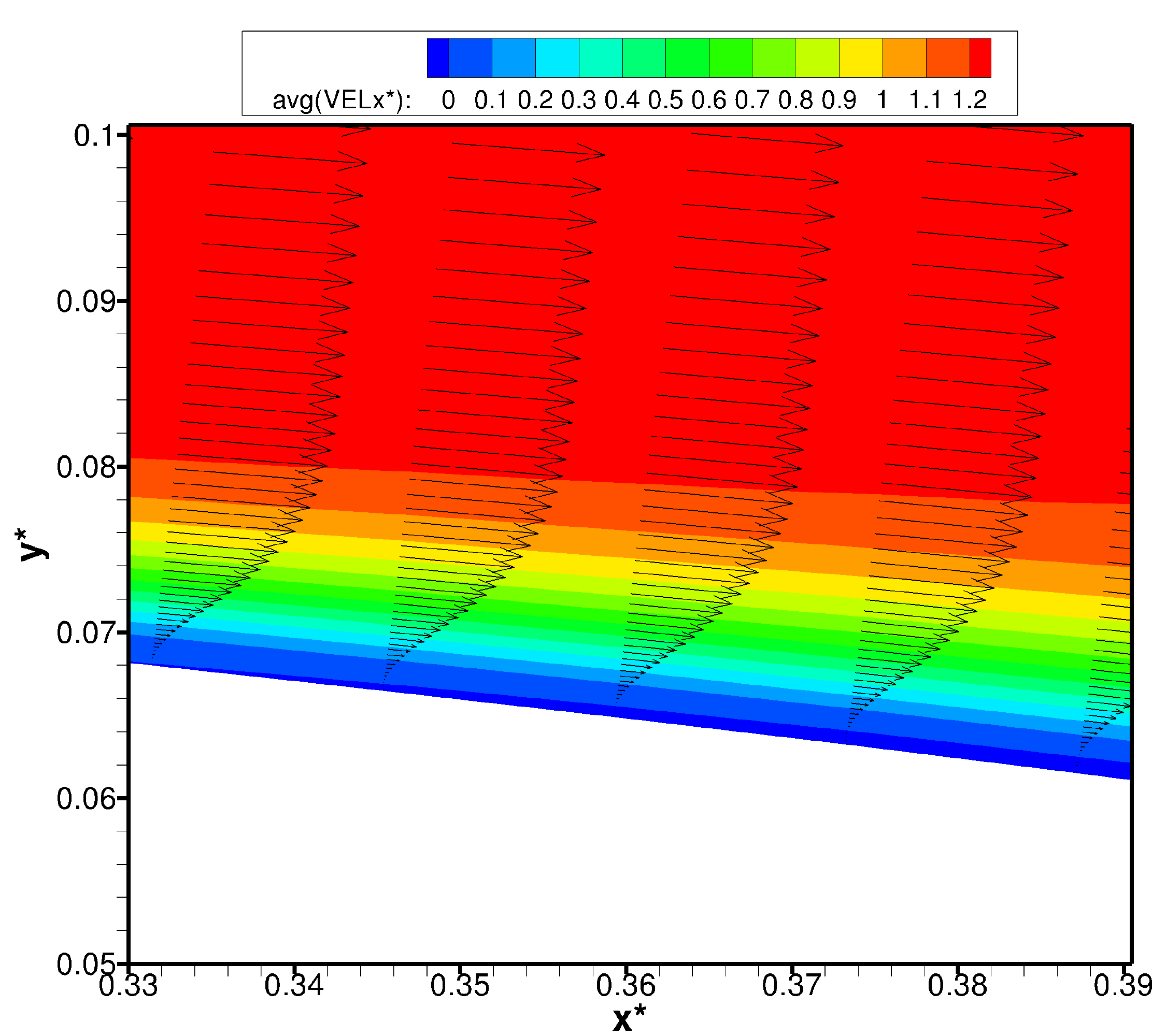} \label{fig:flow_near_profile_medium_grid_e-5_angle_5}}\hfill
	\caption{Time-averaged velocity profiles in the main flow direction for the meshes with an angle of attack of $\alpha=5^\circ$ (results are spatial-averaged in the span-wise direction).}
	\label{fig:velocity_profiles_angle_5}
\end{figure}

  The boundary layer detachment occurs at about $x=0.345c$ for the $m-5-y^+_{med}$ mesh and at about $x=0.331c$ for the $m-5-y^+_{max}$ grid, as illustrated by the velocity profiles in Figs.\ \ref{fig:flow_near_profile_medium_grid_e-6_angle_5} and \ref{fig:flow_near_profile_medium_grid_e-5_angle_5}. The boundary layer is further downstream, at about $x=0.99c$, reattached. Therefore, a separation bubble, i.e$.$ a recirculation zone, is present in the flow around the NACA0012 profile at an angle of attack of $\alpha=5^\circ$ (see Section \ref{sec:flow_summary}). 
\par
\begin{figure}[H]
	\centering
	\subfigure[Time-averaged velocity of the $m-5-y^+_{med}$ mesh for $t^*_{avg}=152$.]{\includegraphics[width=0.49\textwidth]{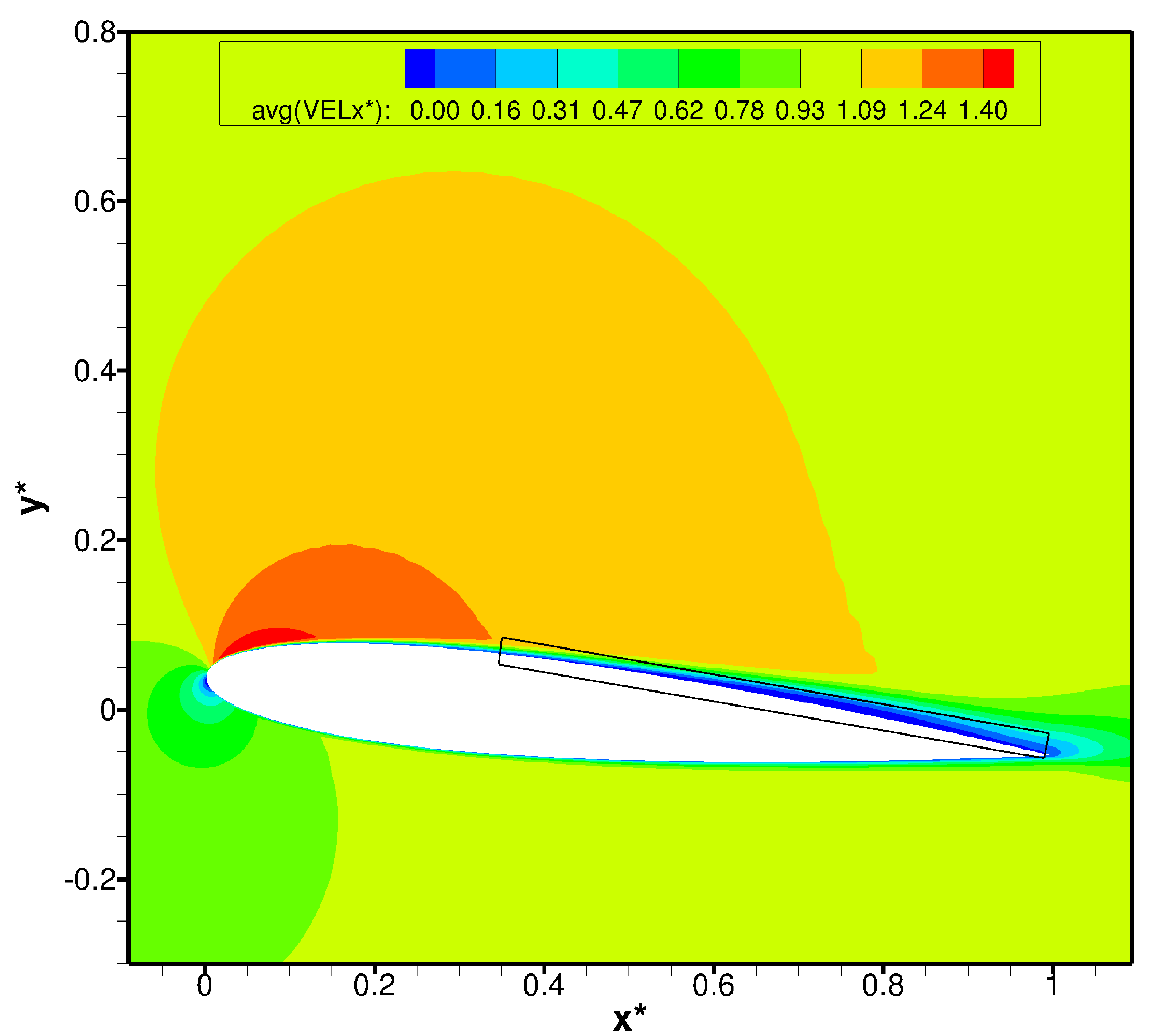}\label{fig:flow_domain_medium_grid_e-6_angle_5_focus_airfoil}}\hfill
	\subfigure[Time-averaged velocity of the $m-5-y^+_{max}$ mesh for $t^*_{avg}=1039$.]{\includegraphics[width=0.49\textwidth]{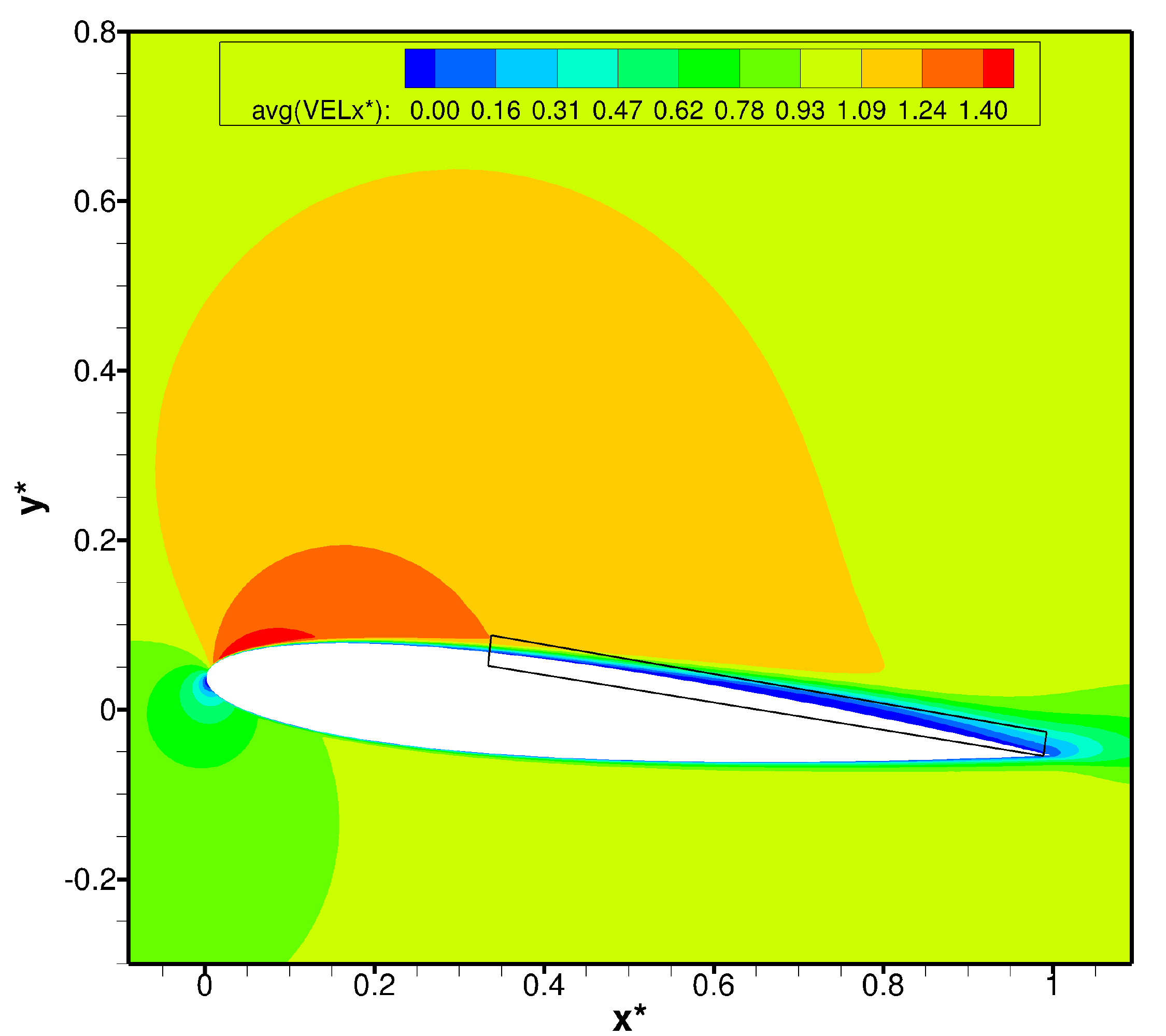} \label{fig:flow_domain_medium_grid_e-5_angle_5_focus_airfoil}}\hfill
	\caption{Time-averaged velocity distribution in the main flow direction for the meshes with an angle of attack of $\alpha=5^\circ$ (results are spatial-averaged in the span-wise direction). Focus on the airfoil and related phenomena.}
	\label{fig:velocities_angle_5_focus_airfoil}
\end{figure}

 The formation of separation bubbles is expected for flow around airfoils at a Reynolds number of $Re=100{,}000$. These are marked by rectangles in Figs.\ \ref{fig:flow_domain_medium_grid_e-6_angle_5_focus_airfoil} and \ref{fig:flow_domain_medium_grid_e-5_angle_5_focus_airfoil} for the $m-5-y^+_{med}$ and $m-5-y^+_{max}$ meshes, respectively. 
\par Since no significant difference is present on the spatial and time-averaged velocity fields of the $m-5-y^+_{med}$ and $m-5-y^+_{max}$ meshes, both can accurately resolve the boundary layer and do not influence the results, that is, both are grid-independent. Nevertheless, investigations of the Reynolds stresses and the formation of separation bubbles are also performed (see Sections \ref{sec:reynolds_stresses} and \ref{sec:flow_summary}).

\subsection{Angle of attack $\alpha=11^\circ$}
 \label{subsec:velocity_analysis_angle_11}

\par The time-averaged dimensionless velocity field in the main flow direction is studied for the $m-11-y^+_{med}$ and $m-11-y^+_{max}$ meshes at $t^*_{avg}=76$ and $t^*_{avg}=454$, respectively.
\par
\begin{figure}[H]
  	\centering
  	\subfigure[Time-averaged velocity of the $m-11-y^+_{med}$ mesh for $t^*_{avg}=76$.]{\includegraphics[width=0.49\textwidth]{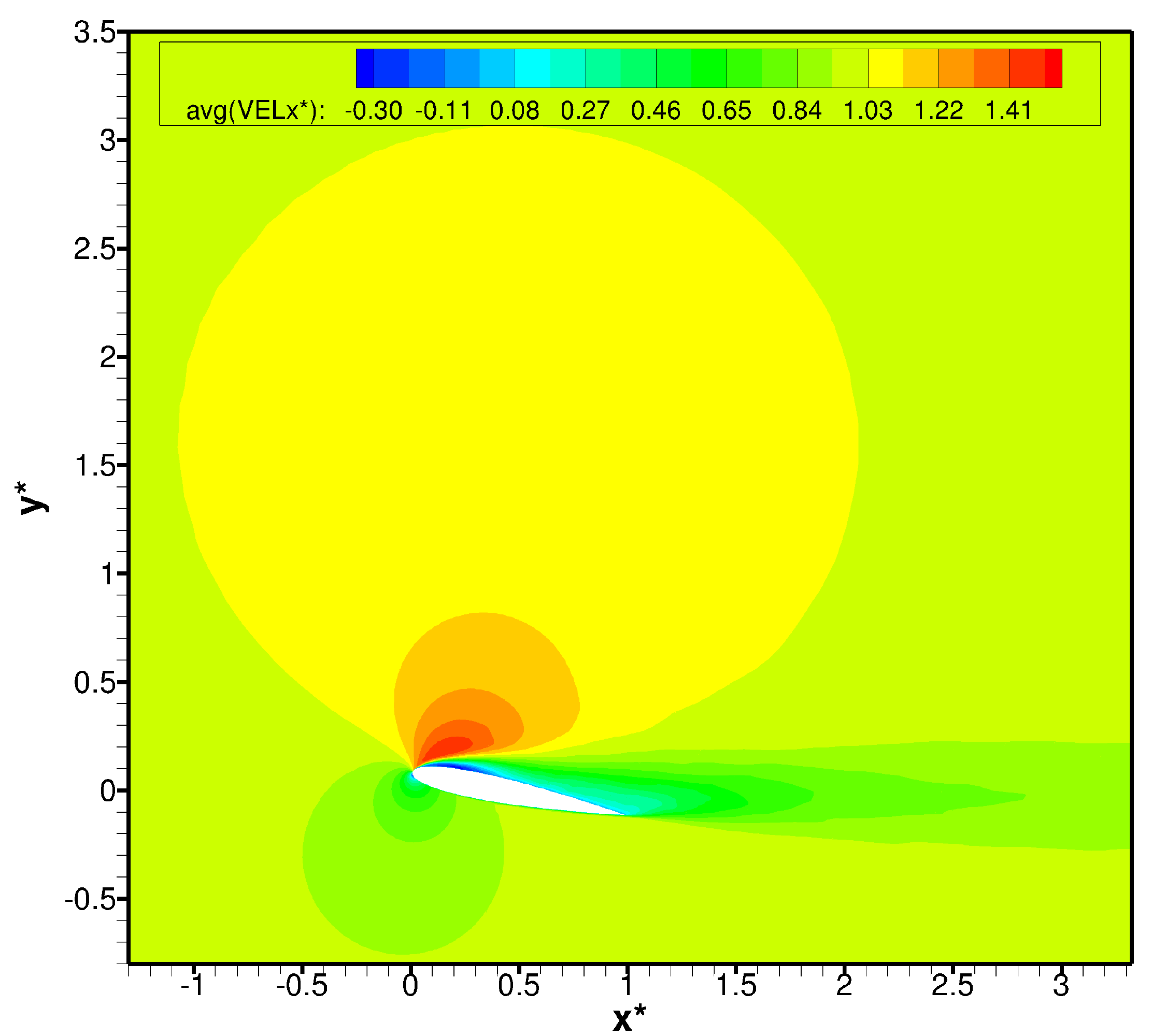}\label{fig:flow_domain_medium_grid_e-6_angle_11}}\hfill
  	\subfigure[Time-averaged velocity of the $m-11-y^+_{max}$ mesh for $t^*_{avg}=454$.]{\includegraphics[width=0.49\textwidth]{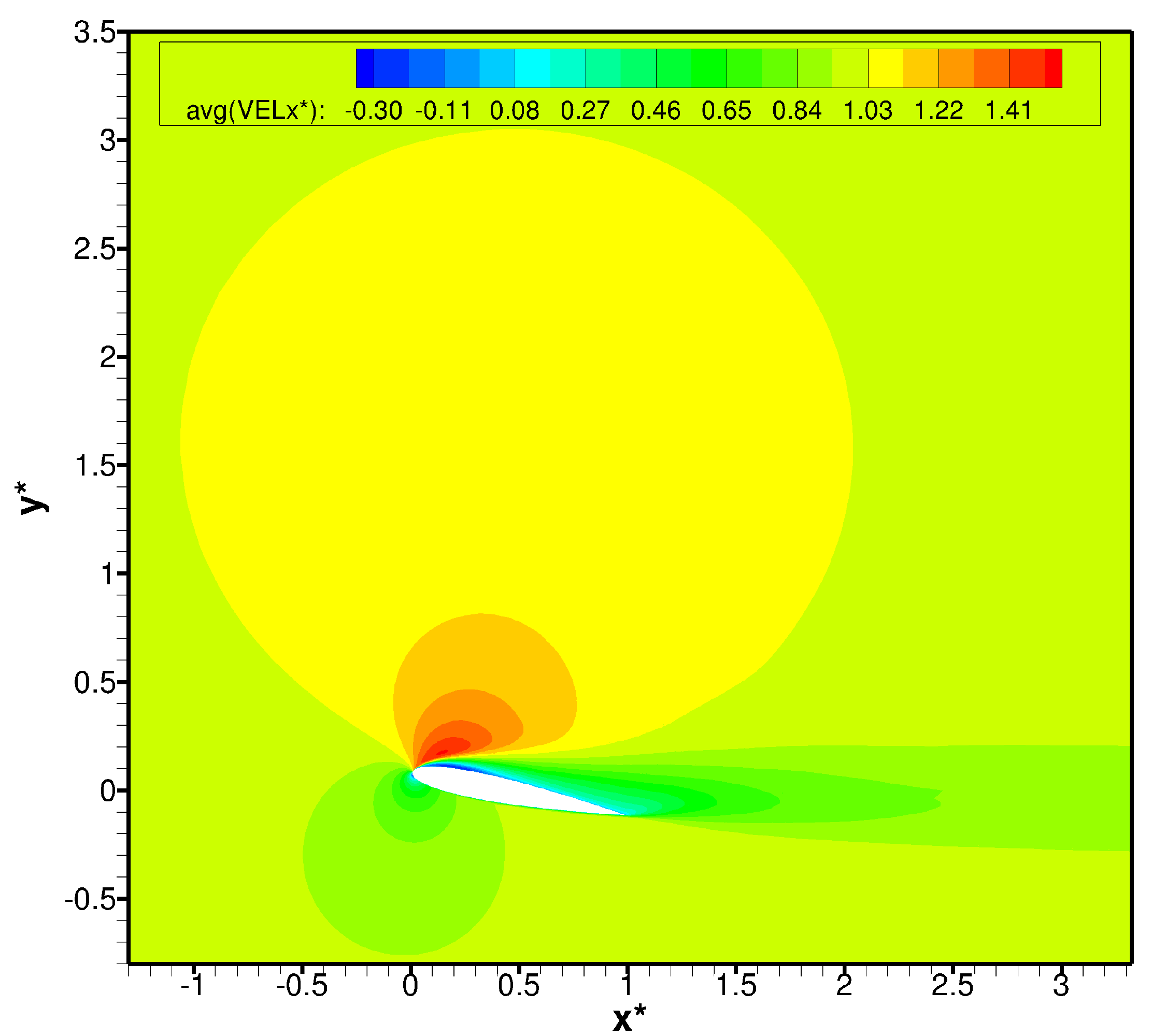} \label{fig:flow_domain_medium_grid_e-5_angle_11}}\hfill
  	\caption{Time-averaged dimensionless velocity distribution in the main flow direction for the meshes with an angle of attack of $\alpha=11^\circ$ (results are spatial-averaged in the span-wise direction).}
  	\label{fig:velocities_angle_11}
  \end{figure}

 The flow around the NACA0012 profile at an angle of attack of $\alpha=11^\circ$ is unsymmetrical and illustrated by Figs.\ \ref{fig:flow_domain_medium_grid_e-6_angle_11} and \ref{fig:flow_domain_medium_grid_e-5_angle_11}, which show the spatial and time-averaged dimensionless velocity fields in the main flow direction of the $m-11-y^+_{med}$ and $m-11-y^+_{max}$ meshes, respectively. A boundary layer detachment near the leading edge and a slightly different wake are recognizable. The latter is caused by the different analyzed dimensionless averaging times.
\par
\begin{figure}[H]
   	\centering
   	\subfigure[$m-11-y^+_{med}$ at $t^*_{avg}=76$.]{\includegraphics[width=0.49\textwidth]{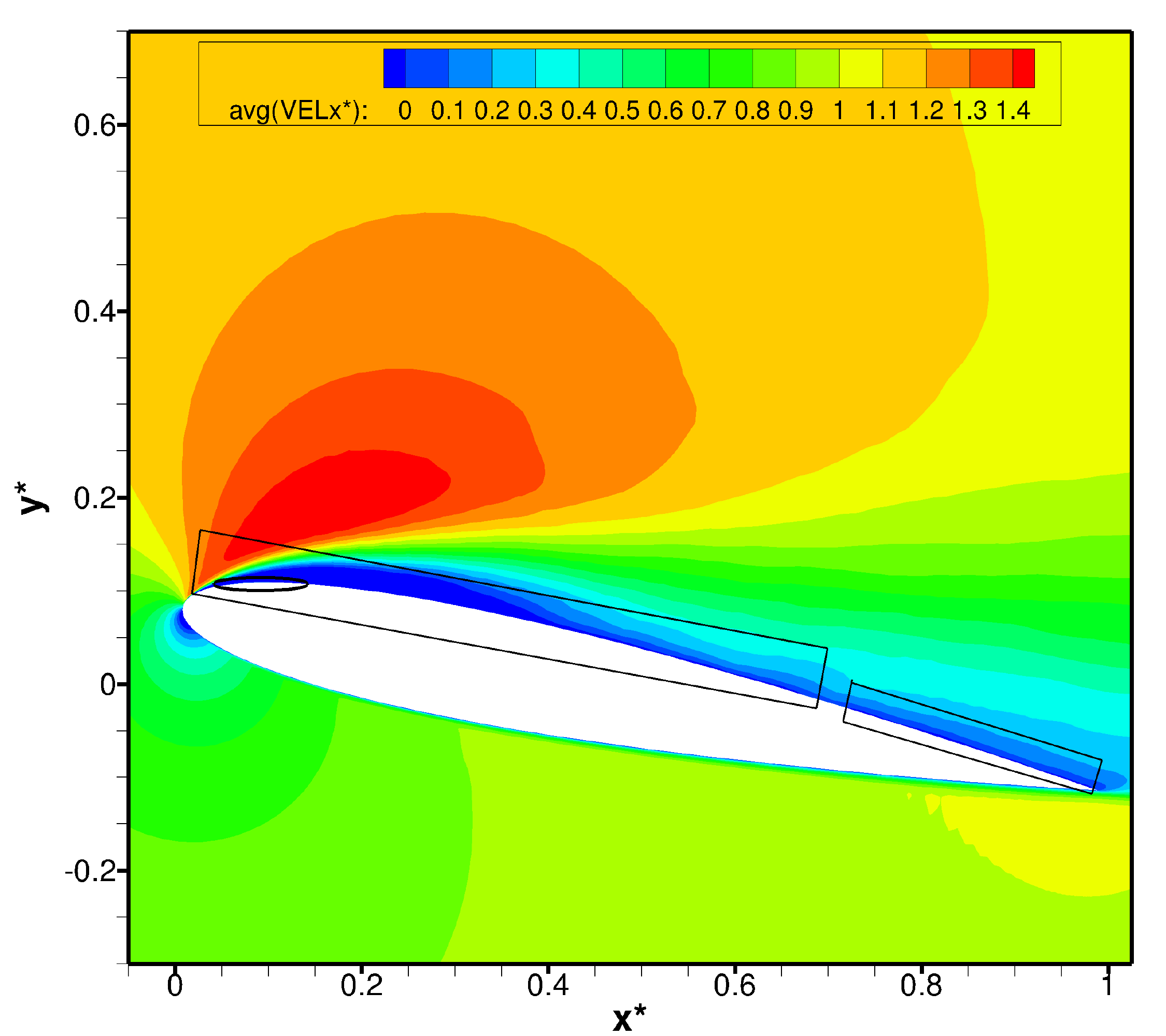}\label{fig:flow_domain_medium_grid_e-5_angle_11_focus_airfoil}}\hfill
   	\subfigure[$m-11-y^+_{max}$ at $t^*_{avg}=454$.]{\includegraphics[width=0.49\textwidth]{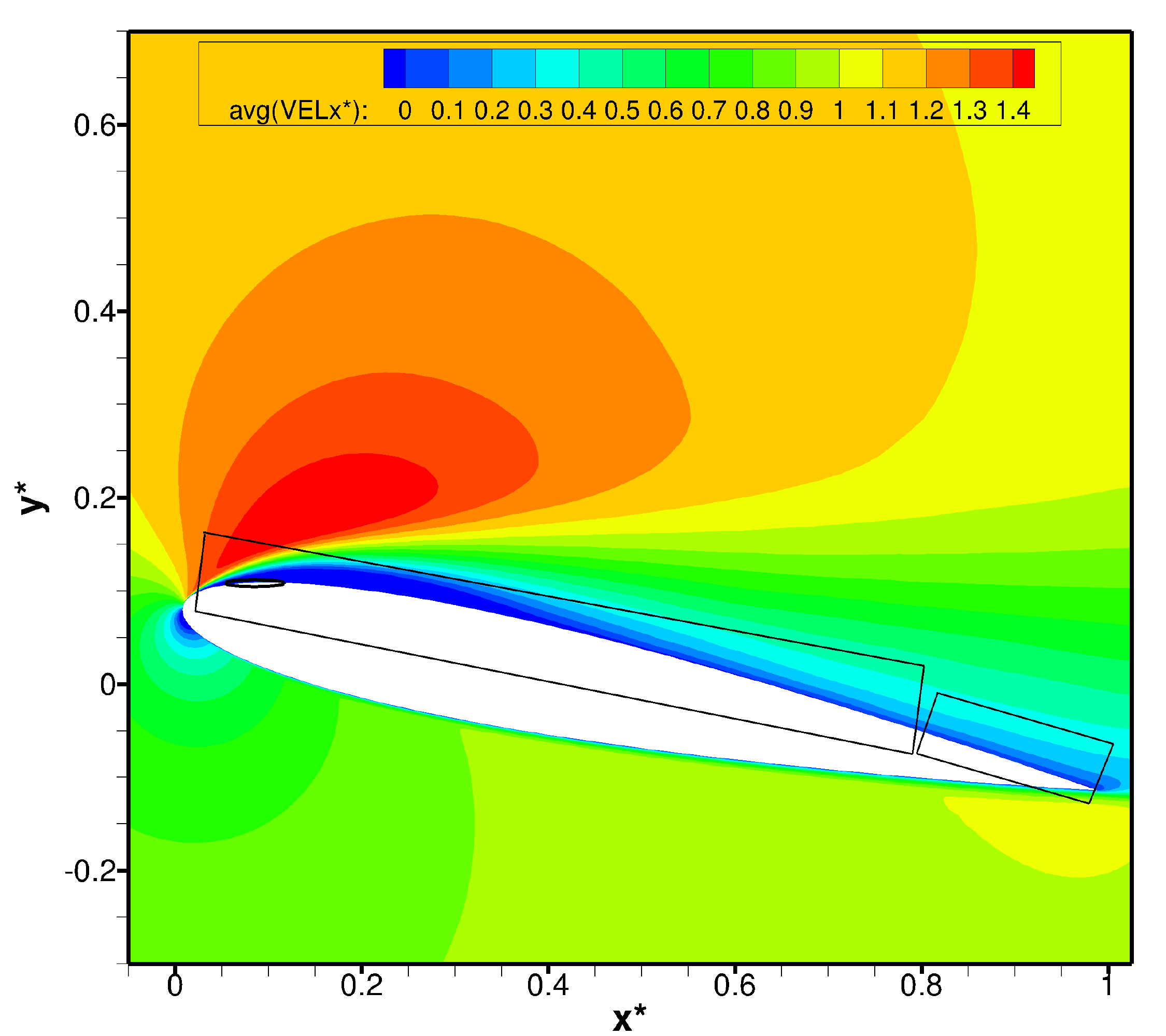}\label{fig:flow_domain_medium_grid_e-6_angle_11_focus_airfoil}}\hfill
\end{figure}

 Figures \ref{fig:flow_domain_medium_grid_e-5_angle_11_focus_airfoil} and \ref{fig:flow_domain_medium_grid_e-6_angle_11_focus_airfoil} illustrate the formation of two recirculation zones on the suction side (marked by rectangles) and also the formation of a counter-wise recirculation zone inside the first recirculation zone, i.e$.$ inside the laminar separation bubble, (marked by ellipses) for the $m-11-y^+_{med}$ and $m-11-y^+_{max}$ meshes, respectively (see \mbox{Section \ref{sec:flow_summary}}).
\par The laminar separation bubble is contained within about $ 0.0208c\leq x \leq 0.6647c$ for the $m-11-y^+_{med}$ mesh and about $ 0.0208c\leq x \leq 0.7721c$ for the $m-11-y^+_{max}$ grid, while the second recirculation zone appears at about $ 0.7184c\leq x \leq 0.9828c$ for the $m-11-y^+_{med}$ mesh and at about $ 0.8090c\leq x \leq 0.9848c$ for the $m-11-y^+_{max}$ grid.
\par
\begin{figure}[H]
   	\subfigure[$m-11-y^+_{med}$ at $t^*_{avg}=76$.]{\includegraphics[width=0.49\textwidth]{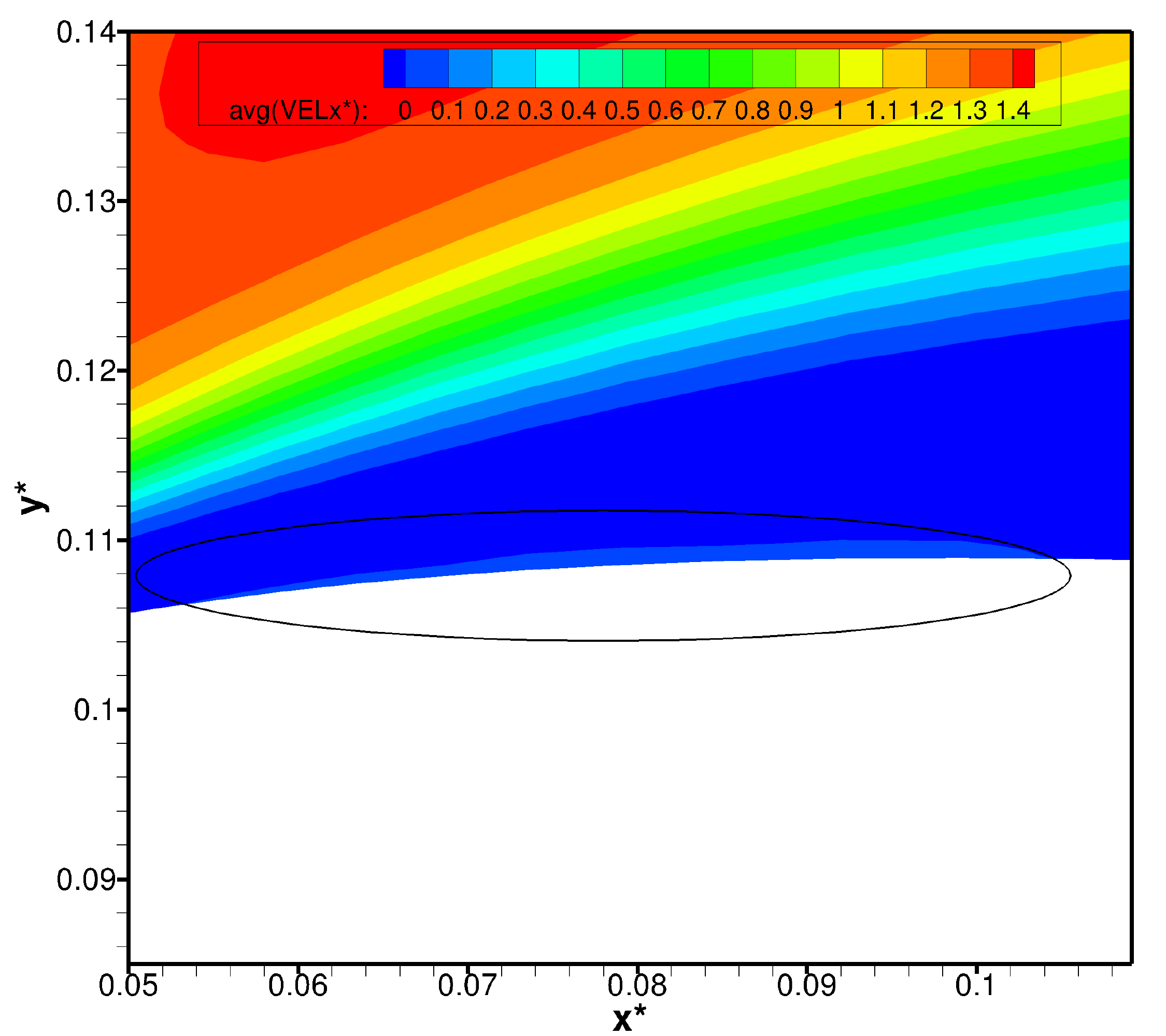}\label{fig:flow_domain_medium_grid_e-5_angle_11_focus_airfoil_zoom}}\hfill
   	\subfigure[$m-11-y^+_{max}$ at $t^*_{avg}=454$.]{\includegraphics[width=0.49\textwidth]{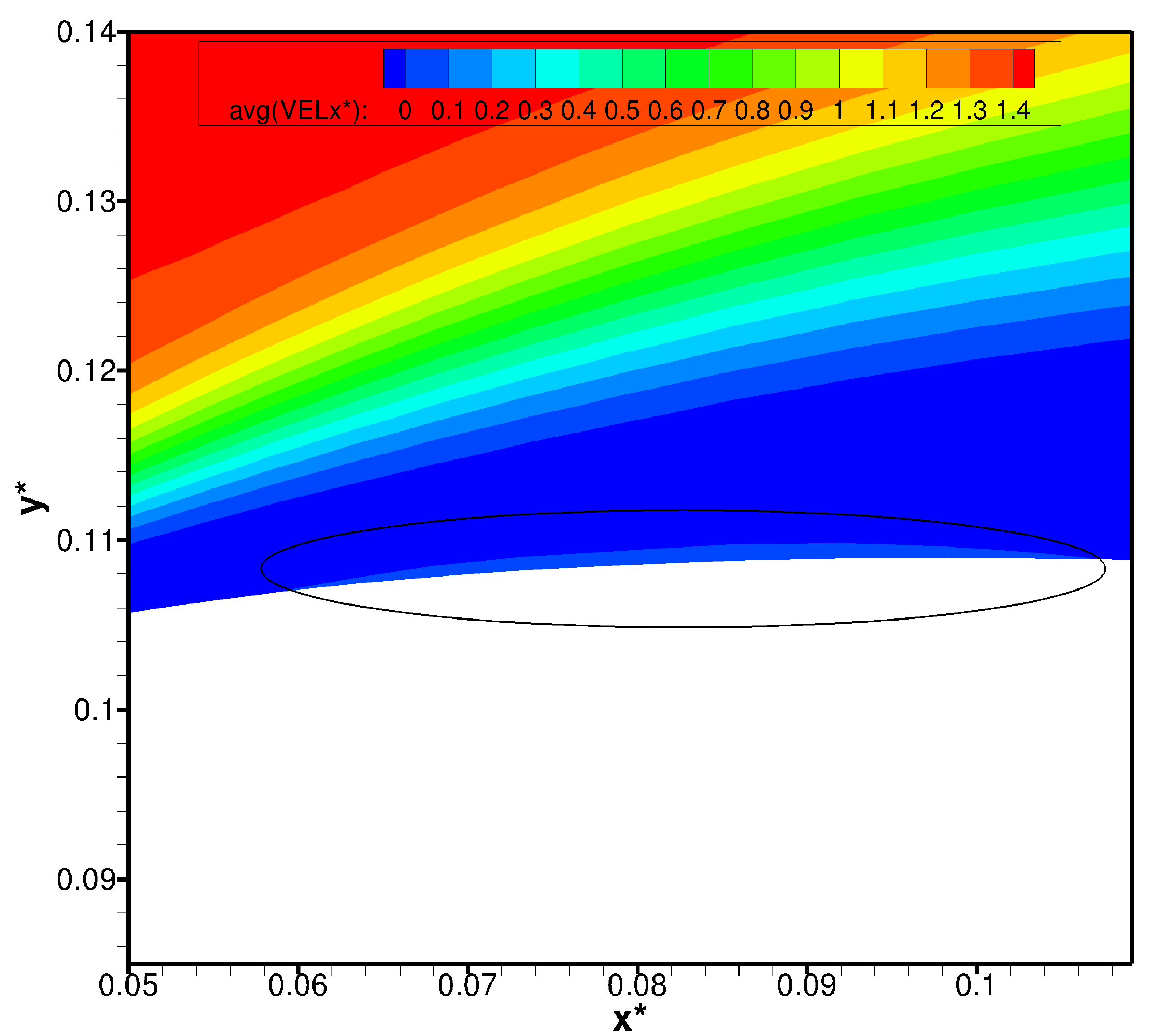}\label{fig:flow_domain_medium_grid_e-6_angle_11_focus_airfoil_zoom}}\hfill
   	\caption{Time-averaged dimensionless velocity distribution in the main flow direction for the meshes with an angle of attack of $\alpha=11^\circ$ (results are spatial-averaged in the span-wise direction). Focus on the airfoil and related phenomena.}
   	\label{fig:velocities_angle_11_focus_airfoil}	
\end{figure}

 The counter-wise recirculation zone formed inside the first laminar separation bubble is shown by Figs.\ \ref{fig:flow_domain_medium_grid_e-5_angle_11_focus_airfoil_zoom} and \ref{fig:flow_domain_medium_grid_e-6_angle_11_focus_airfoil_zoom}. This occurs within about $0.0516c\leq x \leq 0.1068c$ for the $m-11-y^+_{med}$ mesh and about $0.0590c\leq x \leq 0.1068c$ for the $m-11-y^+_{max}$ grid.
\par A divergence in the detachment and reattachment points of the boundary layer is present on the results of both meshes with an angle of attack of $\alpha=11^\circ$. These divergences may be not only influenced by the different analyzed dimensionless averaging times and therefore, the mesh may influence the achieved results.

 

\par Further investigations of the Reynolds stresses and the detachment and reattachment points of the boundary layer are also carried out in order to acquire more information about the divergences in the achieved results for both meshes with $\alpha=11^\circ$ (see \mbox{Sections \ref{sec:reynolds_stresses} and \ref{sec:flow_summary}}).
 
\markboth{CHAPTER 3.$\quad$RES. AND DISC.}{3.3$\quad$AN. OF THE TIME-AVG. STREAMLINES}
\section{Analysis of the time-averaged streamlines}\markboth{CHAPTER 3.$\quad$RES. AND DISC.}{3.3$\quad$AN. OF THE TIME-AVG. STREAMLINES}
\label{sec:flow_summary}

\par The flow around the NACA0012 airfoil at a Reynolds number of $Re=100{,}000$ is laminar and characterized by the detachment of the boundary layer and the formation of recirculation zones. These phenomena are thoroughly investigated based on the streamlines, which are tangent to the spatial and time-averaged velocity vector at every point in the flow at the analyzed dimensionless averaging time $t^*_{avg}$.

\subsection{Angle of attack $\alpha=0^\circ$}
\label{sec:streamlines_angle_0}

\par
\begin{figure}[H]
	\centering
	\subfigure[Overview of the streamlines: $m-0-y^+_{max}$, $t^*_{avg}=1009$.]{\includegraphics[width=0.44\textwidth]{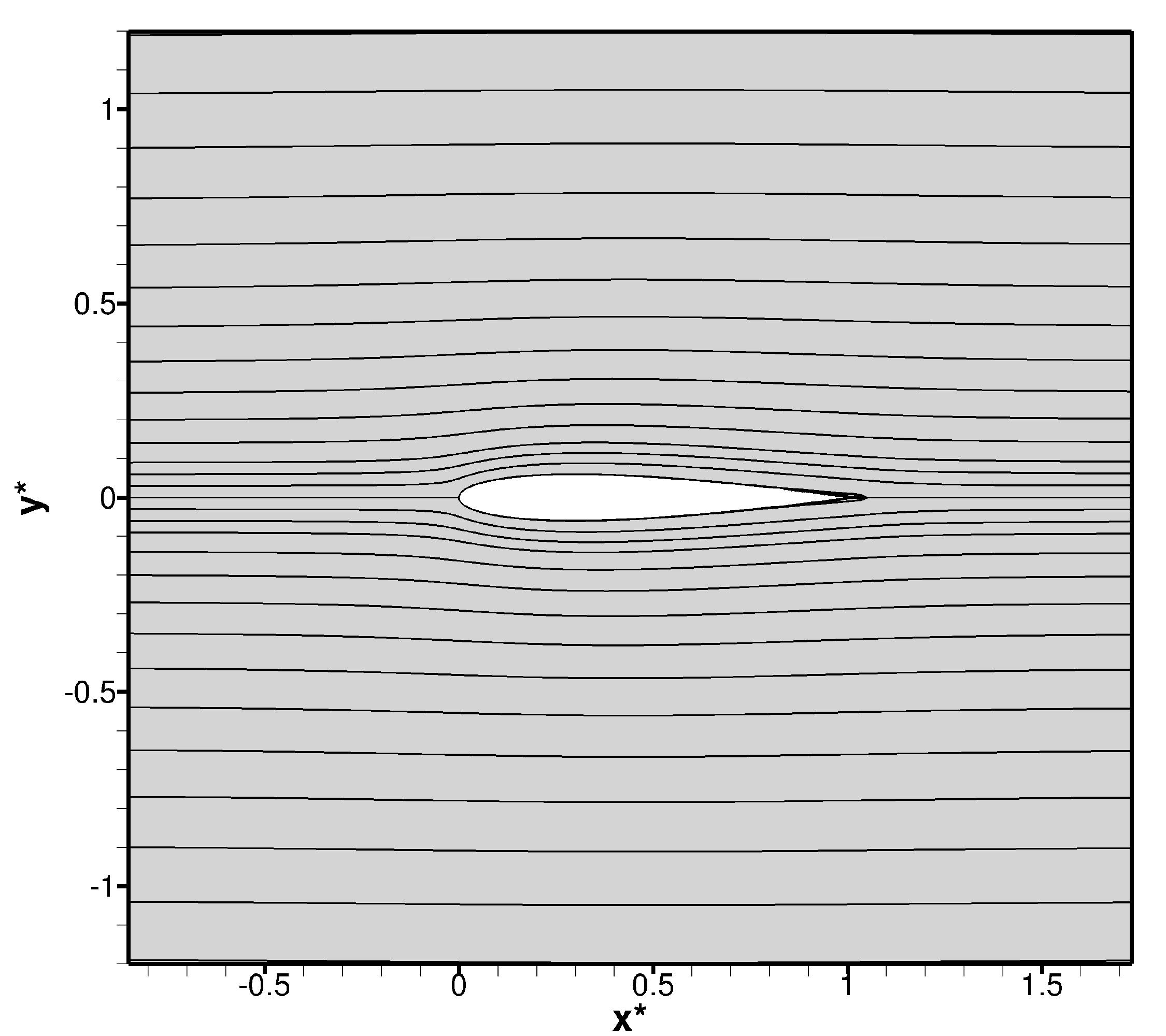}\label{fig:streamlines_e-5_angle_0}}\hfill
	\subfigure[Overview of the streamlines: $m-0-y^+_{med}$, $t^*_{avg}=322$.]{\includegraphics[width=0.44\textwidth]{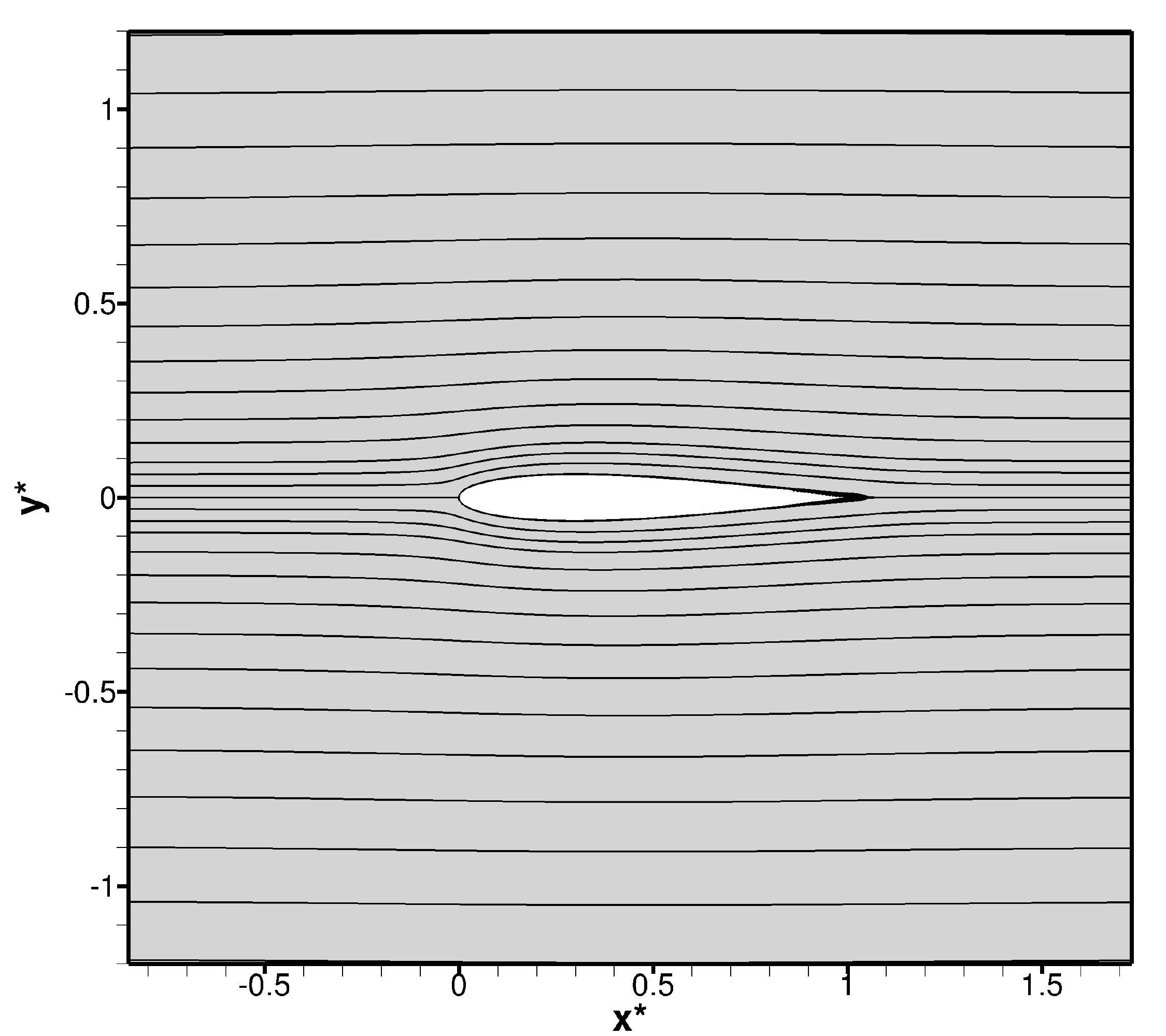}\label{fig:streamlines_e-6_angle_0}}\hfill
	\subfigure[Streamlines focused on the detachment of the boundary layer: $m-0-y^+_{max}$, $t^*_{avg}=1009$.]{\includegraphics[width=0.44\textwidth]{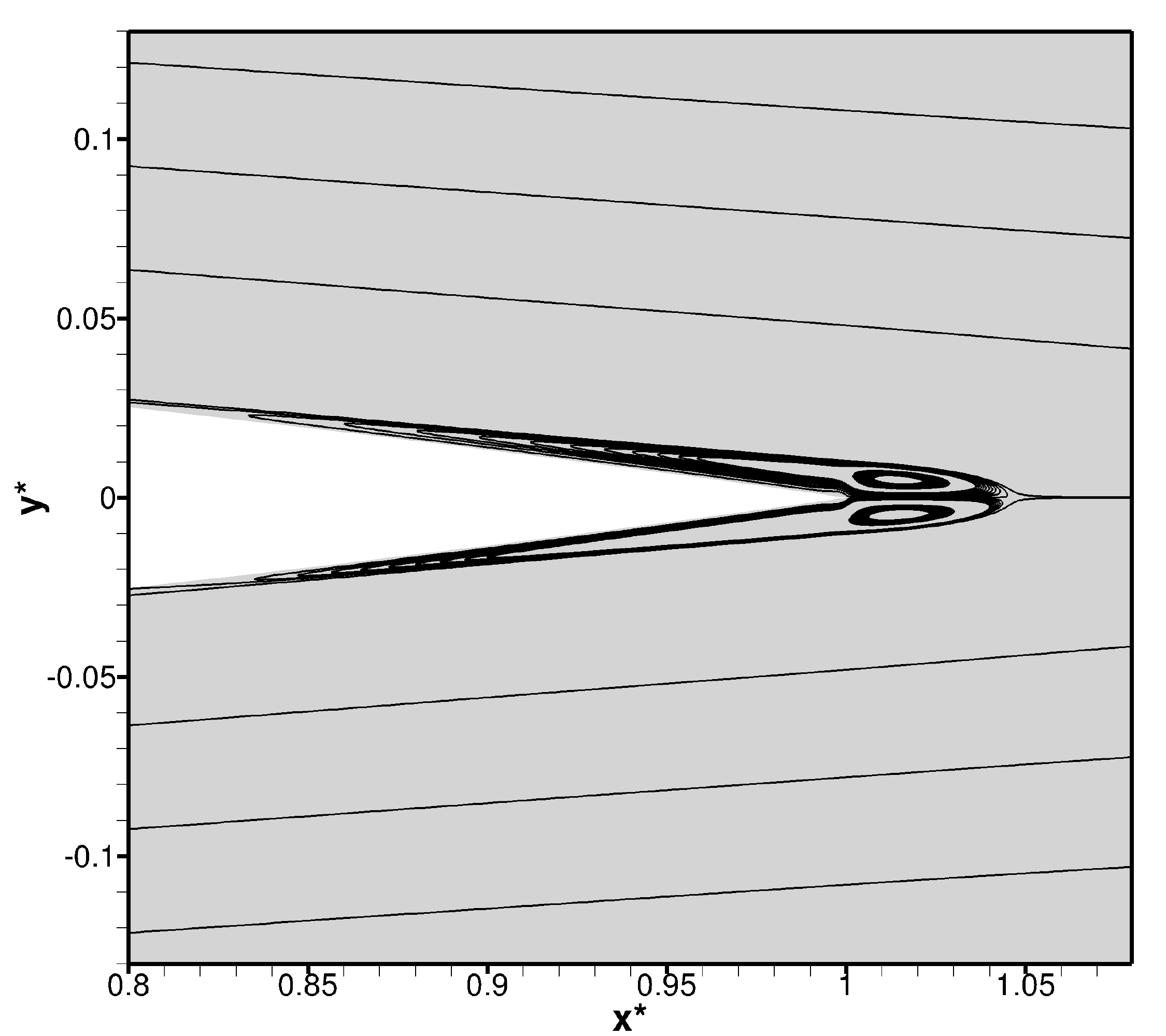}\label{fig:streamlines_e-5_angle_0_zoom}}\hfill
	\subfigure[Streamlines focused on the detachment of the boundary layer: $m-0-y^+_{med}$, $t^*_{avg}=322$.]{\includegraphics[width=0.44\textwidth]{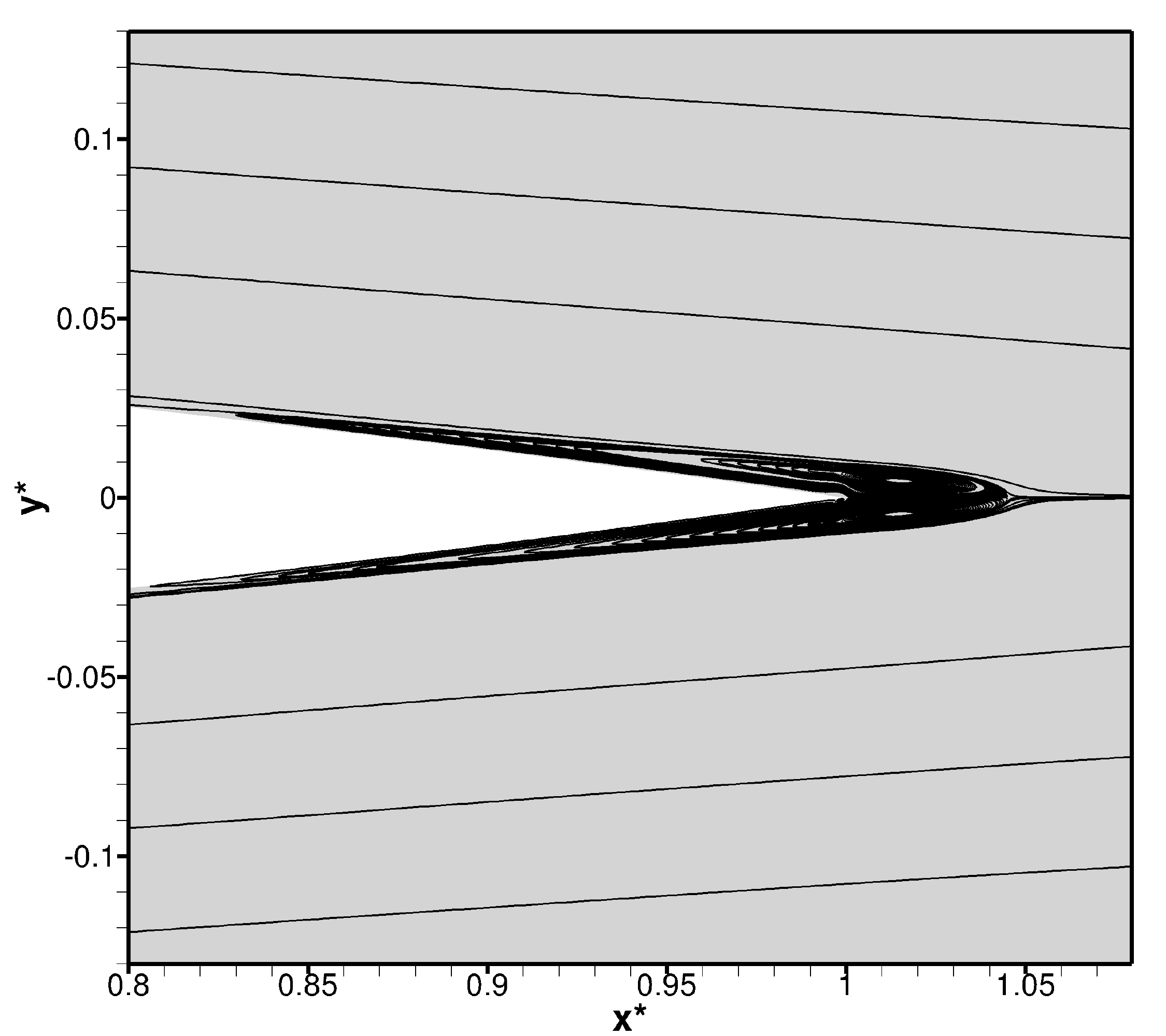}\label{fig:streamlines_e-6_angle_0_zoom}}\hfill	\caption{Time-averaged streamlines of the medium resolution grids at an angle of attack of $\alpha=0^\circ$ (results are spatial-averaged in the span-wise direction).}
	\label{fig:streamlines_angle_of_attack_0}
\end{figure}

 Figure \ref{fig:streamlines_angle_of_attack_0} illustrates the streamlines of the flow around the NACA0012 profile at an angle of attack of $\alpha=0^\circ$ for the $m-0-y^+_{max}$ and $m-0-y^+_{med}$ meshes. The streamlines of the fine resolution mesh, i.e$.$, $f-0-y^+_{min}$, are not investigated due to the missing time-averaged data.
\par The flow on the pressure and suction sides of the airfoil is symmetrical, as already mentioned in Section \ref{sec:simulation_analysis}. A detachment of the boundary layer occurs near the trailing edge on the suction and pressure sides of the airfoil, according to Figs.\ \ref{fig:streamlines_e-5_angle_0_zoom} and \ref{fig:streamlines_e-6_angle_0_zoom} for the $m-0-y^+_{max}$ and $m-0-y^+_{med}$ meshes, respectively. This is not reattached further downstream and is responsible for the creation of the von K\'arm\'an vortex street in the wake (see Section \ref{sec:simulation_analysis_instationary}).
\par Table \ref{table:detachment_points_angle_0} shows the location of the detachment points on the suction ($x_{DP,\,SS}$) and pressure sides ($x_{DP,\,PS}$) of the airfoil for all three grids at an angle of attack of $\alpha=0^\circ$, regarding that the instantaneous data of the $f-0-y^+_{min}$ was previously investigated in Section \ref{sec:simulation_analysis}. The results of this grid also showed the formation of two recirculation zones prior to the detachment of the boundary layer (see Fig.\ \ref{fig:flow_near_profiel_fine_grid_e-6_angle_0}). These zones, however, are attributed to the instantaneous and not yet fully developed state of the fluid and therefore these detachment and reattachment points are not present in \mbox{Table \ref{table:detachment_points_angle_0}}. 
\begin{table}[H]
	\centering
	\begin{tabular}{c c}
		\hline
		\bf{Mesh} & \bf{Location of the detachment points} \tabularnewline \hline
		\multirow{2}{*}{$f-0-y^+_{min}$} & $x_{DP,\,SS}=0.854c$ \tabularnewline
		& $x_{DP,\,PS}=0.863c$ \tabularnewline \cline{2-2}
		\multirow{2}{*}{$m-0-y^+_{med}$} & $x_{DP,\,SS}=0.800c$ \tabularnewline
		& $x_{DP,\,PS}=0.800c$ \tabularnewline \cline{2-2}
		\multirow{2}{*}{$m-0-y^+_{max}$} & $x_{DP,\,SS}=0.800c$ \tabularnewline
		& $x_{DP,\,PS}=0.800c$ \tabularnewline
		\hline	
	\end{tabular}
	\caption{\label{table:detachment_points_angle_0}Location of the detachment points of the grids at an angle of attack of $\alpha=0^\circ$.}
\end{table}
\par The location of the detachment points are only different for the fine resolution mesh, which is also caused by the instantaneous and not yet fully developed state of the fluid. For both medium resolution meshes, however, these points are located at the same position.

\subsection{Angle of attack $\alpha=5^\circ$}
\label{sec:streamlines_angle_5}
\par The location of the detachment and reattachment points of the meshes at an angle of attack of $\alpha=5^\circ$ are investigated according to the streamlines. 
\par
\begin{figure}[H]
	\subfigure[Overview of the streamlines: $m-5-y^+_{max}$, $t^*_{avg}=1039$.]{\includegraphics[width=0.44\textwidth]{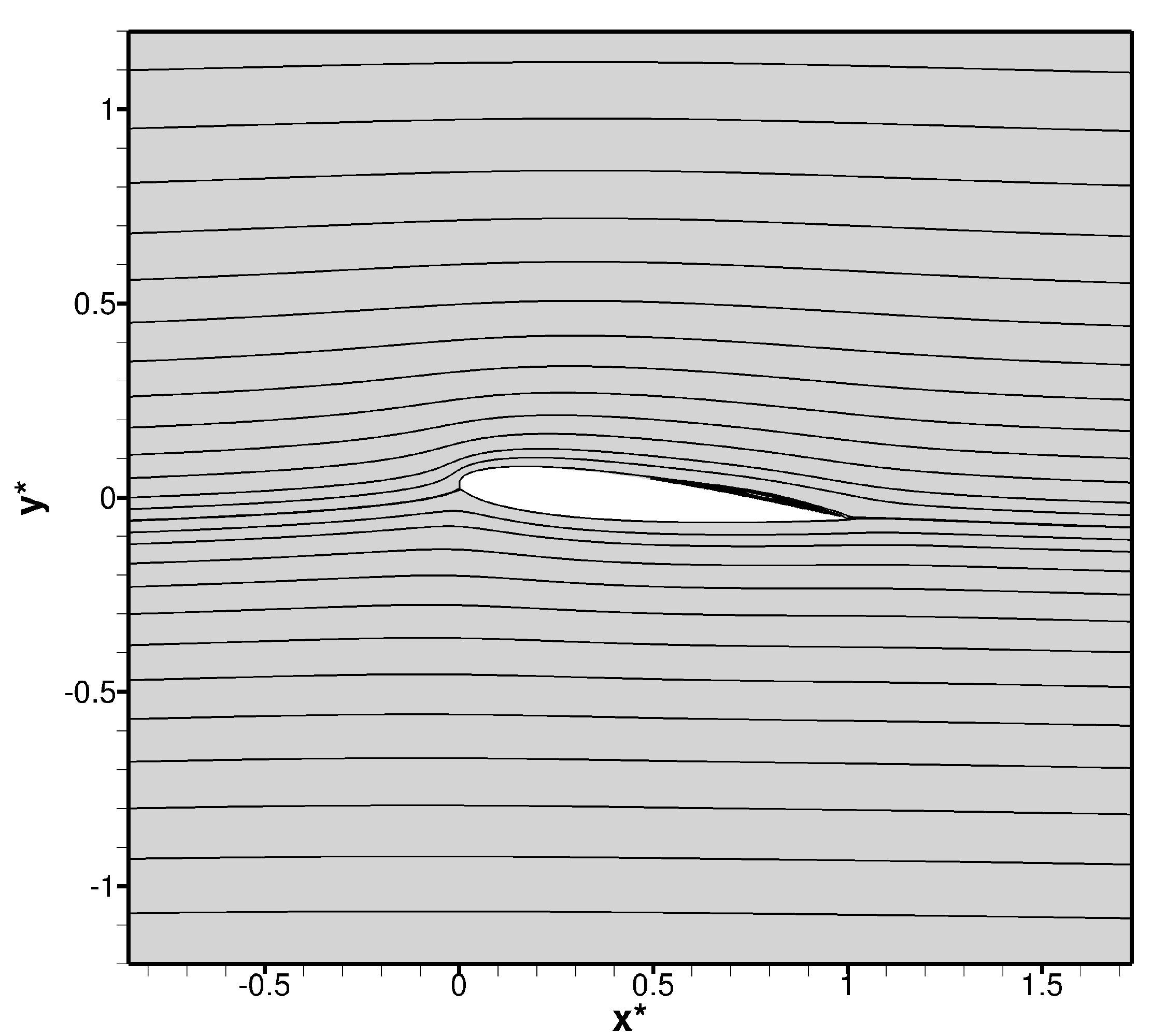}\label{fig:streamlines_e-5_angle_5}}\hfill
	\subfigure[Overview of the streamlines: $m-5-y^+_{med}$, $t^*_{avg}=152$.]{\includegraphics[width=0.44\textwidth]{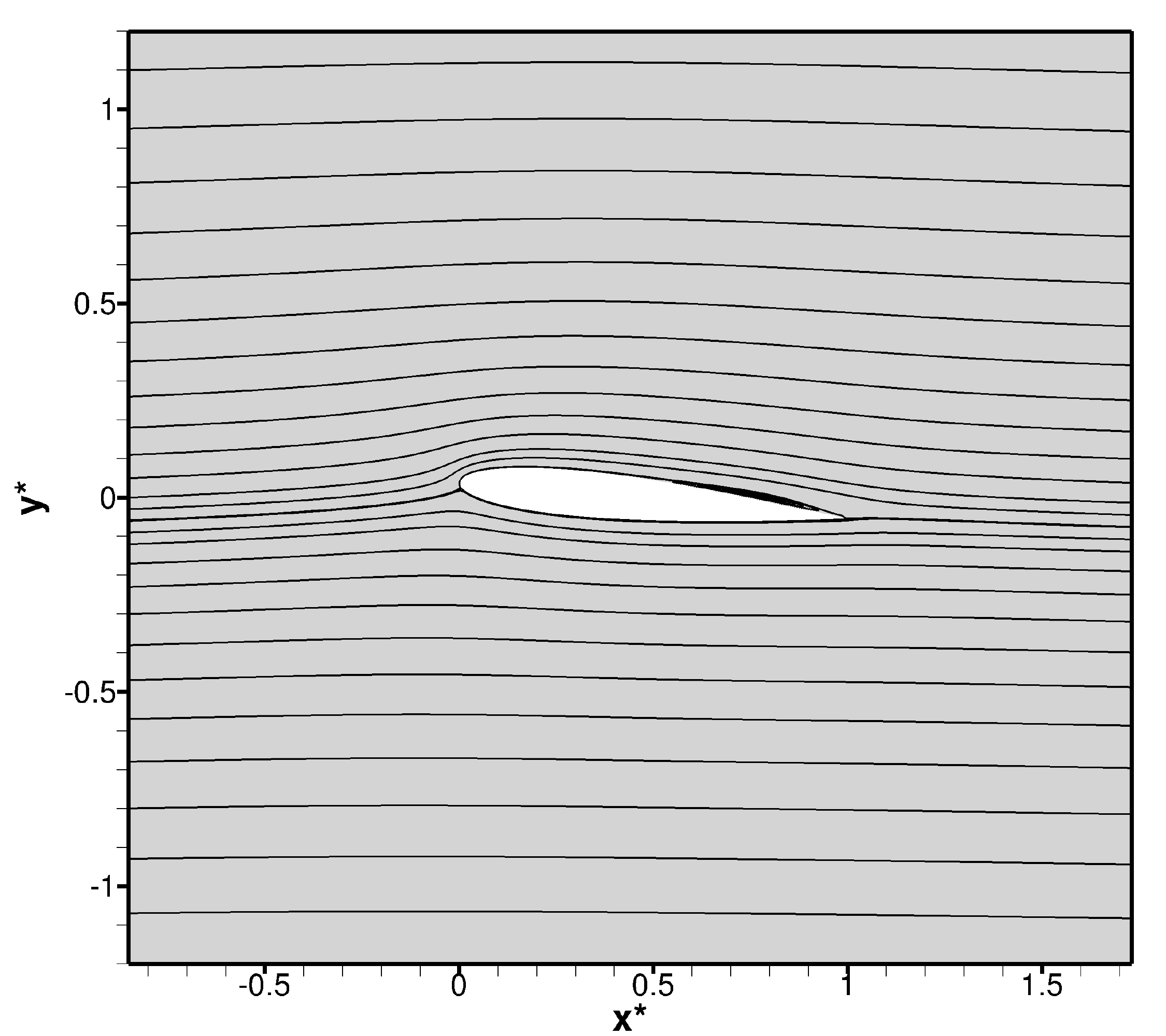}\label{fig:streamlines_e-6_angle_5}}\hfill
\end{figure}

\begin{figure}
	\subfigure[Streamlines focused on the recirculation zone: $m-5-y^+_{max}$, $t^*_{avg}=1039$.]{\includegraphics[width=0.44\textwidth]{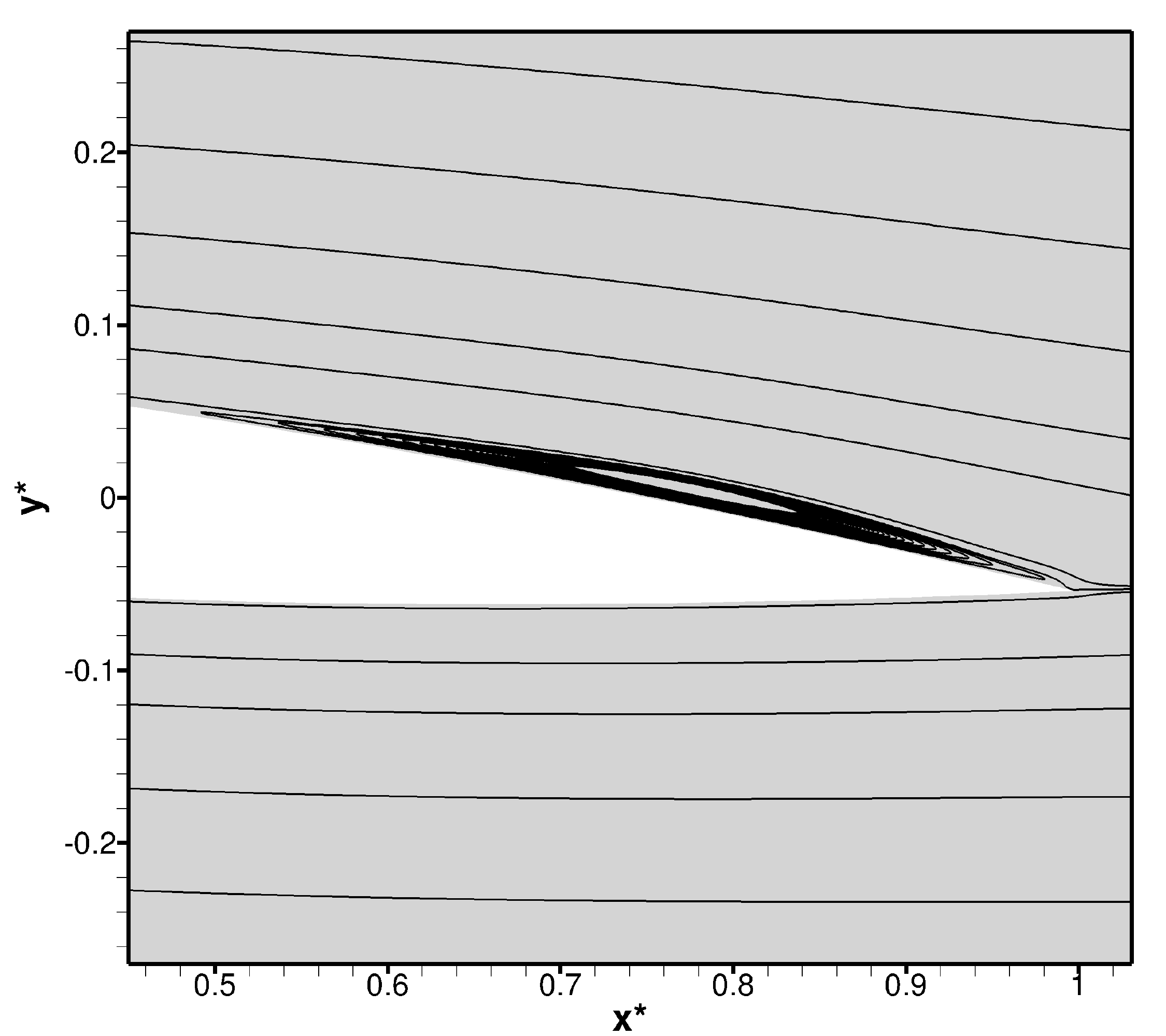}\label{fig:streamlines_e-5_angle_5_zoom}}\hfill
	\subfigure[Streamlines focused on the recirculation zone: $m-5-y^+_{med}$, $t^*_{avg}=152$.]{\includegraphics[width=0.44\textwidth]{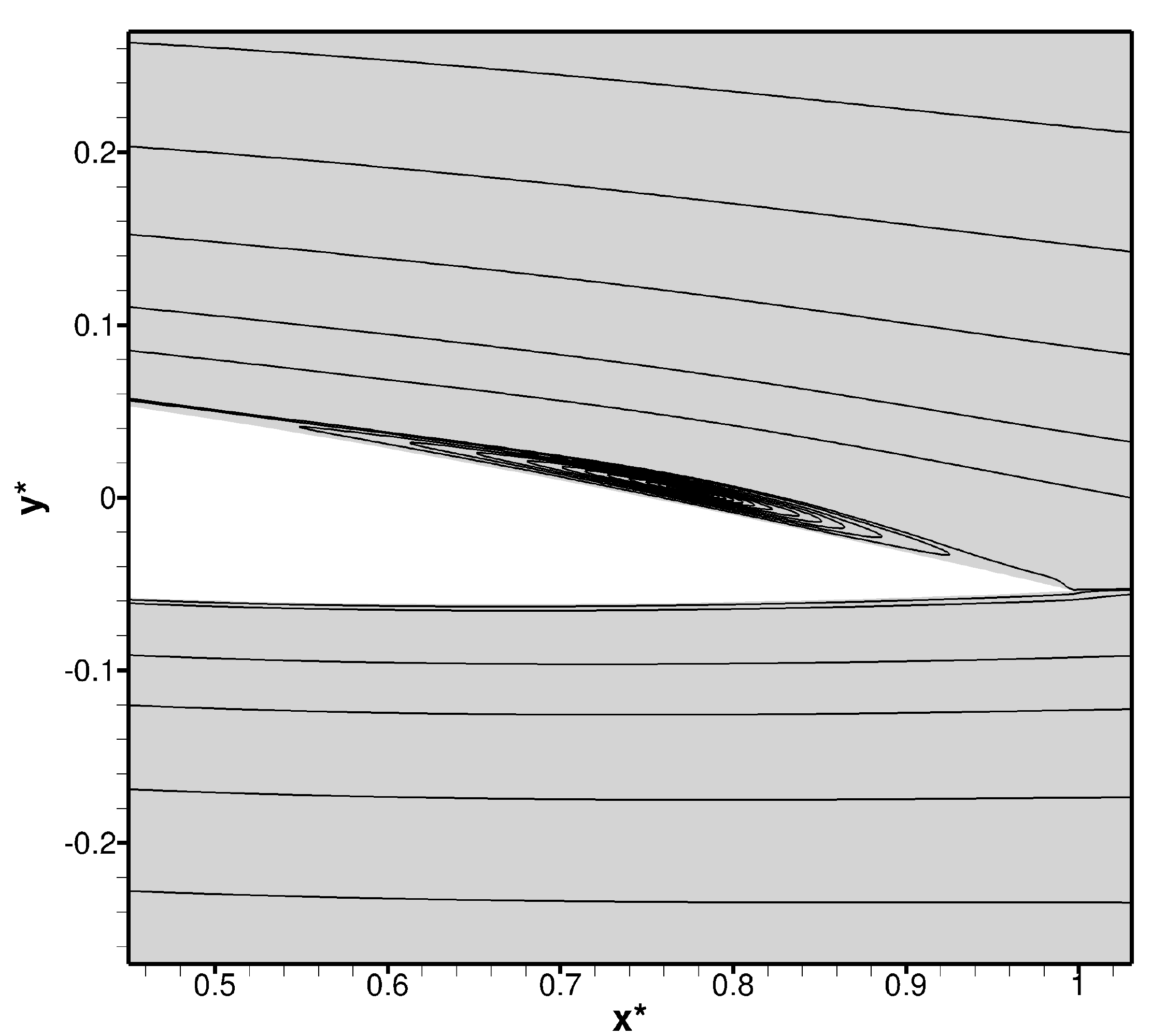}\label{fig:streamlines_e-6_angle_5_zoom}}\hfill
	\caption{Time-averaged streamlines of the medium resolution meshes at an angle of attack of $\alpha=5^\circ$ (results are spatial-averaged in the span-wise direction).}
	\label{fig:streamlines_angle_of_attack_5}
\end{figure}

 Figures \ref{fig:streamlines_e-5_angle_5} and \ref{fig:streamlines_e-6_angle_5} illustrate the overview of the streamlines. A recirculation zone near the trailing edge is present for both meshes on the suction side, as shown in Figs.\ \ref{fig:streamlines_e-5_angle_5_zoom} and \ref{fig:streamlines_e-6_angle_5_zoom}. This zone is characterized by a detachment ($x_{DP,\,SS}$) and a reattachment point ($x_{RP,\,SS}$) on the suction side (see also Section \ref{sec:simulation_analysis}), which are summarized in Table \ref{table:detachment_points_angle_5}. 
\par The location of the detachment point of the boundary layer is slightly different for both investigated meshes. The recirculation zone for the $m-0-y^+_{max}$ grid starts before the recirculation zone for the $m-0-y^+_{med}$ mesh, while it ends at the same position for both meshes.
\begin{table}[H]
	\centering
	\begin{tabular}{c p{4.5cm} p{4.5cm}}
		\hline
		\bf{Mesh} & \centering{\bf{Location of the detachment points}} & \centering{\bf{Location of the reattachment points}} \tabularnewline \hline
		$m-5-y^+_{med}$ & \centering{$x_{DP,\,SS}=0.345c$} & \centering{$x_{RP,\,SS}=0.990c$} \tabularnewline
		$m-5-y^+_{max}$ & \centering{$x_{DP,\,SS}=0.331c$} & \centering{$x_{RP,\,SS}=0.990c$} \tabularnewline
		\hline	
	\end{tabular}
	\caption{\label{table:detachment_points_angle_5}Location of the detachment and reattachment points of the grids at an angle of attack of $\alpha=5^\circ$.}
\end{table}

\subsection{Angle of attack $\alpha=11^\circ$}
\label{sec:streamlines_angle_11}
\par
\begin{figure}[H]
	\centering
	\subfigure[Overview of the streamlines: $m-11-y^+_{max}$, $t^*_{avg}=454$.]{\includegraphics[width=0.435\textwidth]{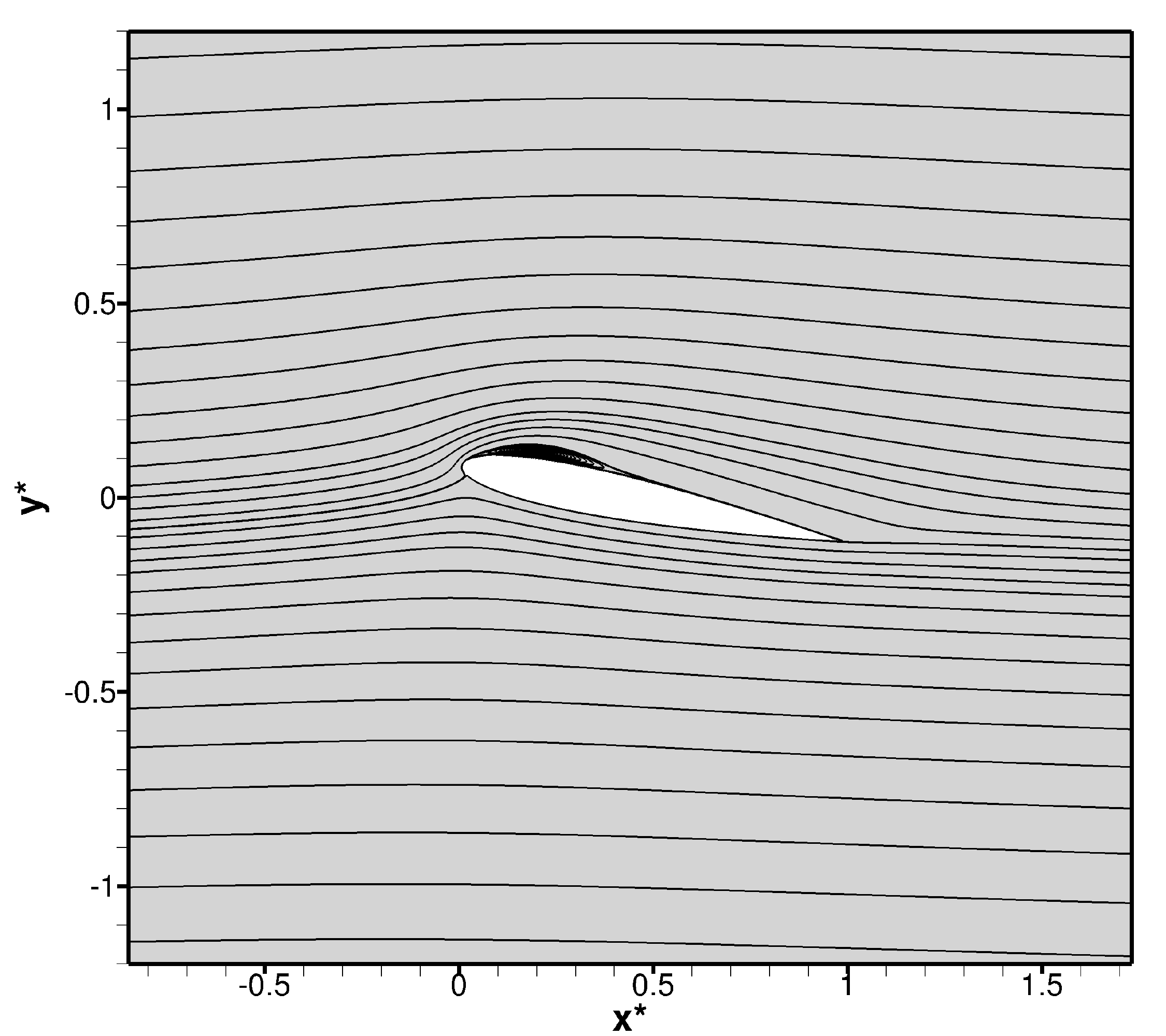}\label{fig:streamlines_e-5_angle_11}}\hfill
	\subfigure[Overview of the streamlines: $m-11-y^+_{med}$, $t^*_{avg}=76$.]{\includegraphics[width=0.435\textwidth]{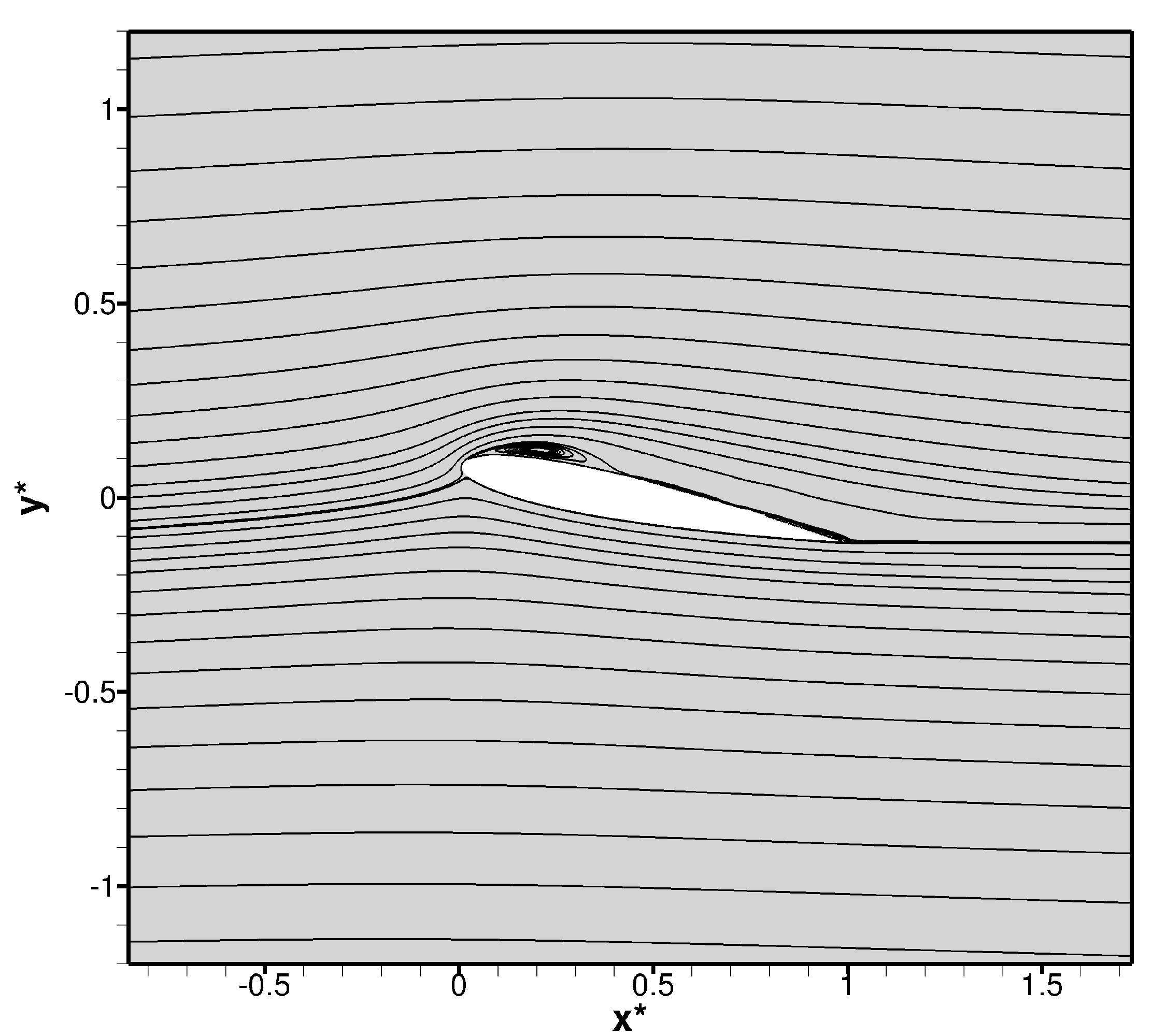}\label{fig:streamlines_e-6_angle_11}}\hfill
	\subfigure[Streamlines focused on the recirculation zones: $m-11-y^+_{max}$, $t^*_{avg}=454$.]{\includegraphics[width=0.435\textwidth]{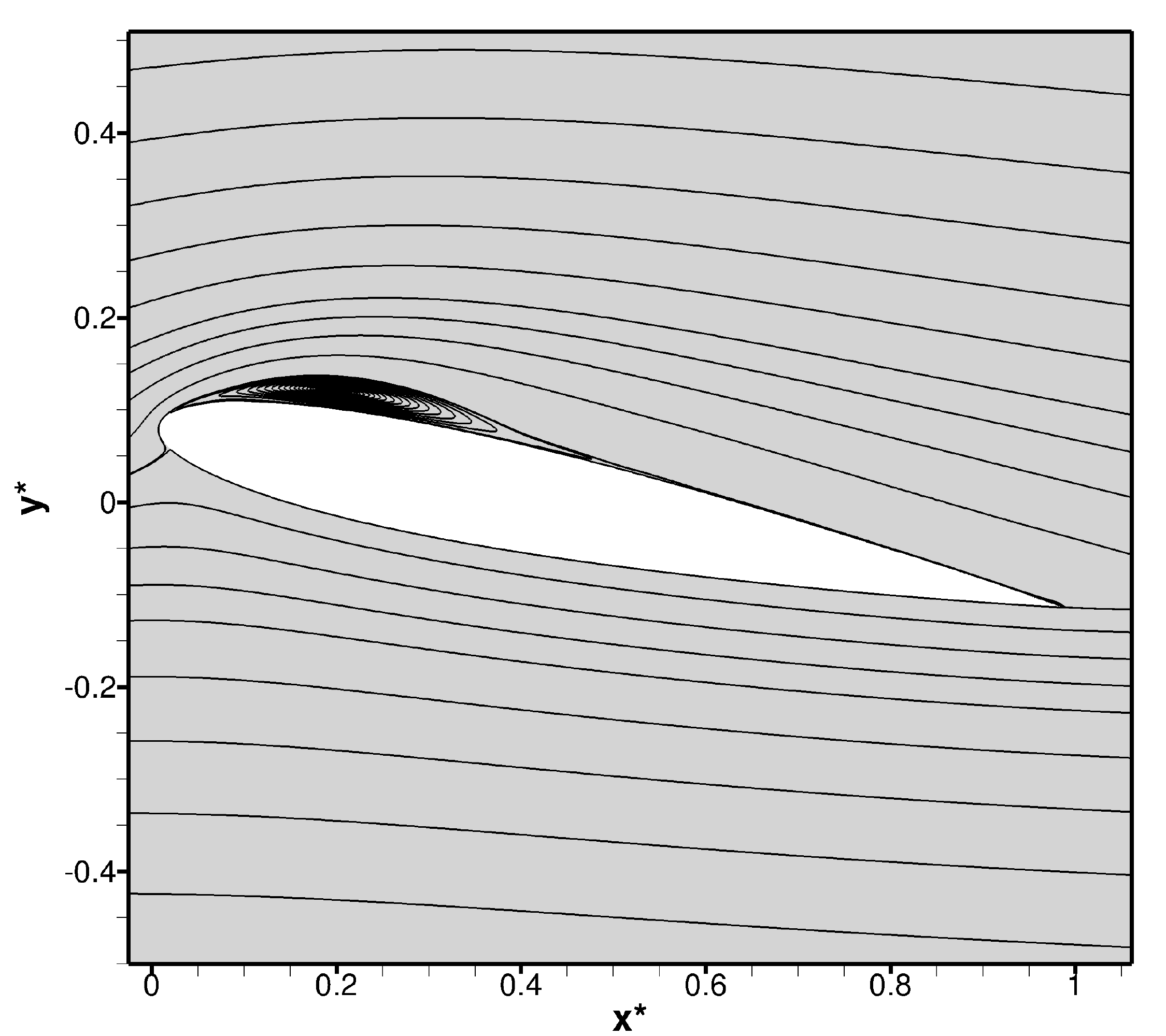}\label{fig:streamlines_e-5_angle_11_zoom}}\hfill
	\subfigure[Streamlines focused on the recirculation zones: $m-11-y^+_{med}$, $t^*_{avg}=76$.]{\includegraphics[width=0.435\textwidth]{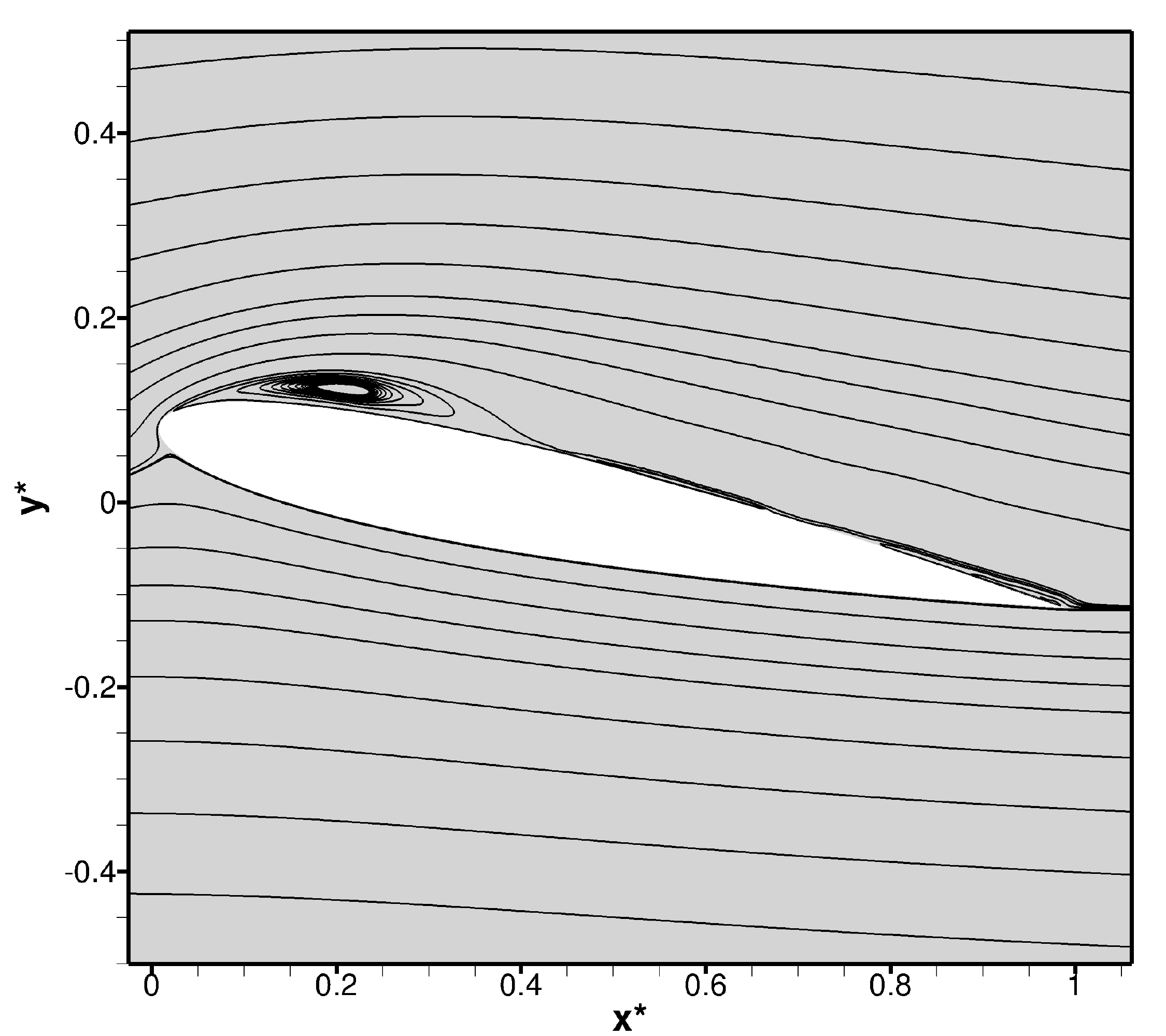}\label{fig:streamlines_e-6_angle_11_zoom}}\hfill
	\subfigure[Streamlines focused on the trailing edge: $m-11-y^+_{max}$, $t^*_{avg}=454$.]{\includegraphics[width=0.435\textwidth]{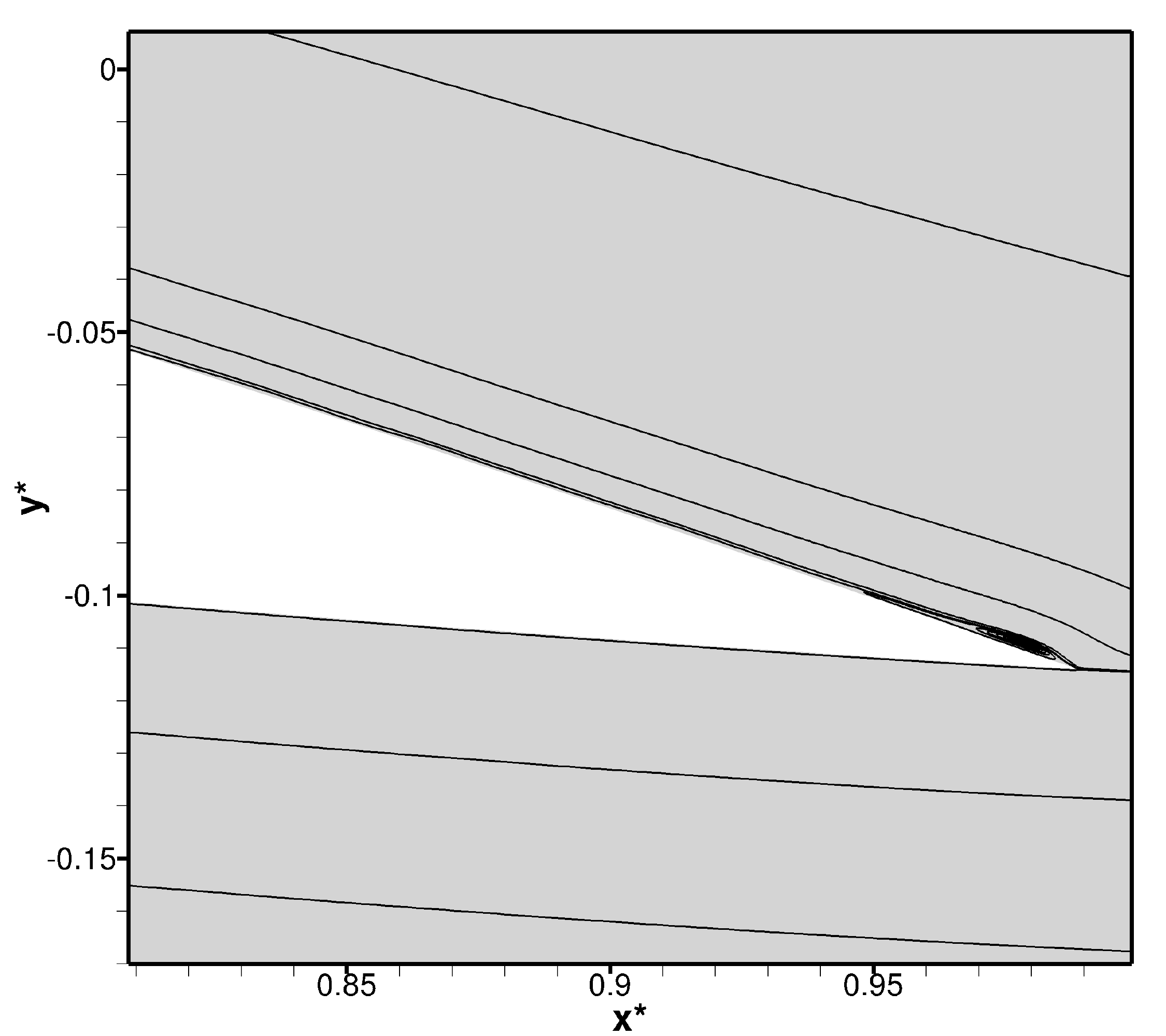}\label{fig:streamlines_e-5_angle_1_zoom_end}}\hfill
	\subfigure[Streamlines focused on the trailing edge: $m-11-y^+_{med}$, $t^*_{avg}=76$.]{\includegraphics[width=0.435\textwidth]{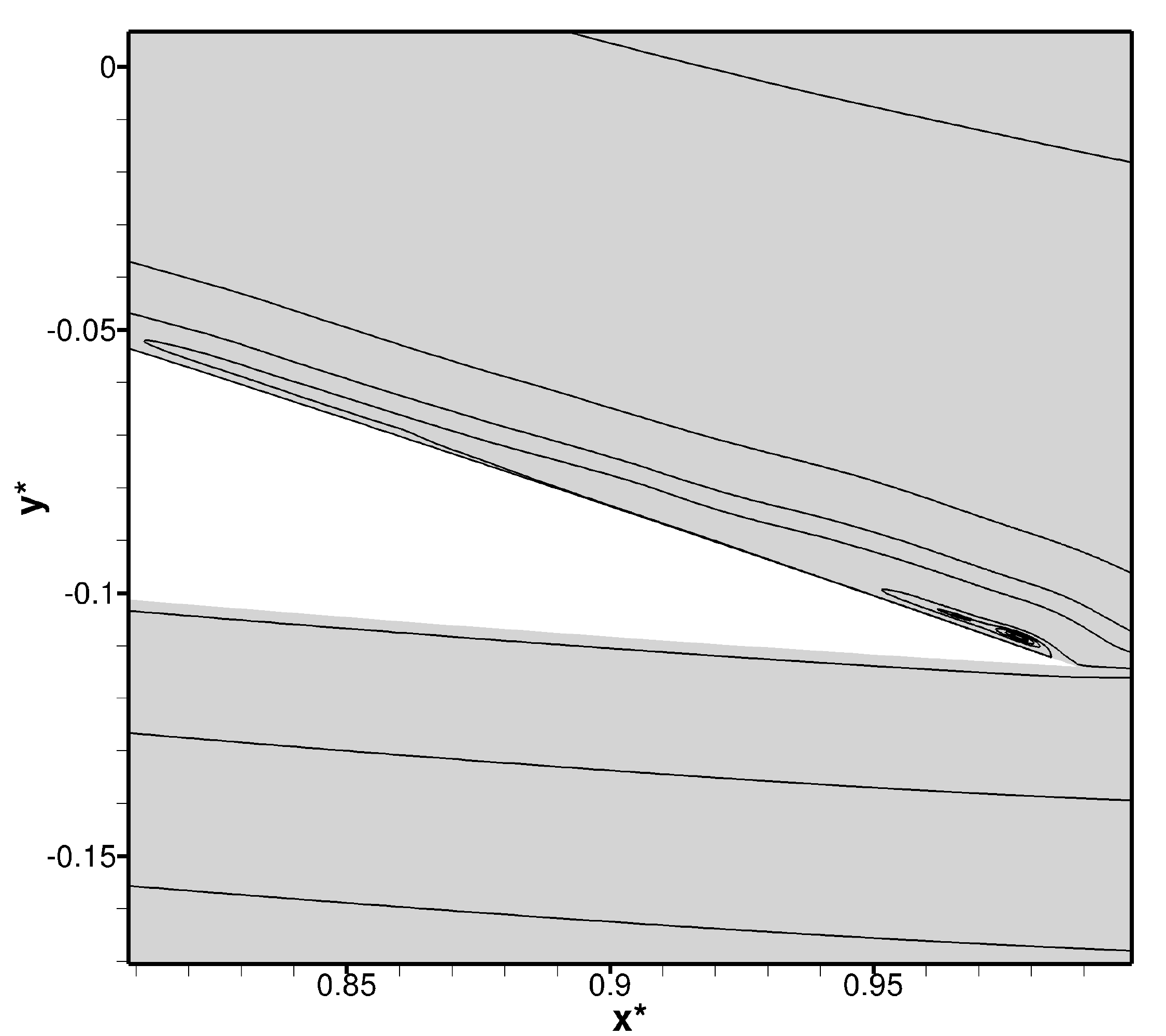}\label{fig:streamlines_e-6_angle_11_zoom_end}}\hfill
\end{figure}

 The flow around the NACA0012 airfoil at an angle of attack of $\alpha=11^\circ$ is characterized by the formation of two main recirculation zones on the suction side: one starts near the leading edge while the other starts on the second half of the profile, as shown in \mbox{Figs.\ \ref{fig:streamlines_e-5_angle_11} through \ref{fig:streamlines_e-6_angle_11_zoom_end}} for the $m-11-y^+_{max}$ and $m-11-y^+_{max}$ meshes. The recirculation zone, i.e$.$, the laminar separation bubble, formed near the leading edge is responsible for the modification of the effective shape of the airfoil and therefore may negatively affect the aerodynamics of this profile.
\par
\begin{figure}[H]
	\subfigure[Streamlines focused on the counter-wise recirculation zone: $m-11-y^+_{max}$, $t^*_{avg}=454$.]{\includegraphics[width=0.435\textwidth]{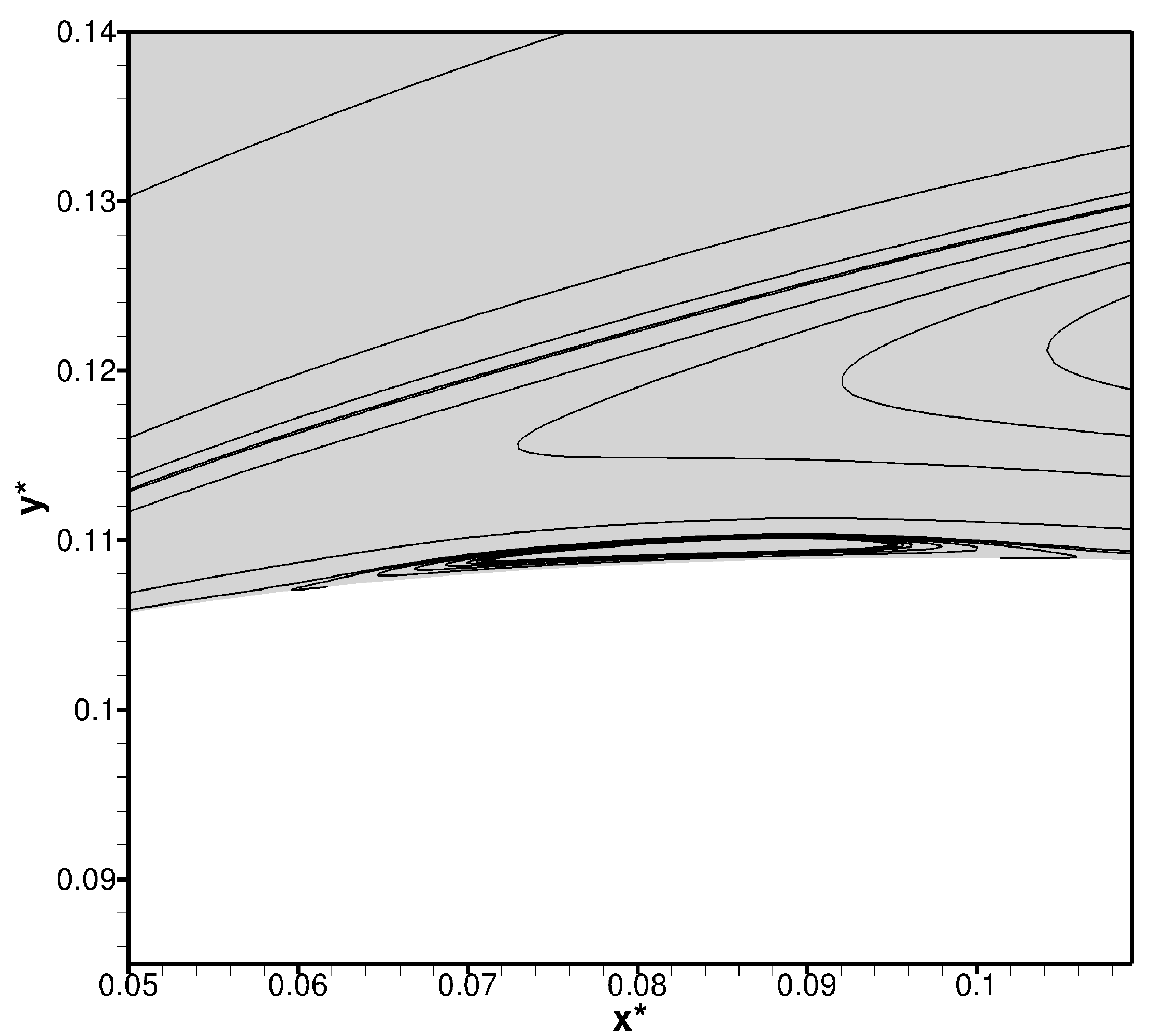}\label{fig:streamlines_e-5_angle_1_zoom_start}}\hfill
	\centering
	\subfigure[Streamlines focused on the counter-wise recirculation zone: $m-11-y^+_{med}$, $t^*_{avg}=76$.]{\includegraphics[width=0.435\textwidth]{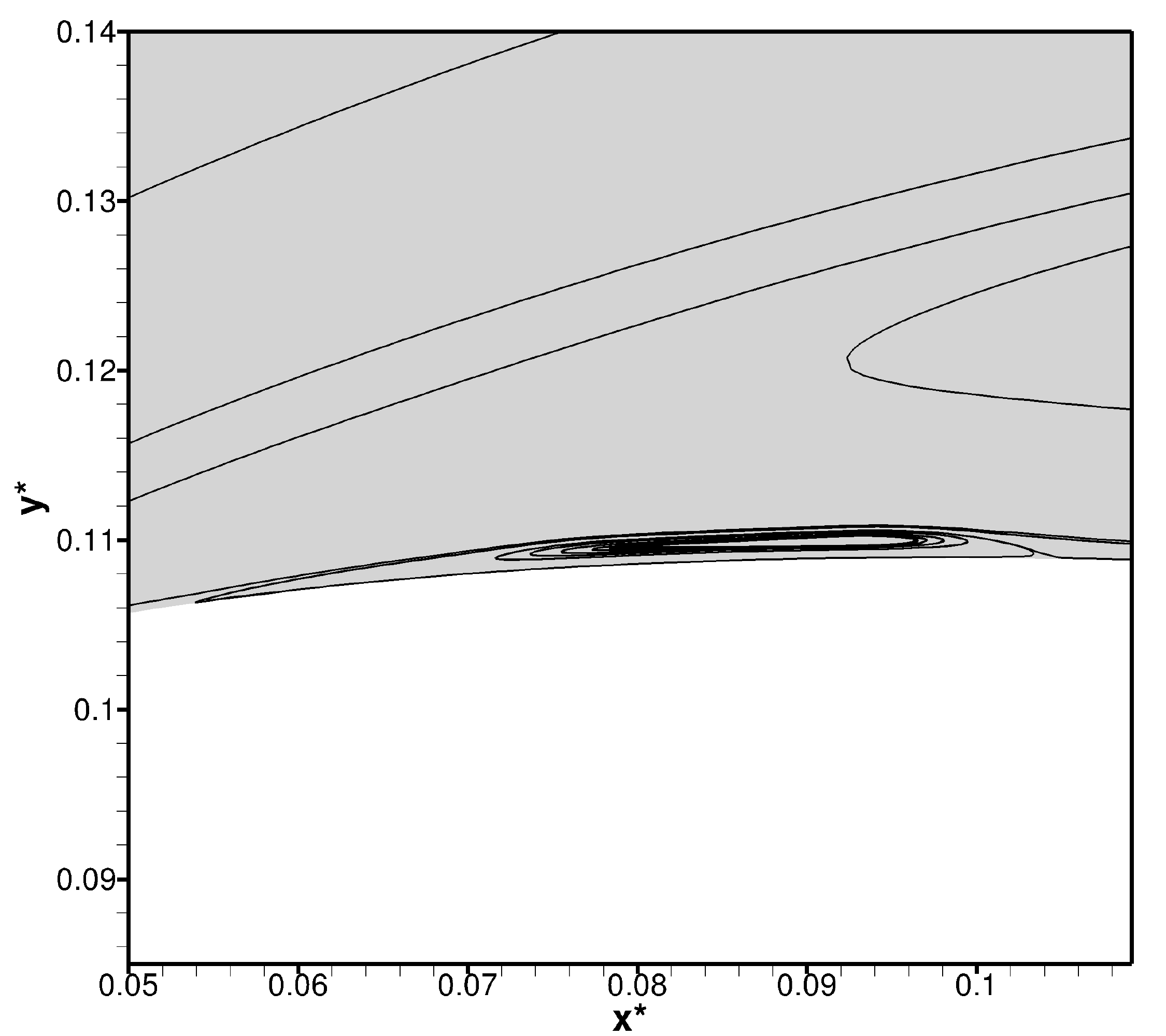}\label{fig:streamlines_e-6_angle_11_zoom_start}}\hfill	
	\caption{Time-averaged streamlines of the medium meshes with $\alpha=11^\circ$ (results are spatial-averaged in the span-wise direction).}
	\label{fig:streamlines_angle_of_attack_11}
\end{figure}

 Inside the laminar separation bubble, a counter-wise recirculation zone is also formed, as illustrated in Figs.\ \ref{fig:streamlines_e-5_angle_1_zoom_start} and \ref{fig:streamlines_e-6_angle_11_zoom_start}. The flow does not suffer any detachment on the pressure side.
\par Aiming at the comparison between the meshes at an angle of attack of $\alpha=11^\circ$, \mbox{Table \ref{table:detachment_points_angle_11}} illustrates the detachment and reattachment points of the laminar separation bubble ($x_{DP1,\,SS}$ and $x_{RP1,\,SS}$), the detachment and reattachment points of the recirculation zone near the trailing edge ($x_{DP2,\,SS}$ and $x_{RP2,\,SS}$), as well as the start and end points of the counter-wise recirculation zone located inside the laminar separation bubble ($x_{CR_S,\,SS}$ and $x_{CR_E,\,SS}$).

\begin{table}[H]
	\centering
	\begin{tabular}{c c c}
		\hline
		\bf{Mesh} & \centering{\bf{$DP1,\,SS$}} & \centering{\bf{$RP1,\,SS$}}  \tabularnewline \hline
		$m-11-y^+_{med}$ & \centering{$x_{DP1,\,SS}=0.0208c$} & \centering{$x_{RP1,\,SS}=0.6647c$} \tabularnewline
		$m-11-y^+_{max}$ & \centering{$x_{DP1,\,SS}=0.0208c$} & \centering{$x_{RP1,\,SS}=0.7721c$} \tabularnewline \cline{2-3}
		& \centering{\bf{$DP2,\,SS$}} & \centering{\bf{$RP2,\,SS$}} \tabularnewline \cline{2-3}
		$m-11-y^+_{med}$ & \centering{$x_{DP2,\,SS}=0.7184c$} & \centering{$x_{RP2,\,SS}=0.9828c$} \tabularnewline
		$m-11-y^+_{max}$ & \centering{$x_{DP2,\,SS}=0.8090c$} & \centering{$x_{RP2,\,SS}=0.9848c$} \tabularnewline \cline{2-3}
		& \centering{\bf{$CR_S,\,SS$}} & \centering{\bf{$CR_E,\,SS$}} \tabularnewline \cline {2-3}
		$m-11-y^+_{med}$ & \centering{$x_{CR_S,\,SS}=0.0516c$} & \centering{$x_{CR_E,\,SS}=0.1068c$} \tabularnewline	
		$m-11-y^+_{max}$ & \centering{$x_{CR_S,\,SS}=0.0590c$} & \centering{$x_{CR_E,\,SS}=0.1068c$} \tabularnewline
		\hline	
	\end{tabular}
	\caption{\label{table:detachment_points_angle_11}Location of the detachment and reattachment points of the grids at an angle of attack of $\alpha=11^\circ$.}
\end{table}

\par The laminar separation bubble starts at the same position for both meshes, while its extension is smaller for the $m-0-y^+_{med}$ mesh. The recirculation zone formed near the trailing edge and the counter-wise recirculation zone formed inside the laminar separation bubble begin first for the $m-0-y^+_{med}$ grid. The end point of these both zones varies only slightly for the former zone and does not vary for the latter zone, when comparing both meshes.

\section{Analysis of the instationary flow}\markboth{CHAPTER 3.$\quad$RESULTS AND DISC.}{3.4$\quad$ANALYSIS OF THE INST. FLOW}
\label{sec:simulation_analysis_instationary}
\par The instationary flow is analyzed according to the instationary velocity field, the pressure fluctuation and the vorticity in the span-wise direction. It aims at a study of the wake regarding vortex sheddings and the formation of von K\'arm\'an vortex streets. Since the flow simulated with the fine resolution mesh ($f-0-y^+_{min}$) is not yet fully developed, this mesh is not taken into account.
\par The pressure fluctuation $p'$ is calculated according to Eq.\ (\ref{eq:pressure_fluctuation}) and the vorticity in the span-wise direction $\omega_z$ is calculated with Eq.\ (\ref{eq:span_wise_vorticity}).
\begin{eqnarray}
p'&=&p\;\;-<p> \label{eq:pressure_fluctuation} \\ 
\omega_z&=&\frac{\partial u_2}{\partial x_1}-\frac{\partial u_1}{\partial x_2} \label{eq:span_wise_vorticity}
\end{eqnarray} 

\subsection{Angle of attack $\alpha=0^\circ$}
\label{subsec:simulation_analysis_instationary_angle_0}

\par
\begin{figure}[H]
	\subfigure[Stream-wise velocity: $m-0-y^+_{med}$ mesh at $t^*=322$. Focus on the boundary layer detachment.]{\includegraphics[width=0.46\textwidth]{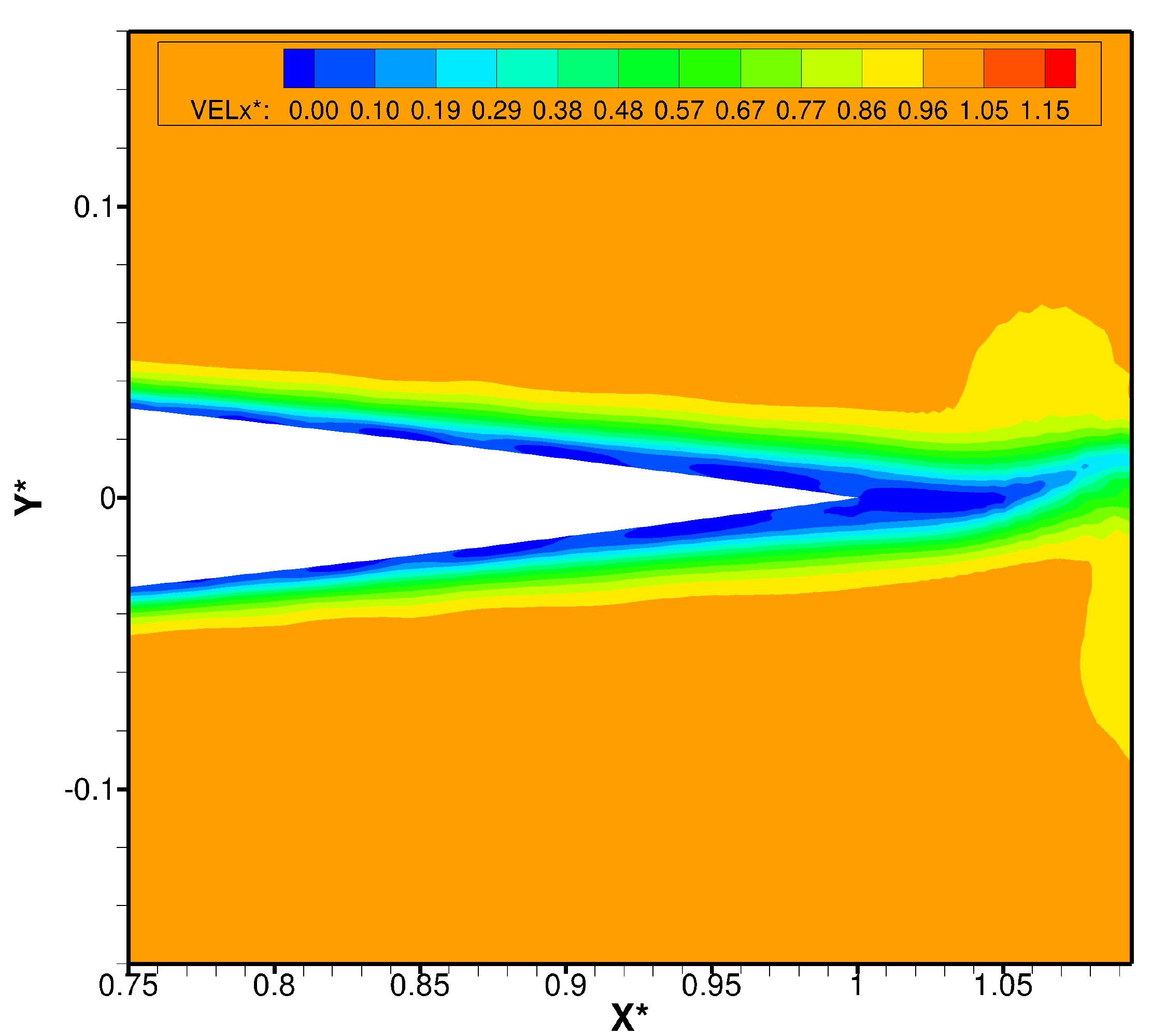}\label{fig:instationary_velocities_e-6_angle_0_zoom}}\hfill
	\subfigure[Stream-wise velocity: $m-0-y^+_{max}$ mesh at $t^*=1009$. Focus on the boundary layer detachment.]{\includegraphics[width=0.46\textwidth]{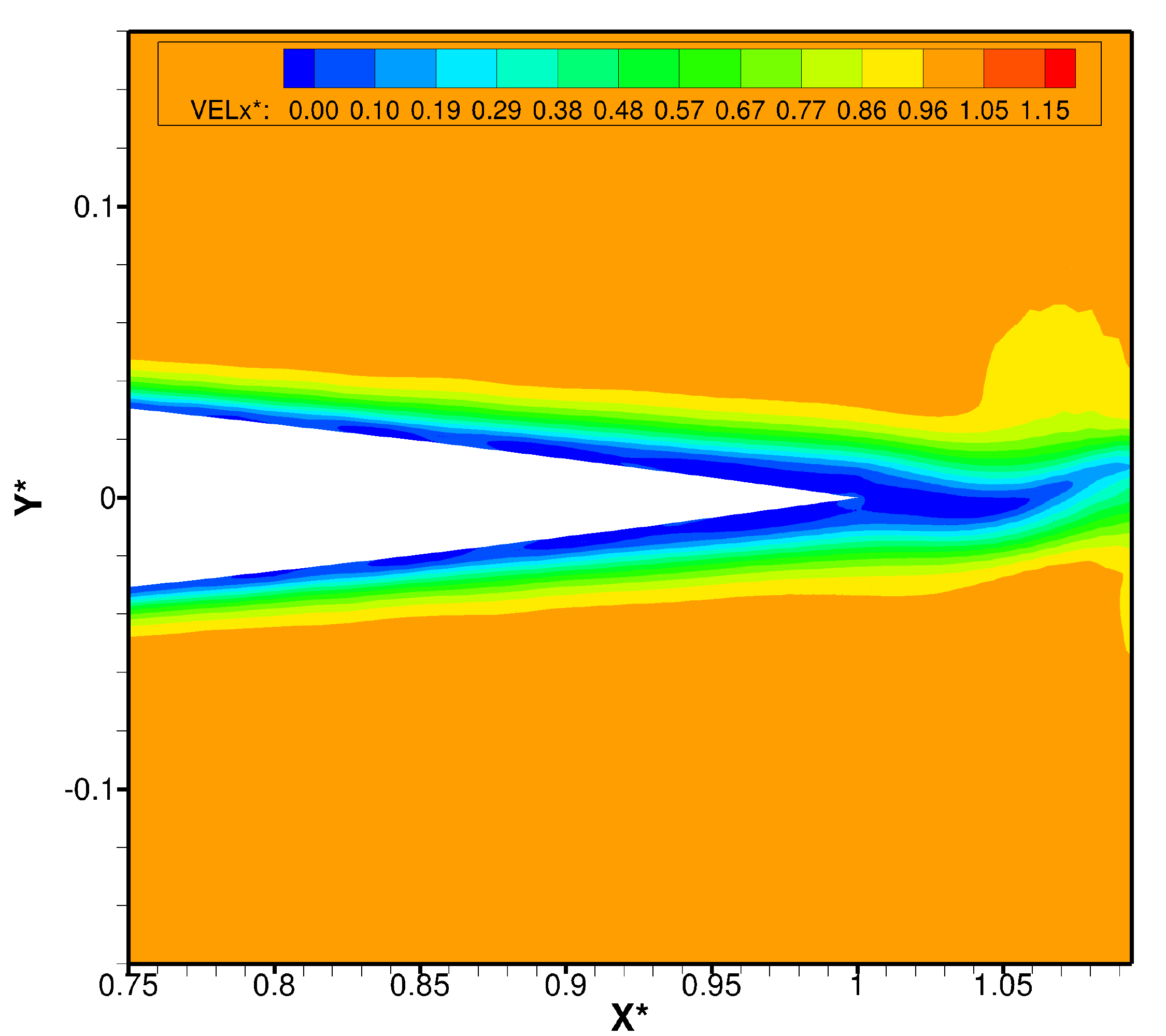}\label{fig:instationary_velocities_e-5_angle_0_zoom}}\hfill
	\caption{Instantaneous stream-wise velocity of the medium meshes at an incidence of $\alpha=0^\circ$.} \label{fig:instationary_velocities_medium_meshes_angle_0}
\end{figure}

 Figure \ref{fig:instationary_velocities_medium_meshes_angle_0} illustrates the instantaneous stream-wise velocity field of the medium meshes at an angle of attack of $\alpha=0^\circ$. Since the studied dimensionless times are different and the instationary velocity is analyzed, both meshes ($m-0-y^+_{med}$ and $m-0-y^+_{max}$) have different patterns for $x>0.8\,c$, point at which the boundary layer is detached (see Section \ref{sec:flow_summary}). Various separation bubbles are formed near the trailing edge for the instantaneous velocity field, while for the time and spatial-averaged velocity field, the flow is detached at $x=0.8\,c$ and does not reattach further downstream (see Section \ref{sec:simulation_analysis}).
\par
\begin{figure}[H]
	\centering
	\subfigure[Pressure fluctuations: $m-0-y^+_{med}$ mesh at $t^*=322$.]{\includegraphics[width=0.46\textwidth]{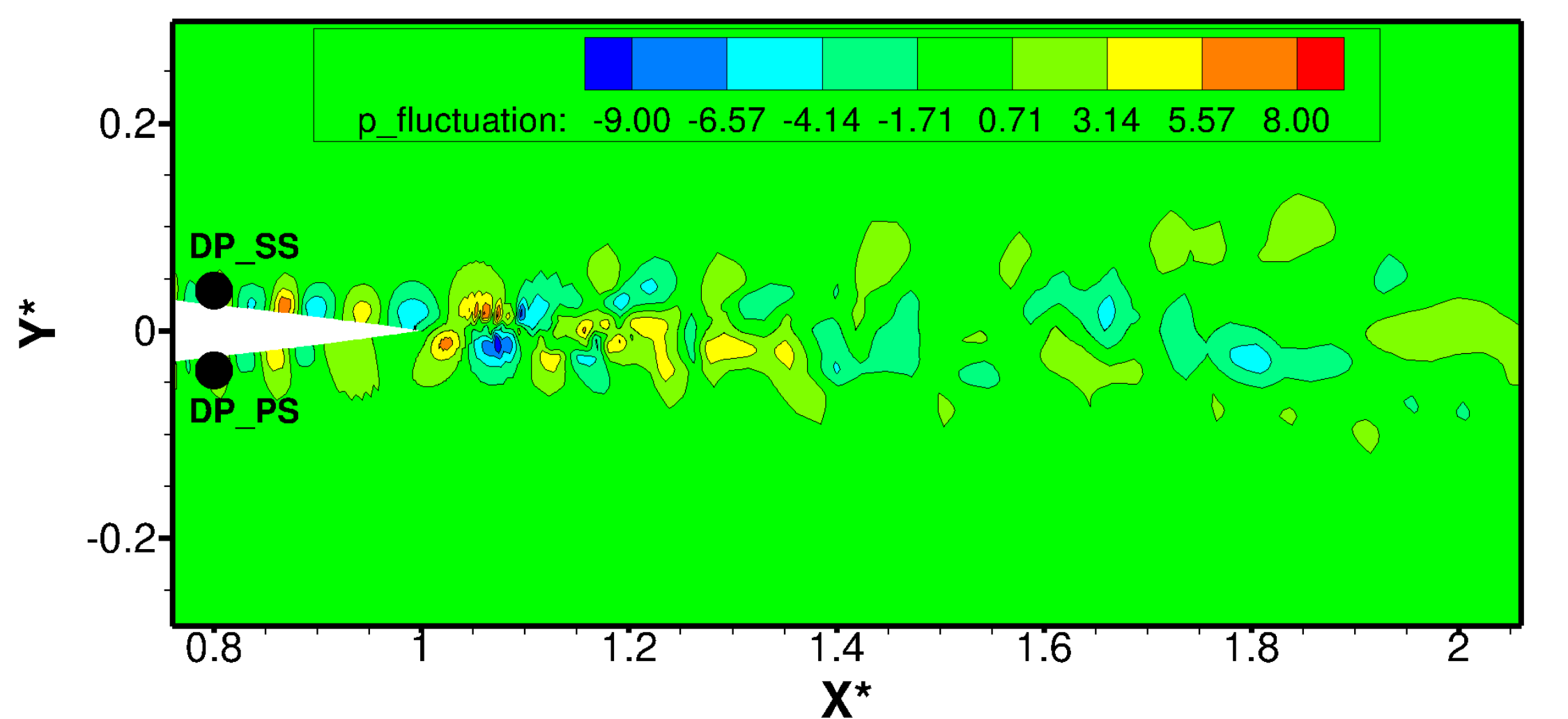}\label{fig:pressure_fluctuations_e-6_angle_0}}\hfill
	\subfigure[Pressure fluctuations: $m-0-y^+_{max}$ mesh at $t^*=1009$.]{\includegraphics[width=0.46\textwidth]{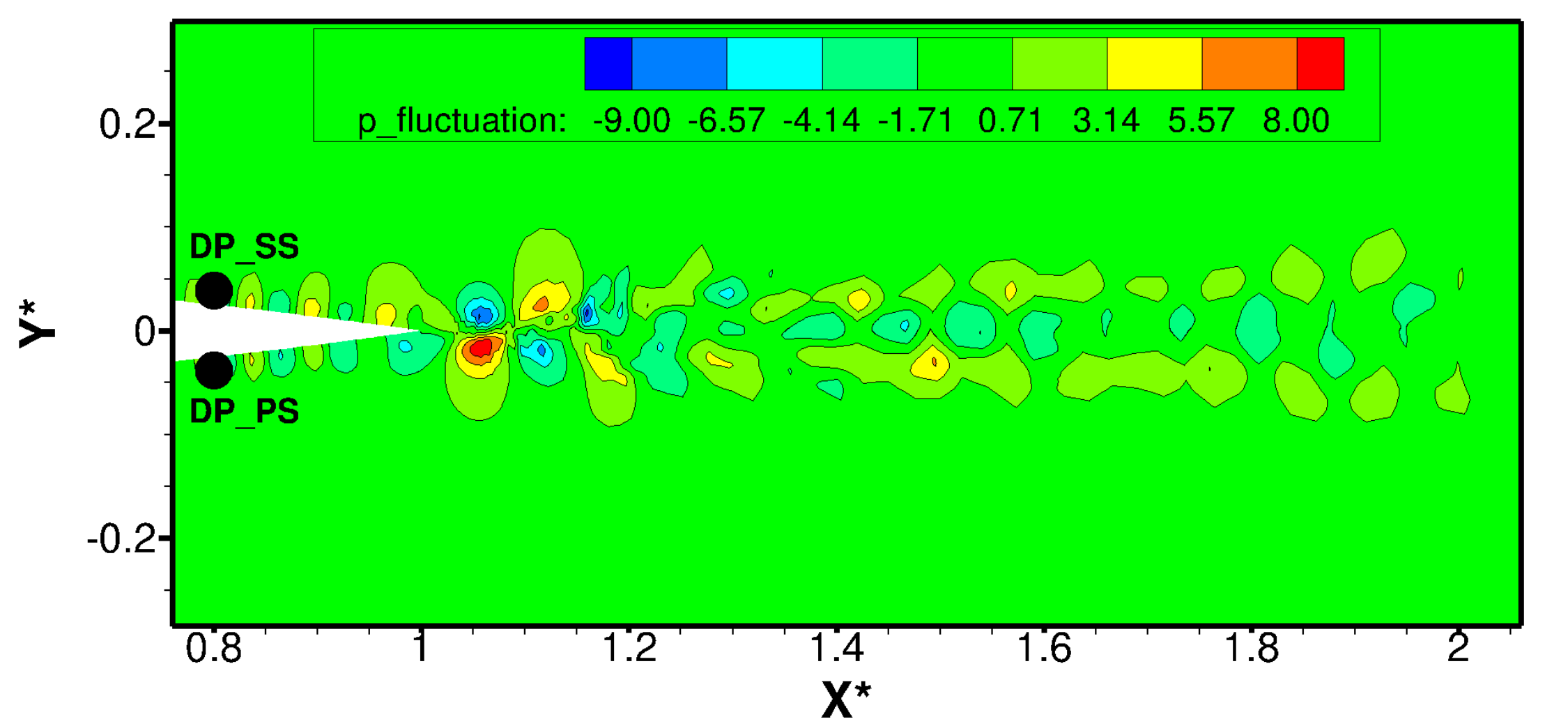}\label{fig:pressure_fluctuations_e-5_angle_0}}\hfill
	\subfigure[Span-wise vorticity: $m-0-y^+_{med}$ mesh at $t^*=322$.]{\includegraphics[width=0.46\textwidth]{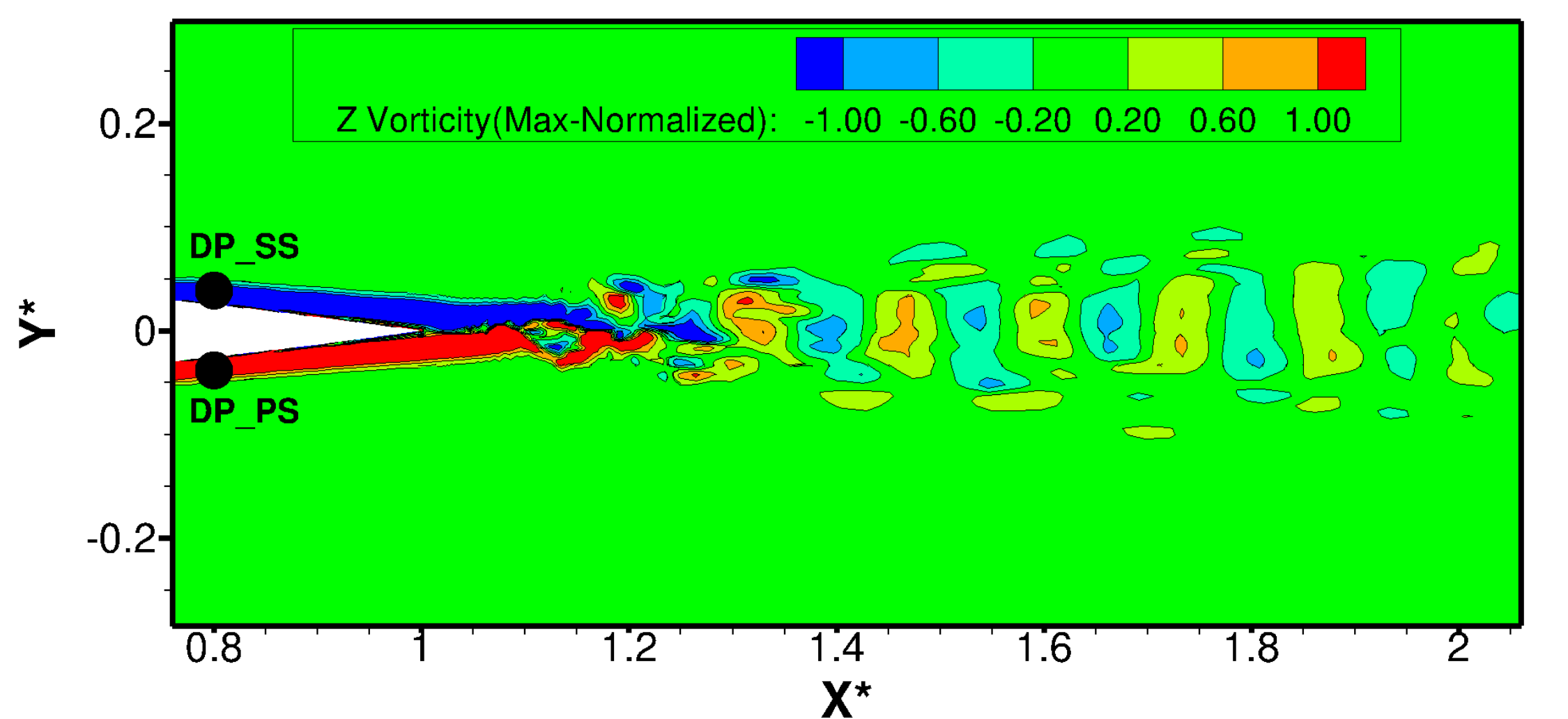}\label{fig:vorticity_e-6_angle_0_zoom}}\hfill
	\subfigure[Span-wise vorticity: $m-0-y^+_{max}$ mesh at $t^*=1009$.]{\includegraphics[width=0.46\textwidth]{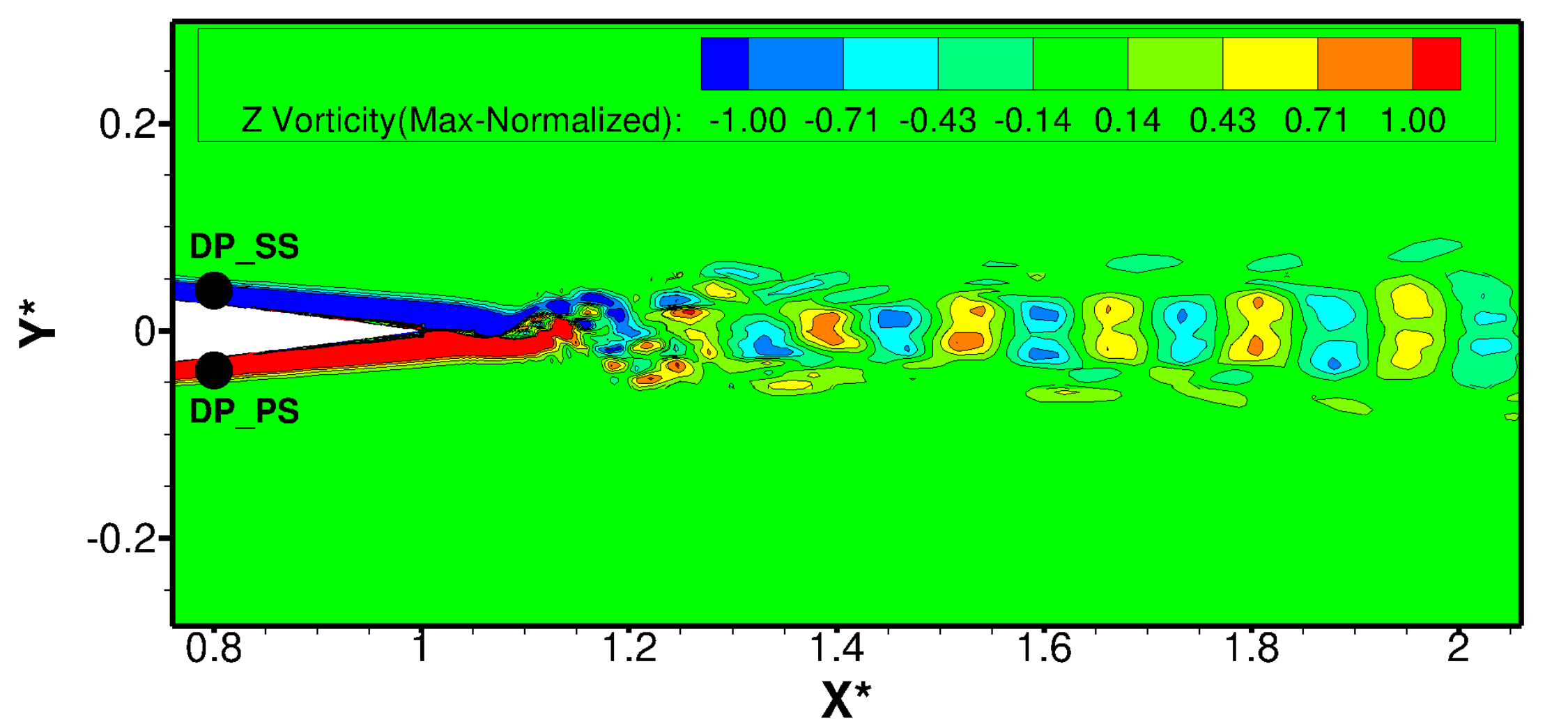}\label{fig:vorticity_e-5_angle_0_zoom}}\hfill
	\caption{Pressure fluctuations and span-wise vorticity of the medium meshes at an incidence of $\alpha=0^\circ$.}\label{fig:pressure_fluctuation_vorticity_medium_meshes_angle_0}
\end{figure}

 A von K\'arm\'an vortex street, characterized by a periodic detachment of pairs of alternate vortices, is present in both wakes. This is illustrated in Fig.\ \ref{fig:pressure_fluctuation_vorticity_medium_meshes_angle_0} by the pressure fluctuations and the vorticity in the span-wise direction. Regions of negative pressure fluctuations correspond to vortex cores whereas regions of positive pressure fluctuation indicate the position of stagnation points in the turbulent velocity field, as stated in the work of Manhart \cite{Manhart_1998}. A negative vorticity indicates that the vortex rolls in the clockwise direction, while a positive vorticity evinces a counter-clockwise rotation.
\begin{figure}[H]
	\centering
	\subfigure[Stream-wise velocity: $m-0-y^+_{med}$ mesh at $t^*=322$.]{\includegraphics[width=0.46\textwidth]{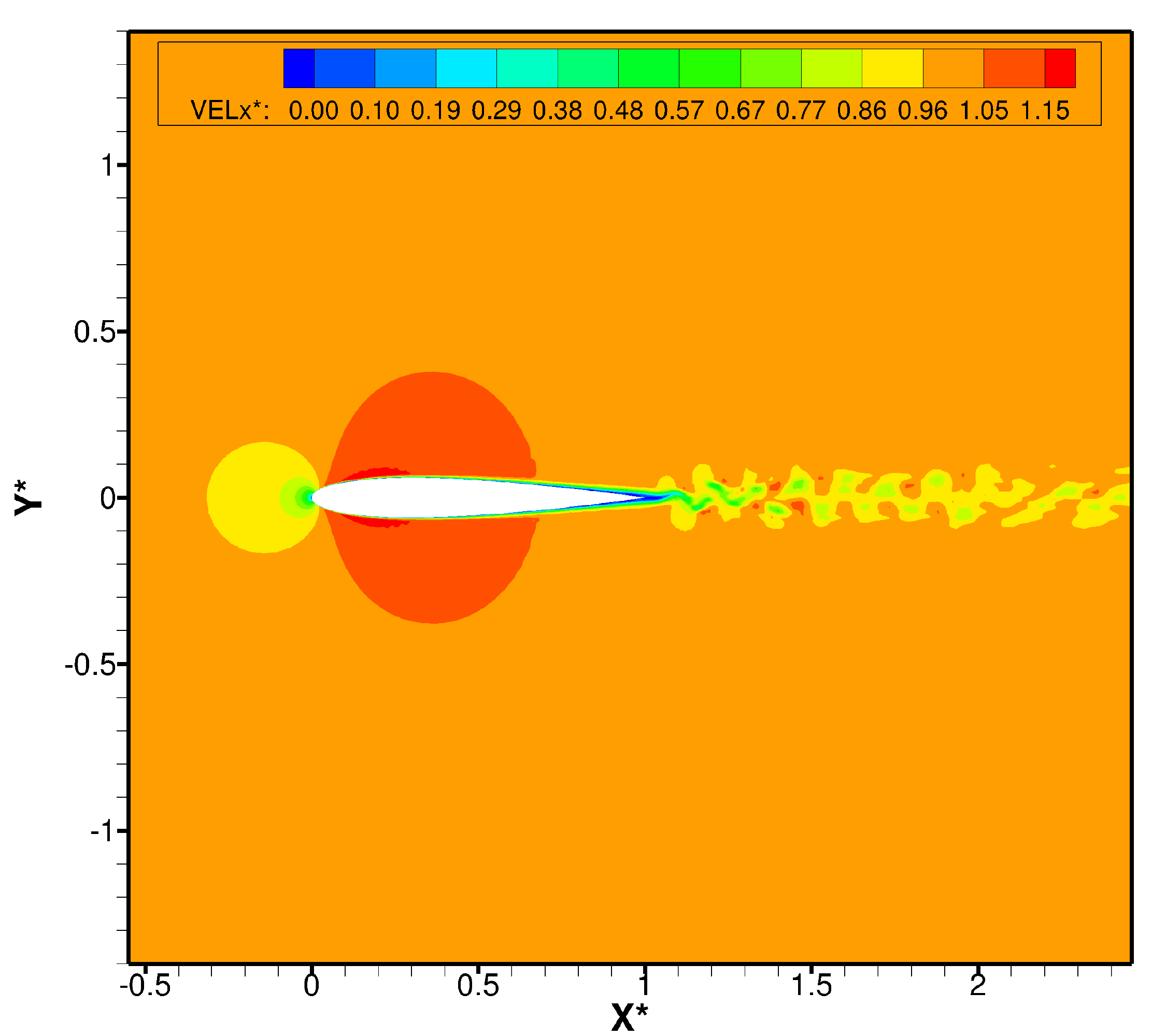}\label{fig:instationary_velocities_e-6_angle_0}}\hfill
	\subfigure[Stream-wise velocity: $m-0-y^+_{max}$ mesh at $t^*=1009$.]{\includegraphics[width=0.46\textwidth]{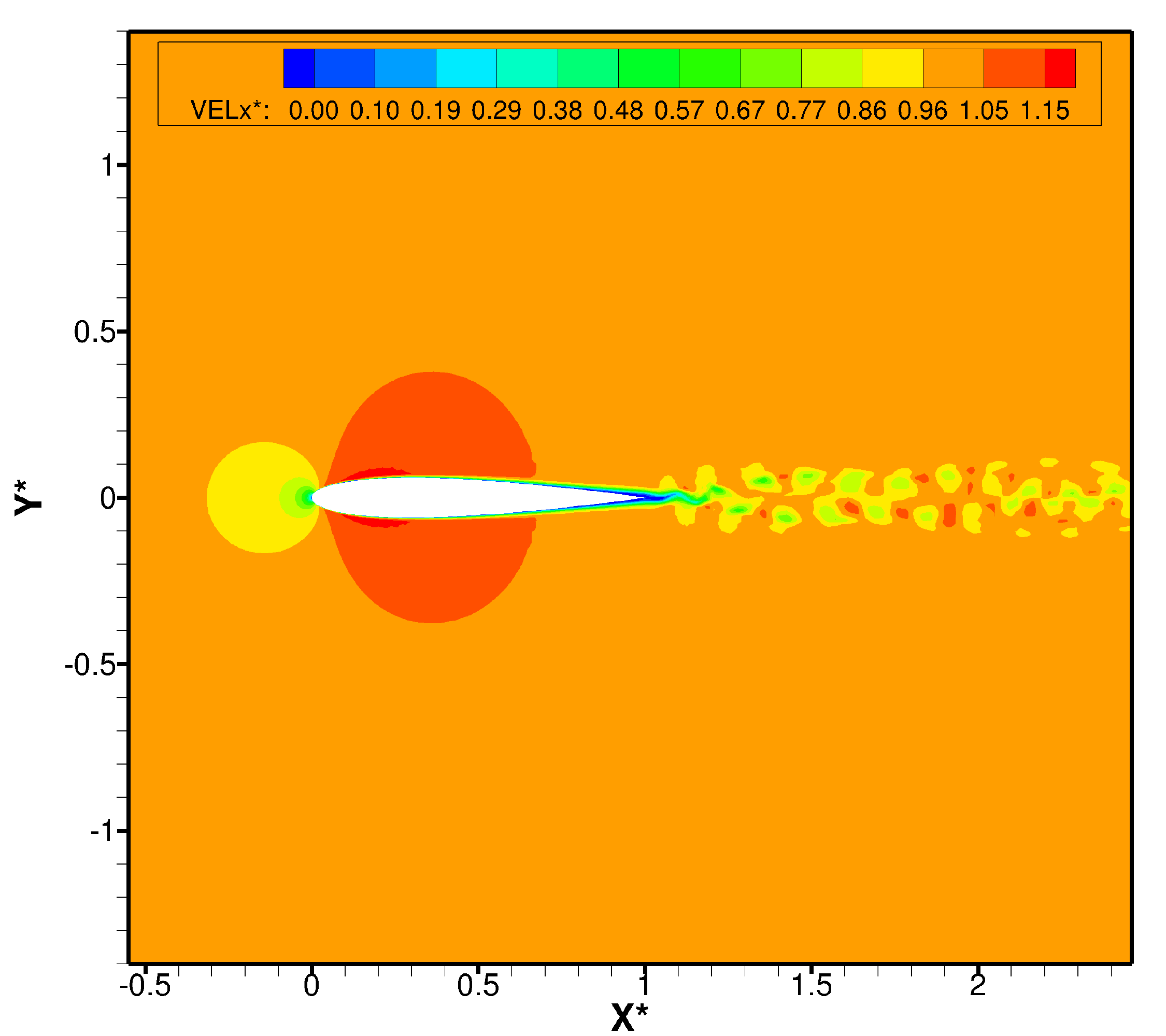}\label{fig:instationary_velocities_e-5_angle_0}}\hfill
\end{figure}

\subsection{Angle of attack $\alpha=5^\circ$}
\label{subsec:simulation_analysis_instationary_angle_5}
\par
\begin{figure}[H]
	\centering
	\subfigure[Stream-wise velocity: $m-5-y^+_{med}$ mesh at $t^*=152$.]{\includegraphics[width=0.46\textwidth]{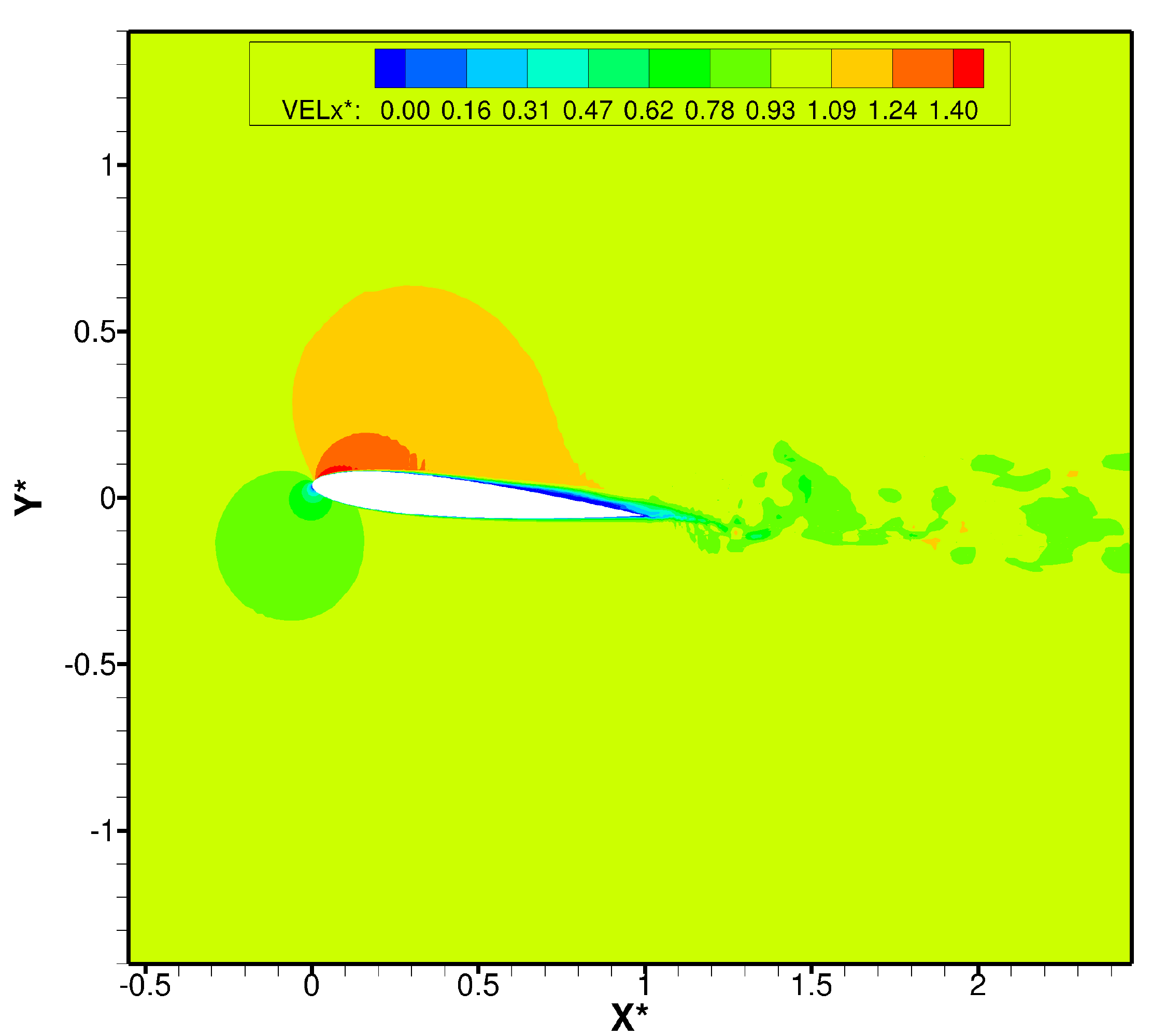}\label{fig:instationary_velocities_e-6_angle_5}}\hfill
	\subfigure[Stream-wise velocity: $m-5-y^+_{max}$ mesh at $t^*=1039$.]{\includegraphics[width=0.46\textwidth]{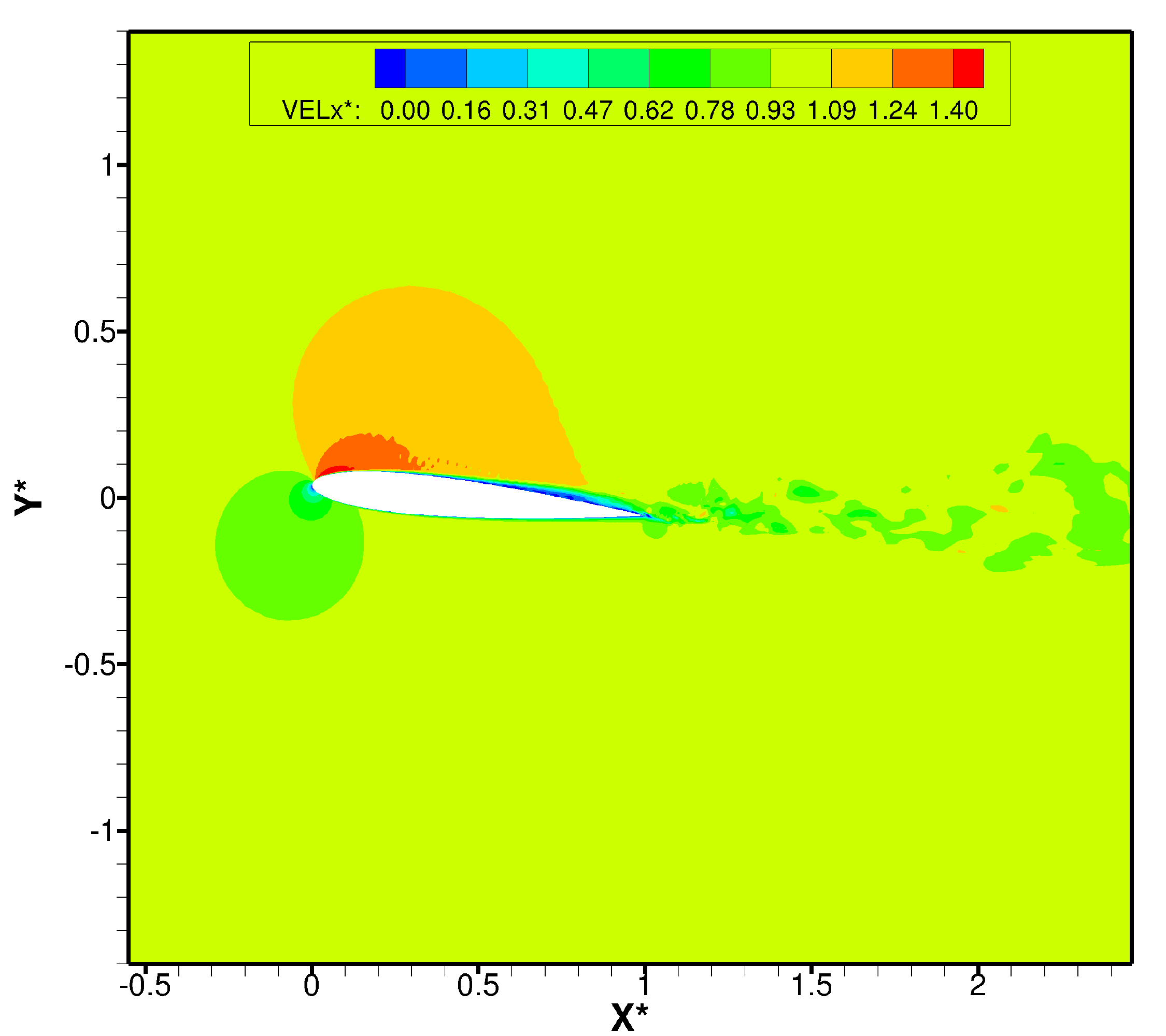}\label{fig:instationary_velocities_e-5_angle_5}}\hfill
	\subfigure[Stream-wise velocity: $m-5-y^+_{med}$ mesh at $t^*=152$.]{\includegraphics[width=0.46\textwidth]{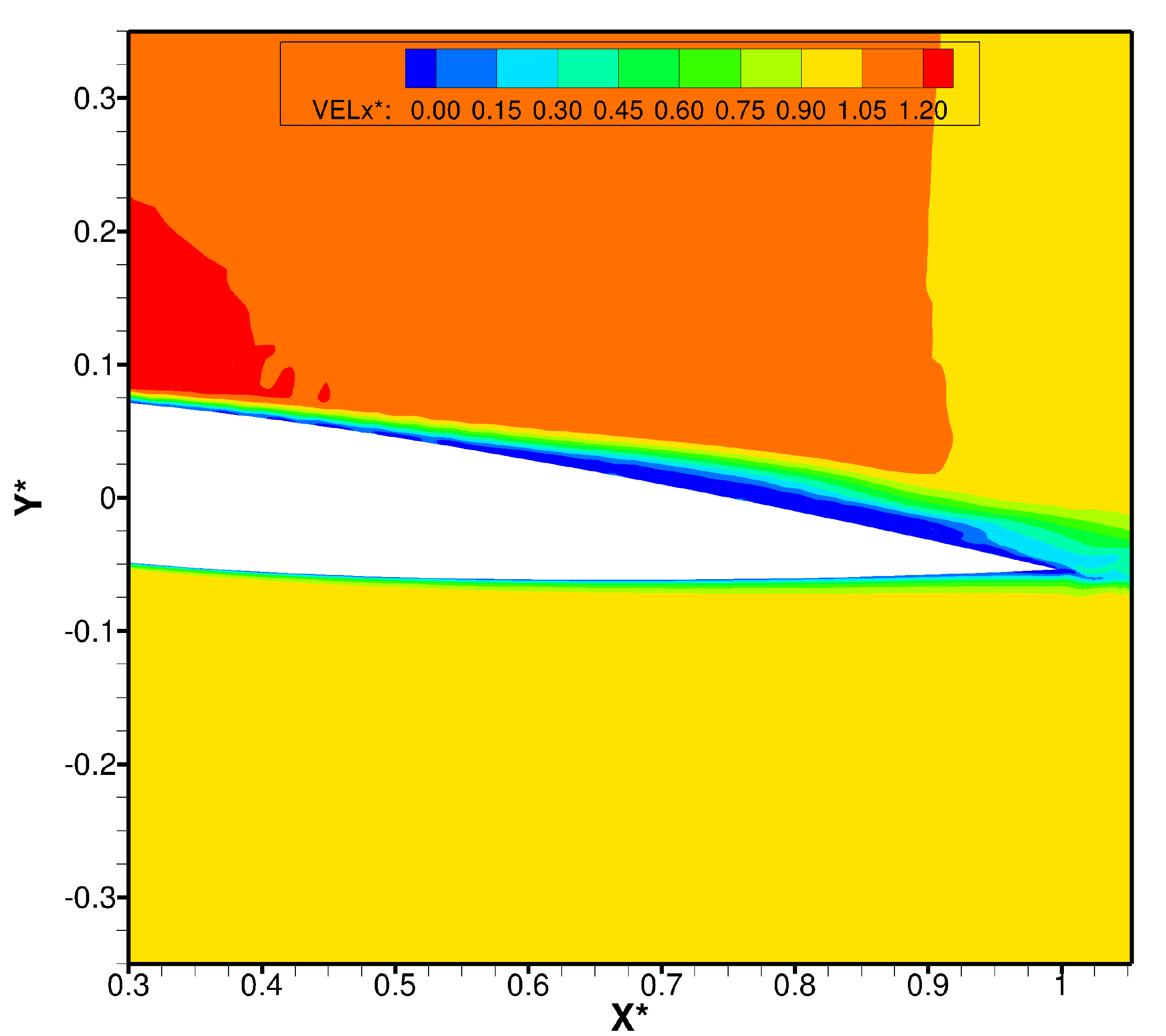}\label{fig:instationary_velocities_e-6_angle_5_zoom}}\hfill	
	\subfigure[Stream-wise velocity: $m-5-y^+_{max}$ mesh at $t^*=1039$.]{\includegraphics[width=0.46\textwidth]{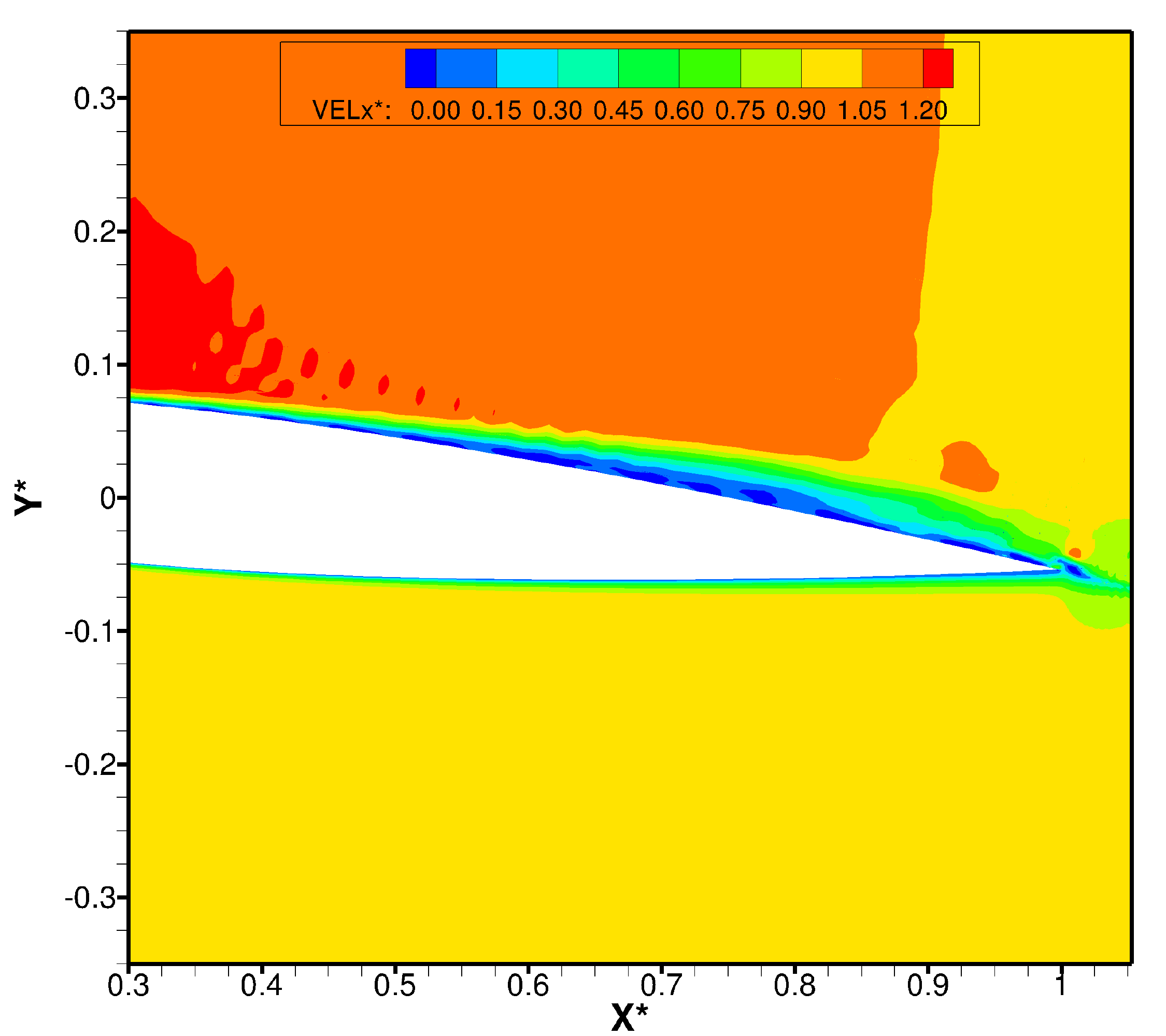}\label{fig:instationary_velocities_e-5_angle_5_zoom}}\hfill
	\caption{Instantaneous stream-wise velocity of the medium meshes at an incidence of $\alpha=0^\circ$.} \label{fig:instationary_velocities_medium_meshes_angle_5}
\end{figure}

 The instantaneous stream-wise velocities are analyzed in Fig.\ \ref{fig:instationary_velocities_medium_meshes_angle_5} for the meshes at an incidence of $\alpha=5^\circ$, i.e$.$, $m-5-y^+_{med}$ and $m-5-y^+_{max}$. Since the investigated dimensionless times ($t^*$) are different and the laminar boundary layer suffer a detachment at $x=0.345\,c$ for the $m-5-y^+_{med}$ mesh and at $x=0.331\,c$ for the $m-5-y^+_{max}$ grid (see Section \ref{sec:flow_summary}), the unsteady velocity field present differences mainly for $x>0.33\,c$. Numerical oscillations caused by the utilized second-order accuracy central differencing scheme and the resolution of the mesh are present in both simulated grids (see \mbox{Breuer \cite{Breuer_2013}}). A refinement of the grids are, however, not appropriate, since it would compromise the required computational time.  
\par The instationary velocity field of the $m-5-y^+_{med}$ mesh shows various short laminar separation bubbles and a long separation bubble on the suction side, while the results of the $m-5-y^+_{max}$ grid shows  various short separation bubbles. Both of these patterns are in agreement with the time-averaged velocity field (see Section \ref{sec:simulation_analysis}), which illustrates the formation of a long laminar separation bubble already in the first half of the suction side. 
\par
\begin{figure}[H]
	\centering
	\subfigure[Pressure fluctuations: $m-5-y^+_{med}$ mesh at $t^*=152$.]{\includegraphics[width=0.46\textwidth]{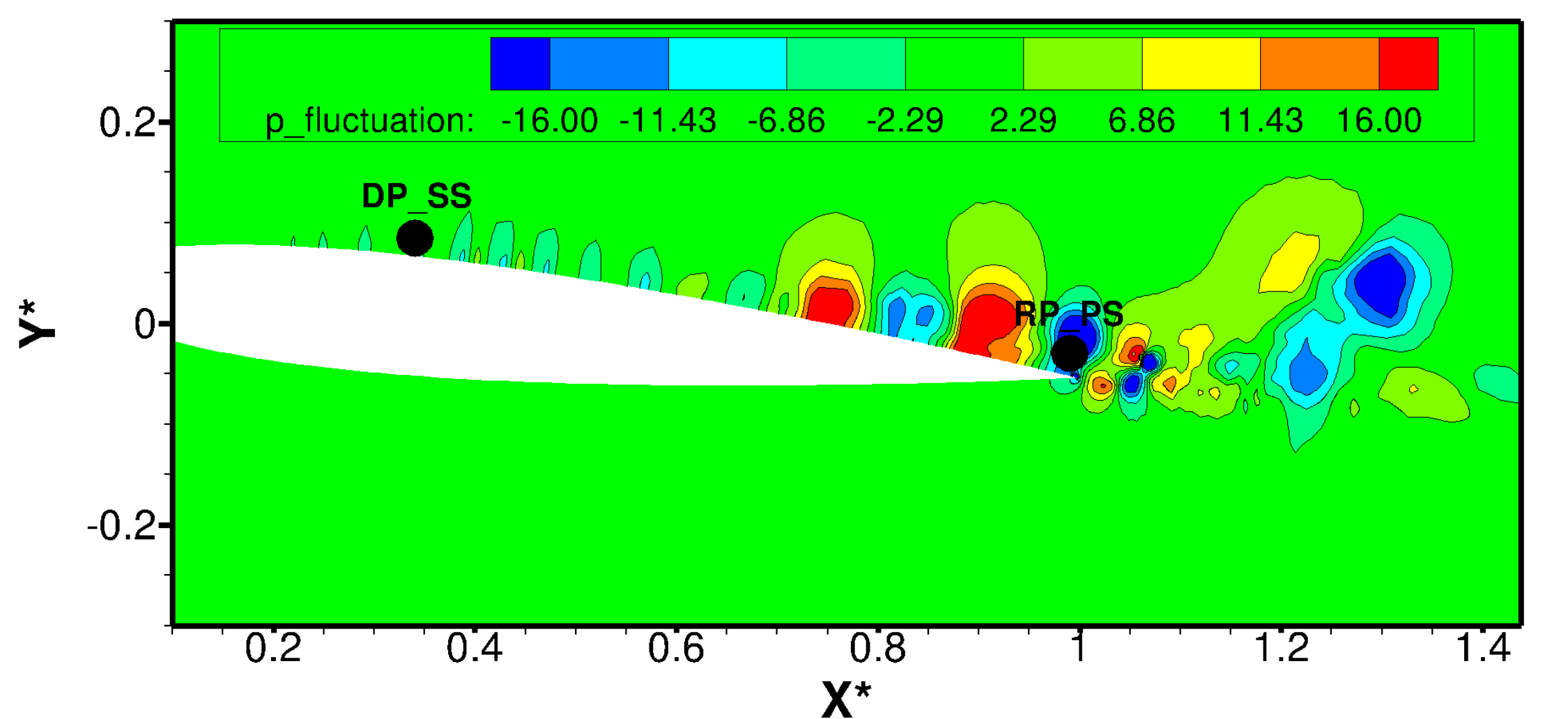}\label{fig:pressure_fluctuation_e-6_angle_5}}\hfill
	\subfigure[Pressure fluctuations: $m-5-y^+_{max}$ mesh at $t^*=1039$.]{\includegraphics[width=0.46\textwidth]{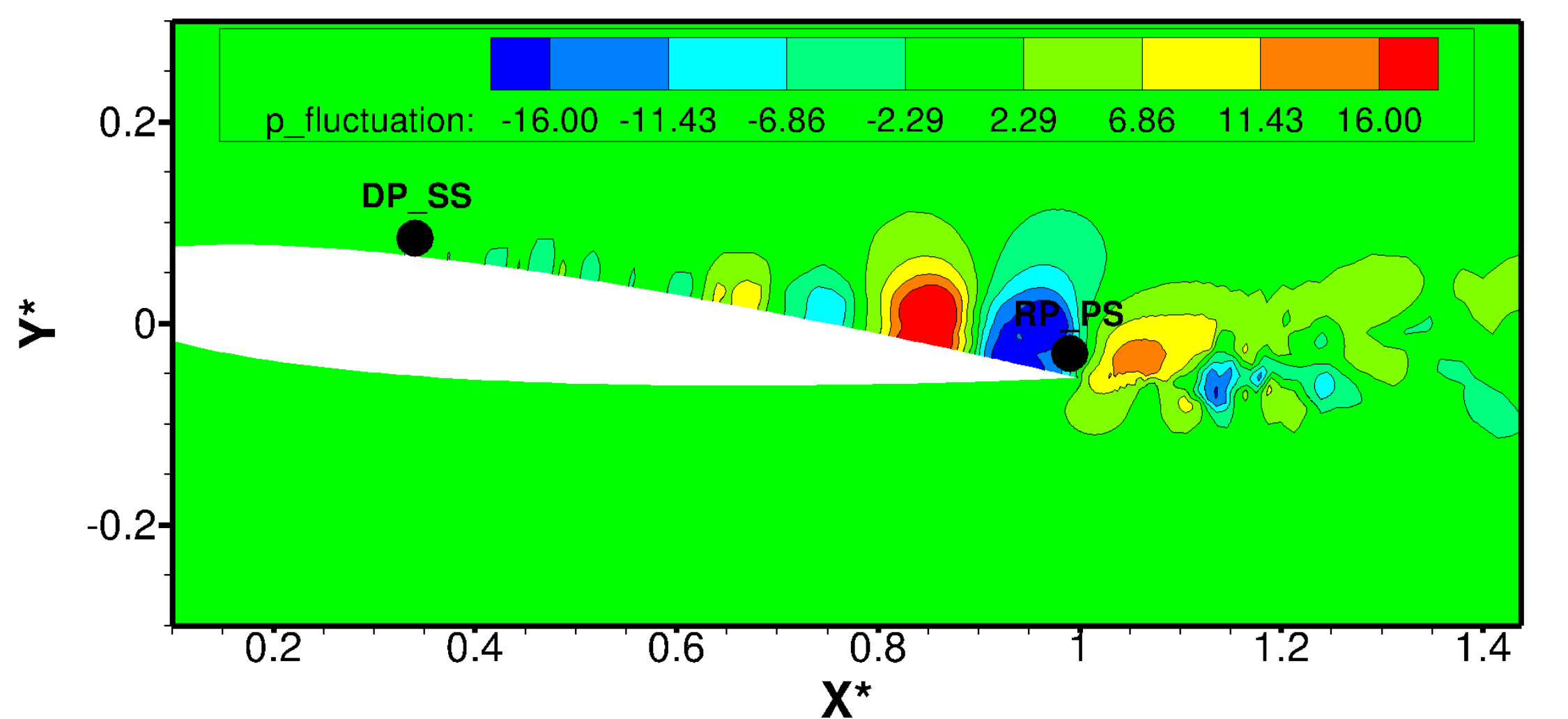}\label{fig:pressure_fluctuation_e-5_angle_5}}\hfill	
	\subfigure[Span-wise vorticity: $m-5-y^+_{med}$ mesh at $t^*=152$.]{\includegraphics[width=0.46\textwidth]{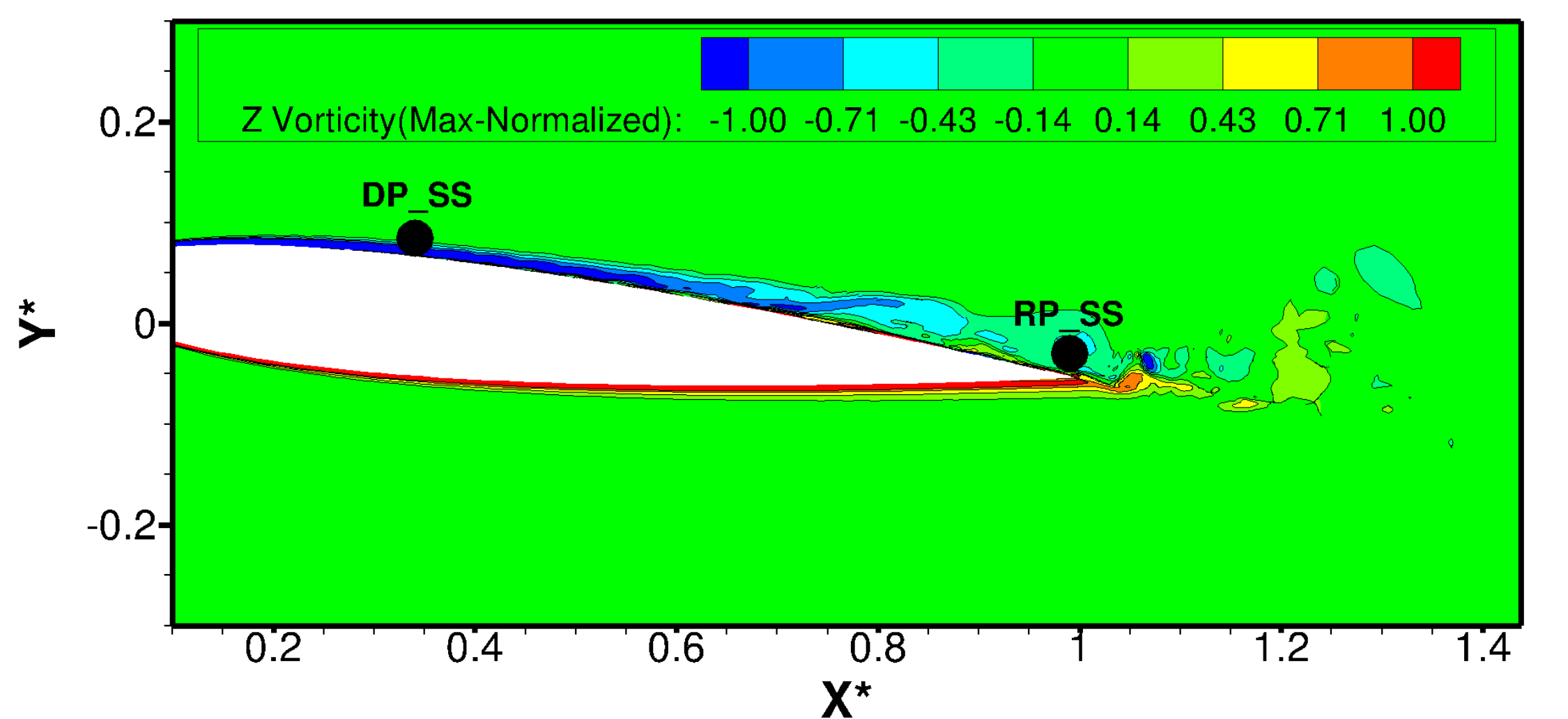}\label{fig:vorticity_e-6_angle_5}}\hfill
	\subfigure[Span-wise vorticity: $m-5-y^+_{max}$ mesh at $t^*=1039$.]{\includegraphics[width=0.46\textwidth]{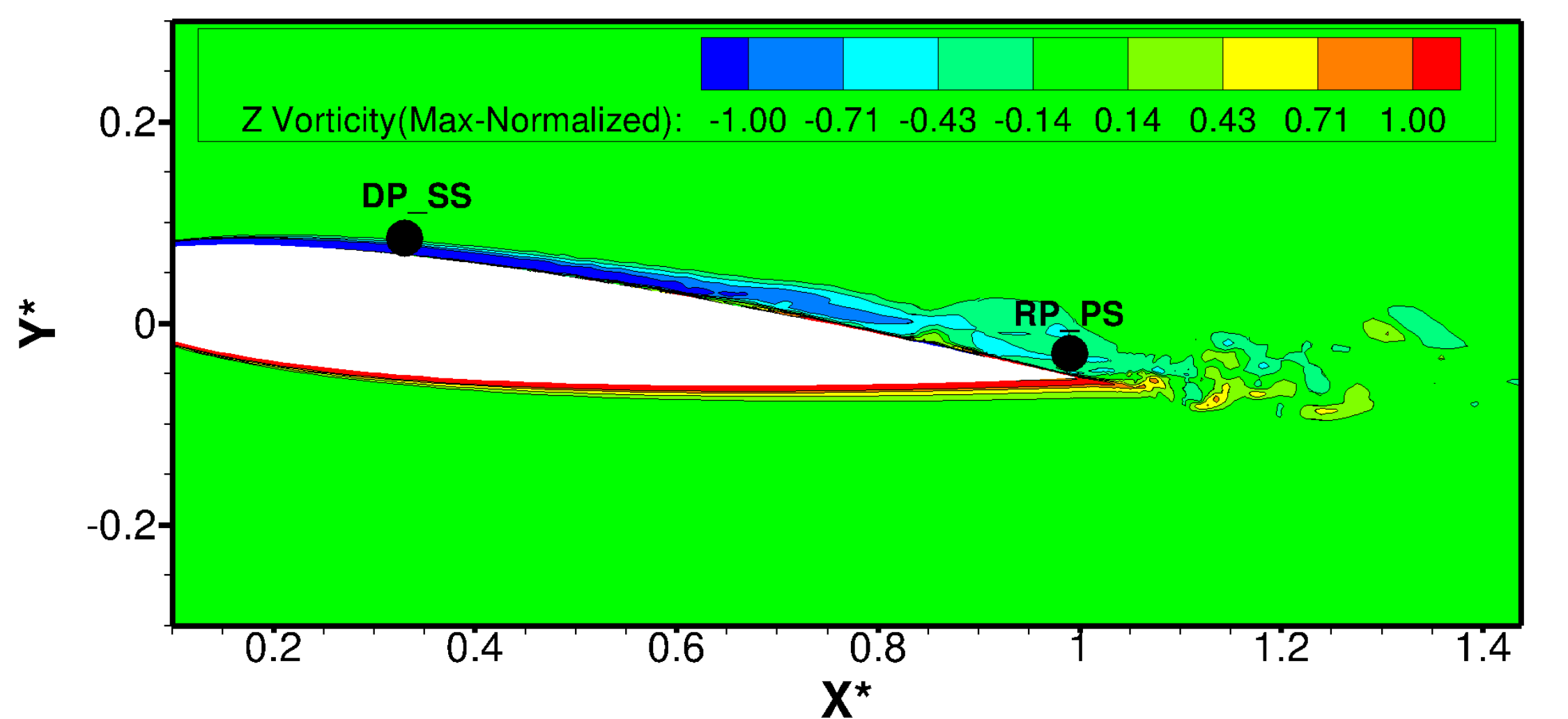}\label{fig:vorticity_e-5_angle_5}}\hfill
	\caption{Pressure fluctuations and span-wise vorticity of the medium meshes at an incidence of $\alpha=5^\circ$.}\label{fig:pressure_fluctuation_vorticity_medium_meshes_angle_5}
\end{figure}

 The formation of vortices is studied through pressure fluctuations and the vorticity in the span-wise direction, as illustrated in Fig.\ \ref{fig:pressure_fluctuation_vorticity_medium_meshes_angle_5}. A series of clockwise vortices are formed on the suction side of the airfoil, while in the wake clockwise and counter-clockwise vortices are present.

\subsection{Angle of attack $\alpha=11^\circ$}
\label{subsec:simulation_analysis_instationary_angle_11}
\par
\begin{figure}[H]
	\centering
	\subfigure[Stream-wise velocity: $m-11-y^+_{med}$ mesh at $t^*=76$.]{\includegraphics[width=0.46\textwidth]{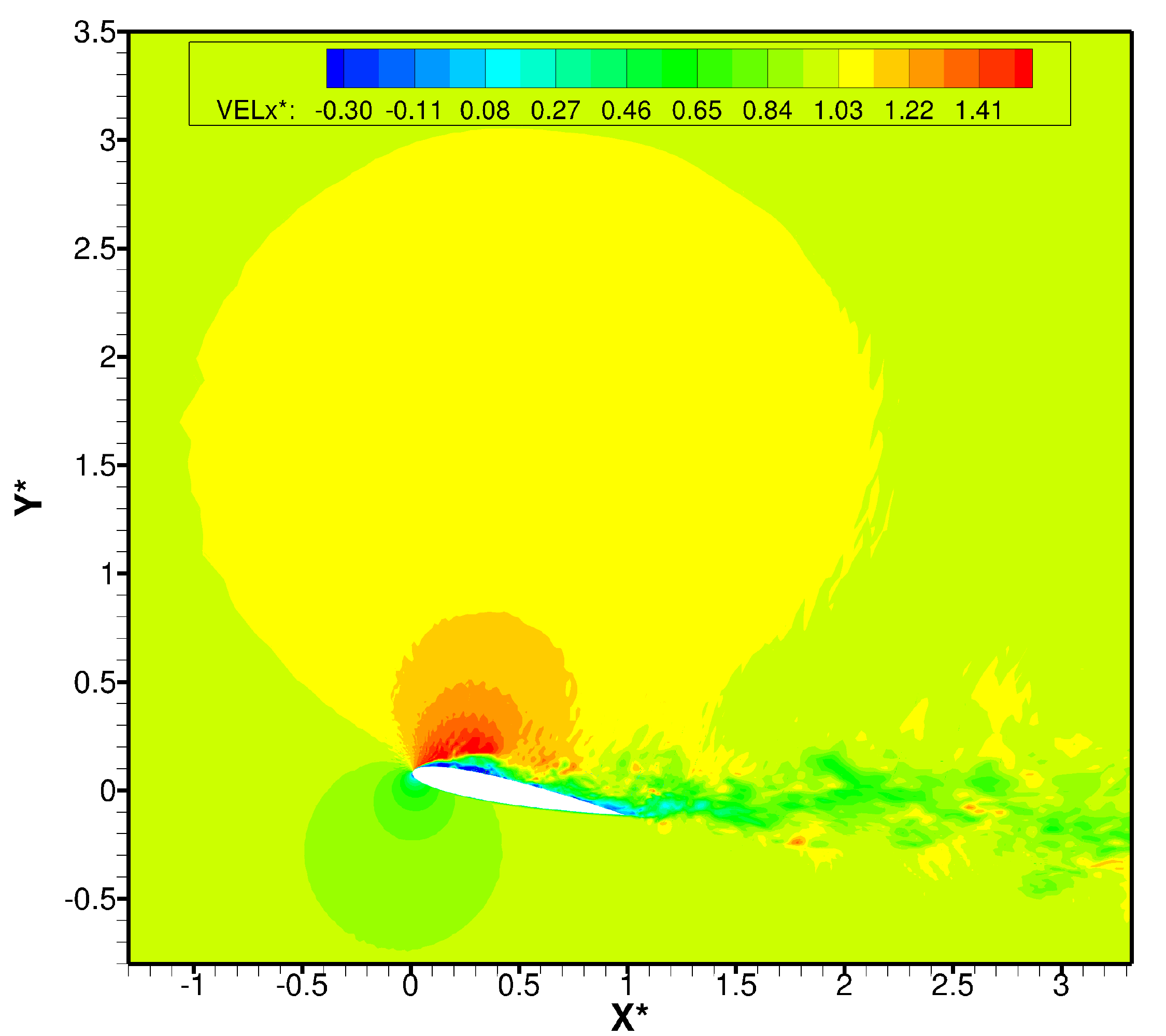}\label{fig:instationary_velocities_e-6_angle_11}}\hfill
	\subfigure[Stream-wise velocity: $m-11-y^+_{max}$ mesh at $t^*=454$.]{\includegraphics[width=0.46\textwidth]{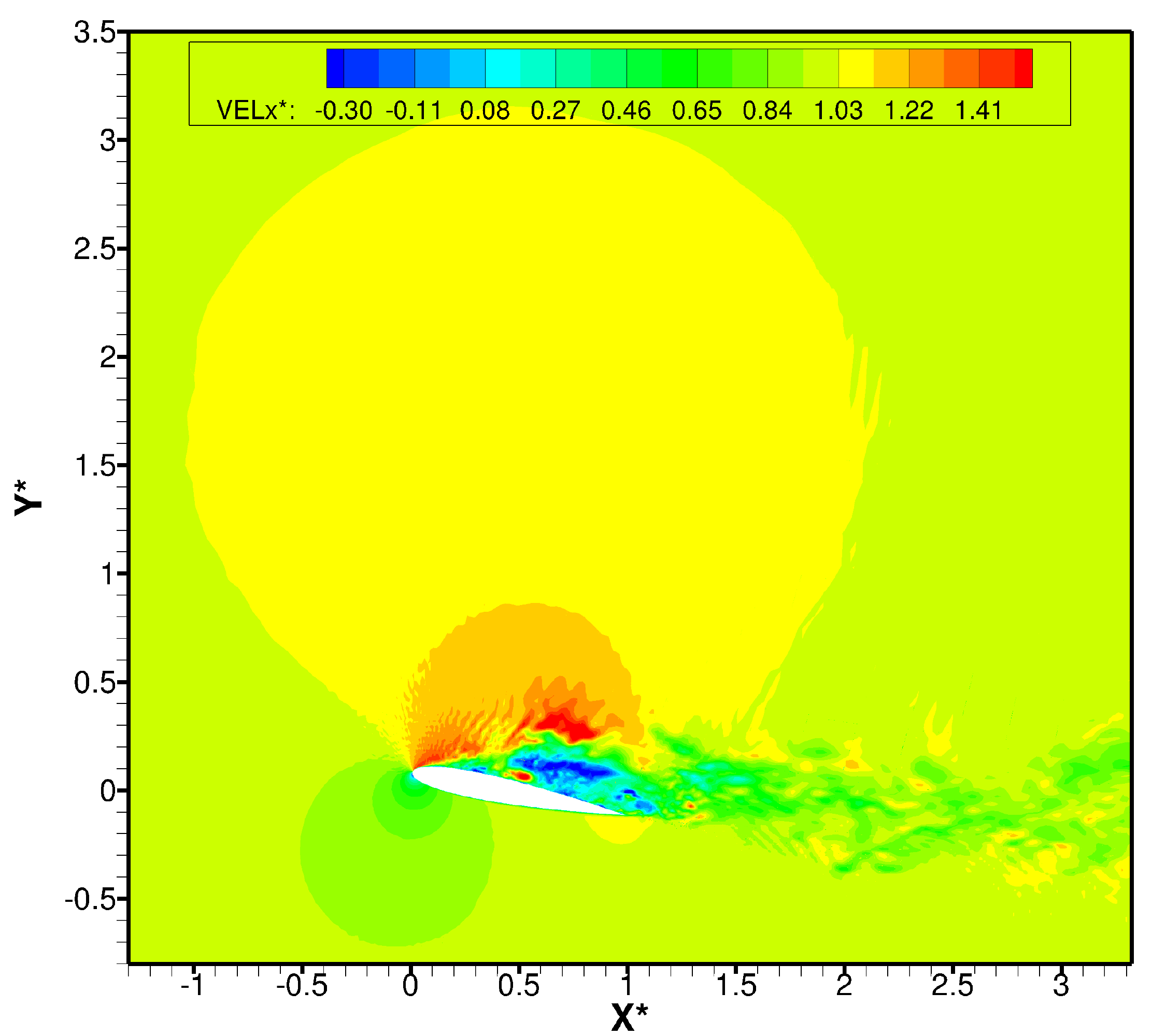}\label{fig:instationary_velocities_e-5_angle_11}}\hfill
	\subfigure[Stream-wise velocity: $m-11-y^+_{med}$ mesh at $t^*=76$.]{\includegraphics[width=0.46\textwidth]{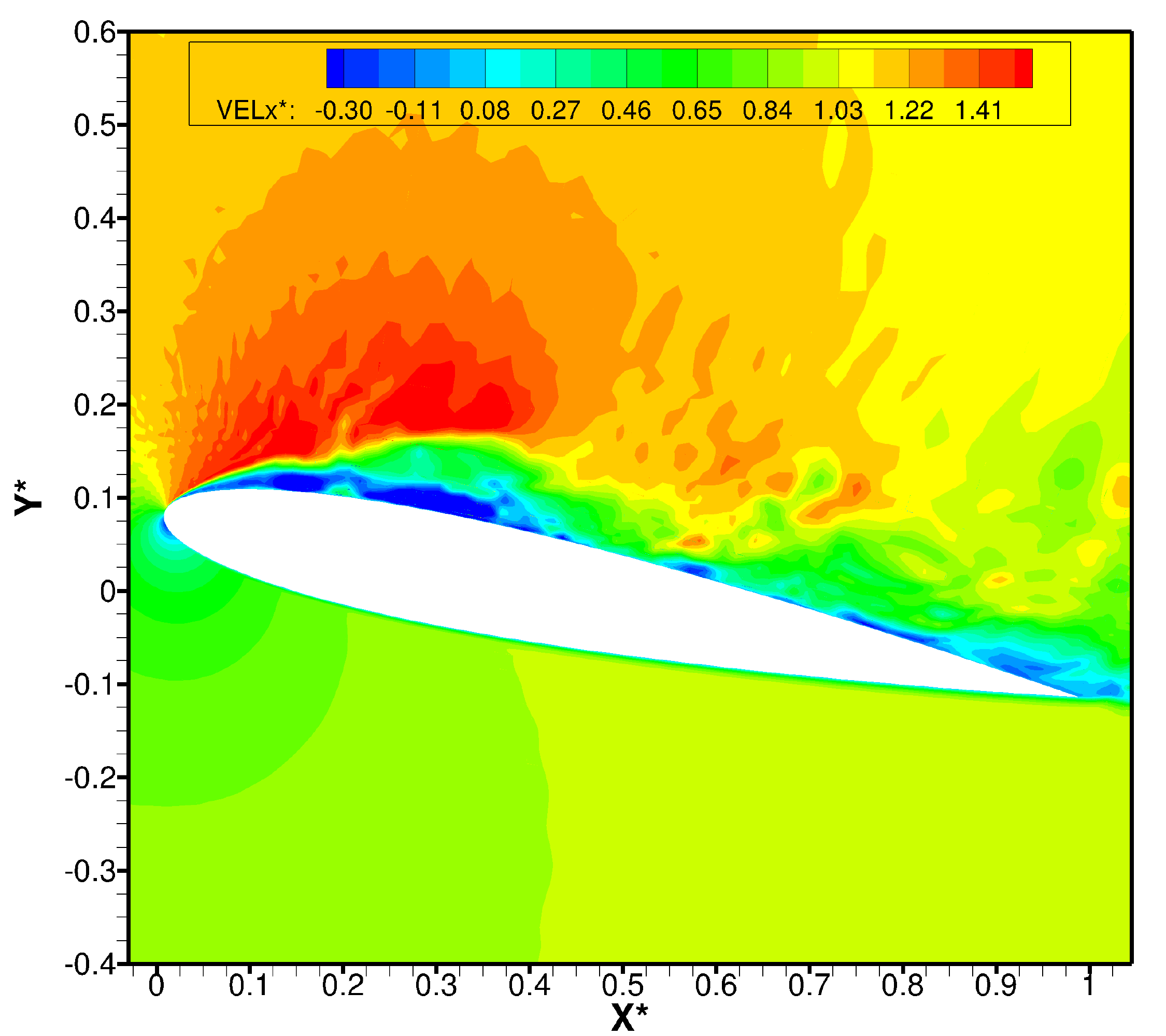}\label{fig:instationary_velocities_e-6_angle_11_zoom}}\hfill	
	\subfigure[Stream-wise velocity: $m-11-y^+_{max}$ mesh at $t^*=454$.]{\includegraphics[width=0.46\textwidth]{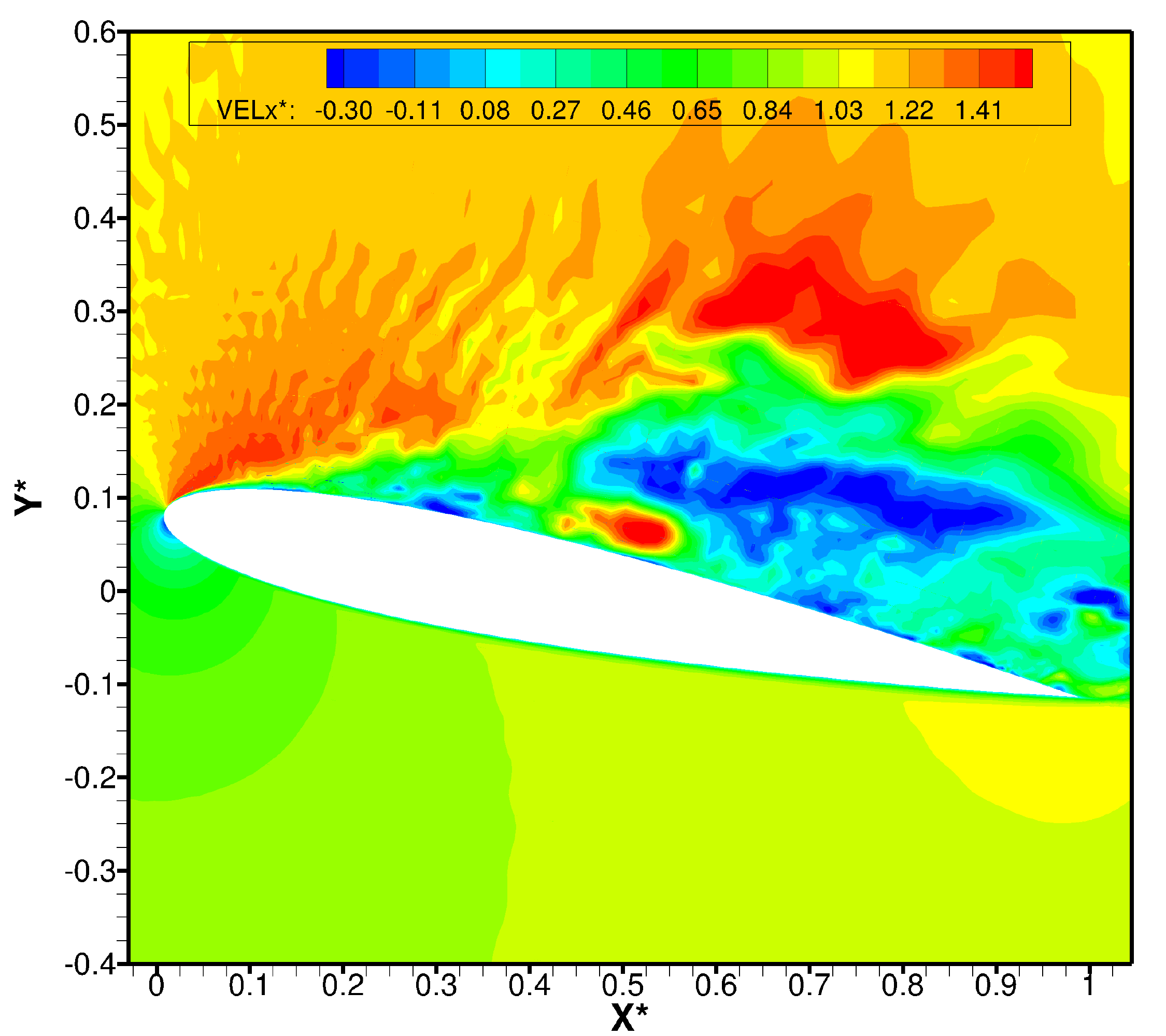}\label{fig:instationary_velocities_e-5_angle_11_zoom}}\hfill
	\caption{Instantaneous stream-wise velocity of the medium meshes at an incidence of \mbox{$\alpha=11^\circ$}.} \label{fig:instationary_velocities_medium_meshes_angle_11}	
\end{figure}

 Figure \ref{fig:instationary_velocities_medium_meshes_angle_11} illustrates the instantaneous velocity field of the large-eddy simulations for the grids at an incidence of $\alpha=11^\circ$. Numerical oscillations are present in both velocity fields as a result of the utilized second-order accuracy central differencing scheme and the mesh resolution. Both grids ($m-11-y^+_{med}$ and $m-11-y^+_{max}$) are studied at different dimensionless times ($t^*$), which results in different flow patterns for $x>0.02\,c$, point at which the laminar boundary layer is detached (see Section \ref{sec:flow_summary}). 
\par The instantaneous velocity for the $m-11-y^+_{med}$ mesh shows the formation of a long laminar separation bubble on the suction side near the leading edge and various short separation bubbles in the second half of the profiles leeward side. For the $m-11-y^+_{max}$ grid, short laminar separation bubbles are formed on the first half of the profiles suction side and a large recirculation region is formed on the second half of the airfoils leeward side. No recirculation zone is present on the pressure side for both meshes.
\par
\begin{figure}[H]
	\centering
	\subfigure[Stream-wise vorticity: $m-11-y^+_{med}$ mesh at $t^*=76$.]{\includegraphics[width=0.46\textwidth]{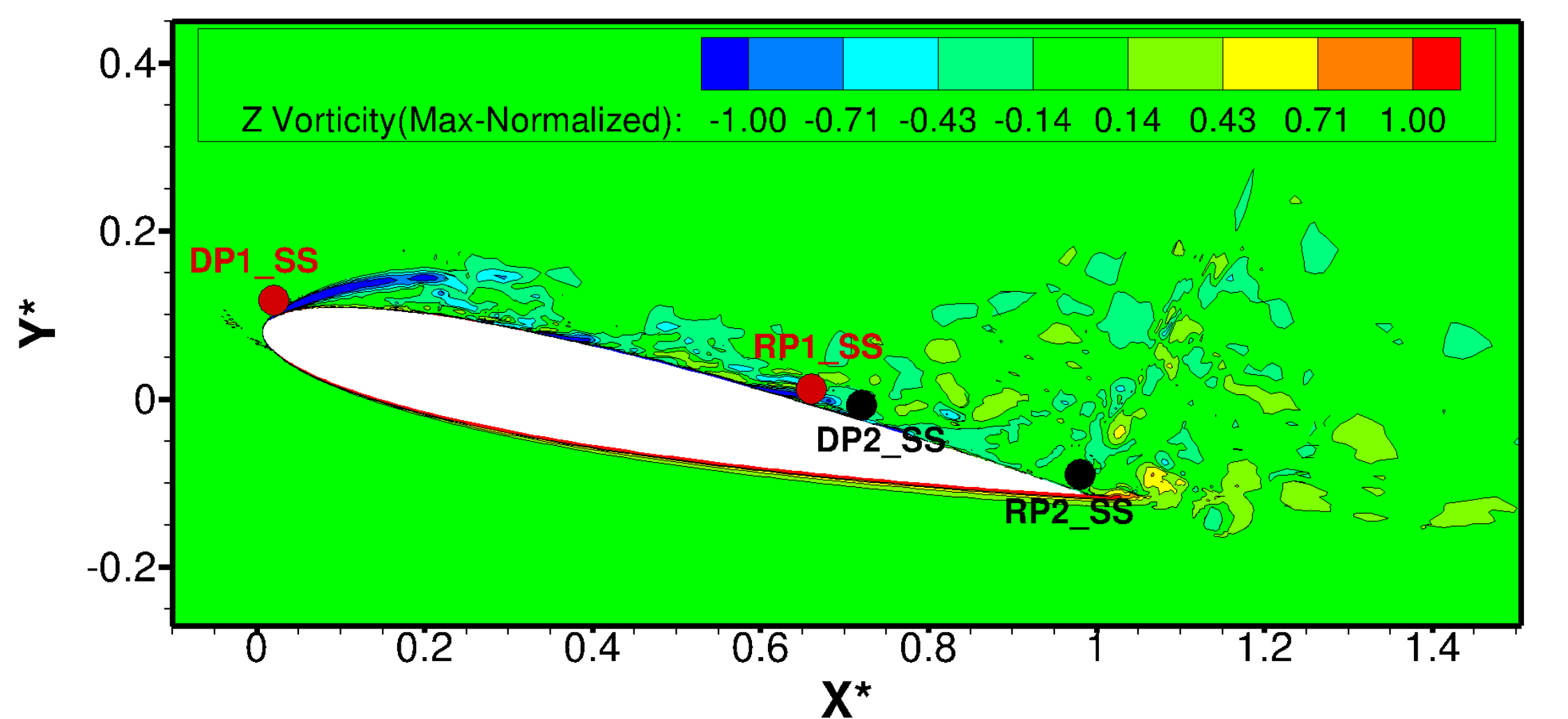}\label{fig:vorticity_e-6_angle_11}}\hfill
	\subfigure[Span-wise vorticity: $m-11-y^+_{max}$ mesh at $t^*=454$.]{\includegraphics[width=0.46\textwidth]{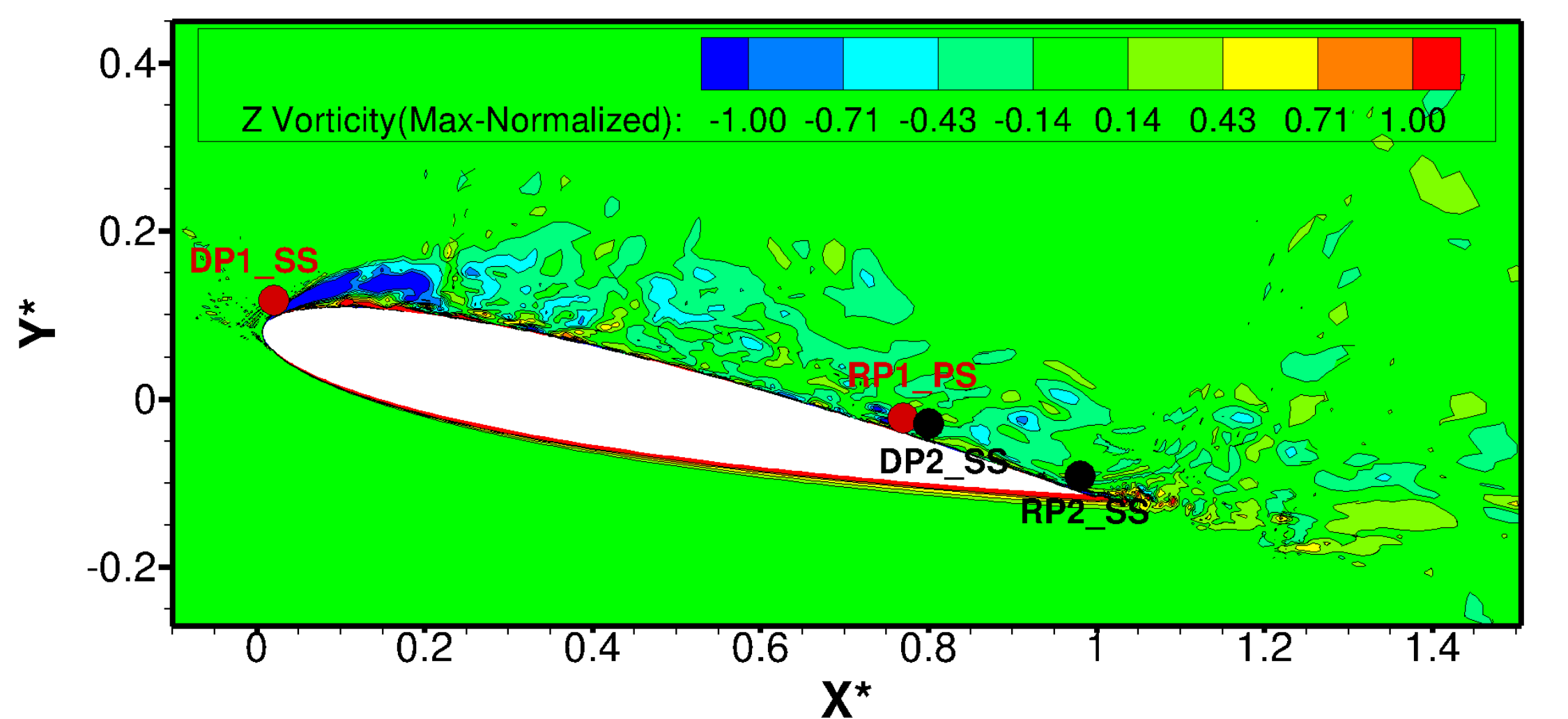}\label{fig:vorticity_e-5_angle_11}}\hfill
	\caption{Pressure fluctuations and span-wise vorticity of the medium meshes at an incidence of $\alpha=11^\circ$.} \label{fig:pressure_fluctuation_vorticity_medium_meshes_angle_11}
\end{figure}

 The analysis of the pressure fluctuations together with the vorticity in the span-wise direction (see Fig.\ \ref{fig:pressure_fluctuation_vorticity_medium_meshes_angle_11}) indicates the presence of clockwise vortices on the leeward side of the airfoil and clockwise and counter-clockwise vortices in the wake. Furthermore, a shear layer, i.e$.$, a region with a significant velocity gradient, is formed near the leading edge and seems to be thicker for the $m-11-y^+_{max}$ mesh.
\begin{figure}[H]
	\centering
	\subfigure[Pressure fluctuations: $m-11-y^+_{med}$ mesh at $t^*=76$.]{\includegraphics[width=0.46\textwidth]{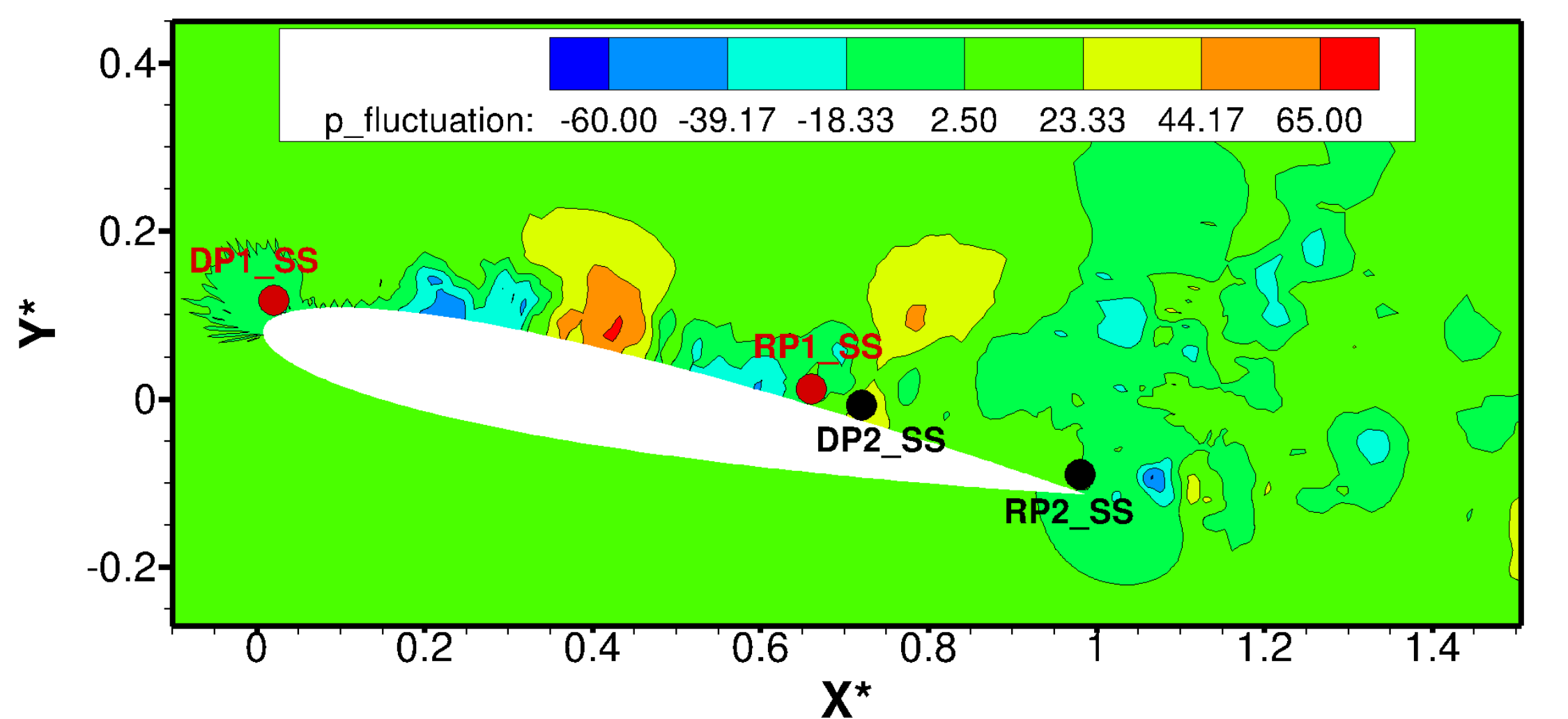}\label{fig:pressure_fluctuation_e-6_angle_11}}\hfill
	\subfigure[Pressure fluctuations: $m-11-y^+_{max}$ mesh at $t^*=454$.]{\includegraphics[width=0.46\textwidth]{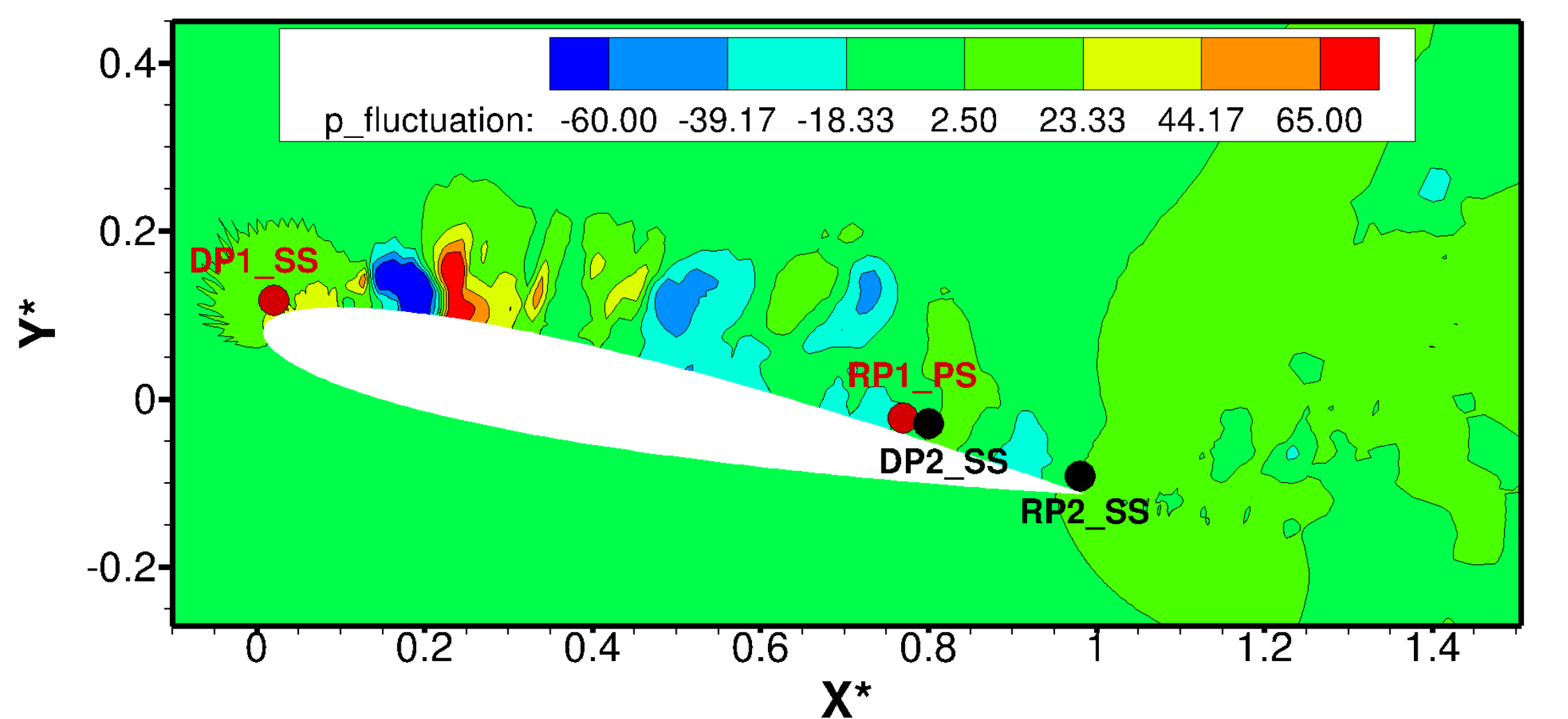}\label{fig:pressure_fluctuation_e-5_angle_11}}\hfill
\end{figure}
\markboth{\MakeUppercase{chapter 3.$\quad$results and discussion}}{\MakeUppercase{3.5$\quad$Rey. Stresses Analysis}}
\section{Reynolds stresses analysis} \markboth{\MakeUppercase{chapter 3.$\quad$results and discussion}}{\MakeUppercase{3.5$\quad$Rey. Stresses Analysis}}
\label{sec:reynolds_stresses}
\par The dimensionless Reynolds stress tensor is based on the time-averaging process and is calculated according to Eq.\ (\ref{eq:dimensionless_reynolds_stresses}). Equation (\ref{eq:time_averaging_reynolds}) represents the fundamental of the time-averaging process, in which the velocity, for instance, is divided into a mean value $\widetilde{u_i}$ plus a fluctuation $u_i'$. The former is calculated according to Eq.\ (\ref{eq:time_averaging_reynolds_2}), in which $t_{avg}$ is the time period utilized to average the instantaneous values. This must be sufficiently long in order to guarantee stable statistics, that is, to average the fluctuations to zero ($\widetilde{u_i'}=0$).
\begin{eqnarray}
u_i&=&\widetilde{u_i}+u_i'
\label{eq:time_averaging_reynolds} \\
\widetilde{u_i}&\approx&\lim_{t_{avg}\to\infty}\;\frac{1}{t_{avg}}\int\limits_{t}^{t_{avg}}u_i\,\mathrm{d}t
\label{eq:time_averaging_reynolds_2}
\end{eqnarray}
\par The analysis of this stress tensor is important in order to acquire more information about the turbulent flow and consequently provide a thorough investigation of the meshes. 
\markboth{\MakeUppercase{chapter 3.$\quad$results and discussion}}{\MakeUppercase{3.5$\quad$Rey. Stresses Analysis}}
\subsection{Angle of attack $\alpha=0^\circ$}
\label{subsec:reynolds_analysis_angle_0}
\par The components $\tau_{11}^{turb^*}$, $\tau_{12}^{turb^*}$ and $\tau_{22}^{turb^*}$ are studied for the $m-0-y^+_{med}$ and $m-0-y^+_{max}$ meshes at $t^*_{avg}=322$ and $t^*_{avg}=1009$, respectively. The rest of the components are not analyzed, since the flow around the airfoil is considered almost two-dimensional. Due to the missing time-averaged data, the fine resolution mesh is not investigated.  
\par
\begin{figure}[H]
	\centering
	\subfigure[$\tau_{11}^{turb^*}$: $m-0-y^+_{med}$ mesh for $t^*_{avg}=322$.]{\includegraphics[width=0.455\textwidth]{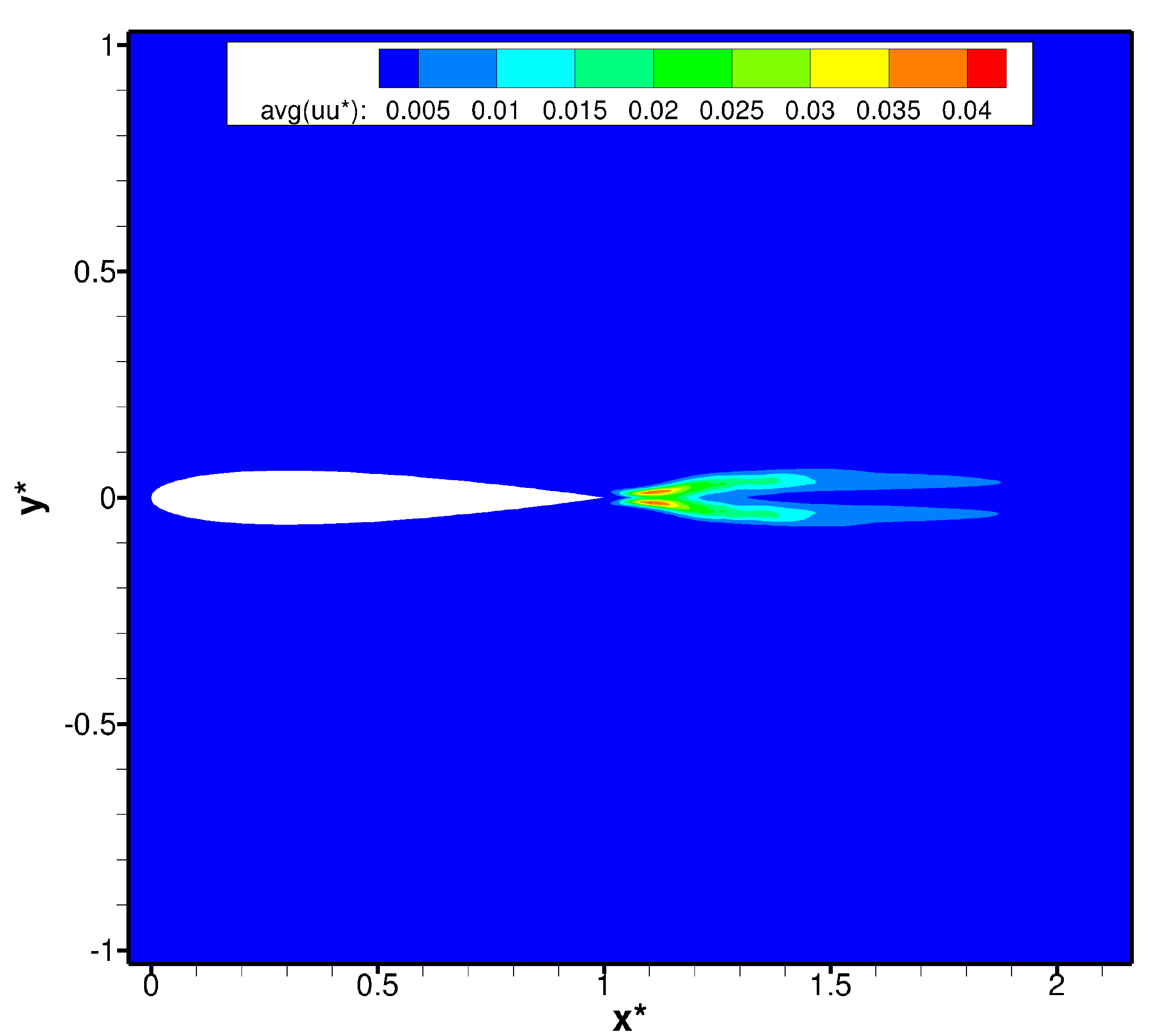}\label{fig:avg_uu_m_0_y+_med}}\hfill
	\subfigure[$\tau_{11}^{turb^*}$: $m-0-y^+_{max}$ mesh for $t^*_{avg}=1009$.]{\includegraphics[width=0.455\textwidth]{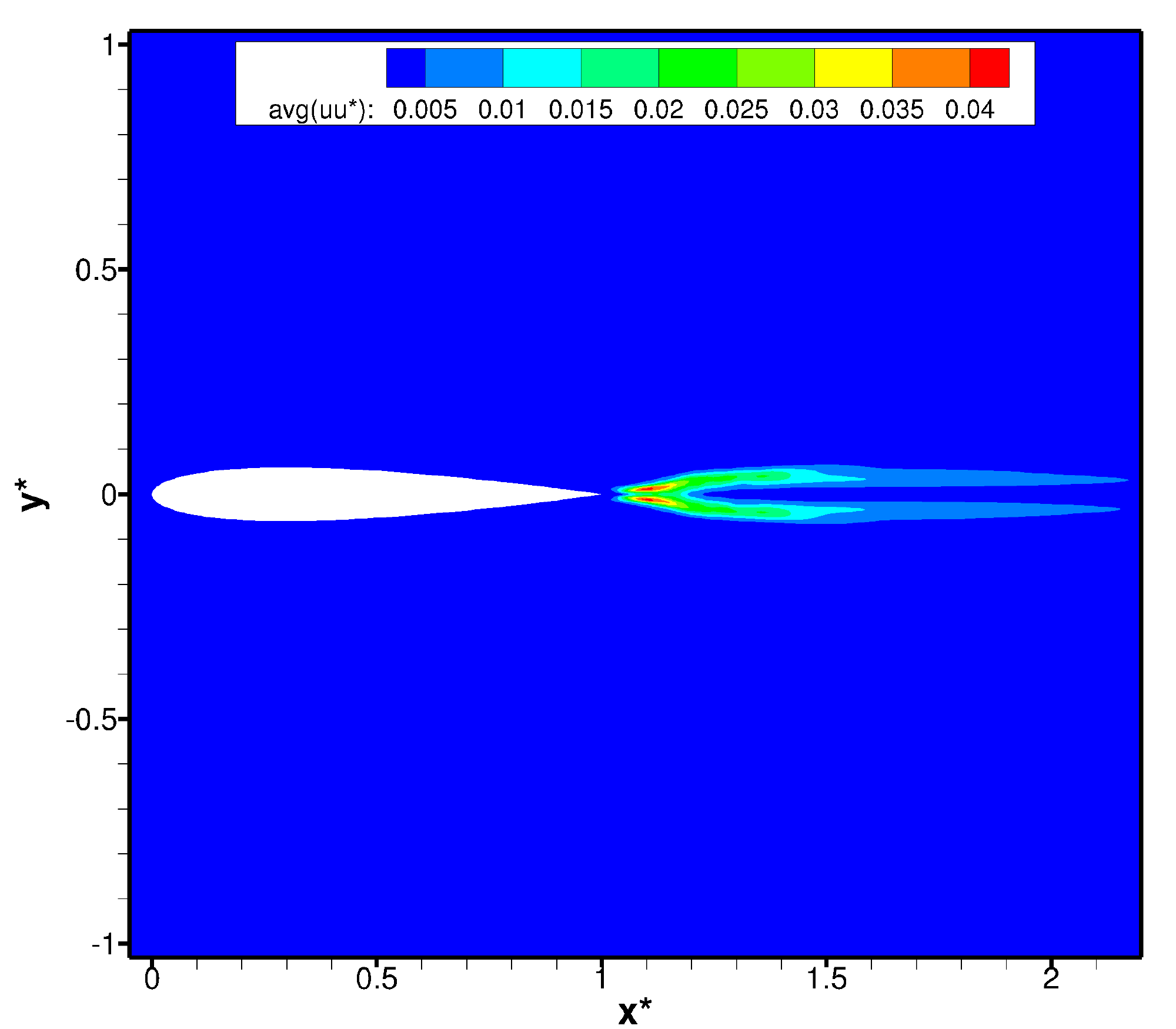}\label{fig:avg_uu_m_0_y+_max}}\hfill
	\centering
	\subfigure[$\tau_{12}^{turb^*}$: $m-0-y^+_{med}$ mesh for $t^*_{avg}=322$.]{\includegraphics[width=0.455\textwidth]{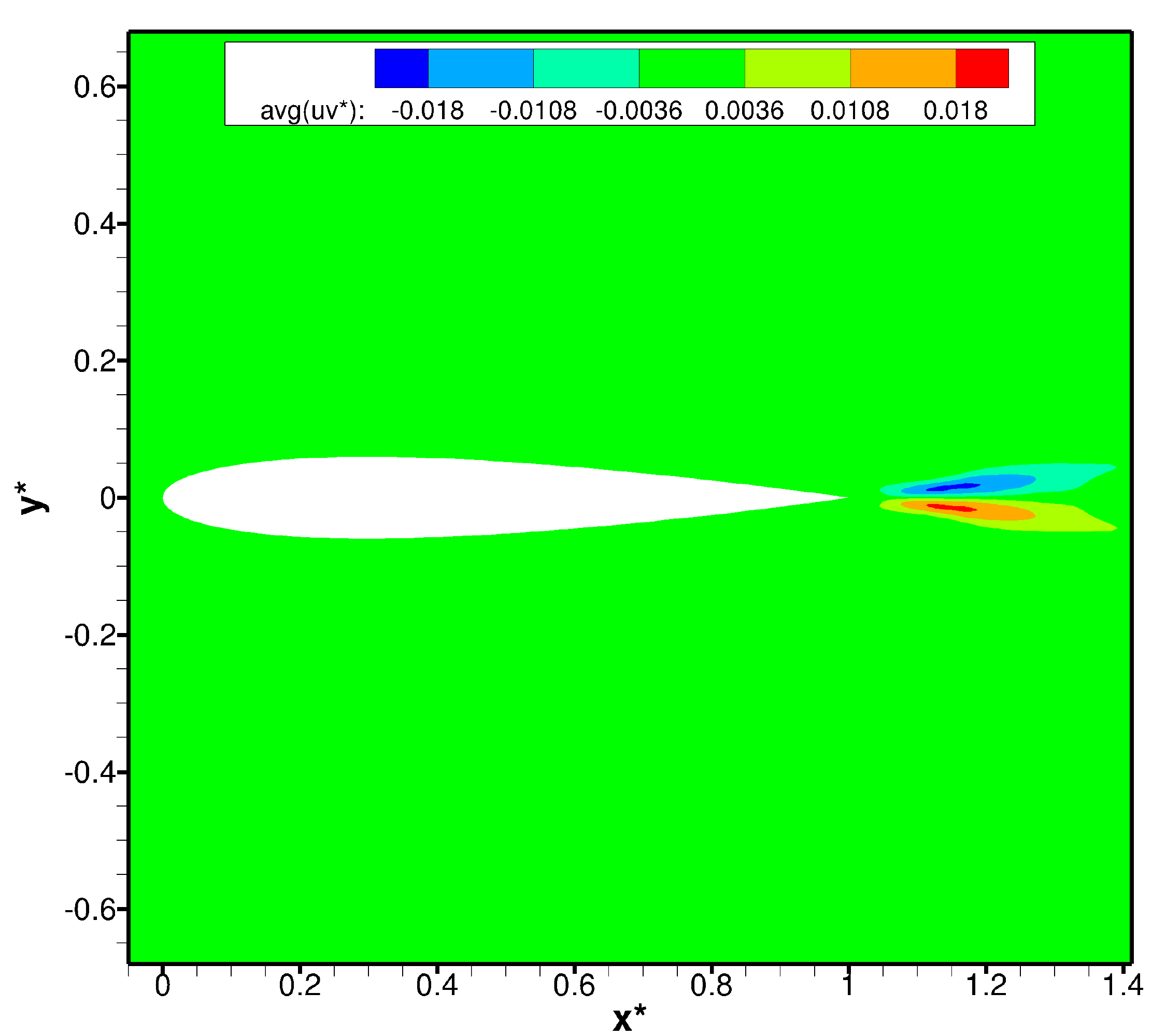}\label{fig:avg_uv_m_0_y+_med}}\hfill
	\subfigure[$\tau_{12}^{turb^*}$: $m-0-y^+_{max}$ mesh for $t^*_{avg}=1009$.]{\includegraphics[width=0.455\textwidth]{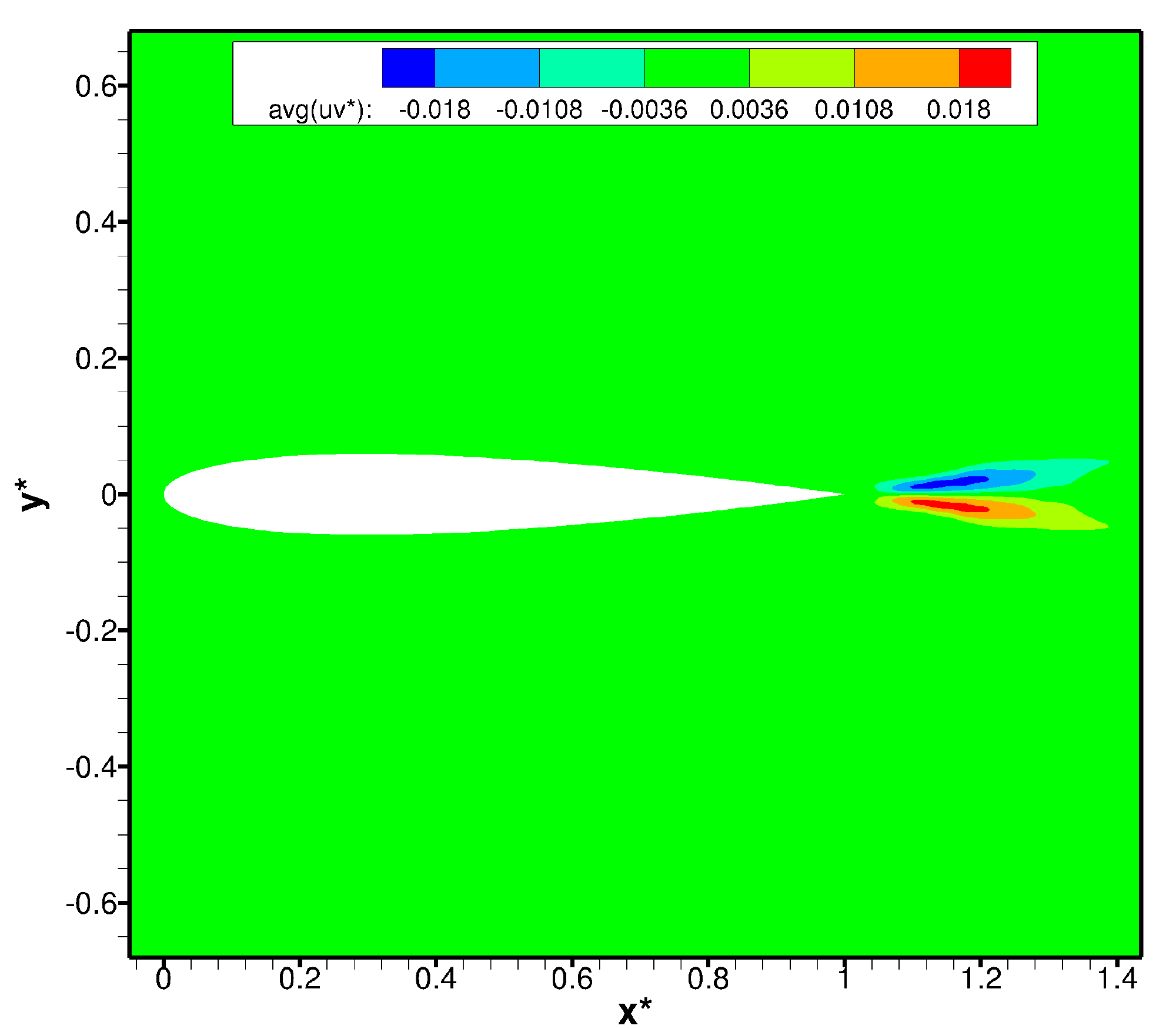}\label{fig:avg_uv_m_0_y+_max}}\hfill
	\centering
	\subfigure[$\tau_{22}^{turb^*}$: $m-0-y^+_{med}$ mesh for $t^*_{avg}=322$.]{\includegraphics[width=0.455\textwidth]{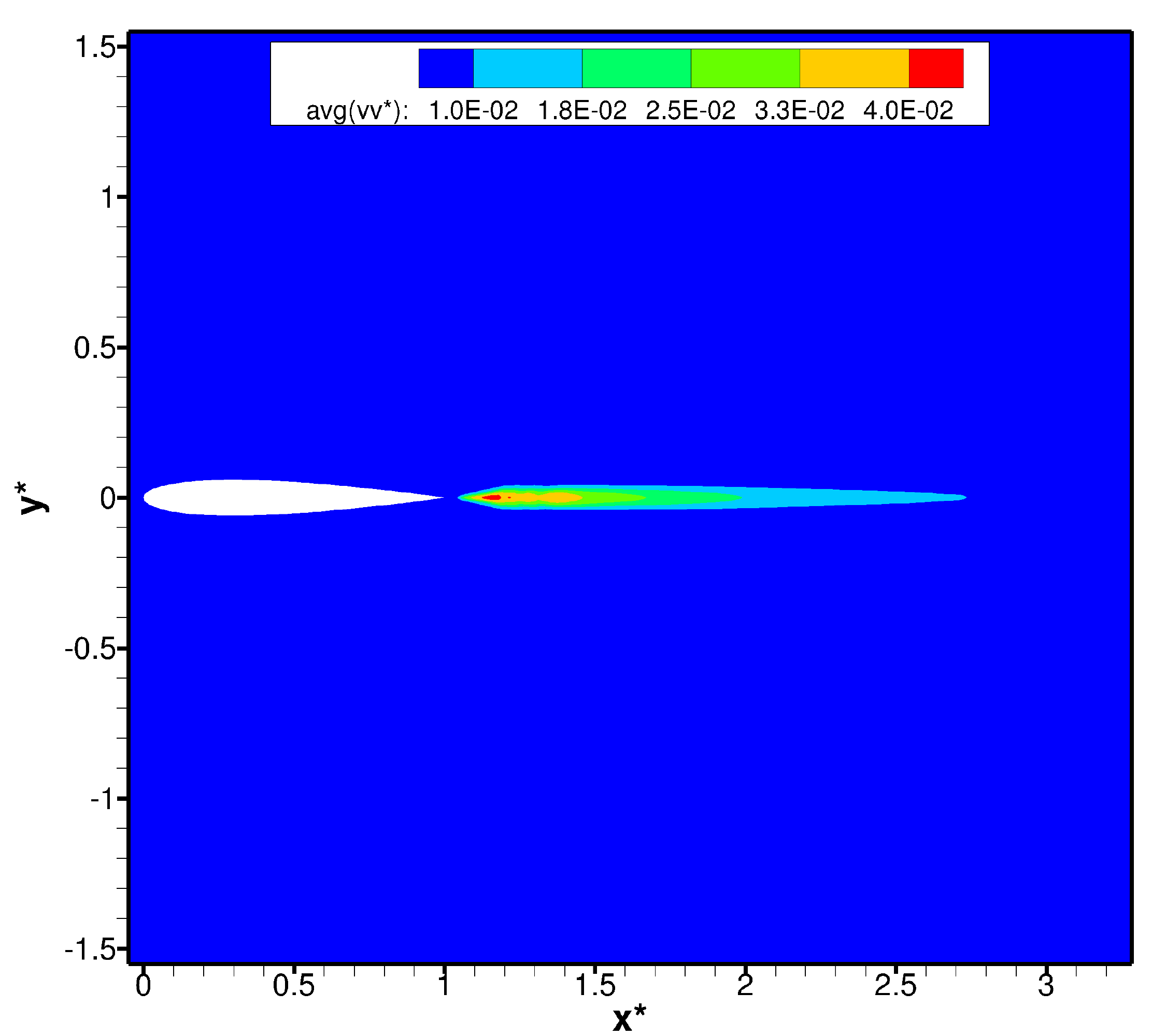}\label{fig:avg_vv_m_0_y+_med}}\hfill
	\subfigure[$\tau_{22}^{turb^*}$: $m-0-y^+_{max}$ mesh for $t^*_{avg}=1009$.]{\includegraphics[width=0.455\textwidth]{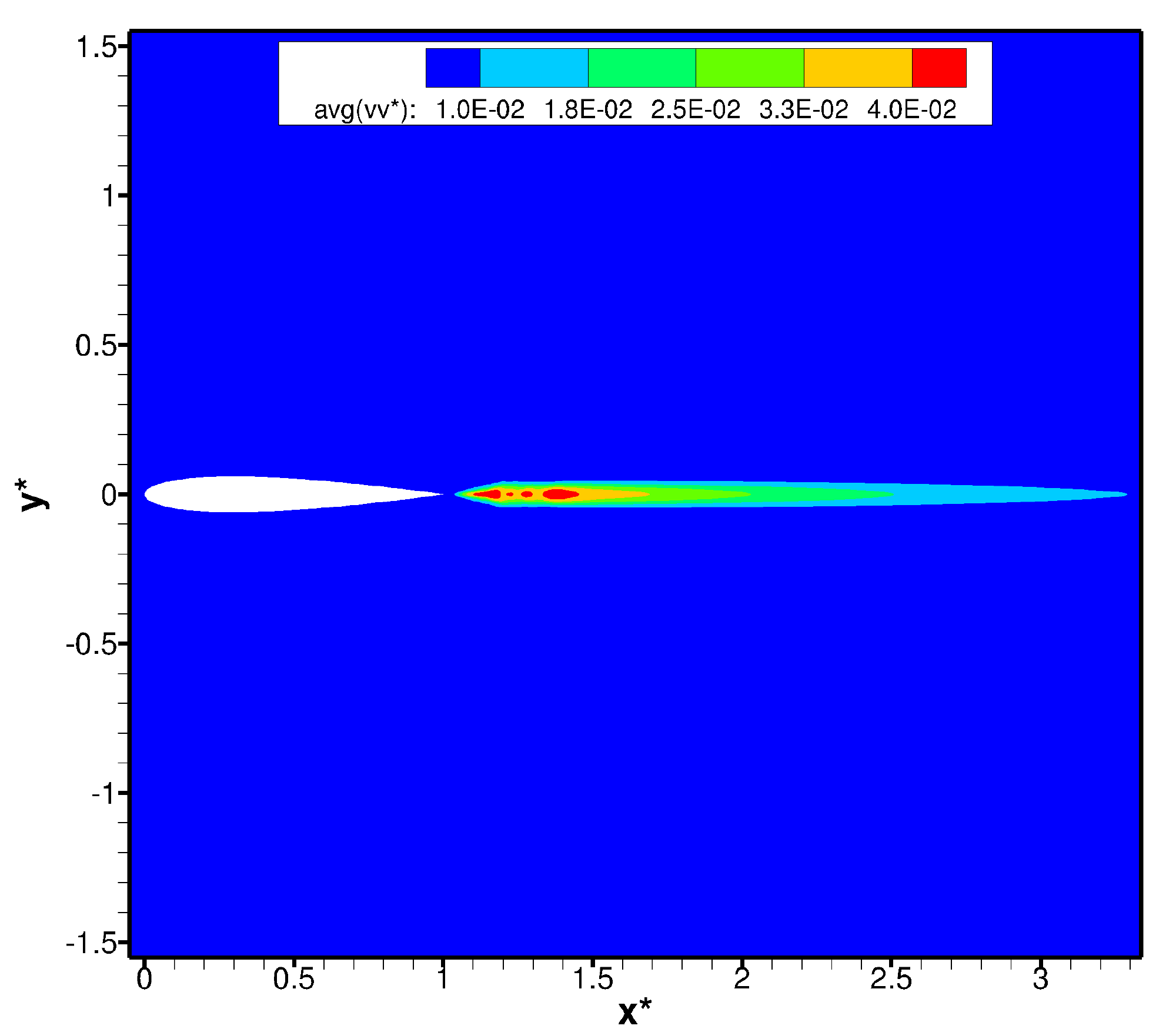}\label{fig:avg_vv_m_0_y+_max}}\hfill
	\caption{Dimensionless Reynolds stresses $\tau_{11}^{turb^*}$, $\tau_{12}^{turb^*}$ and $\tau_{22}^{turb^*}$ for the meshes with $\alpha=0^\circ$ (results are spatial-averaged in the span-wise direction).}
	\label{fig:avg_reynolds_stresses_angle_0}
\end{figure}

 Figures \ref{fig:avg_uu_m_0_y+_med} and \ref{fig:avg_uu_m_0_y+_max} illustrate $\tau_{11}^{turb^*}$ for the $m-0-y^+_{med}$ and $m-0-y^+_{max}$ meshes, respectively. This varies within $0 \leq \tau_{11}^{turb^*} \leq 0.0419$ for the former grid, and within $0 \leq \tau_{11}^{turb^*} \leq 0.0444$ for the latter, therefore some small discrepancies are present in the figures. The distribution of this Reynolds stresses are also symmetrical, which is caused by the symmetrical airfoil at an angle of attack of $\alpha=0^\circ$. 
\par The $\tau_{12}^{turb^*}$ component of the Reynolds stress tensor is shown in \mbox{Figs$.$ \ref{fig:avg_uv_m_0_y+_med} and \ref{fig:avg_uv_m_0_y+_max}} for the $m-0-y^+_{med}$ and $m-0-y^+_{max}$ grids, respectively. While the former has values varying within $-0.0208 \leq \tau_{12}^{turb^*} \leq 0.0209$, the latter is contained within \mbox{$-0.0236 \leq \tau_{12}^{turb^*} \leq 0.0236$}.  $\tau_{12}^{turb^*}$ is negative for $y\geq0$ and positive for $y\leq 0$. This is caused by the different directions of the velocity $u_2$ at the end of the airfoil for the suction and pressure side. 
\par The $\tau_{22}^{turb^*}$ component is illustrated in Figs$.$ \ref{fig:avg_vv_m_0_y+_med} and \ref{fig:avg_vv_m_0_y+_max} for respectively the $m-0-y^+_{med}$ and $m-0-y^+_{max}$ grids. It varies within $0 \leq \tau_{22}^{turb^*} \leq 0.0466$ for the former mesh, and between $0 \leq \tau_{22}^{turb^*} \leq 0.0493$ for the latter. A discrepancy caused by the different analyzed values of the dimensionless averaging time is present in the airfoil wake for the $\tau_{22}^{turb^*}$ component.
\par Some patterns in the values of the studied components of the Reynolds stresses are observed: Firstly, the Reynolds stresses are only present in the airfoil wake since the flow around the NACA0012 profile at a Reynolds number of $Re=100{,}000$ is laminar. Secondly, the modulus of the Reynolds stresses are symmetrical with the symmetry plane located at $y=0c$. This happens because the airfoil is symmetrical and has an angle of attack of $\alpha=0^\circ$. Finally, the range of this stresses is always slightly greater for the $m-0-y^+_{max}$ mesh, which is caused by the different analyzed dimensionless times. These deviations are negligible, therefore the investigation of the Reynolds stresses also indicate a mesh independence, as already concluded in Sections \ref{sec:analysis_dimensionless_wall_distance} and \ref{sec:simulation_analysis}.

\subsection{Angle of attack $\alpha=5^\circ$}
\label{subsec:reynolds_analysis_angle_5}

\par The components $\tau_{11}^{turb^*}$, $\tau_{12}^{turb^*}$ and $\tau_{22}^{turb^*}$ of the dimensionless Reynolds stress tensor are studied for the medium resolution meshes at an angle of attack of $\alpha=5^\circ$. The investigation is based only on the previous described components due to the almost two-dimensional flow around the NACA0012 airfoil. A velocity component in the span-wise direction is generated due to turbulence, but this is considerably smaller compared to the components in chord-wise and wall-normal directions.

\par
\begin{figure}[H]
	\centering
	\subfigure[$\tau_{11}^{turb^*}$: $m-5-y^+_{med}$ mesh for $t^*_{avg}=152$.]{\includegraphics[width=0.455\textwidth]{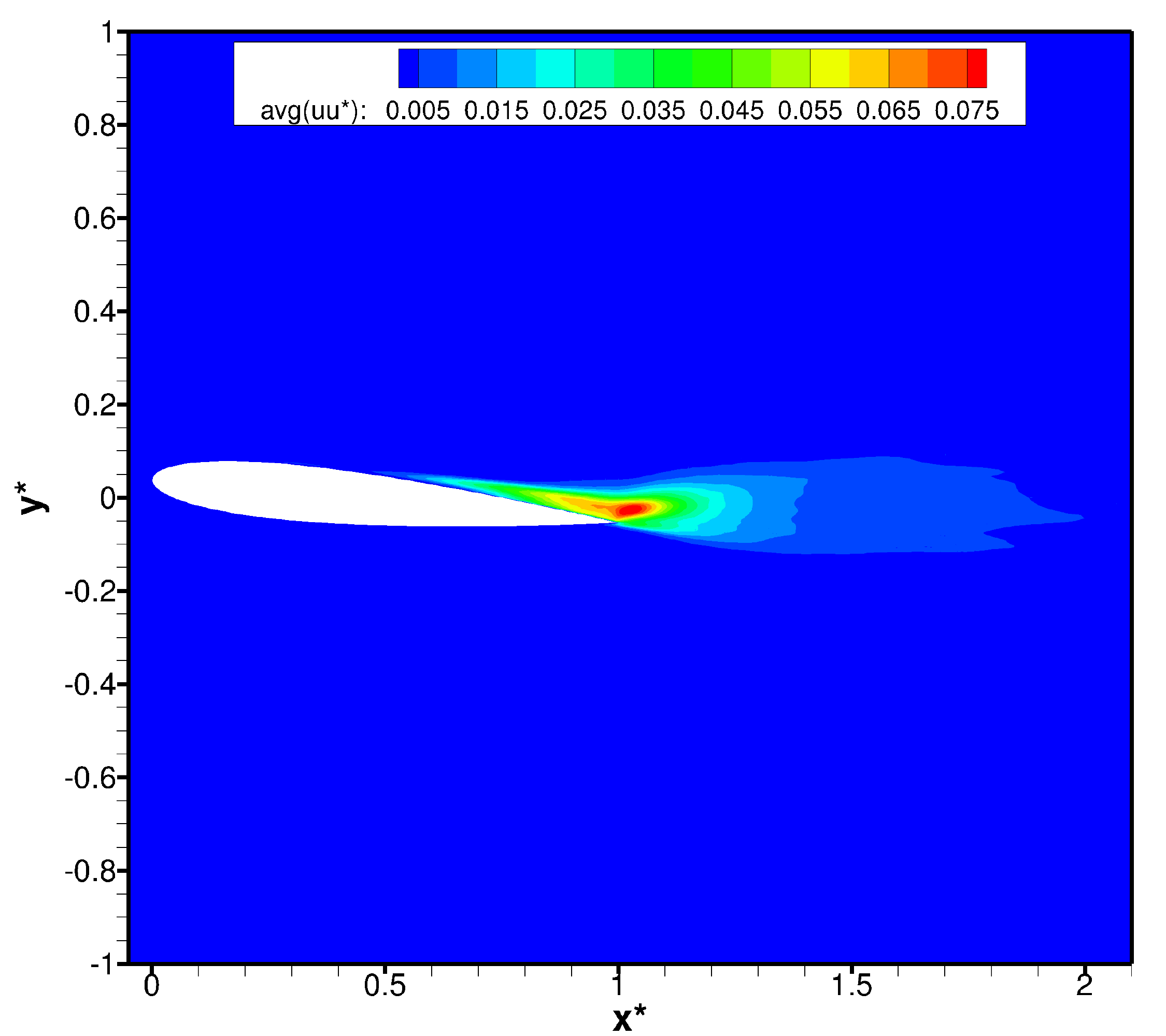}\label{fig:avg_uu_m_5_y+_med}}\hfill
	\subfigure[$\tau_{11}^{turb^*}$: $m-5-y^+_{max}$ mesh for $t^*_{avg}=1009$.]{\includegraphics[width=0.455\textwidth]{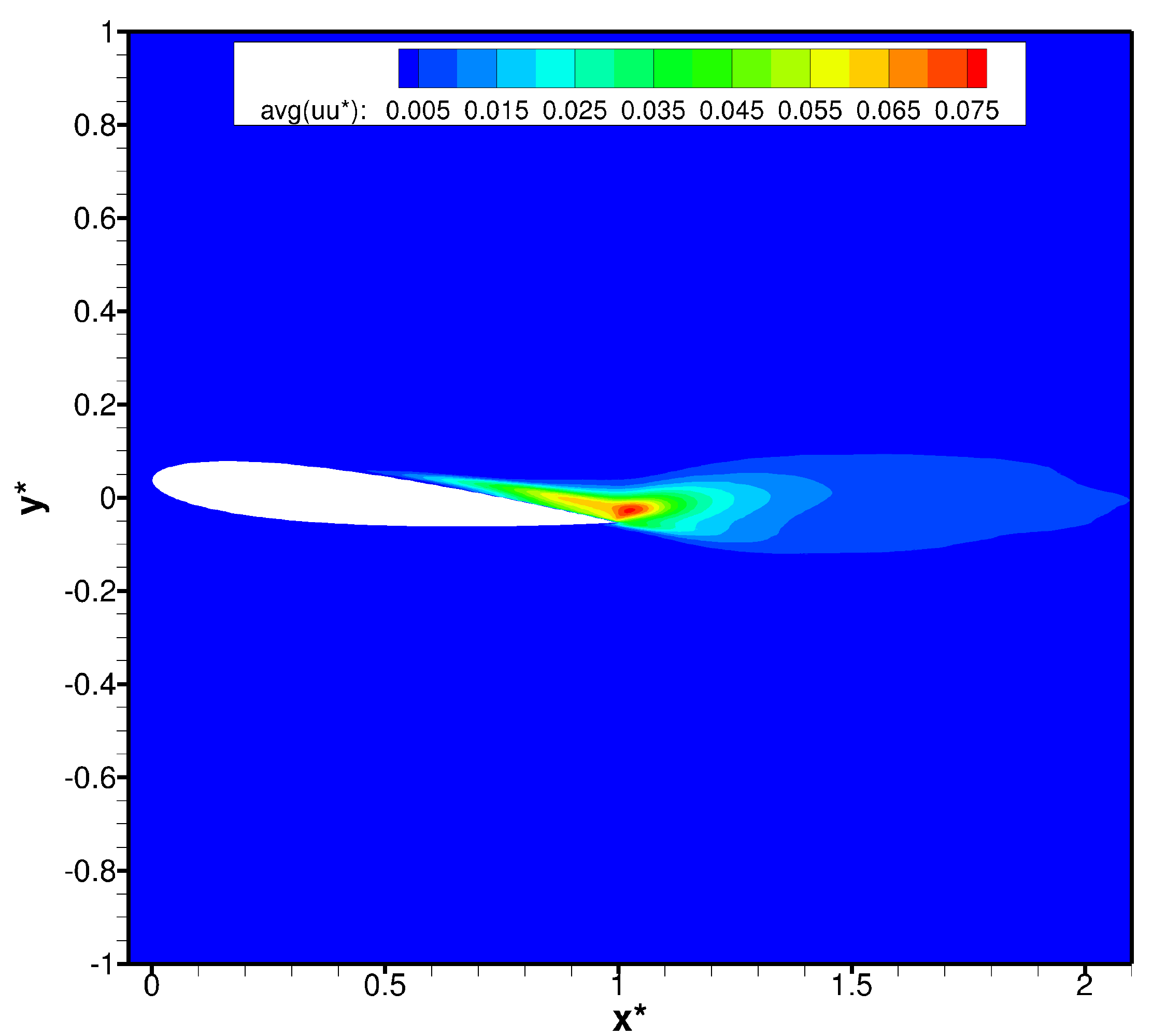}\label{fig:avg_uu_m_5_y+_max}}\hfill
	\centering
	\subfigure[$\tau_{12}^{turb^*}$: $m-5-y^+_{med}$ mesh for $t^*_{avg}=152$.]{\includegraphics[width=0.455\textwidth]{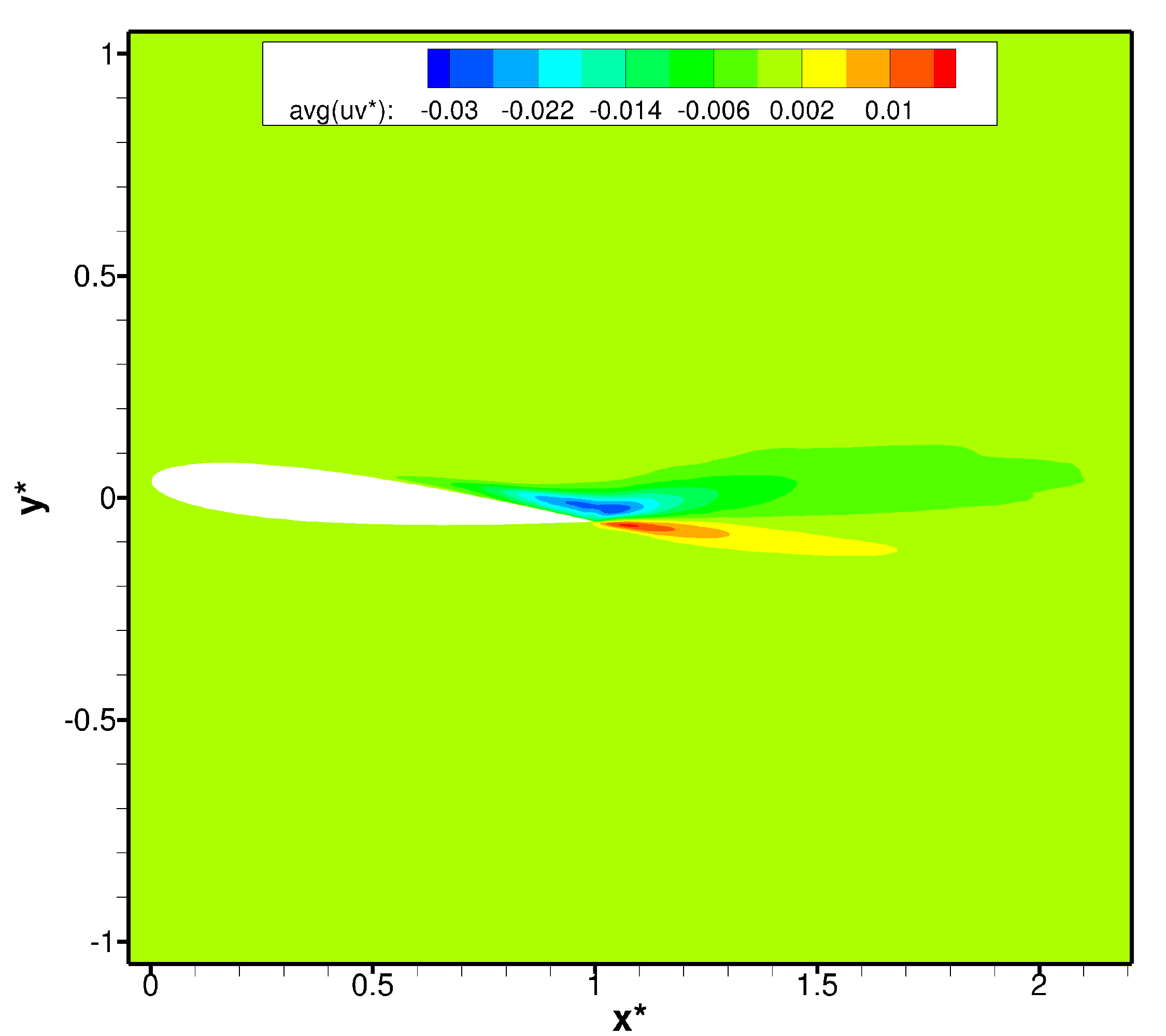}\label{fig:avg_uv_m_5_y+_med}}\hfill
	\subfigure[$\tau_{12}^{turb^*}$: $m-5-y^+_{max}$ mesh for $t^*_{avg}=1009$.]{\includegraphics[width=0.455\textwidth]{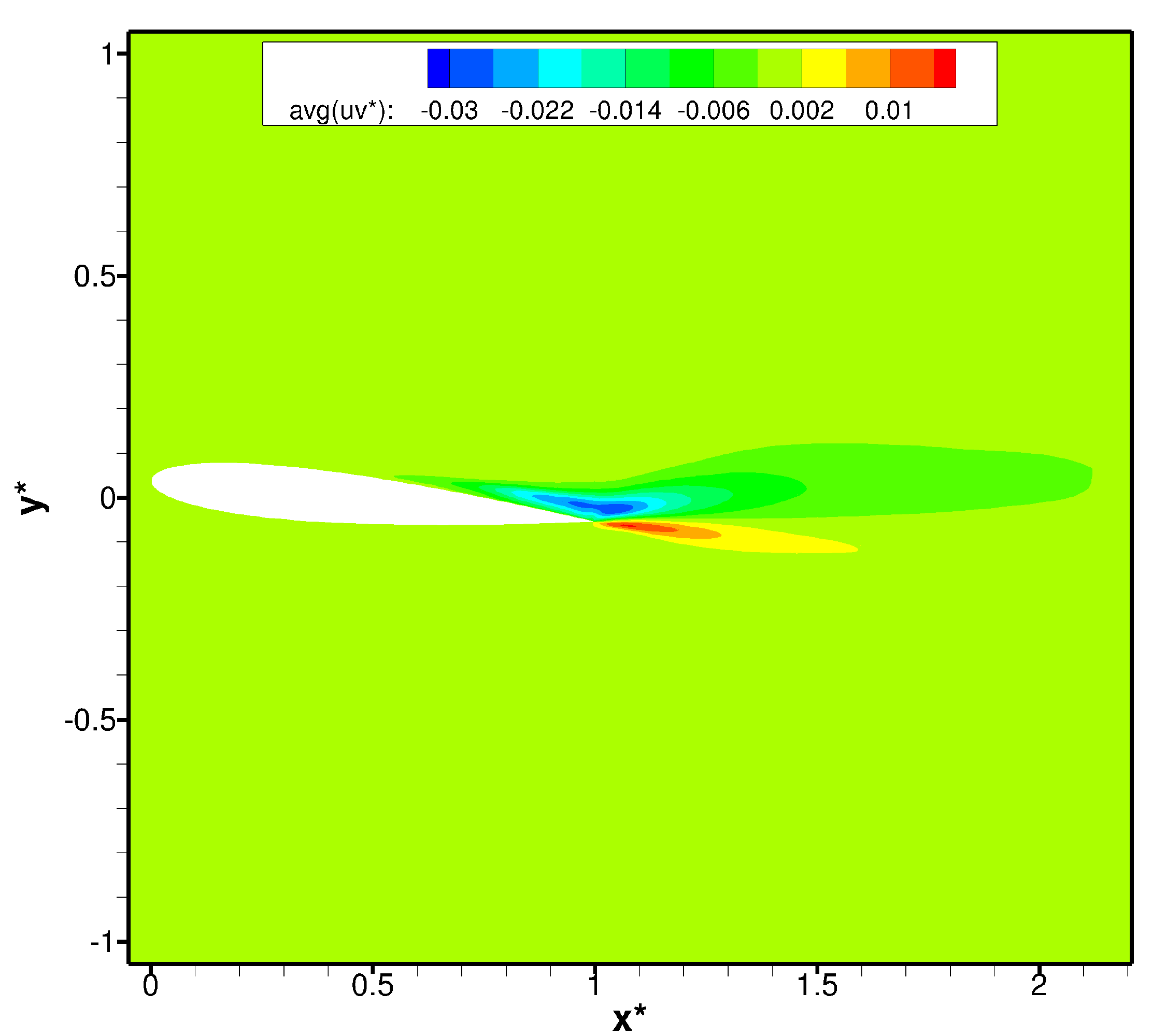}\label{fig:avg_uv_m_5_y+_max}}\hfill
	\centering
	\subfigure[$\tau_{22}^{turb^*}$: $m-5-y^+_{med}$ mesh for $t^*_{avg}=152$.]{\includegraphics[width=0.455\textwidth]{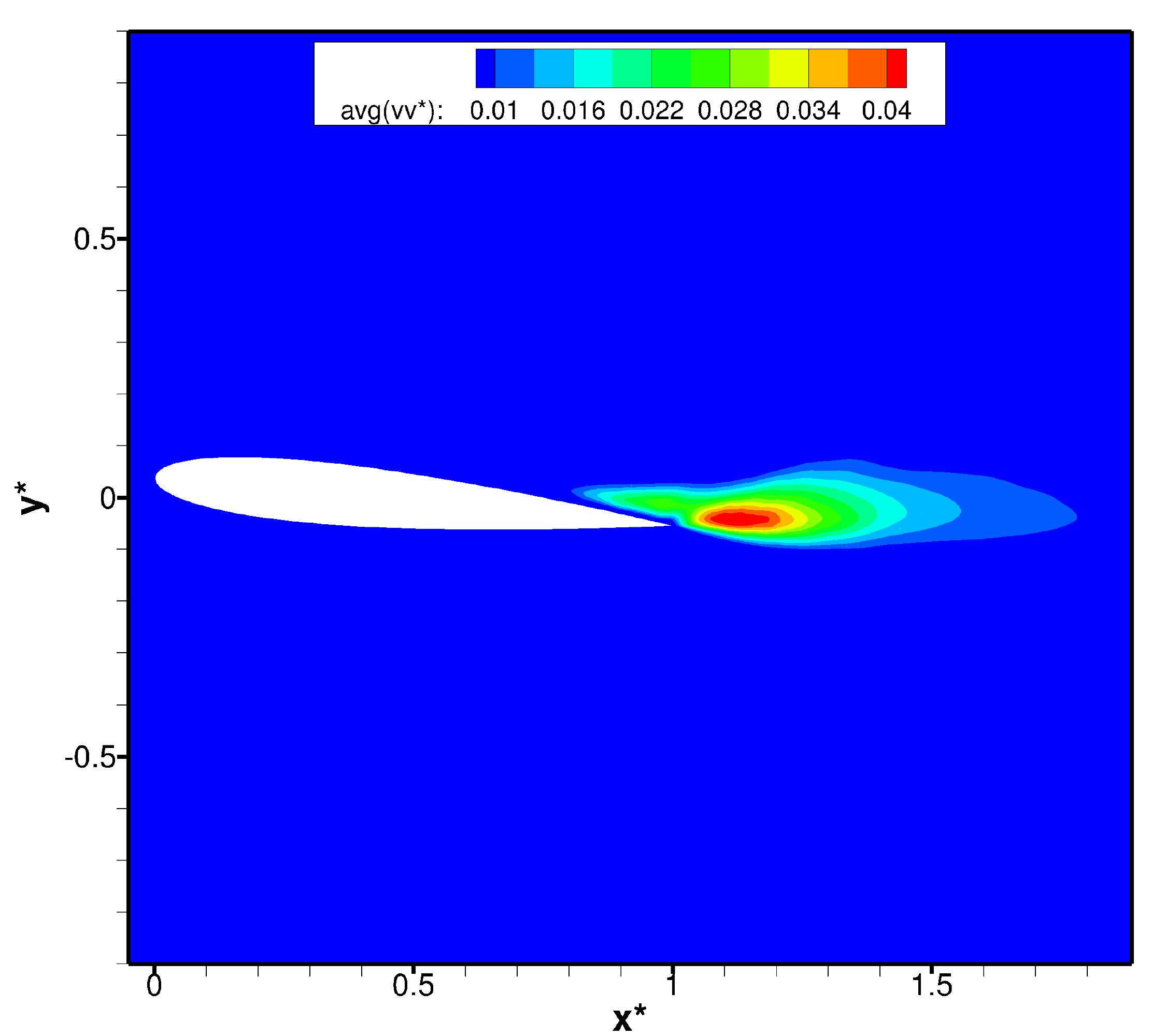}\label{fig:avg_vv_m_5_y+_med}}\hfill
	\subfigure[$\tau_{22}^{turb^*}$: $m-5-y^+_{max}$ mesh for $t^*_{avg}=1009$.]{\includegraphics[width=0.455\textwidth]{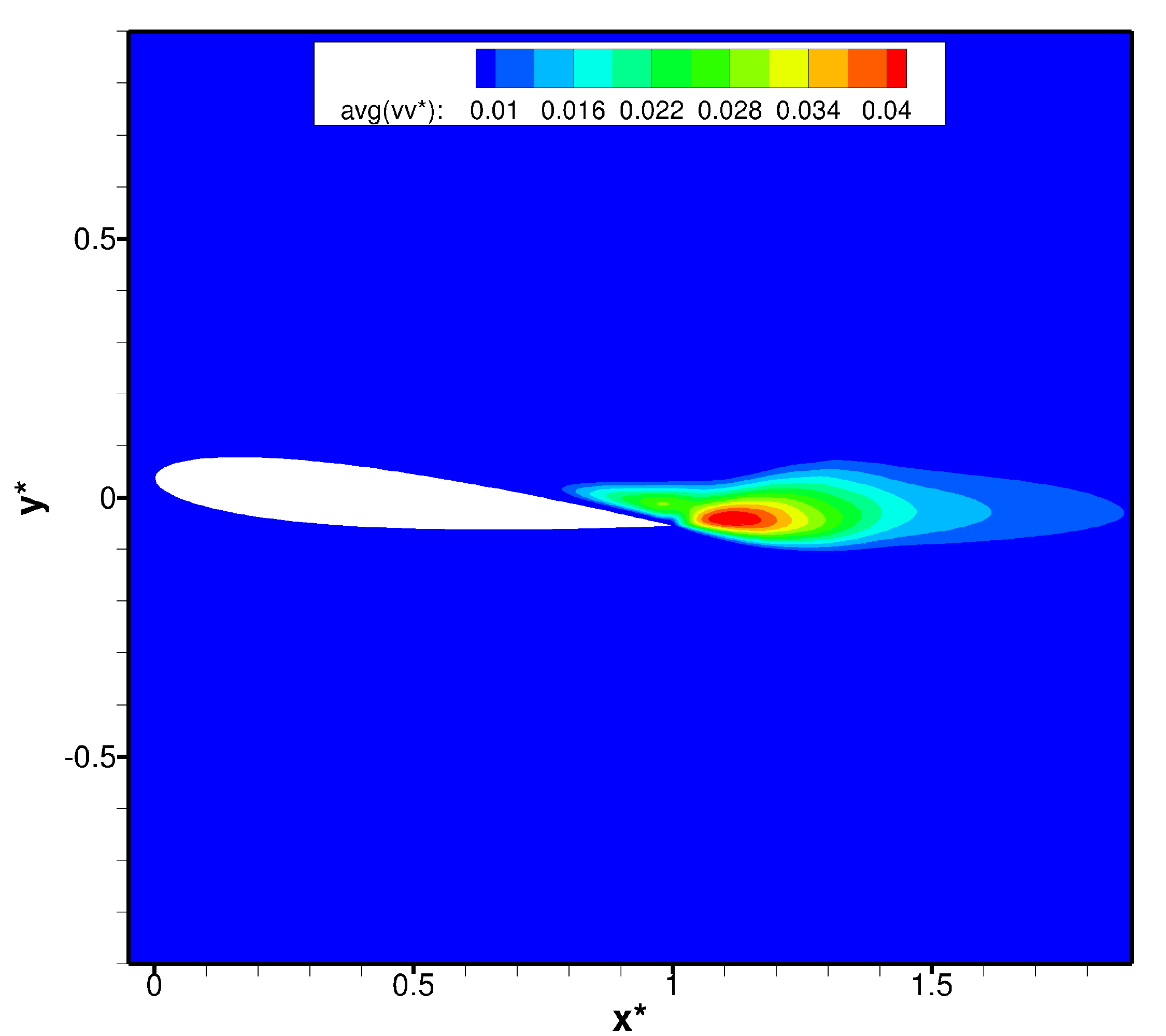}\label{fig:avg_vv_m_5_y+_max}}\hfill
	\caption{Dimensionless Reynolds stresses $\tau_{11}^{turb^*}$, $\tau_{12}^{turb^*}$ and $\tau_{22}^{turb^*}$ for the meshes with $\alpha=5^\circ$ (results are spatial-averaged in the span-wise direction).}
	\label{fig:avg_reynolds_stresses_angle_5}
\end{figure}

 Figures \ref{fig:avg_uu_m_5_y+_med} and \ref{fig:avg_uu_m_5_y+_max} show the $\tau_{11}^{turb^*}$ component of the stress tensor for the $m-5-y^+_{med}$ and $m-5-y^+_{max}$ meshes, respectively. This varies within $0 \leq \tau_{11}^{turb^*} \leq 0.0906$ for the former grid, and within $0 \leq \tau_{11}^{turb^*} \leq 0.0811$ for the latter. Besides the different intervals, the region affected by this stresses is larger for the $m-5-y^+_{max}$ mesh. The discrepancies present in both results are, however, caused by the differences in the analyzed dimensionless averaging times.

\par The $\tau_{12}^{turb^*}$ component of the Reynolds stress tensor is illustrated in Figs.\ \ref{fig:avg_uv_m_5_y+_med} and \ref{fig:avg_uv_m_5_y+_max} for the $m-5-y^+_{med}$ and $m-5-y^+_{max}$ grids, respectively. While the former has values varying between $-0.0330 \leq \tau_{12}^{turb^*} \leq 0.0162$, the latter is within \mbox{$-0.0305 \leq \tau_{12}^{turb^*} \leq 0.0149$}. $\tau_{12}^{turb^*}$ is negative near the suction side and positive near the pressure side. This phenomenon is caused by the flow following the airfoil profile, i.e$.$, by the opposite directions of the velocity $u_2$ near the pressure and the suction sides. 
\par The $\tau_{22}^{turb^*}$ component is presented in Figs.\ \ref{fig:avg_vv_m_5_y+_med} and \ref{fig:avg_vv_m_5_y+_max} for respectively the $m-5-y^+_{med}$ and $m-5-y^+_{max}$ grids. It varies within $0 \leq \tau_{22}^{turb^*} \leq 0.0476$ for the former mesh, and between $0 \leq \tau_{22}^{turb^*} \leq 0.0443$ for the latter. In addition to the divergence in the range of the component $\tau_{22}^{turb^*}$ of the Reynolds stresses, the region affected by this component is also larger for the $m-5-y^+_{max}$ mesh, which is caused by the different dimensionless averaging times.
\par As for the meshes with an angle of attack of $\alpha=0^\circ$, some patterns in the values of the studied components of the Reynolds stresses are observed: Firstly, no Reynolds stresses are present on the pressure side, since the flow around the profile is laminar and it does not suffer a boundary layer detachment on this side. Secondly, on the suction side, Reynolds stresses are present due to the detachment of the boundary layer and the formation of a recirculation zone (see Sections \ref{sec:simulation_analysis} and \ref{sec:flow_summary}). Finally, the interval of this tensor values is always slightly longer for the $m-5-y^+_{med}$ mesh, which is analyzed for a smaller dimensionless averaging time. Since these values tend to decrease with the increase in the dimensionless averaging time, the present deviations are caused by the different investigated dimensionless averaging time. Therefore, the investigation of the Reynolds stresses also indicate a mesh independence, as already concluded in \mbox{Sections \ref{sec:analysis_dimensionless_wall_distance} and \ref{sec:simulation_analysis}}. 

\subsection{Angle of attack $\alpha=11^\circ$}
\label{subsec:reynolds_analysis_angle_11}

\par The dimensionless Reynolds stress tensor of the meshes at an angle of attack of \mbox{$\alpha=11^\circ$}, i.e$.$, $m-11-y^+_{med}$ and $m-11-y^+_{max}$, are analyzed with help of the components $\tau_{11}^{turb^*}$, $\tau_{12}^{turb^*}$ and $\tau_{22}^{turb^*}$. Although the results are already spatial and time-averaged, fluctuations are still present in the values of the $m-11-y^+_{med}$ mesh due to its larger required computational time and consequently small achieved dimensionless averaging time, i.e$.$, $t^*_{avg}=76$.

\par
\begin{figure}[H]
	\centering
	\subfigure[$\tau_{11}^{turb^*}$: $m-11-y^+_{med}$ mesh for $t^*_{avg}=76$.]{\includegraphics[width=0.455\textwidth]{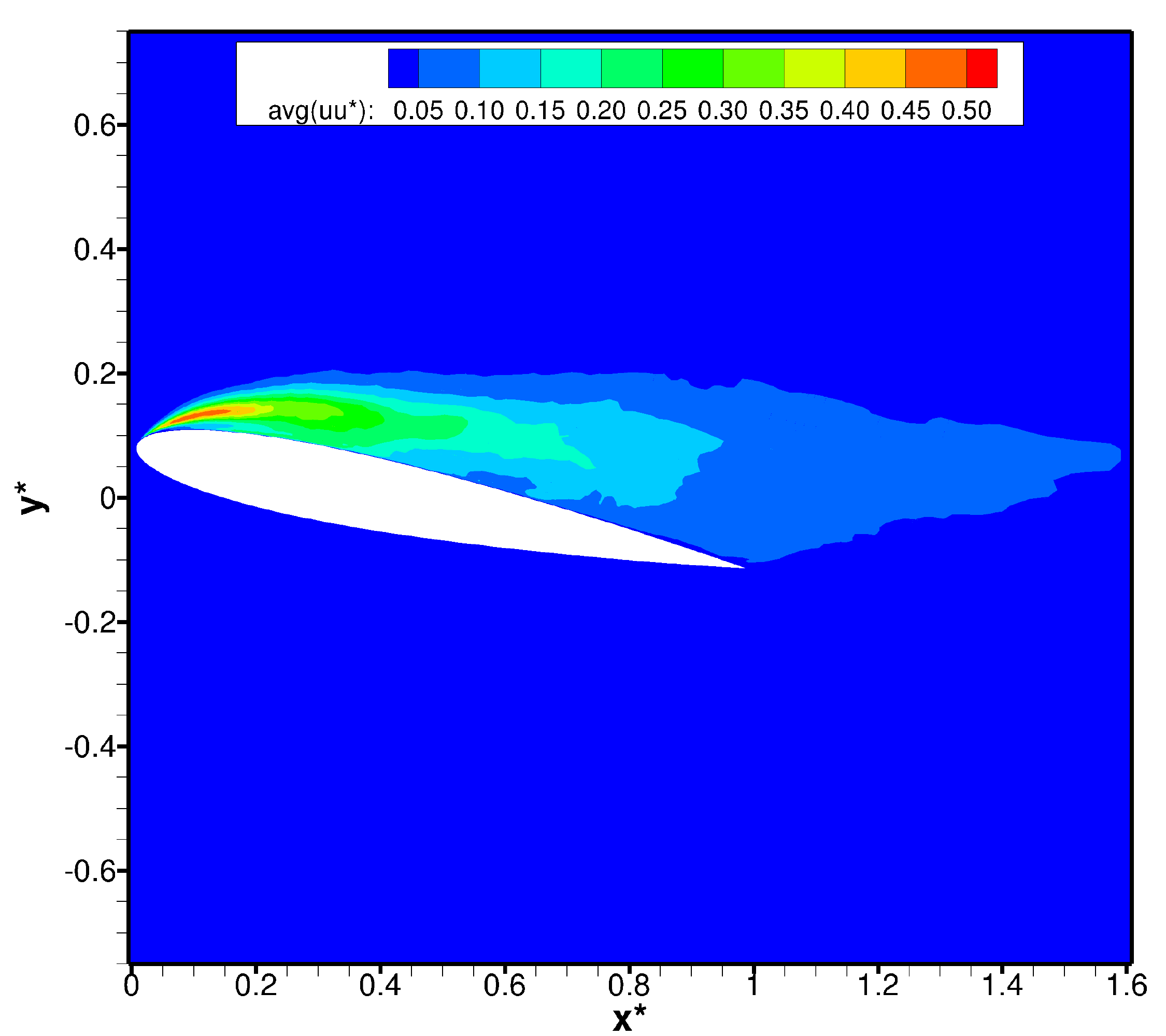}\label{fig:avg_uu_m_11_y+_med}}\hfill
	\subfigure[$\tau_{11}^{turb^*}$: $m-11-y^+_{max}$ mesh for $t^*_{avg}=454$.]{\includegraphics[width=0.455\textwidth]{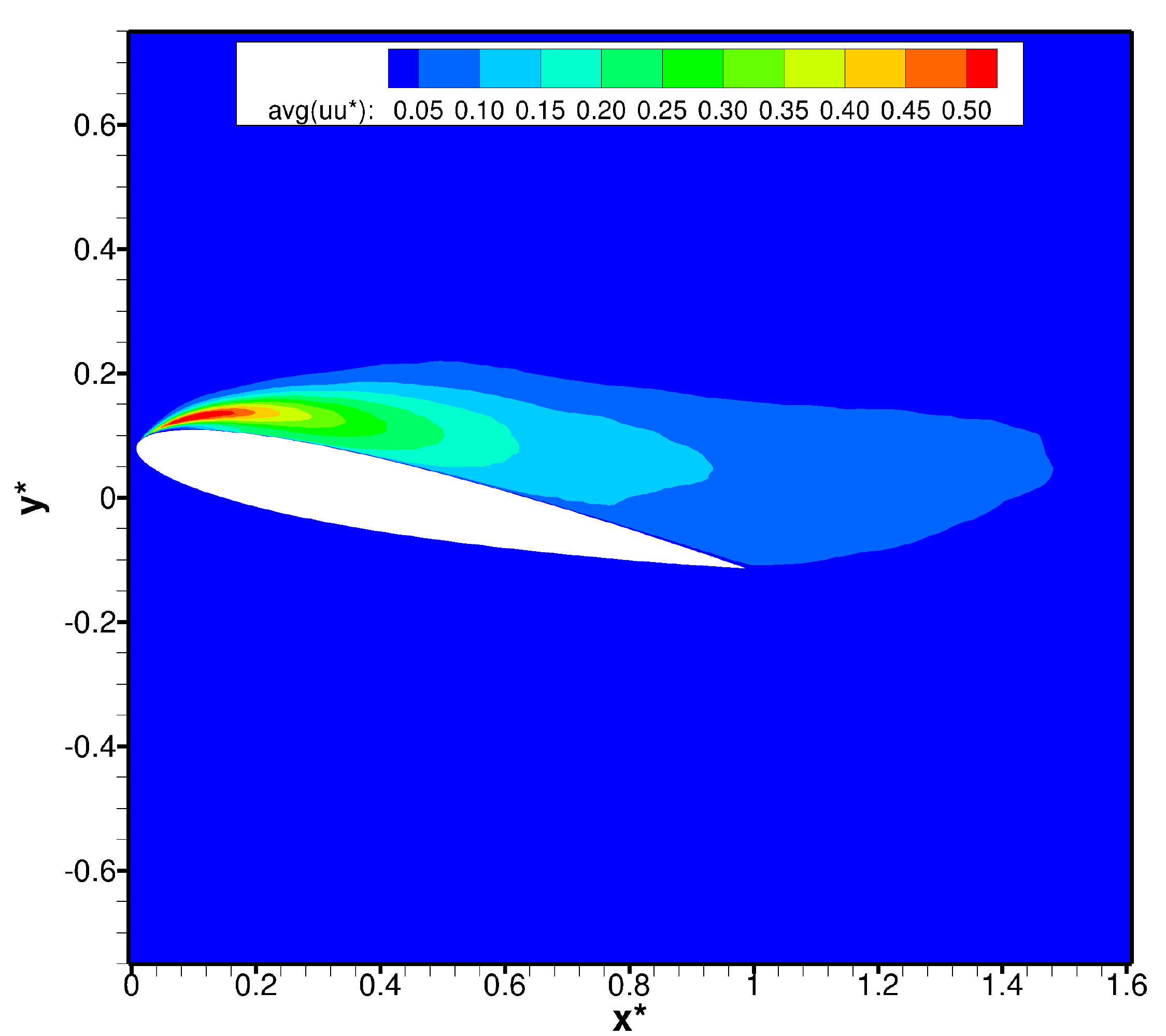}\label{fig:avg_uu_m_11_y+_max}}\hfill
	\centering
	\subfigure[$\tau_{12}^{turb^*}$: $m-11-y^+_{med}$ mesh for $t^*_{avg}=76$.]{\includegraphics[width=0.455\textwidth]{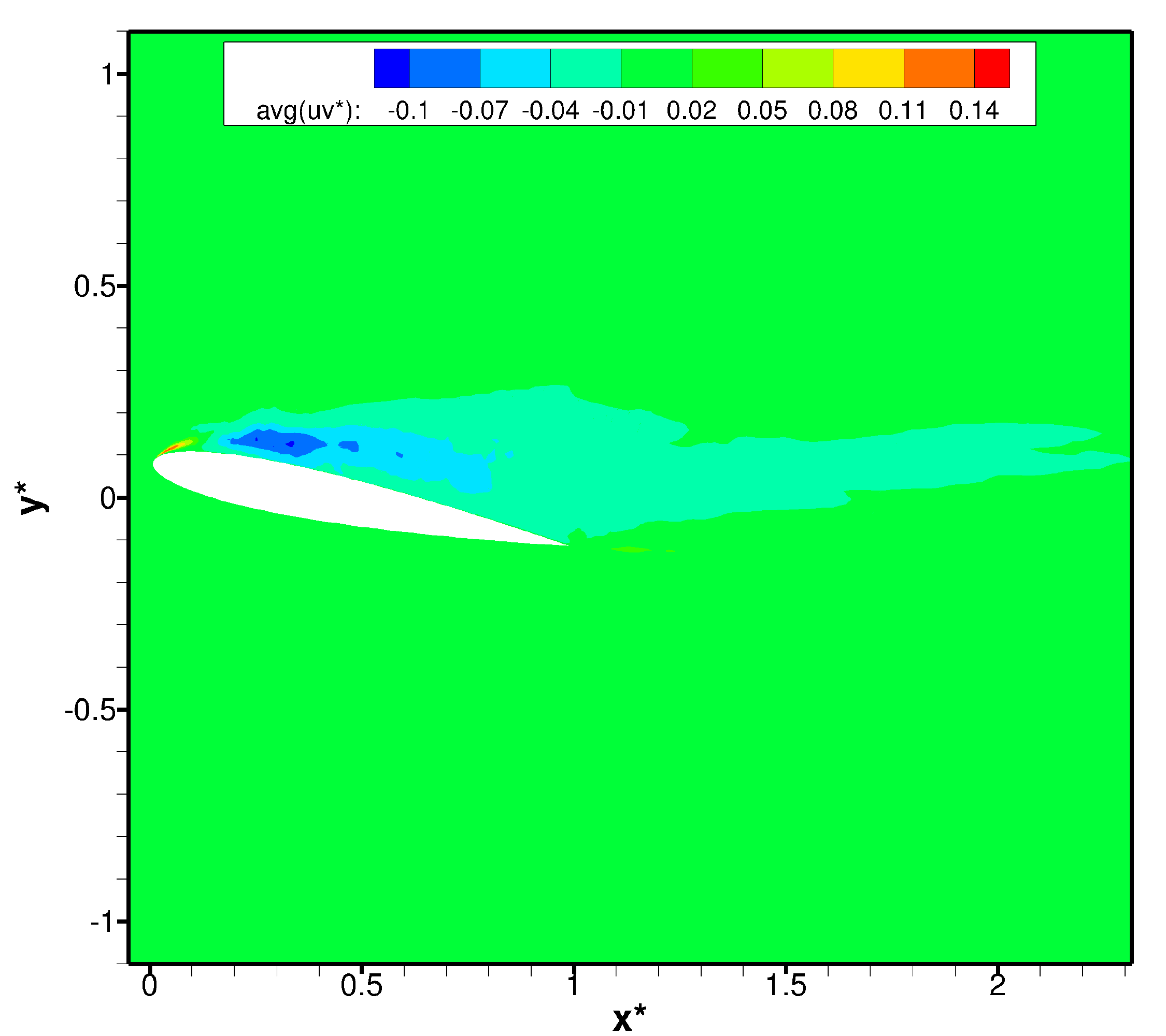}\label{fig:avg_uv_m_11_y+_med}}\hfill
	\subfigure[$\tau_{12}^{turb^*}$: $m-11-y^+_{max}$ mesh for $t^*_{avg}=454$.]{\includegraphics[width=0.455\textwidth]{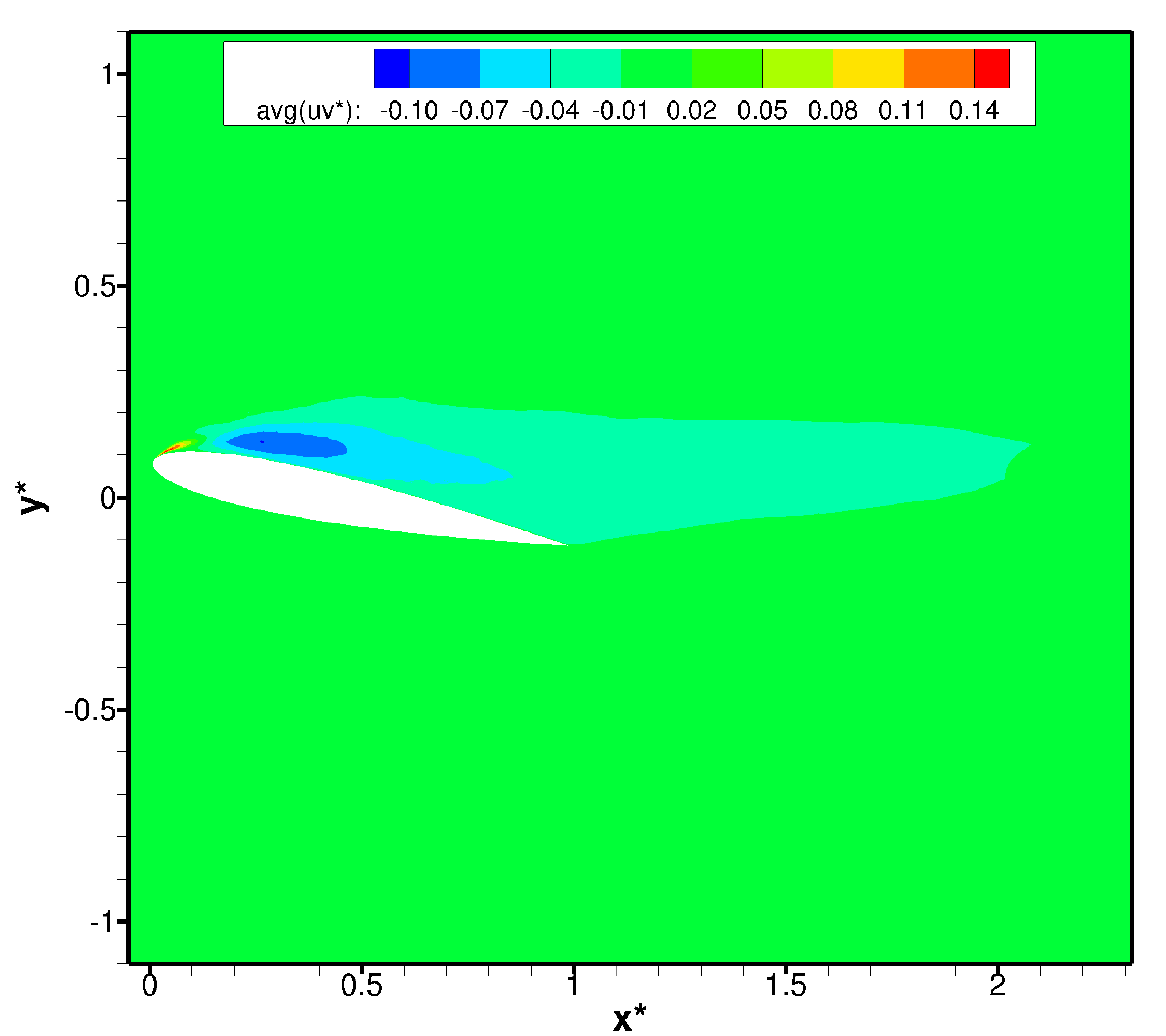}\label{fig:avg_uv_m_11_y+_max}}\hfill
\end{figure}

 Figures \ref{fig:avg_uu_m_11_y+_med} and \ref{fig:avg_uu_m_11_y+_max} show the $\tau_{11}^{turb^*}$ component of the stress tensor for the $m-11-y^+_{med}$ and $m-11-y^+_{max}$ meshes, respectively. This varies within $0 \leq \tau_{11}^{turb^*} \leq 0.7440$ for the former grid, and within $0 \leq \tau_{11}^{turb^*} \leq 0.6122$ for the latter. The region with stresses greater than $\tau_{11}^{turb^*}\geq 0.5$ are located near the leading edge and are larger for the $m-11-y^+_{max}$ grid, while the region affected by this stress component is larger for the $m-11-y^+_{med}$ mesh. The differences on both results may not be only caused by the difference in the analyzed dimensionless times and therefore, the grid may influence the computation of the flow.  

\par The $\tau_{12}^{turb^*}$ component of the Reynolds stress tensor is illustrated in Figs$.$ \ref{fig:avg_uv_m_11_y+_med} and \ref{fig:avg_uv_m_11_y+_max} for the $m-11-y^+_{med}$ and $m-11-y^+_{max}$ grids, respectively. While the former has values varying within $-0.1375 \leq \tau_{12}^{turb^*} \leq 0.2130$, the latter is contained within $-0.1049 \leq \tau_{12}^{turb^*} \leq 0.1689$. The highest values of this stress component, i.e$.$, $\tau_{12}^{turb^*}\geq 0.14$, are present near the leading edge for both grids. The area influenced by this Reynolds stress component is larger for the $m-11-y^+_{med}$ mesh.

\par
\begin{figure}[H]
	\centering
	\subfigure[$\tau_{22}^{turb^*}$: $m-11-y^+_{med}$ mesh for $t^*_{avg}=76$.]{\includegraphics[width=0.455\textwidth]{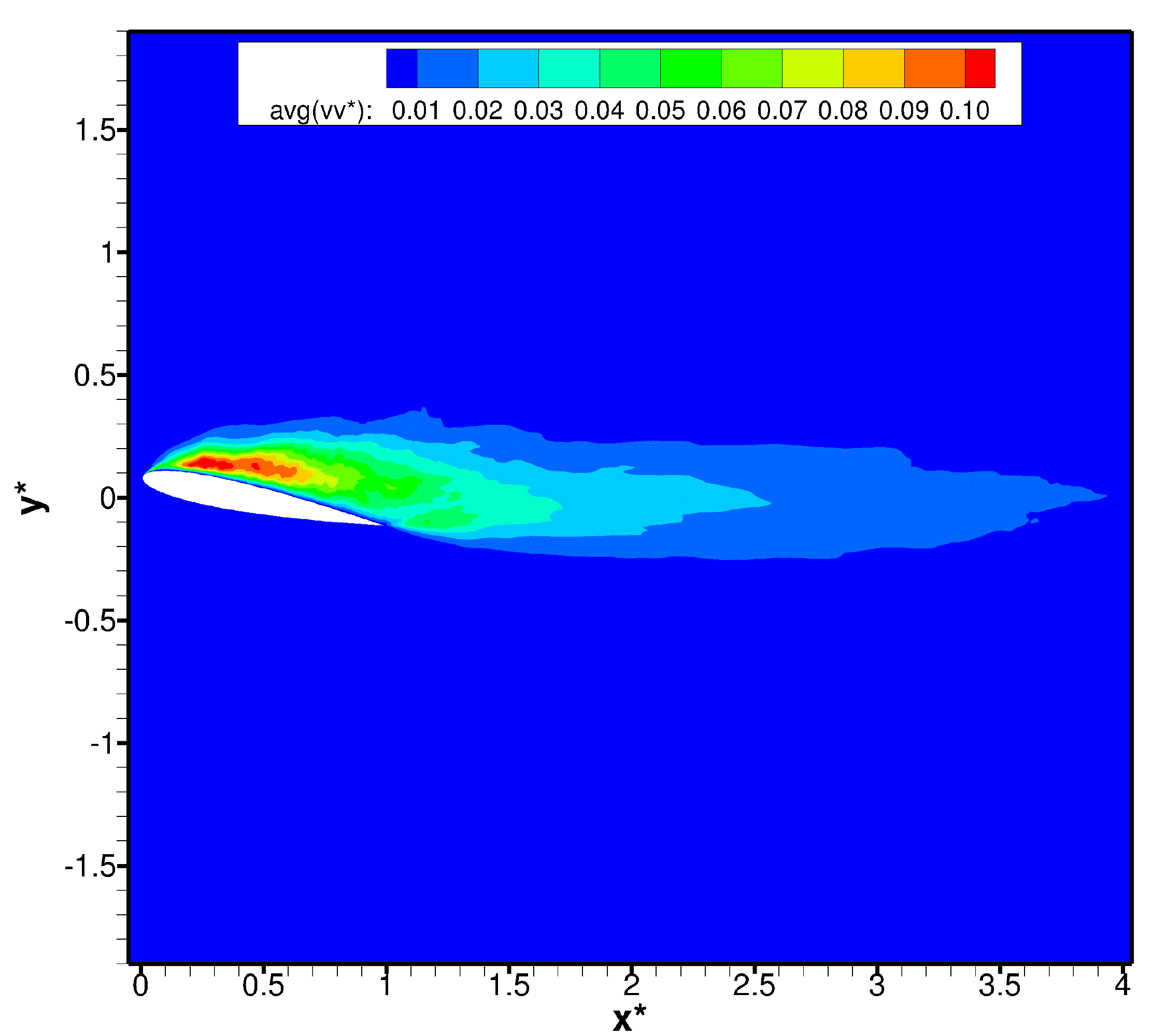}\label{fig:avg_vv_m_11_y+_med}}\hfill
	\subfigure[$\tau_{22}^{turb^*}$: $m-11-y^+_{max}$ mesh for $t^*_{avg}=454$.]{\includegraphics[width=0.455\textwidth]{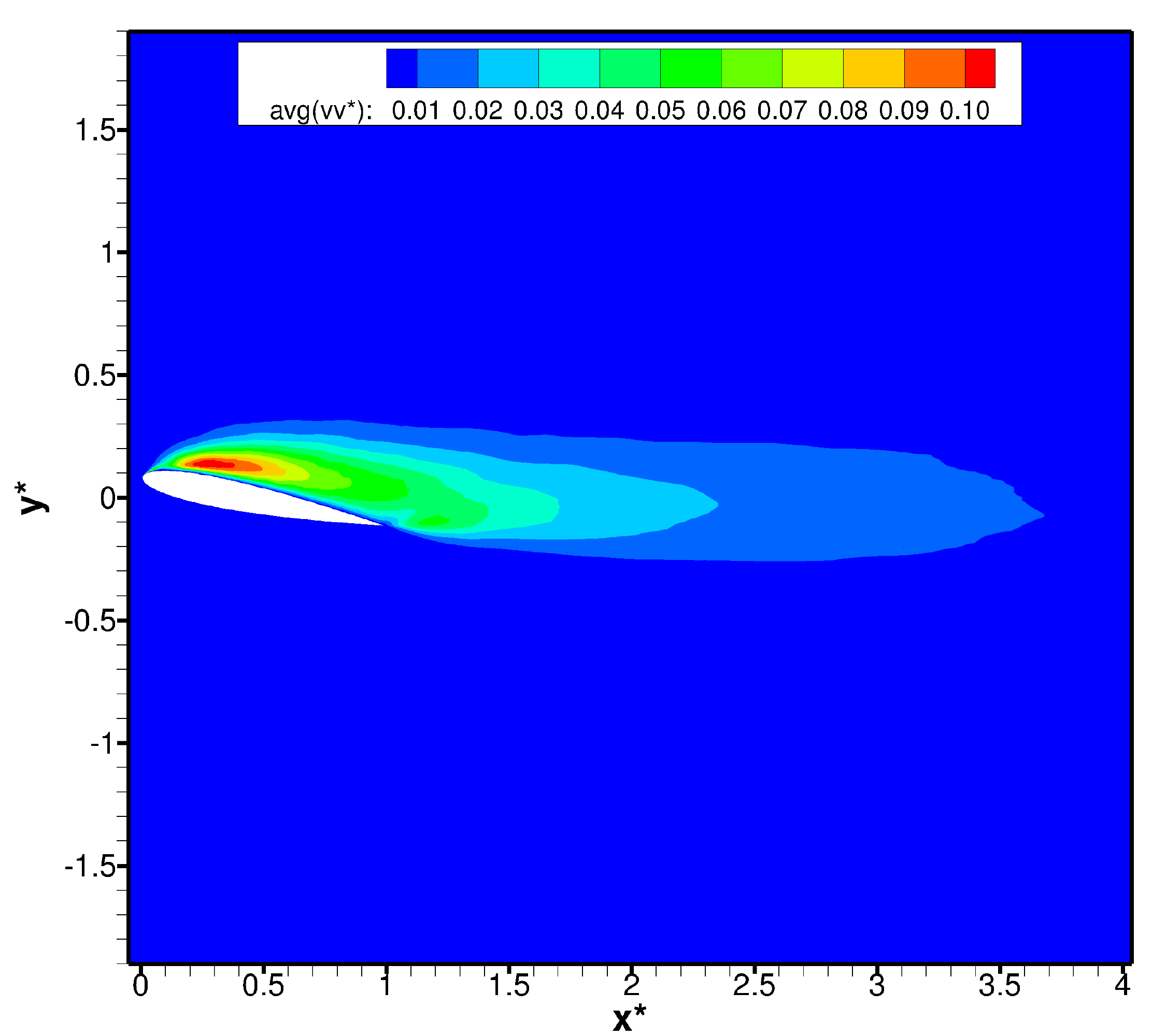}\label{fig:avg_vv_m_11_y+_max}}\hfill
	\caption{Dimensionless Reynolds stresses $\tau_{11}^{turb^*}$, $\tau_{12}^{turb^*}$ and $\tau_{22}^{turb^*}$ for the meshes with $\alpha=\nobreak 11^\circ$ (results are spatial-averaged in the span-wise direction).}
	\label{fig:avg_reynolds_stresses_angle_11}
\end{figure}

 The $\tau_{22}^{turb^*}$ component is presented in Figs$.$ \ref{fig:avg_vv_m_11_y+_med} and \ref{fig:avg_vv_m_11_y+_max} for respectively the $m-11-y^+_{med}$ and $m-11-y^+_{max}$ grids. It varies within $0 \leq \tau_{22}^{turb^*} \leq 0.1350$ for the former mesh, and between $0 \leq \tau_{22}^{turb^*} \leq 0.1153$ for the latter. The highest values of this component are located in the first half of the profile for the results of both meshes. 

\par The recognizable patterns on the Reynolds stresses of the meshes at an angle of attack of $\alpha=11^\circ$ are: Firstly, the achieved dimensionless averaging time is not enough to acquire data without fluctuations for the $m-11-y^+_{med}$ mesh. Secondly, the pressure side does not contain Reynolds stresses due to the fact that the flow around the NACA0012 profile is laminar at $Re=100{,}000$ and that this side do not suffer any boundary layer detachment. On the suction side, however, the boundary layer is detached near the leading edge (see Sections \ref{sec:simulation_analysis} and \ref{sec:flow_summary}). Therefore, Reynolds stresses are present on this side. Moreover, the largest values of the analyzed components are present in the first half of the profile. Finally, the interval of the values of the analyzed Reynolds stresses components are always greater for the $m-11-y^+_{med}$ mesh and tend to decrease with the increase in the dimensionless averaging time. However, since divergences in the intervals of $28\%$ are present, the differences on the Reynolds stresses fields may be not only caused by the different dimensionless averaging times. Therefore, a mesh independence cannot be guaranteed.

\par In the case of the fluid-structure interaction between the NACA0012 airfoil and the flow, the greater the achieved rotation angle due to the flutter, the greater are the influence of the meshes on the results. For a regime inducing maximal rotations of $\theta=5^\circ$, the medium grid with the large wall distance can be used without problem. However, when the airfoil is subjected to greater angle of attack due to flutter rotations ($\theta>5^\circ$), further investigations are required to measure the impact of the grid resolution on the fluid forces and consequently on the structural response.


\section{Aerodynamic coefficients}\markboth{\MakeUppercase{chapter 3.$\quad$results and discussion}}{\MakeUppercase{3.6$\quad$Aerodynamic Coeff.}}
\label{sec:drag_lift_coefficients}

\par The relative movement between the NACA0012 profile and air generates aerodynamic forces due to the pressure on the surface of the body and due to the viscosity of the air. The aerodynamic coefficients are a function of these forces, as well as of the angle of attack and the Reynolds number. These are frequently used to compare the aerodynamic performance of different bodies.
\par The aerodynamic coefficients are analyzed only for the medium resolution meshes, since the results for the fine resolution mesh are still not yet available.

\subsection{Pressure coefficient $C_p$}
\label{subsec:pressure_coefficient}

\par As the NACA0012 airfoil moves through the air, the fluid velocity varies producing a variation of pressure, which is responsible for the generation of a pressure force $F_p$ that acts on the center of pressure. The pressure variation on the airfoil is characterized by the pressure coefficient $C_p$, calculated according to Eq.\ (\ref{eq:pressure_coefficient}).
\begin{equation}
C_p=\frac{2\,<p-p_{\infty}>}{\rho_f\, u_{1,\,in}^2} \label{eq:pressure_coefficient}
\end{equation}
\par $\rho_f$ and $u_{1,\,in}$ are the air density and the free stream velocity. $p$ and $p_\infty$ represent the static pressure at the point at which the pressure coefficient is evaluated and the free stream static pressure, respectively. 
\subsubsection{Angle of attack $\alpha=0^\circ$}
\label{subsubsec:pressure_coefficient_angle_0}
\par
\begin{figure}[H]
	\centering
	\centering
	\includegraphics[width=0.56\textwidth,draft=\drafttype]{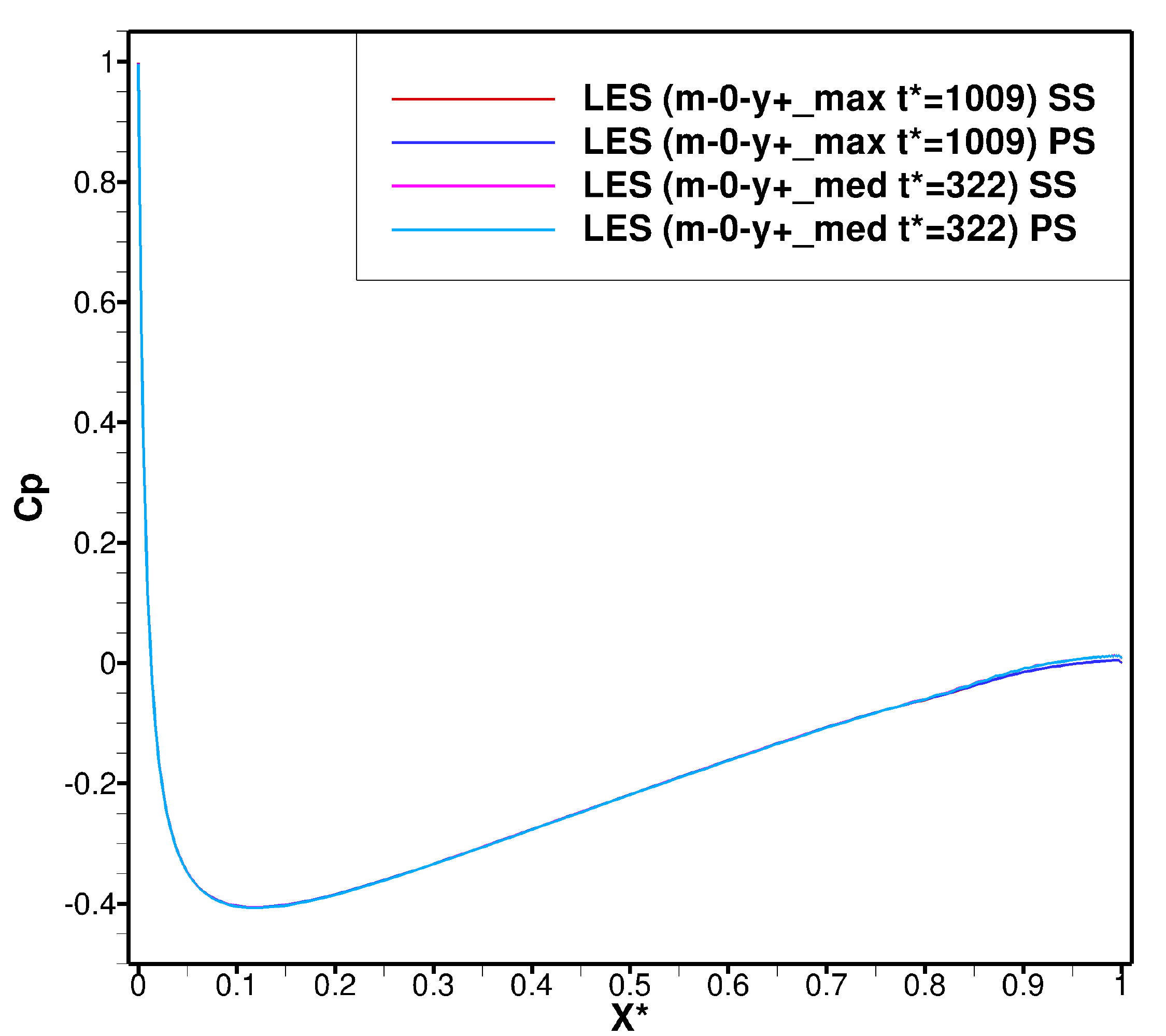}
	\caption{\label{fig:cp_medium_meshes_angle_0}Distribution of the time-averaged pressure coefficient at the airfoil for the medium meshes at an angle of attack of $\alpha=0^\circ$ (results are spatial-averaged in the span-wise direction).}
\end{figure}

 The pressure coefficient distribution for the medium meshes at an angle of attack of $\alpha=0^\circ$ are illustrated in Fig.\ \ref{fig:cp_medium_meshes_angle_0}. 

\par Since the pressure coefficient at $x=0$ is $C_p=1$ a stagnation point, i.e$.$ a point where the local velocity of the fluid is zero, is present at this location. At about $x=0.01\,c$ the value of the static pressure $p$ is the same as the free stream pressure $p_\infty$ and therefore the pressure coefficient is $C_p=0$. A minimal value of the pressure coefficient, i.e$.$, \mbox{$C_p=-0.406$}, is achieved at about $x=0.12\,c$.
\par The pressure coefficient distribution is the same for the pressure and suction sides due to the symmetrical profile submitted to an angle of attack of $\alpha=0^\circ$. Since the static pressure is inversely proportional to the velocity, the pressure coefficient firstly decrease and then increases. A minimal difference in the pressure coefficients of the $m-0-y^+_{med}$ and $m-0-y^+_{max}$ is noticed for $x>0.8\,c$.
\par
\begin{figure}[H]
	\centering
	\centering
	\includegraphics[width=0.56\textwidth,draft=\drafttype]{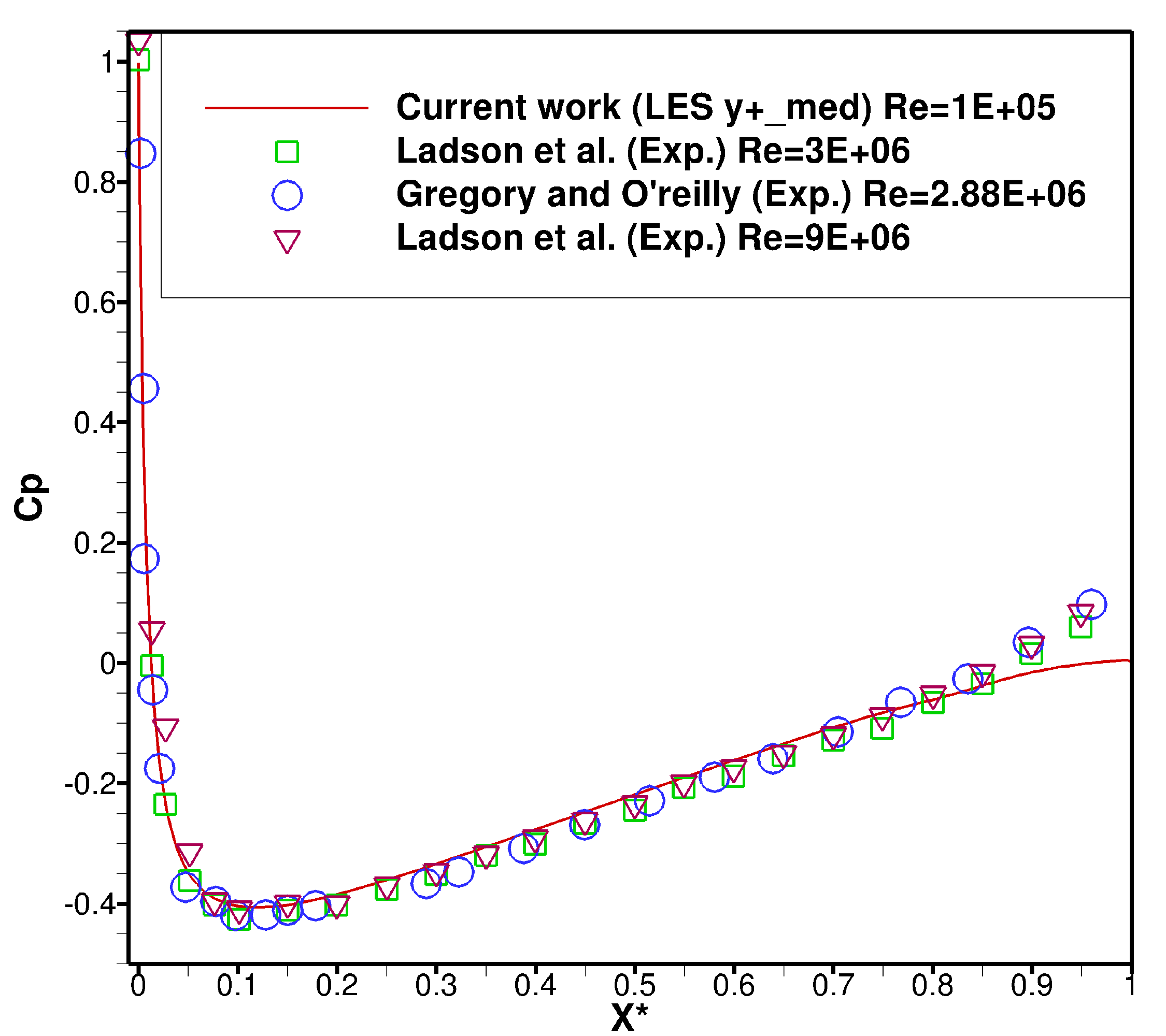}
	\caption{\label{fig:cp_comparison_angle_0}Time-averaged pressure coefficients comparison on the suction side of the NACA0012 profile at an angle of attack of $\alpha=0^\circ$ (simulation results are spatial-averaged in the span-wise direction).}
\end{figure}

 In Fig.\ \ref{fig:cp_comparison_angle_0}, the large-eddy simulation results for the $m-0-y^+_{med}$ mesh on the suction side are compared to the experiments performed by \mbox{Ladson et al.\ \cite{Ladson_1987a}} for a NACA0012 profile at a range of Reynolds numbers of \mbox{$3 \cdot 10^6 \leq Re \leq 45 \cdot 10^6$}, as well as with the experimental results of \mbox{Gregory and O'reilly \cite{Gregory_1973}} for a NACA0012 airfoil at \mbox{$Re=1.44\cdot 10^6$} and \mbox{$Re=2.88\cdot 10^6$}.
\par Despite of the difference in the analyzed Reynolds number, the same pattern can be observed until $x=0.8\,c$, which indicates that the pressure coefficient remains almost constant for $Re\geq100{,}000$. Pattern differences appears at $x\geq 0.8\,c$, point at which a detachment of the boundary layer occurs for the large-eddy simulation (see \mbox{Section \ref{sec:streamlines_angle_0}}). For the case of Gregory and O'reilly's \cite{Gregory_1973} tests, a wind tunnel boundary layer control through the blowing technique is utilized. This re-energizes the boundary layer, avoiding a boundary layer detachment. Therefore, no separation is present at $Re=2{,}880{,}000$ and $\alpha=0^\circ$. Regarding the Ladson's et al.\ \cite{Ladson_1987a} results, the cryogenic tunnel also has a boundary layer control system, described by Ladson and Ray \cite{Ladson_1987b}, which aims at the reduction of the thickness of the boundary layer in the test section region, reducing the possibility of boundary layer separation. Thus, no detachment of the boundary layer is present at an incidence of $\alpha=0^\circ$, for both analyzed cases, i.e$.$, $Re=3{,}000{,}000$ at $Ma=0.3$ and $Re=9{,}000{,}000$ at $Ma=0.4$. 

\subsubsection{Angle of attack $\alpha=5^\circ$}
\label{subsubsec:pressure_coefficient_angle_5}

\par
\begin{figure}[H]
	\centering
	\centering
	\includegraphics[width=0.56\textwidth,draft=\drafttype]{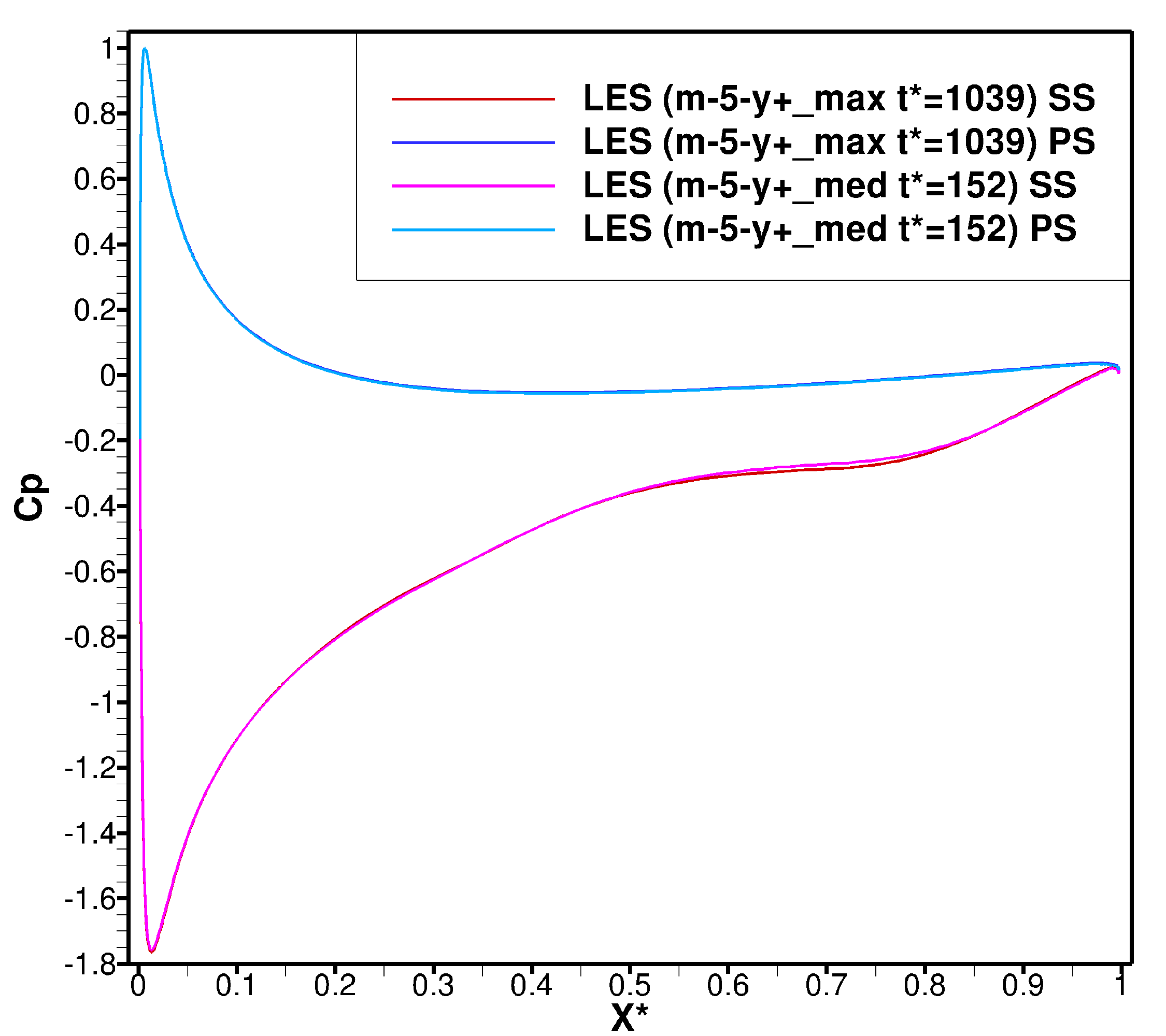}
	\caption{\label{fig:cp_medium_meshes_angle_5}Distribution of the time-averaged pressure coefficient at the airfoil for the medium meshes at an angle of attack of $\alpha=5^\circ$ (results are spatial-averaged in the span-wise direction).}
\end{figure}

 The distribution of the pressure coefficient for the meshes at an incidence of $\alpha=5^\circ$ is illustrated in Fig.\ \ref{fig:cp_medium_meshes_angle_5}. A stagnation point is present on the pressure side at $x=0.006\,c$ since $C_p=1$. A point in which the local static pressure is equal the free stream static pressure, i.e$.$, $C_p=0$, is located on the pressure side at about $x=0.002\,c$. The minimum of the pressure coefficient $C_{p_{min}}=-1.76$ is achieved on the suction side of the airfoil at $x=0.014\,c$. The pressure coefficient distribution shows minimal discrepancies on the suction sides of both analyzed meshes at an incidence of $\alpha=5^\circ$. 
\par
\begin{figure}[H]
	\centering
	\centering
	\includegraphics[width=0.56\textwidth,draft=\drafttype]{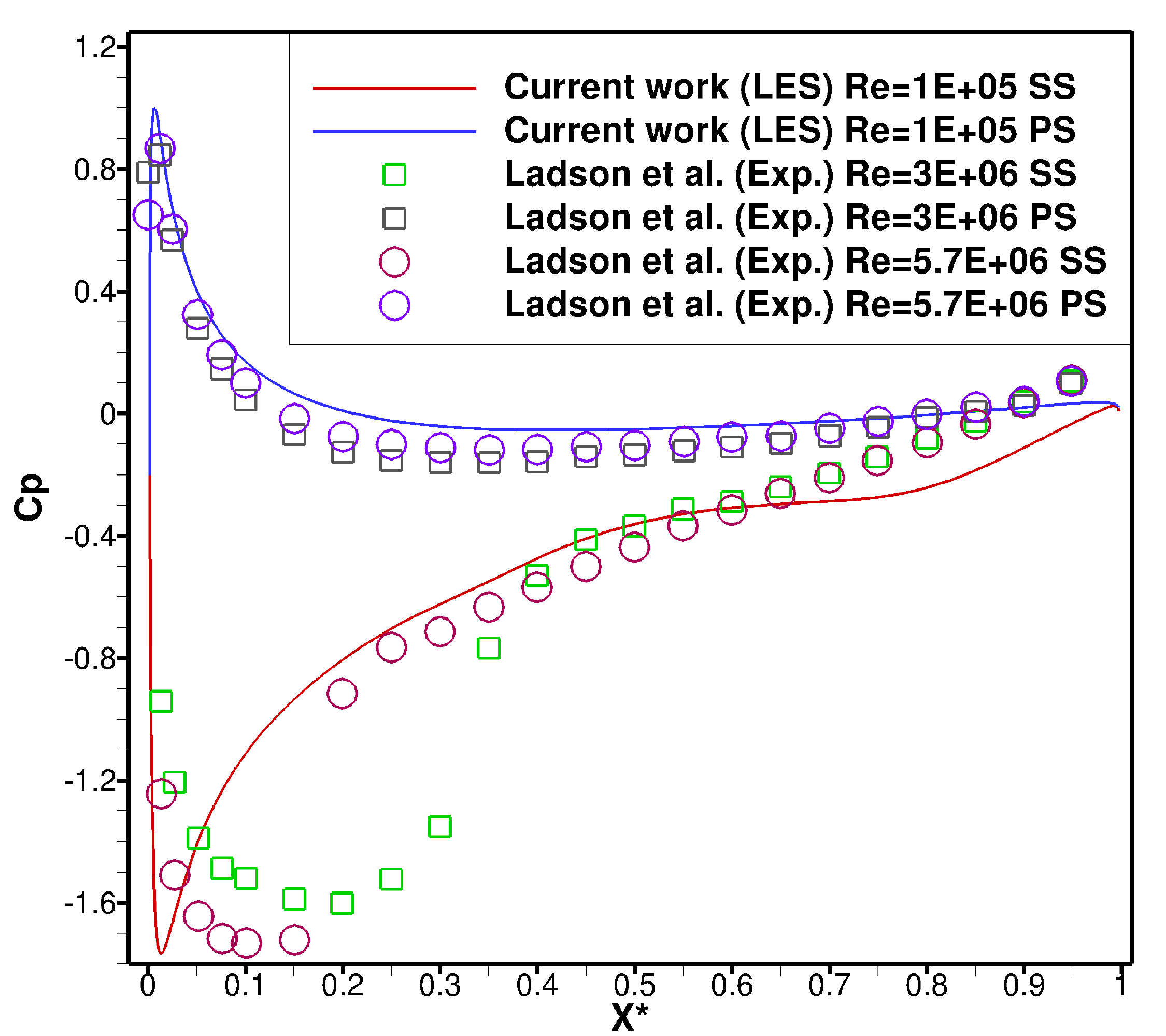}
	\caption{\label{fig:cp_comparison_angle_5}Time-averaged pressure coefficients comparison on the suction (SS) and pressure (PS) sides of the NACA0012 profile at an angle of attack of $\alpha=5^\circ$ (simulation results are spatial-averaged in the span-wise direction).}
\end{figure}

 The pressure coefficients for the large-eddy simulation of the $m-5-y^+_{max}$ mesh are compared to the experimental data, i.e$.$, $Re=3{,}000{,}000$ at $Ma=0.7$ and $Re=5{,}7000{,}000$ at $Ma=0.65$, of Ladson et al.\ \cite{Ladson_1987a} in Fig.\ \ref{fig:cp_comparison_angle_5}. The pressure coefficients on the pressure side show almost the same pattern, since the flow on this side does not suffer any detachment. The coefficients on the suction side, however, are strongly dissimilar. This occurs because of the different analyzed Reynolds number, which causes a different flow behavior, that is, the presence and location of detachment and reattachment points change with the variation of the Reynolds number.

\subsubsection{Angle of attack $\alpha=11^\circ$}
\label{subsubsec:pressure_coefficient_angle_11}
\par
\begin{figure}[H]
	\centering
	\centering
	\includegraphics[width=0.56\textwidth,draft=\drafttype]{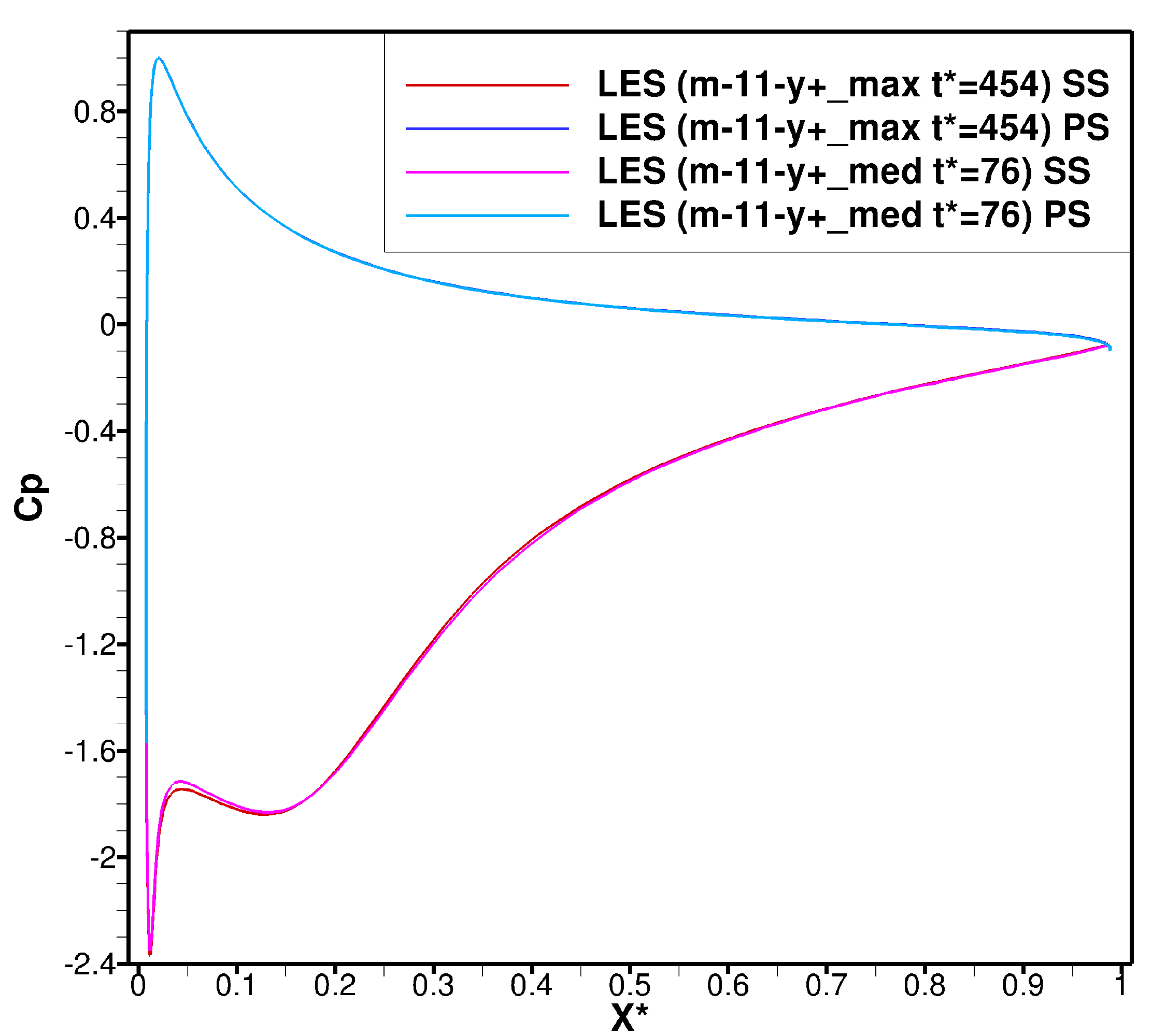}
	\caption{\label{fig:cp_medium_meshes_angle_11}Distribution of the time-averaged pressure coefficient at the airfoil for the medium meshes at an angle of attack of $\alpha=11^\circ$ (results are spatial-averaged in the span-wise direction).}
\end{figure}

 The pressure coefficients at the airfoil at an incidence of $\alpha=11^\circ$ are shown in \mbox{Fig.\ \ref{fig:cp_medium_meshes_angle_11}} for the $m-11-y^+_{med}$ and $m-11-y^+_{max}$ meshes. A point at which the local static pressure is the same as the free stream pressure, i.e$.$, $C_p=0$, is located on the pressure side at $x=0.0085\,c$. The minimum value of this coefficient is $C_{p_{min}}=-2.36$ located on the suction side at $x=0.011\,c$. A stagnation point is present on the pressure side at $x=0.021\,c$ since the pressure coefficient is $C_p=1$. A very small divergence on the data is available for the suction side of the analyzed meshes.
\begin{figure}[H]
	\centering
	\centering
	\includegraphics[width=0.56\textwidth,draft=\drafttype]{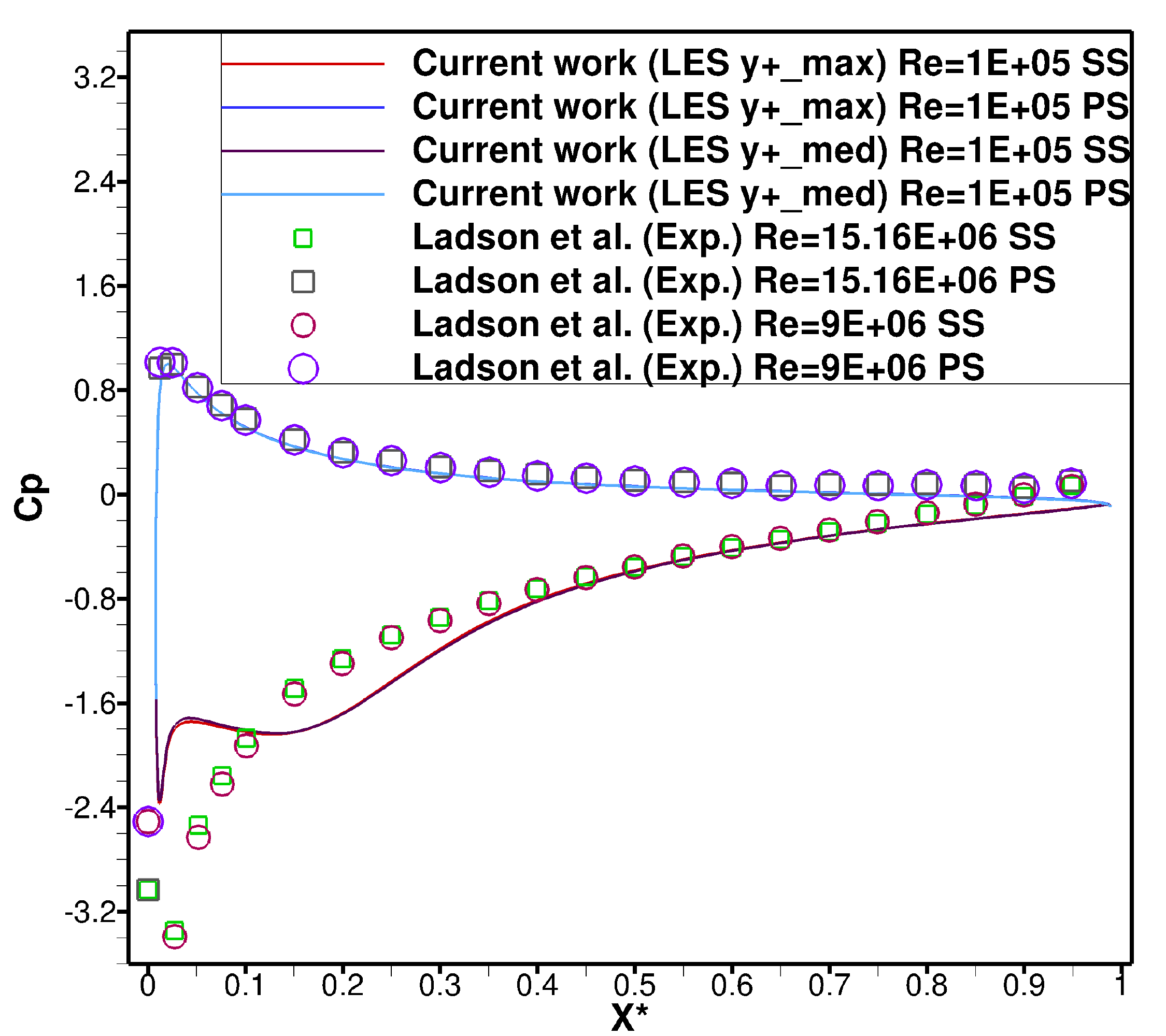}
	\caption{\label{fig:cp_comparison_angle_11}Time-averaged pressure coefficients comparison on the suction (SS) and pressure (PS) sides of the NACA0012 profile at an angle of attack of $\alpha=11^\circ$ (simulation results are spatial-averaged in the span-wise direction).}
\end{figure}
\par The pressure coefficients for the NACA0012 profile at an angle of attack of $\alpha=11^\circ$ are compared for both large-eddy simulations and the experimental results of \mbox{Ladson et al.\ \cite{Ladson_1987a}} for $Re=15{,}160{,}000$ at $Ma=0.3$ and for $Re=9{,}000{,}000$ at $Ma=0.4$. Although the compared Reynolds numbers are different, the coefficients of the pressure side show a great similarity, since the flow on this side does not suffer any separation. On the suction side, however, these values are strongly dissimilar. This occurs because the flow on this side suffers with boundary layer detachments and reattachments, which location and formation are affected by the free stream velocity and therefore by the Reynolds number.

\subsection{Friction coefficient $C_f$}
\label{subsec:friction_coefficient}

\par Skin friction is caused by the interaction of the viscous fluid with the airfoil profile that results in the formation of a wall shear stress $\tau_w$, which is also influenced by the surface roughness. In the present work the simulated airfoil is aerodynamically smooth.
\par The friction coefficient $C_f$ is used to illustrate the distribution of the wall shear stress on the pressure and suction sides of the profile and is calculated in accordance with \mbox{Eq.\ (\ref{eq:friction_coefficient})}.
\begin{equation}
C_f=\frac{2\,\tau_w}{\rho_f\,u_{1,\,in}^2} \label{eq:friction_coefficient}
\end{equation} 
\par $\rho_f$ and $u_{1,\,in}$ are the fluid viscosity and the free stream velocity, respectively. The wall shear stress $\tau_w$ is approximated with \mbox{Eq.\ (\ref{eq:wall_shear_stress_approximation})}.
\begin{equation}
\tau_w=\frac{\mu_f\;\,sgn(u_1)\;\,\sqrt{\sum_{i=1}^{3}<u_i^2>}}{\Delta y} \label{eq:wall_shear_stress_approximation}
\end{equation} 
\par $\mu_f$, $u_i$ and $\Delta y$ are the air dynamic viscosity, the local velocity and the distance from the first cell middle point to the airfoil, respectively. $sgn(u_1)$ is the signum function of the stream-wise velocity $u_1$.

\subsubsection{Angle of attack $\alpha=0^\circ$}
\label{subsubsec:friction_coefficient_angle_0}
\par
\begin{figure}[H]
	\centering
	\centering
	\includegraphics[width=0.56\textwidth,draft=\drafttype]{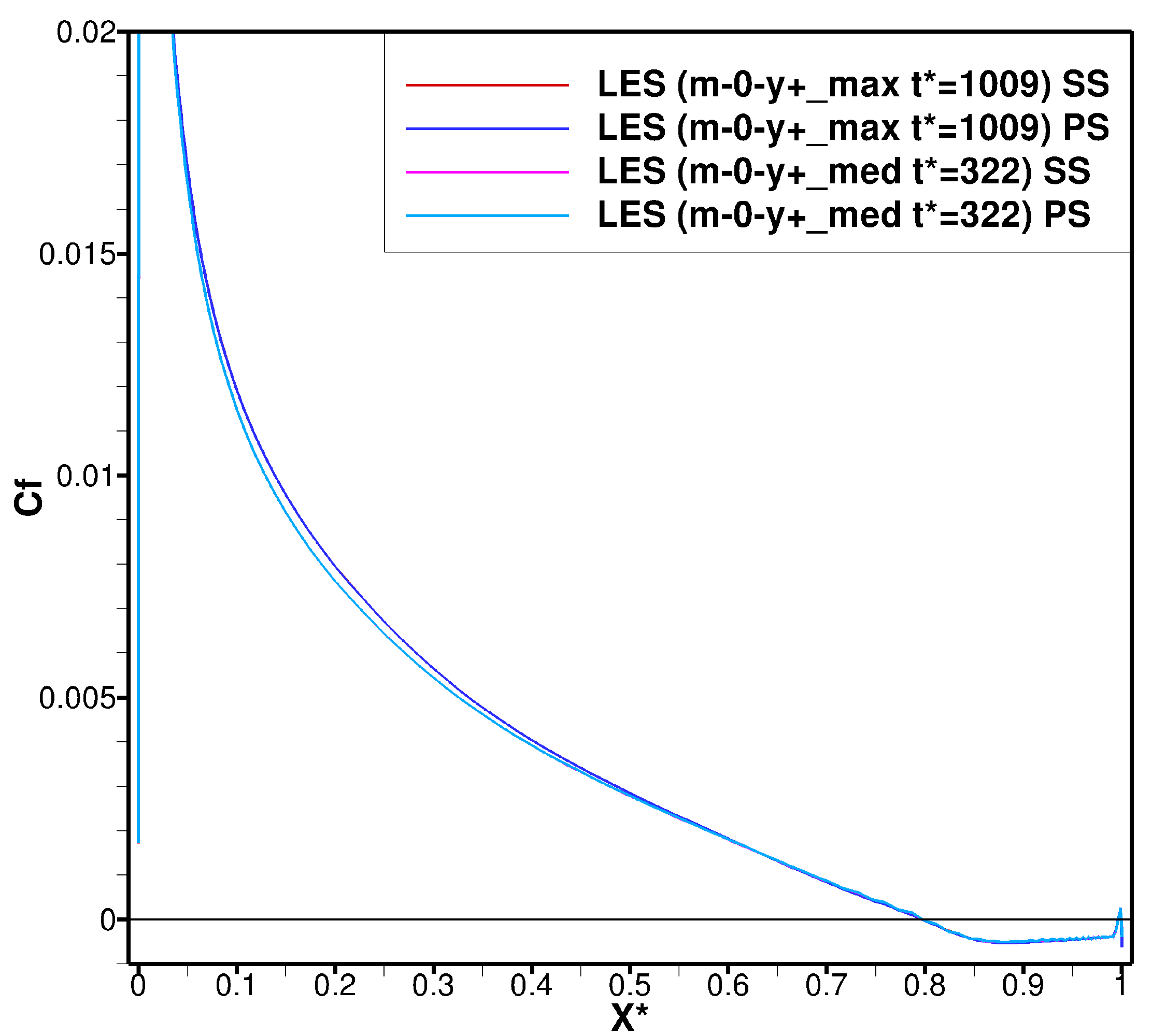}
	\caption{\label{fig:cf_medium_meshes_angle_0}Distribution of the time-averaged friction coefficient at the airfoil for the medium meshes at an angle of attack of $\alpha=0^\circ$ (results are spatial-averaged in the span-wise direction).}
\end{figure}

 The skin friction coefficients of the meshes at an incidence of $\alpha=0^\circ$ are illustrated in \mbox{Fig.\ \ref{fig:cf_medium_meshes_angle_0}}. Since the airfoil is symmetrical and submitted to an angle of attack of $\alpha=0^\circ$ the skin friction coefficient is the same for both pressure and suction sides. The spatial evolution of the skin friction coefficients is very similar for both meshes.
\par A maximal skin friction coefficient of $C_{p_{max}}=0.036$ is located at $x=0.0046\,c$, which also indicates the point of maximal velocity. The boundary layer is detached at $x=0.8\,c$ (see Section \ref{sec:streamlines_angle_0}), since the skin friction coefficient at this point is $C_f=0$.

\subsubsection{Angle of attack $\alpha=5^\circ$}
\label{subsubsec:friction_coefficient_angle_5}

\par
\begin{figure}[H]
	\centering
	\centering
	\includegraphics[width=0.56\textwidth,draft=\drafttype]{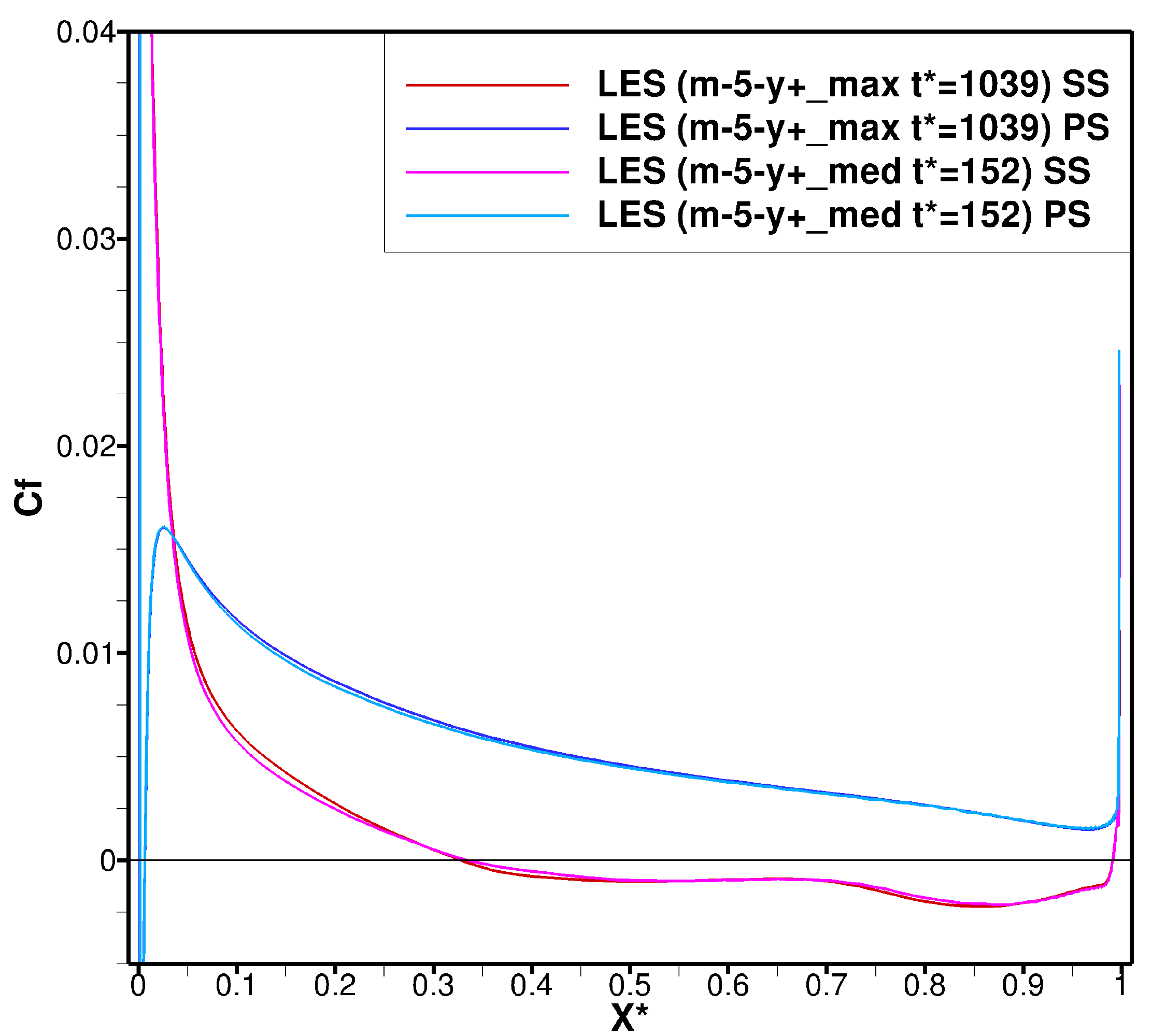}
	\caption{\label{fig:cf_medium_meshes_angle_5}Distribution of the time-averaged friction coefficient at the airfoil for the medium meshes at an angle of attack of $\alpha=5^\circ$ (results are spatial-averaged in the span-wise direction).}
\end{figure}

 The skin friction coefficient for the meshes at an angle of attack of $\alpha=5^\circ$ are investigated through Fig.\  \ref{fig:cf_medium_meshes_angle_5}. 

\par The maximal value of the skin friction coefficient is $C_{p_{max}}=0.083$, located on the suction side at $x=0.003\,c$. This is also the point with the maximum achieved local velocity. Two points with $C_f=0$ are present, representing the detachment and reattachment points of the boundary layer (see also Section \ref{sec:streamlines_angle_5}). The former is located at about $x=0.34\,c$ for the $m-5-y^+_{med}$ mesh and at about $x=0.33\,c$ for the $m-5-y^+_{max}$ mesh. The latter is located at $x=0.99\,c$ for both meshes. The distribution of the coefficients remains very similar for both investigated meshes.

\subsubsection{Angle of attack $\alpha=11^\circ$}
\label{subsubsec:friction_coefficient_angle_11}

\par
\begin{figure}[H]
	\centering
	\centering
	\includegraphics[width=0.56\textwidth,draft=\drafttype]{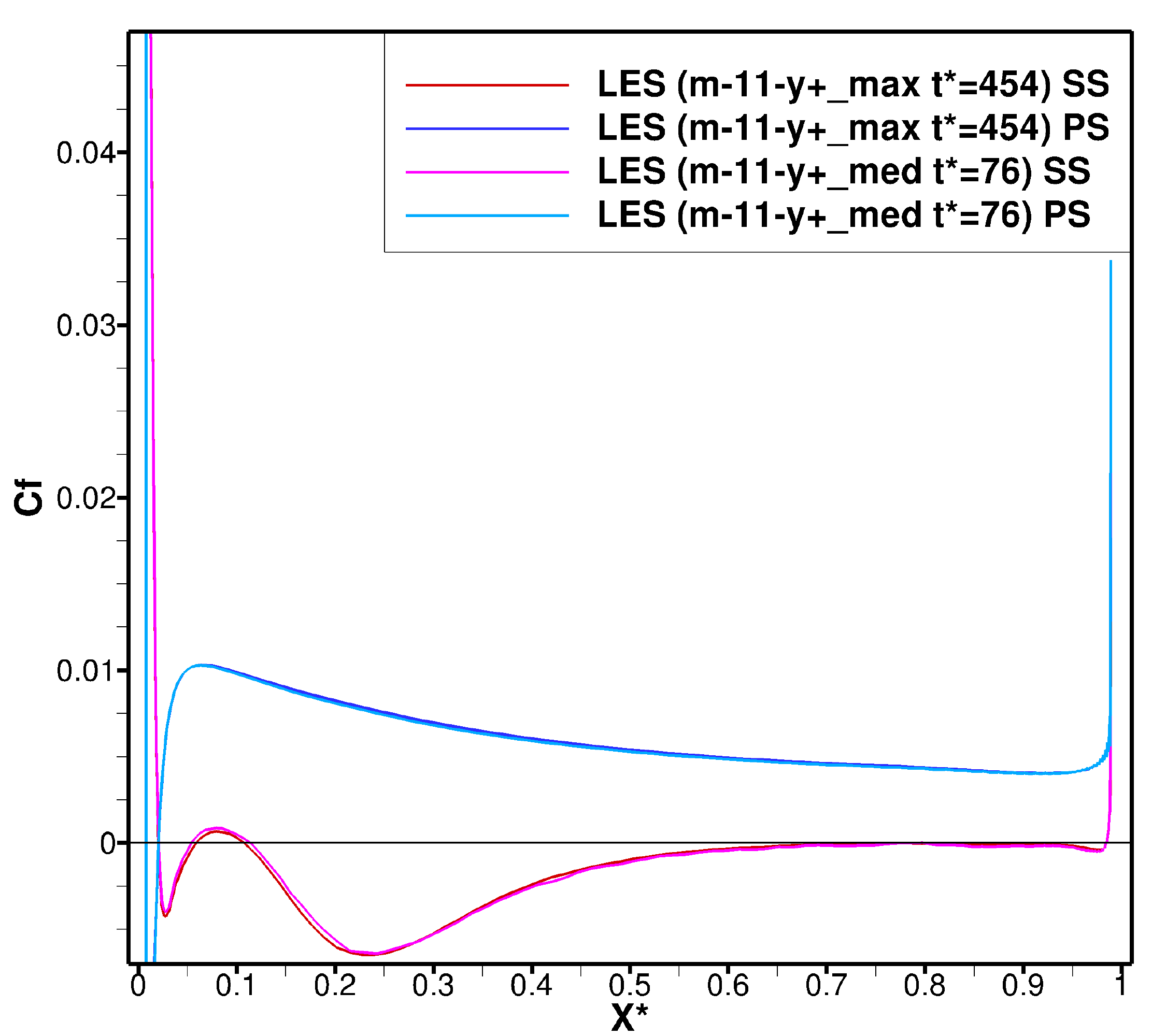}
	\caption{\label{fig:cf_medium_meshes_angle_11}Distribution of the time-averaged friction coefficient at the airfoil for the medium meshes at an angle of attack of $\alpha=11^\circ$ (results are spatial-averaged in the span-wise direction).}
\end{figure}

 The coefficients of skin friction at the airfoil for the meshes at an angle of attack of $\alpha=11^\circ$ are illustrated in Fig.\ \ref{fig:cf_medium_meshes_angle_11}.

\par The maximal skin friction coefficient is achieved at the point of maximal velocity, i.e$.$ $x=0.003\,c$. On the suction side, six points with $C_f\approx0$ are present, which indicate boundary layer detachments and reattachments. The location of these points are summarized in Section \ref{sec:streamlines_angle_11} and are slightly different for both meshes. However, the distribution of the skin friction coefficients are nearly identical for the two medium meshes.

\subsection{Drag coefficient $C_D$}
\label{subsec:drag_coefficient}

\par The drag coefficient $C_D$ is a sum of the skin friction drag and the pressure drag coefficients, which is calculated according to Eq.\ (\ref{eq:drag_coefficient}). $\rho_f$ is the fluid density, $u_{1,\,in}$ is the free stream velocity and $S$ is the reference area, which is calculated through Eq.\ (\ref{eq:drag_coefficient_area}). $c$ is the airfoil chord ($c=0.1\,m$) and $L_z$ is the span-wise length ($L_z=0.25\,c=0.025\,m$).
\begin{eqnarray}
C_D&=&\frac{2<F_D>}{\rho_f\;S\;u_{1,\,in}^2} \label{eq:drag_coefficient} \\
S&=&c\,L_z \label{eq:drag_coefficient_area}
\end{eqnarray}
\par The drag force on the wall $F_D$ has an opposite direction of the motion, i.e$.$, negative stream-wise direction, and is a sum of the frictional $F_{\tau_{11}}$, i.e$.$, shear, and the pressure $F_{p_{x_1}}$ forces parallel to the motion, as demonstrated in Eq.\ (\ref{eq:drag}).  
\begin{equation}
F_D=F_{\tau_{11}}+F_{p_{x_1}} \label{eq:drag}
\end{equation}
\par The shear force $F_{\tau_{11}}$ is calculated according to Eq.\ (\ref{eq:shear_force_drag}). $\tau_{11}$ and $A_{\tau}$ are the shear stress in the negative stream-wise direction and the area submitted to this shear stress, respectively. The former is proportional to the dynamic viscosity of the fluid $\mu_f$, as stated in Eq.\ (\ref{eq:shear_stress_drag}).  
\begin{eqnarray}
F_{\tau_{11}}&=&\tau_{11}\; A_{\tau} \label{eq:shear_force_drag} \\
\tau_{11}&=&\mu_f \,\frac{\partial u_1}{\partial x_1} \label{eq:shear_stress_drag} \\
\end{eqnarray}
\par The pressure force $F_{p_{x_1}}$ is a function of the pressure in the chord-wise direction $p_{x_1}$ and the area normal to this pressure $A_{p}$, as described in Eq.\ (\ref{eq:pressure_force_drag}).
\begin{equation}
F_{p_{x_1}}=p_{x_1} \; A_p \label{eq:pressure_force_drag}
\end{equation} 
\par The drag coefficients and the standard deviations for the medium resolution meshes are illustrated in Table \ref{table:drag_coefficients}. The standard deviations are relatively small and indicates a small drag amplitude oscillation in time. Moreover, these amplitudes are smaller for the $m-0-y^+_{max}$ mesh.
\begin{table}[H]
	\centering
	\begin{tabular}{c c c}
		\hline
		\bf{Mesh} & \centering{\bf{Drag coefficient}} &\centering{\bf{Standard deviation}} \tabularnewline \hline
		$m-0-y^+_{med}$ & $C_D=0.015$ & $\sigma_{C_D}=0.020$ \tabularnewline
		$m-0-y^+_{max}$ & $C_D=0.015$ & $\sigma_{C_D}=0.000$ \tabularnewline
		$m-5-y^+_{med}$ & $C_D=0.020$ & $\sigma_{C_D}=0.010$ \tabularnewline
		$m-5-y^+_{max}$ & $C_D=0.019$ & $\sigma_{C_D}=0.001$ \tabularnewline
		$m-11-y^+_{med}$ & $C_D=0.047$ & $\sigma_{C_D}=0.008$ \tabularnewline
		$m-11-y^+_{max}$ & $C_D=0.043$ & $\sigma_{C_D}=0.001$ \tabularnewline
		\hline		
	\end{tabular}
	\caption{\label{table:drag_coefficients}Time-averaged drag coefficients and standard deviations of the medium resolution meshes at $Re=100{,}000$.}
\end{table}
\par
\begin{figure}[H]
	\centering
	\centering
	\includegraphics[width=0.56\textwidth,draft=\drafttype]{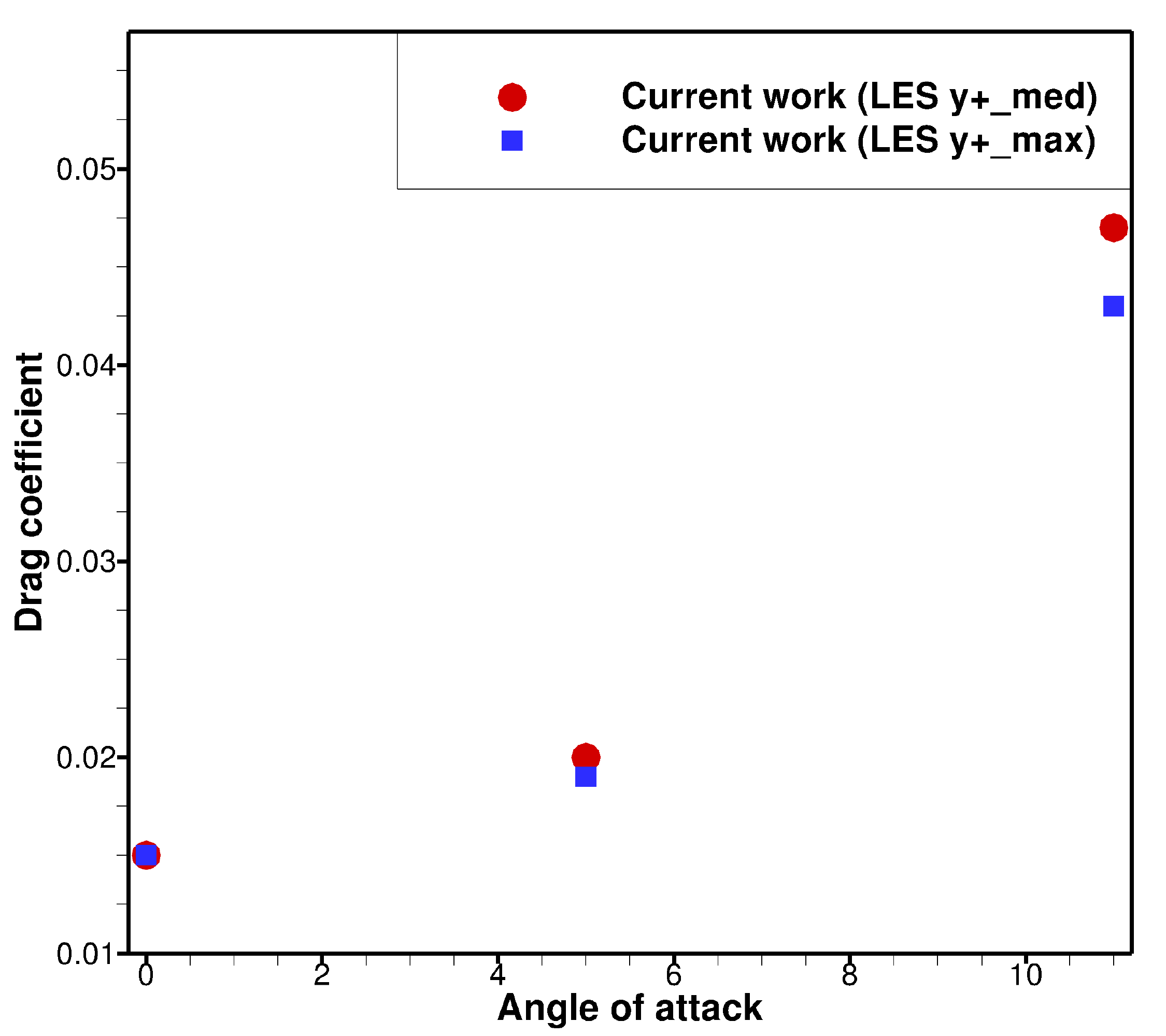}
	\caption{\label{fig:cd_comparison_angles}Variation of the time-averaged drag coefficients with incidence at $Re=100{,}000$.}
\end{figure}

 An increase in the incidence results in an exponential growth of the drag coefficient, since the pressure drag rises exponentially while the skin friction drag stays almost constant, as stated in the work of Jones \cite{Jones_2008}. This growth is illustrated in Fig.\ \ref{fig:cd_comparison_angles} for the medium meshes with the maximum and the medium first cell wall distances. The meshes with a medium and maximal wall distances of the first cell at angles of attack of $\alpha=5^\circ$ and $\alpha=11^\circ$ present different values for the drag coefficients, presenting a variation of $5\%$ for the former grids and of $8.51\%$ for the latter. 
\par
\begin{figure}[H]
	\centering
	\subfigure[Qualitative comparison of drag coefficients of current work, Ladson et al.\ \cite{Ladson_1987a}, Almutari \cite{Almutari_2010} and Jones \cite{Jones_2008}.]{\includegraphics[width=0.46\textwidth,draft=\drafttype]{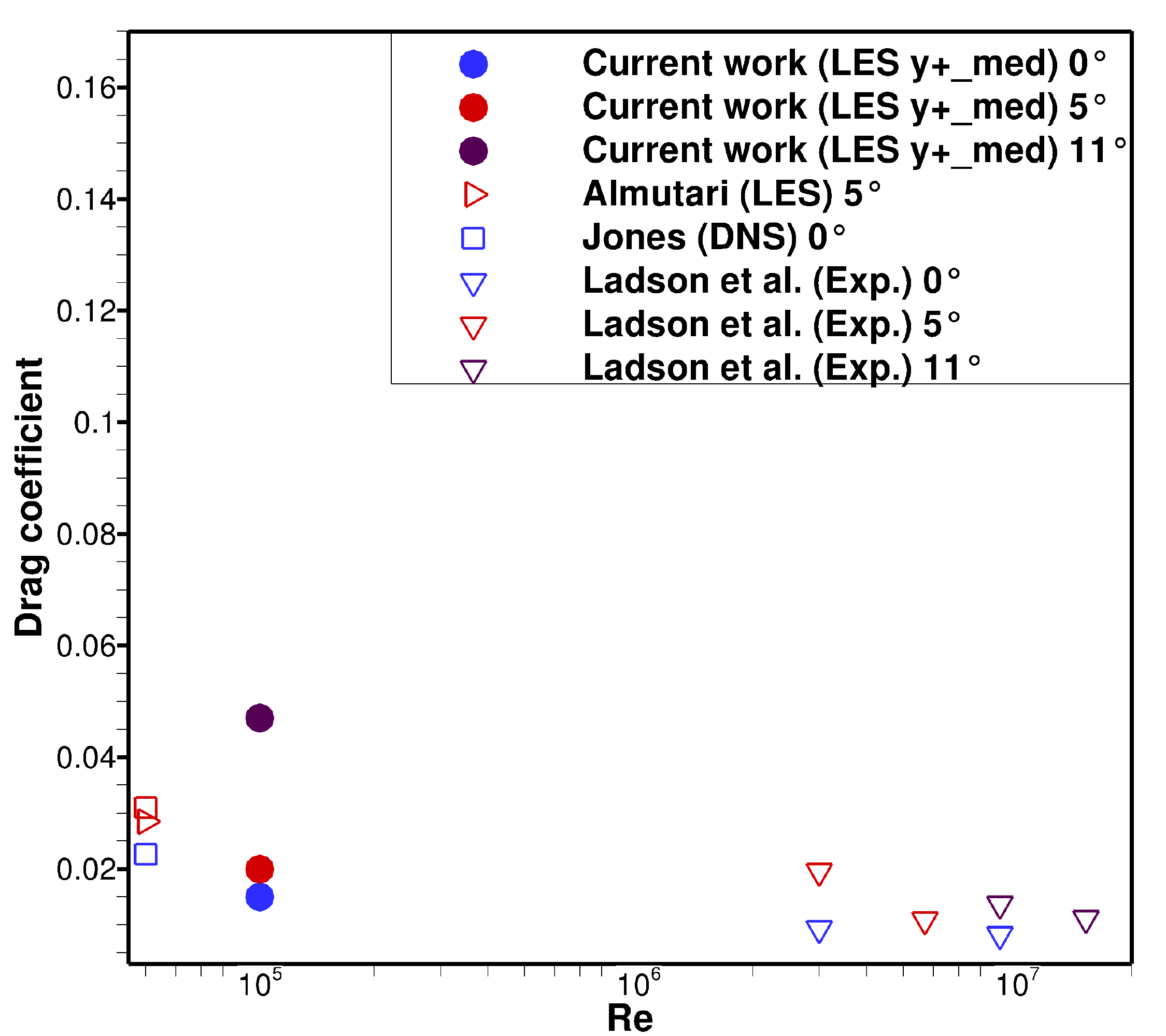}\label{fig:cd_comparison_angles_reynolds_a}} \hfill
	\subfigure[Quantitative comparison of drag coefficients of current work and Sheldahl and Klimas \cite{Sheldahl_1981} experiments on a NACA0012 Eppler model profile.]{	\includegraphics[width=0.46\textwidth,draft=\drafttype]{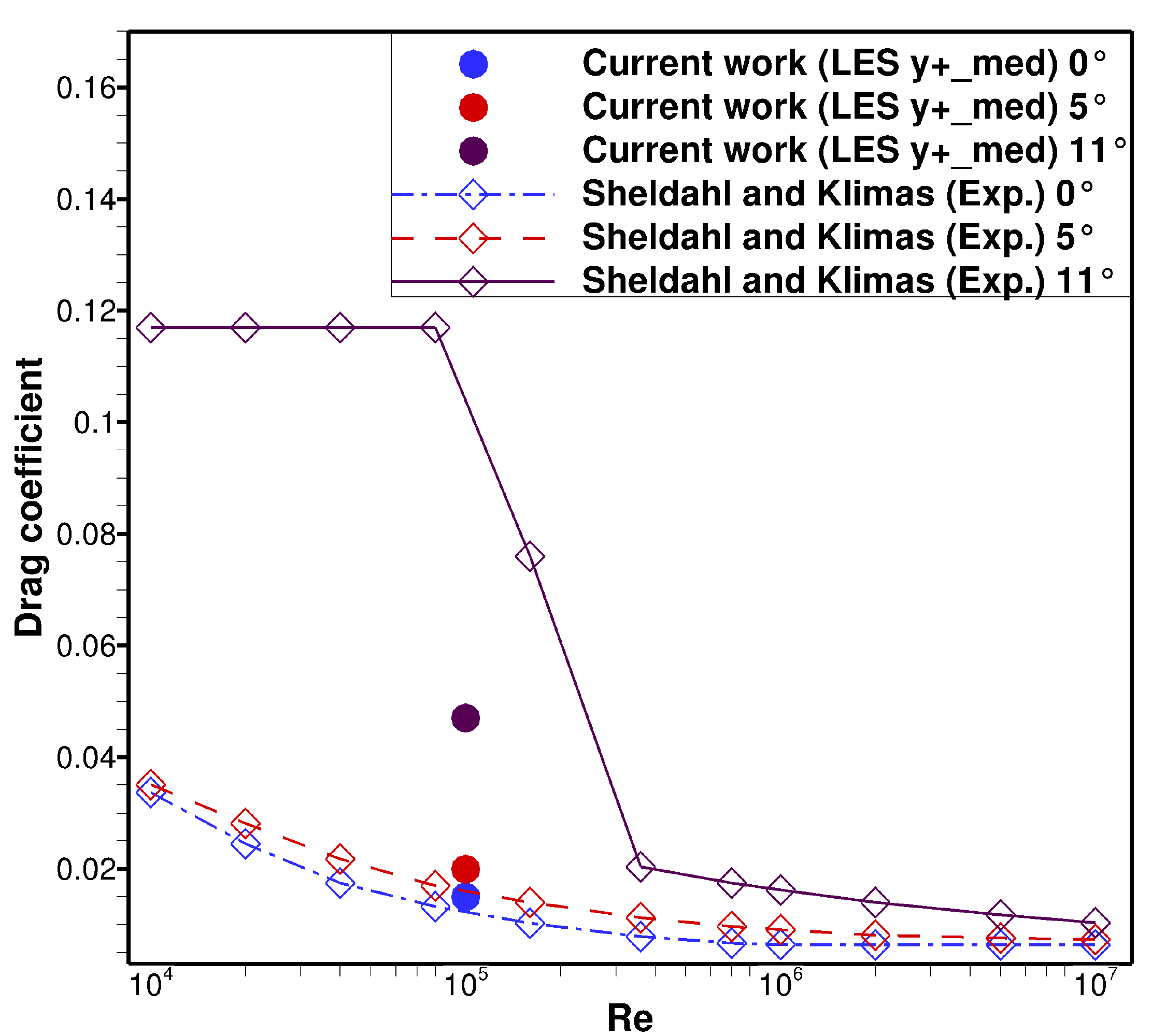}\label{fig:cd_comparison_angles_reynolds_b}}	\hfill
	\caption{\label{fig:cd_comparison_angles_reynolds}Comparison of the time-averaged drag coefficients of the present large-eddy simulations and the data available in the literature.}
\end{figure}

 The drag coefficients for the meshes with a medium first cell wall distance, i.e$.$, $m-\nobreak 0-\nobreak y^+_{med}$, $m-5-y^+_{med}$ and $m-11-y^+_{med}$, are compared to the results of the experiments performed by Ladson et al.\ \cite{Ladson_1987a}, the large-eddy simulations carried out by Almutari \cite{Almutari_2010} and the direct-numerical simulations performed by \mbox{Jones \cite{Jones_2008}}, as illustrated in Fig.\ \ref{fig:cd_comparison_angles_reynolds_a}. This evaluation is qualitative, since the conditions of the experiments and simulations (see Table \ref{table:drag_coefficients_conditions}) differ from the current work simulated conditions. A tendency to decrease the drag coefficient with the increase in the Reynolds number is present and therefore the coefficients computed by the present large-eddy simulations are plausible.

\par
\begin{figure}[H]
	\centering
	\subfigure[NASA NACA0012 \cite{NASA_2016} profile from current work and NACA0012 Eppler model profile from Sheldahl and Klimas \cite{Sheldahl_1981}.]{	\includegraphics[width=0.84\textwidth,draft=\drafttype]{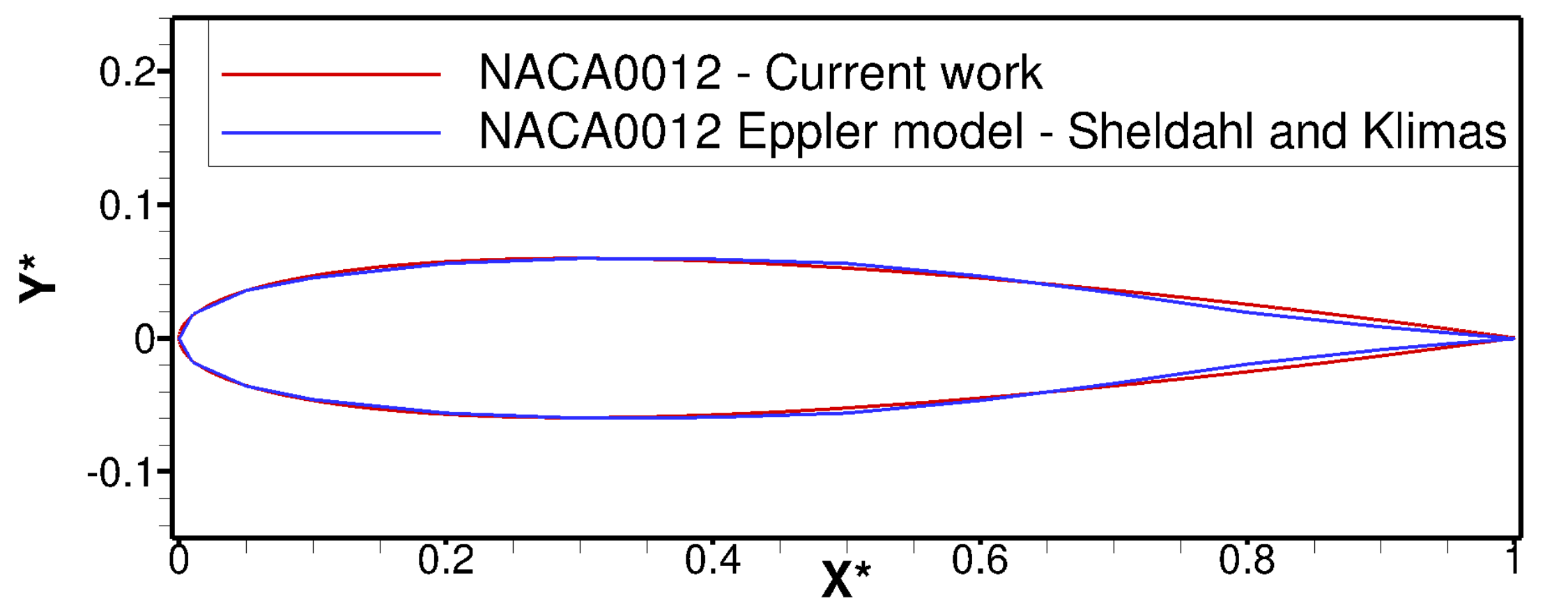}\label{fig:geometry_NACA0012_NACA0012_Eppler}} \hfill
\end{figure}

 In Fig.\ \ref{fig:cd_comparison_angles_reynolds_b}, a quantitative comparison of the data for the simulated meshes with a medium wall distance of the first cell is carried out with the available experimental data of \mbox{Sheldahl and Klimas \cite{Sheldahl_1981}}. The experimental data are for a NACA0012 Eppler model profile at a Reynolds number range of $1\cdot 10^{4} \leq Re \leq 1\cdot 10^{7}$ and an incidence range of $0^\circ \leq \nobreak \alpha \leq \nobreak 180^\circ$. This profile is, however, slightly different from the NACA0012 profile from NASA \cite{NASA_2016} used in the present work, according to Fig.\ \ref{fig:geometry_NACA0012_NACA0012_Eppler}. Therefore, it could explain the difference observed in the drag coefficients between the performed large-eddy simulations and the experiments from Sheldahl and Klimas \cite{Sheldahl_1981}. 

\par The geometry of the airfoil has a direct influence on the critical angle of attack, at which the airfoil stalls. For the case of the NACA0012 Eppler model airfoil the critical angle of attack is $\alpha_{crit}>11^\circ$. Thus, the transition between the stall and non-stall region occurs at a higher Reynolds number than for the simulated NACA0012 profile, resulting in a smaller drag coefficient computed by the large-eddy simulation at and incidence of $\alpha=11^\circ$.

\par
\begin{figure}[H]
	\centering
	\subfigure[NASA NACA0012 \cite{NASA_2016} profile from current work and NACA0012H profile from Sheldahl and Klimas \cite{Sheldahl_1981}]{\includegraphics[width=0.84\textwidth,draft=\drafttype]{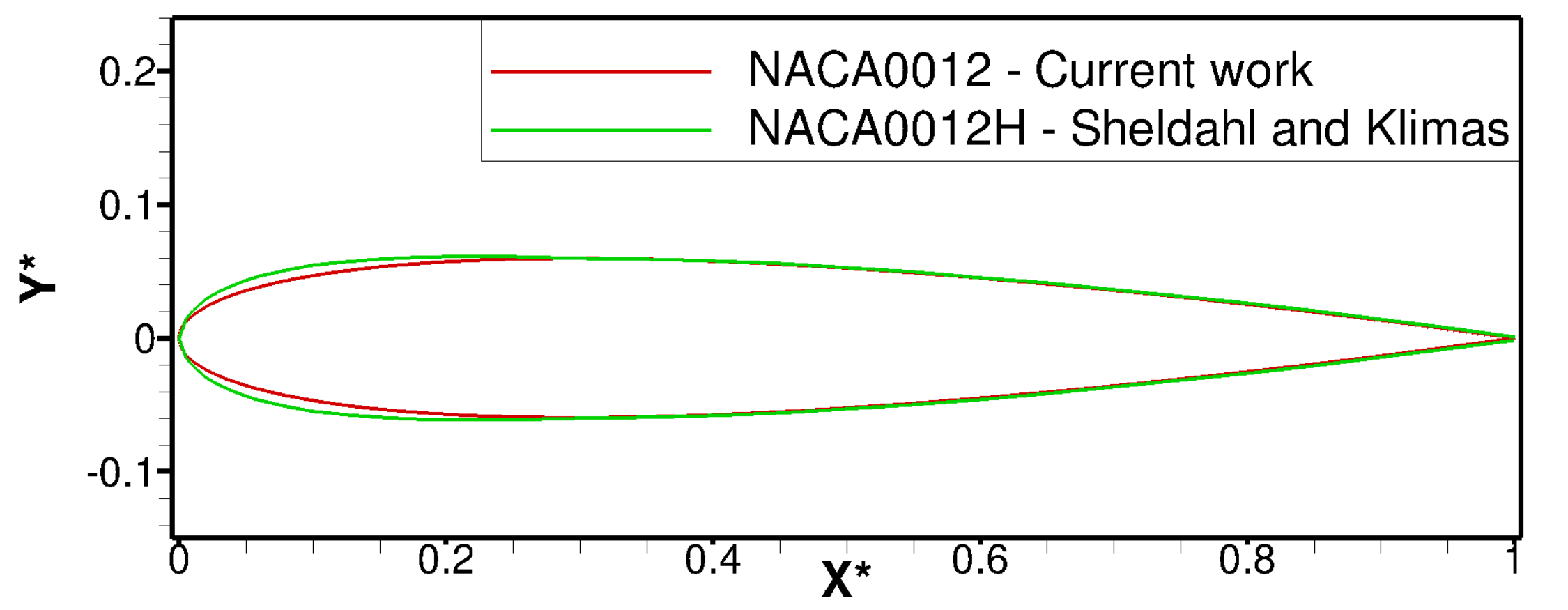}\label{fig:geometry_NACA0012_NACA0012H}} \hfill
	\caption{\label{fig:comparison_NACA0012_profiles}Comparison of the NACA0012 profiles utilized by the present large-eddy simulations and by the experiments of Sheldahl and Klimas \cite{Sheldahl_1981}.}
\end{figure}

 Sheldahl and Klimas \cite{Sheldahl_1981} also compared the NACA0012 Eppler model profile to a NACA0012H profile, which is modified in order to improve the aerodynamic performance. This comparison showed that the drag coefficients for the NACA0012H profile is greater than the ones for the NACA0012 Eppler model profile and also that the critical angle of attack was improved to $\alpha_{crit}>13^\circ$. A comparison between the geometry of this improved airfoil and the NACA0012 airfoil from NASA \cite{NASA_2016}, i.e$.$, the profile utilized in the present work, is shown in Fig.\ \ref{fig:geometry_NACA0012_NACA0012H}. When comparing the NACA0012 from NASA \cite{NASA_2016}, the NACA0012 Eppler model \cite{Sheldahl_1981} and the NACA0012H \cite{Sheldahl_1981} profiles, it is noticeable, that the simulated NACA0012 profile  has approximately the geometry of the leading edge of the NACA0012 Eppler model and the geometry of the trailing edge of the NACA0012H airfoil. Therefore, the performance of this airfoil, i.e$.$, the drag and lift coefficients, may also be increased in relation to the NACA0012 Eppler model profile for incidences of $\alpha<\alpha_{crit}$.  

\begin{table}[H]
	\centering
	\begin{tabular}{c c l l l l}
		\hline
		\bf{Literature} & \multicolumn{1}{c}{\bf{Type}} & \multicolumn{1}{c}{$\alpha$} & \multicolumn{1}{c}{$Re$} & \multicolumn{1}{c}{$Ma$} \tabularnewline \hline
		Almutari & LES & $5^\circ$ & $\quad\;\;\,50{,}000$ & \multicolumn{1}{c}{--} & \tabularnewline
		Jones & DNS & $0^\circ$ & $\quad\;\;\,50{,}000$ & \multicolumn{1}{c}{--} & \tabularnewline
		Jones & DNS & $5^\circ$ & $\quad\;\;\,50{,}000$ & \multicolumn{1}{c}{--} & \tabularnewline
		Ladson et al.\ & Experiment & $0^\circ$ & $\;\,3{,}000{,}000$ & $0.30$ \tabularnewline
		Ladson et al.\ & Experiment & $0^\circ$ & $\;\,9{,}000{,}000$ & $0.40$ \tabularnewline
		Ladson et al.\ & Experiment & $5^\circ$ & $\;\,3{,}000{,}000$ & $0.70$ \tabularnewline
		Ladson et al.\ & Experiment & $5^\circ$ & $\;\,5{,}730{,}000$ & $0.65$ \tabularnewline
		Ladson et al.\ & Experiment & $11^\circ$ & $\;\,9{,}000{,}000$ & $0.40$ \tabularnewline
		Ladson et al.\ & Experiment & $11^\circ$ & $15{,}150{,}000$ & $0.30$ \tabularnewline
		\hline		
	\end{tabular}
	\caption{\label{table:drag_coefficients_conditions}Conditions for the experiments and simulations performed by Ladson et al.\ \cite{Ladson_1987a}, Almutari \cite{Almutari_2010} and Jones \cite{Jones_2008} which are used to compare with the time-averaged drag coefficients of the present large-eddy simulations.}
\end{table}

\subsection{Lift coefficient $C_L$}
\label{subsec:lift_coefficient}
\par The lift coefficient $C_L$ is calculated according to Eq.\ (\ref{eq:lift_coefficient}). $\rho_f$, $u_{1,\,in}$ and $S$ are the density of the fluid, the free stream velocity and the reference area (see Eq.\ (\ref{eq:drag_coefficient_area})), respectively. 
\begin{equation}
C_L=\frac{2<F_L>}{\rho_f\;S\;u_{1,\,in}^2} \label{eq:lift_coefficient}
\end{equation}
\par The lift force on the wall $F_L$ is perpendicular to the motion and calculated as the sum of the shear force $F_{\tau_{22}}$ and the pressure force $F_{p_{x_2}}$, as stated in Eq.\ (\ref{eq:lift_force}). The shear $F_{\tau_{22}}$ and pressure $F_{p_{x_2}}$ forces are calculated similarly to the $F_{\tau_{11}}$ and $F_{p_{x_1}}$ forces (see \mbox{Eqs.\ (\ref{eq:shear_force_drag}) through (\ref{eq:pressure_force_drag})}). 
\begin{equation}
F_L=F_{\tau_{22}}+F_{p_{x_2}} \label{eq:lift_force}
\end{equation}
\par The lift coefficients and the standard deviations of the medium resolution meshes are illustrated in Table \ref{table:lift_coefficients}. The meshes with medium and maximal first cell wall distances present different values for this coefficient, with greater values for the meshes with a maximal wall distance of the first cell. The standard deviations are relatively small, indicating a small amplitude of the lift oscillations in time.
\begin{table}[H]
	\centering
	\begin{tabular}{c c c}
		\hline
		\bf{Mesh} & \centering{\bf{Lift coefficient}} & \centering{\bf{Standard deviation}} \tabularnewline \hline
		$m-0-y^+_{med}$ & $C_L=0.003$ & $\sigma_{C_L}=0.004$ \tabularnewline
		$m-0-y^+_{max}$ & $C_L=0.004$ & $\sigma_{C_L}=0.013$ \tabularnewline
		$m-5-y^+_{med}$ & $C_L=0.589$ & $\sigma_{C_L}=0.011$ \tabularnewline
		$m-5-y^+_{max}$ & $C_L=0.606$ & $\sigma_{C_L}=0.003$ \tabularnewline
		$m-11-y^+_{med}$ & $C_L=0.951$ & $\sigma_{C_L}=0.086$ \tabularnewline
		$m-11-y^+_{max}$ & $C_L=1.091$ & $\sigma_{C_L}=0.004$ \tabularnewline
		\hline		
	\end{tabular}
	\caption{\label{table:lift_coefficients}Time-averaged lift coefficients and standard deviations of the medium resolution meshes at $Re=100{,}000$.}
\end{table}
\par
\begin{figure}[H]
	\centering
	\centering
	\includegraphics[width=0.56\textwidth,draft=\drafttype]{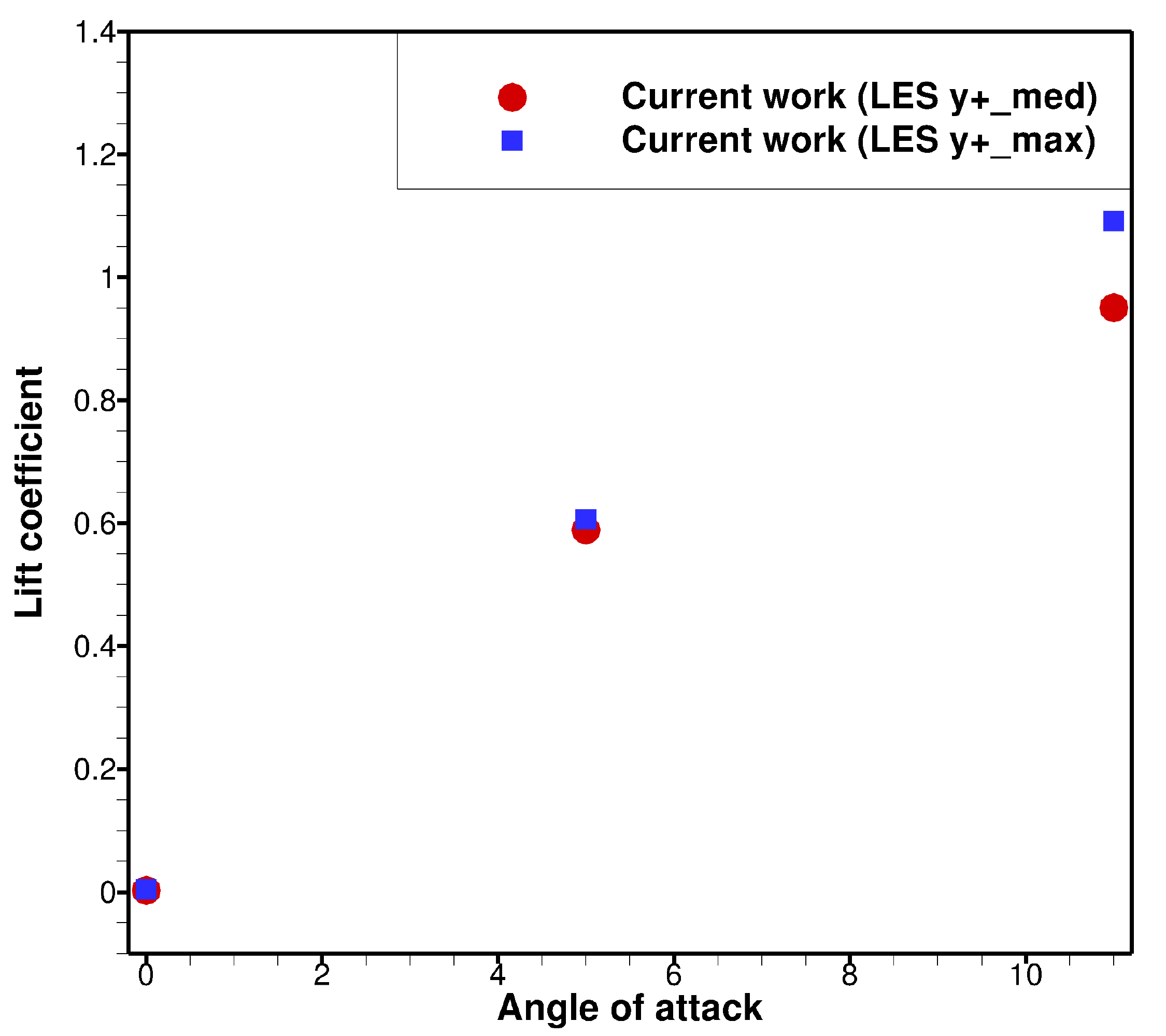}
	\caption{\label{fig:cl_comparison_angles}Variation of the time-averaged lift coefficients with incidence at $Re=100{,}000$.}
\end{figure}

 Since the NACA0012 airfoil is symmetrical, the lift coefficient at an angle of attack of $\alpha=0^\circ$ must be $C_L=0$, which is not the case of both simulated meshes. An increase in the incidence results in a lift increase until a critical angle, at which the flow is disrupted and the airfoil stalls, is achieved. This lift coefficient increase with the incidence is shown in Fig.\ \ref{fig:cl_comparison_angles} for both medium and maximal first cell wall distances.

\par
\begin{figure}[H]
	\centering
	\subfigure[Comparison of lift coefficients of current work, Ladson \cite{Ladson_1988}, Almutari \cite{Almutari_2010} and Jones \cite{Jones_2008}.]{\includegraphics[width=0.46\textwidth,draft=\drafttype]{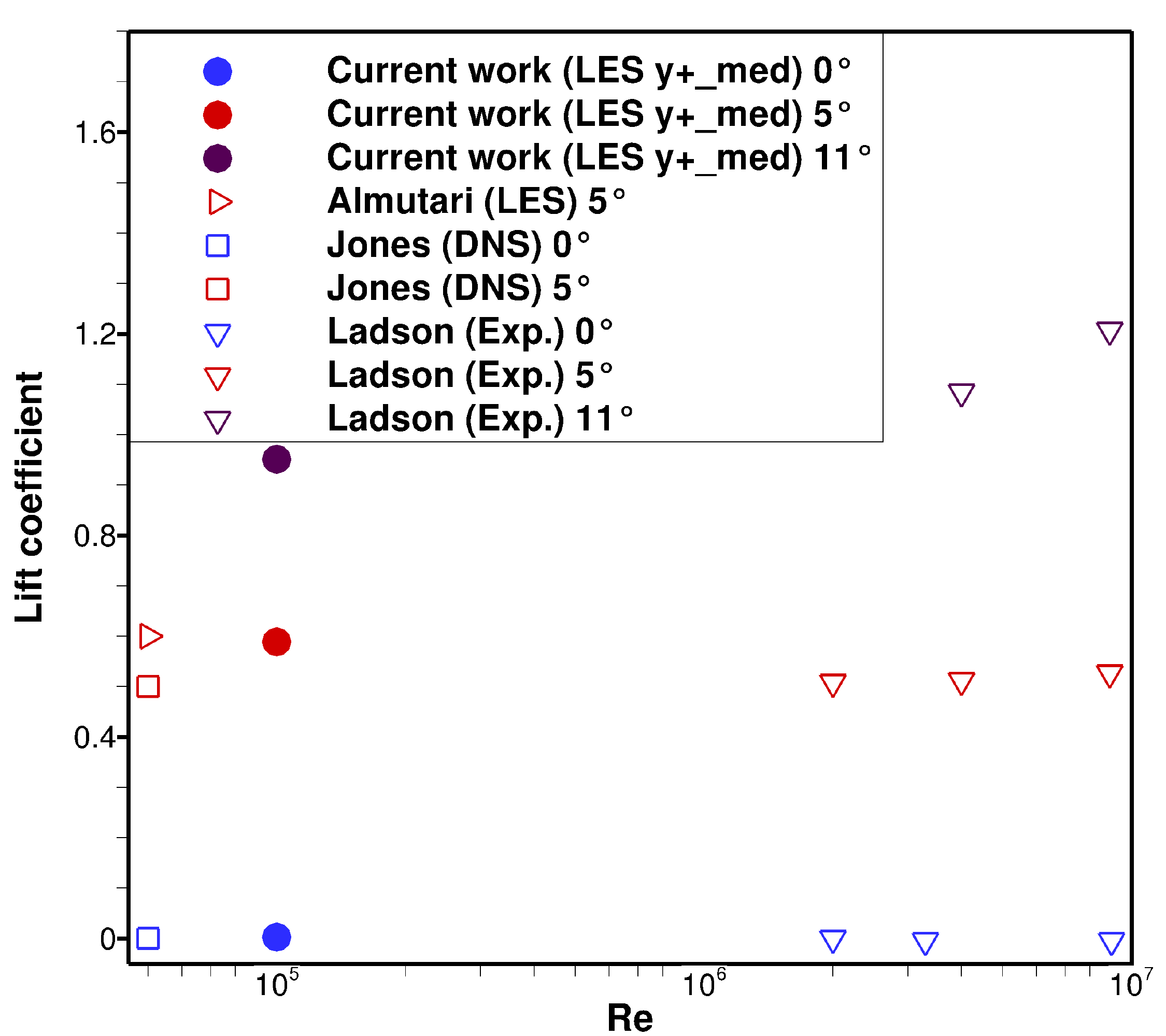}\label{fig:cf_comparison_angles_reynolds_a}} \hfill
	\subfigure[Comparison of lift coefficients of current work and Sheldahl and Klimas \cite{Sheldahl_1981}.]{	\includegraphics[width=0.46\textwidth,draft=\drafttype]{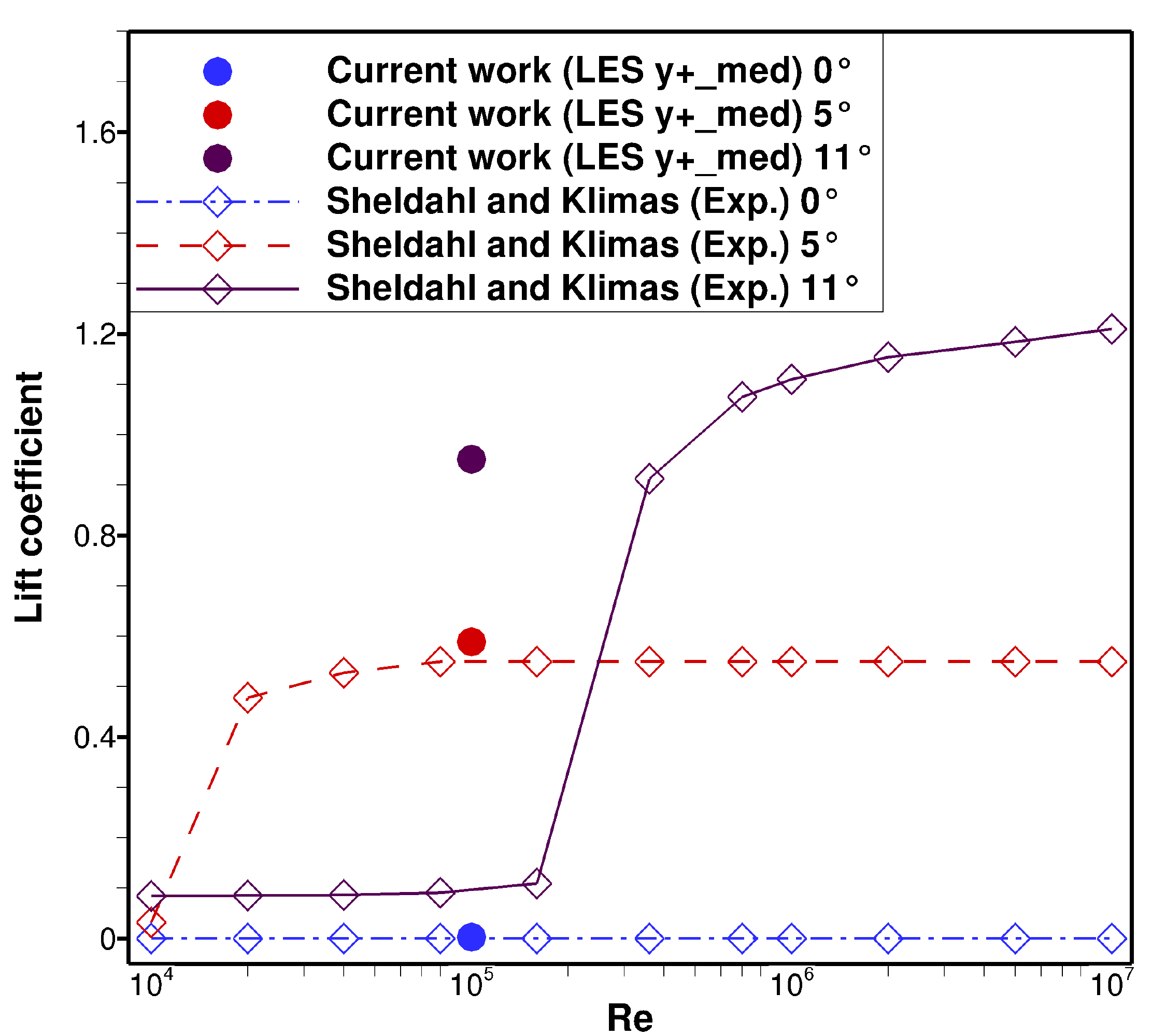}\label{fig:cf_comparison_angles_reynolds_b}}	\hfill
	\caption{\label{fig:cf_comparison_angles_reynolds}Comparison of the time-averaged lift coefficients of the present large-eddy simulations and the data available in the literature.}
\end{figure}

 In Fig.\ \ref{fig:cf_comparison_angles_reynolds_a}, the computed lift coefficients of the meshes with a medium wall distance of the first cell are qualitatively compared with the experimental data of Ladson \cite{Ladson_1988}, the large-eddy simulations data of Almutari \cite{Almutari_2010} and the direct-numerical simulations data of Jones \cite{Jones_2008}. The analyzed lift coefficients at angles of attack of $\alpha=0^\circ$, $\alpha=5^\circ$ and $\alpha=11^\circ$ for diverse conditions of Reynolds and Mach numbers (see Table \ref{table:lift_coefficients_conditions}) indicate that at $\alpha=0^\circ$ the lift coefficient is about $C_L=0$ and that at $\alpha=11^\circ$ the lift coefficient increases with the Reynolds number (for $100{,}000\leq Re\leq 8{,}900{,}000$). For an incidence of $\alpha=5^\circ$, the lift coefficient of the direct numerical simulations performed by Jones \cite{Jones_2008} indicates a smaller value than the large-eddy simulations performed by Almutari \cite{Almutari_2010} for the same investigated condition, i.e$.$, $Re=50{,}000$. The experimental data of Ladson \cite{Ladson_1988} indicate a slight increase in the lift coefficient with the Reynolds number.   
\par Figure \ref{fig:cf_comparison_angles_reynolds_b} illustrates the lift coefficients computed by the large-eddy simulations and measured by the experiments of Sheldahl and Klimas \cite{Sheldahl_1981}. As already stated in Section \ref{subsec:drag_coefficient}, the experimented NACA0012 Eppler model profile \cite{Sheldahl_1981} is slightly different from the NASA NACA0012 profile \cite{NASA_2016}, which is used in the present work. Therefore, the overestimation of the computed coefficients could be caused by the difference of the investigated geometries. Specially in the case of an incidence of $\alpha=11^\circ$, the influence of the geometry on the critical angle of attack could change the location of the transition between stall and non-stall region, which occurs firtly for the \mbox{NASA NACA0012 airfoil \cite{NASA_2016}} computed by the present large-eddy simulation.

\begin{table}[H]
	\centering
	\begin{tabular}{c c l l l l}
		\hline
		\bf{Literature} & \multicolumn{1}{c}{\bf{Type}} & \multicolumn{1}{c}{$\alpha$} & \multicolumn{1}{c}{$Re$} & \multicolumn{1}{c}{$Ma$} \tabularnewline \hline
		Almutari & LES & $5^\circ$ & $\quad\;\;\,50{,}000$ & \multicolumn{1}{c}{--} & \tabularnewline
		Jones & DNS & $0^\circ$ & $\quad\;\;\,50{,}000$ & \multicolumn{1}{c}{--} & \tabularnewline
		Jones & DNS & $5^\circ$ & $\quad\;\;\,50{,}000$ & \multicolumn{1}{c}{--} & \tabularnewline
		Ladson et al.\ & Experiment & $0^\circ$ & $\;\,2{,}000{,}000$ & $0.15$ \tabularnewline
		Ladson et al.\ & Experiment & $0^\circ$ & $\;\,3{,}290{,}000$ & $0.25$ \tabularnewline
		Ladson et al.\ & Experiment & $0^\circ$ & $\;\,8{,}920{,}000$ & $0.25$ \tabularnewline
	\end{tabular}
\end{table}
\begin{table}[H]
	\centering
	\begin{tabular}{c c l l l l}
		Ladson et al.\ & Experiment & $5^\circ$ & $\;\,2{,}000{,}000$ & $0.15$ \tabularnewline
		Ladson et al.\ & Experiment & $5^\circ$ & $\;\,4{,}000{,}000$ & $0.20$ \tabularnewline
		Ladson et al.\ & Experiment & $5^\circ$ & $\;\,8{,}900{,}000$ & $0.25$ \tabularnewline
		Ladson et al.\ & Experiment & $11^\circ$ & $\;\,4{,}000{,}000$ & $0.15$ \tabularnewline
		Ladson et al.\ & Experiment & $11^\circ$ & $8{,}900{,}000$ & $0.20$ \tabularnewline
		\hline		
	\end{tabular}
	\caption{\label{table:lift_coefficients_conditions}Conditions for the experiments and simulations performed by Ladson \cite{Ladson_1988}, Almutari \cite{Almutari_2010} and Jones \cite{Jones_2008} which are used to compare with lift coefficients of the present large-eddy simulations.}
\end{table}

%% file: conclusion.tex
\chapter*{Conclusions and outlook\markboth{CONCLUSIONS AND OUTLOOK}{}}
\label{general_conclusion}

\addcontentsline{toc}{chapter}{Conclusions and outlook}

\par Wall-resolved large-eddy simulations on a fixed NACA0012 profile at various angles of attack at a Reynolds number of $Re=100{,}000$ are performed aiming at future studies of the fluid-structure interaction between the fluid flow and an elastically mounted airfoil.
\par Firstly, the problem is studied in order to generate high quality meshes for wall-resolved LES. The fluid domain and the number of nodes per edge are initially established based on the work of Almutari \cite{Almutari_2010}, while the wall distance of the first cell is originally determined according to the work of Kasibhotla \cite{Kasibhotla_2014}. The fluid domain is established in order to generate a block structured C-grid and has a wake length of $W=5\,c$, a domain radius of $R=7.3\,c$ and a span-wise length of $L_z=0.25\,c$. The first cell wall distance is primarily set to $ y_{first\,cell}=4.4\cdot10^{-6}\,m$. The stretching factor is adapted for every edge in order to guarantee a mesh with smooth transitions. This varies within $1.02 \leq q \leq 1.06$ for the cells near the airfoil profile.
\par A first mesh for an angle of attack of $\alpha=0^\circ$, i.e$.$, $f-0-y^+_{min}$, is generated in ANSYS ICEM CFD. Because of its high number of control volumes and therefore high required computational time, another two medium resolution meshes, i.e$.$, $m-0-y^+_{med}$ and $m-0-y^+_{max}$, varying according to first cell wall distances are generated. These have almost a quarter of the control volumes of the fine resolution mesh and the first cell wall distances of $y_{first\,cell}=9.0\cdot10^{-6}\,m$ and $y_{first\,cell}=1.8\cdot10^{-5}\,m$, respectively.
\par In order to achieve angles of attack of $\alpha=5^\circ$ and $\alpha=11^\circ$ the generated medium resolution meshes are adapted based on an enhanced inverse distance weighting interpolation method described by Sen et al.\ \cite{Sen_2017}.
\par Seven high quality block-structured meshes are generated in total: $f-0-y^+_{min}\,$, $m-0-y^+_{med}\,$, $m-\nobreak 0-\nobreak y^+_{max}\,$, $m-5-y^+_{med}\,$, $m-5-y^+_{max}\,$, $m-11-y^+_{med}\,$ and $m-11-y^+_{max}\,$. These vary in the resolution, the first cell wall distance and the angle of attack.
\par The computational setup is defined and the simulation of the incompressible flow around the NACA0012 profile is carried out by the in-house software FASTEST-3D. A finite-volume method is utilized for the spatial discretization, while a predictor-corrector method, with an explicit low-storage three sub-steps Runge-Kutta scheme as predictor and a Poisson equation as corrector, is used for the temporal discretization.   
\par The turbulence spectrum is divided by an implicit spatial filter. The large scale is solved according to the LES governing equations, while the subgrid-scales are modeled based on the Smagorinsky model combined with the Van-Driest damping function.
\par The initial condition of the velocity is determined in order to achieve a Reynolds number of $Re=100{,}000$. The inlet, the outlet, and the fluid domain top and bottom edges boundary conditions are given by Dirichlet, convective and symmetry boundary conditions, respectively. A no-slip wall is set for the rigid airfoil and a periodic boundary condition is established in the homogeneous span-wise direction.
\par A time-step for all seven meshes is determined in order to maintain the simulation stable. This varies for the meshes and is related mainly to the first cell wall distance.
\par A time-averaging process is required in order to obtain statistical data, enabling thorough investigations of the results. This is started after a long initialization phase, characterized by a dimensionless time $t^*_{init}\geq50$.
\par With available time-averaged data for all meshes except the $f-\nobreak 0-\nobreak y^+_{min}$ grid, the study of the results can be finally executed. This is performed in order to analyze the solution of the boundary layer, the velocity distribution and the Reynolds stresses.
\par The dimensionless wall distance investigation assure that all meshes contain cells within the viscous sublayer. For the $m-11-y^+_{max}$ mesh, however, only two control volumes are always contained within $0\leq y^+_{visc} \leq 5$. Therefore, an accurate solution of the flow in the immediate vicinity of the profile cannot be guaranteed for this mesh.
\par The study of the spatial and time-averaged velocities and streamlines is performed aiming at a mesh comparison. The flow itself, as well as the detachment of the boundary layer and the formation of separation bubbles is analyzed. The results of the meshes for an angle of attack of $\alpha=0^\circ$ are similar, as well as the results for the meshes for an angle of attack of $\alpha=5^\circ$. Therefore, a mesh independence for these five meshes can be assured. For the grids with an angle of attack of $\alpha=11^\circ$ some discrepancies are present, such as a difference in the velocity distribution and the formation and location of the separation bubbles. Since the dimensionless times utilized to investigate the velocities are different for both meshes, these discrepancies can be caused either due to this difference, or due to a mesh influence in the computation of the test case.
\par The analysis of the instationary flow field aims at the investigation of the instantaneous velocity field regarding the detachment and reattachment of the boundary layer. Moreover, the vortex shedding and the formation of von K\'arm\'an vortex streets are studied through the pressure fluctuations and the vorticity in the span-wise direction. The meshes at an angle of attack of $\alpha=0^\circ$ show the formation of various separation bubbles near the trailing edge, as well as the formation of a von K\'arm\'an vortex street in the wake. At incidences of $\alpha=5^\circ$ $\alpha=11^\circ$, various separation bubbles are formed on the suction side and small numerical oscillations caused by mesh resolution and the utilized second-order accuracy central differencing scheme of the finite volume method are present in the instantaneous velocity field. Furthermore, vortex shedding is present on the suction side of the airfoil and in the wake. A shear layer near the leading edge is also formed for an angle of attack of $\alpha=11^\circ$.
\par The investigation of the Reynolds stresses provides information about the turbulence present in the flow. Due to the almost two-dimensional flow, only three components of the stress tensor, i.e$.$, $\tau_{11}^{turb^*}$, $\tau_{12}^{turb^*}$ and $\tau_{22}^{turb^*}$, are studied. While the meshes with angles of attack of $\alpha=0^\circ$ and $\alpha=5^\circ$ present no significant discrepancies, the meshes with an angle of attack of $\alpha=11^\circ$ show differences that may be not caused only by the different dimensionless times. Therefore, a conclusion about the independence of both meshes for an angle of attack of $\alpha=11^\circ$ cannot be established.
\par Finally, the aerodynamic coefficients are investigated and compared to the available data in the literature. The pressure coefficients at the airfoil at an incidence of $\alpha=0^\circ$ are equal for the pressure and suction sides since the profile is symmetrical. Furthermore, both analyzed medium meshes show a great agreement with the experimental data of Ladson et al.\ \cite{Ladson_1987a} and Gregory and O'reilly \cite{Gregory_1973}, although these experiments were performed for different Reynolds numbers. A deviation is present near the trailing edge, since a boundary layer detachment occurs for the large-eddy simulations and is not present on the the experiments of Ladson et al.\ \cite{Ladson_1987a} and Gregory and O'reilly \cite{Gregory_1973} due to the usage of boundary layer control techniques. At an incidence of $\alpha=5^\circ$ and $\alpha=11^\circ$, small pressure coefficient discrepancies are present on the suction side of both simulated grids. A comparison with the experiments performed by Ladson et al.\ \cite{Ladson_1987a} shows strong deviations of the results achieved on the suction side. These are caused by the different analyzed Reynolds numbers, which influences the formation and location of detachment and reattachment points. 
\par The skin friction coefficients caused by the interaction of the viscous fluid with the airfoil are studied regarding the comparison between the meshes and the location of the detachment and reattachment points. These show minimal deviations for the grids at an incidence of $\alpha=0^\circ$, $\alpha=5^\circ$ and $\alpha=11^\circ$, which may be caused by the different analyzed dimensionless averaging times $t^*_{avg}$. Furthermore, the location of the detachment and reattachment points analyzed with the skin friction coefficients and with the spatial and time-averaged velocity field and streamlines are consonant. 
\par The investigation of the drag and lift coefficients shows some deviations for the computed grids at a same incidence. These are greater for the meshes at an incidence of $\alpha=11^\circ$, which may indicate an influence of the mesh in the computation. The results of the meshes with the medium wall distance of the first cell are qualitatively compared to the results achieved by the large-eddy simulations of Almutari \cite{Almutari_2010}, the direct-numerical simulations of Jones \cite{Jones_2008}, and the experiments of Ladson et al.\ \cite{Ladson_1987a} and Ladson \cite{Ladson_1988}, indicating plausible results. A quantitatively comparison to the results of the experiments of Sheldahl and Klimas \cite{Sheldahl_1981} shows some deviations, which may be caused by the slightly different experimented profiles, since the geometry has a direct influence on the flow field and on the critical angle of attack.  
\par In the near future fluid-structure interactions between NACA0012 airfoil and flow at $Re=100{,}000$ will be investigated utilizing the present work as basis for the grid and time step selection. For instance, for FSI studies inducing rotations of $\theta\leq5^\circ$, the medium grid with the larger first cell wall distance, i.e$.$, $y^+_{max}$, will be used as it provides the best compromise between accuracy of the results and the CPU-time consumption. Since the independence of both generated meshes has not been granted for angles of attack of $\alpha=11^\circ$, for FSI induced rotations of $\theta>5^\circ$ further investigations on the impact of the grid resolution on the fluid forces and consequently on the structural responses are required.